\documentclass[a4paper,twoside,11pt]{report}
\pdfoutput=1
\usepackage{amsmath}
\usepackage{amssymb}
\usepackage{graphicx}
\usepackage{cancel} 

\renewcommand{\[}{\begin{equation}}
\renewcommand{\]}{\end{equation}}
\newcommand{\st}{\begin{eqnarray}}
\newcommand{\en}{\end{eqnarray}}

\newcommand{\prt}{\partial}
\newcommand{\darrow}{\LARGE{\nearrow\!\!\!\!\!\!\searrow}}
\newcommand{\ddarrow}{\LARGE{\nwarrow\!\!\!\!\!\!\swarrow}}
\newcommand{\ndiv}{\mid\!\!\!\not\,\,\,}

\newcommand{\D}{\mathcal{D}}
\newcommand{\e}{\textrm{\large{\textit{e}}}}

\begin{document}

\begin{titlepage}
\begin{center}
{ \phantom{.} \vspace{43mm} }
\Huge\textbf{Non-Compact Geometries \\ in \\ 2D Euclidean Quantum Gravity}

\LARGE
\line(1,0){250}
\\[15mm]
\LARGE 
J\large ENS \LARGE A. \LARGE G\large ESSER\\
\vspace{90mm}
\large
Submitted for the degree of 
\\
Ph.D. in Physics
\\
from
\\
The Niels Bohr Institute
\\
Faculty of Science
\\
University of Copenhagen
\\[5mm]
June, 2008
\end{center}
\newpage
\thispagestyle{empty}
\mbox{}
\end{titlepage}

\begin{abstract}
This thesis discusses the worldsheet geometries in the $(p,q)$ minimal model coupled to 2D euclidean Quantum Gravity with particular focus on non-compact geometries. We first calculate the FZZT-FZZT cylinder amplitudes for all pairs of Cardy matter states thereby generalizing the result obtained in arXiv:hep-th/0406030 and discuss the decompositions of the cylinder amplitudes in terms of solutions to the homogeneous Wheeler-DeWitt equation.  We then show, that the principal ZZ boundary conditions in Liouville theory can be viewed as effective boundary conditions obtained by integrating out the matter degrees of freedom on the worldsheet. We now consider the $(2,2m-1)$ minimal model coupled to 2D euclidean Quantum Gravity obtained in the scaling limit of dynamical triangulations, in which the $m^{\text{th}}$ multi-critical hyper-surface is approached and the conformal background is turned on. We show, that only one concrete realization of matter boundary condition, the $(1,1)$ Cardy boundary condition, is obtained in this scaling limit. Finally, we study the cylinder amplitude with fixed distance and provide some evidence of a transition from a FZZT-brane to a ZZ-brane on the exit loop in the limit, where the distance approaches infinity, and for particular values of the exit-boundary cosmological constant.
\end{abstract}

\tableofcontents

\chapter*{Preface}
\addcontentsline{toc}{chapter}{Preface} 

In the twentieth century Physics was revolutionized by two very different theories, 
Quantum Mechanics and General Relativity.  
General Relativity was developed by Albert Einstein in 1915/1916 
and completely altered our understanding of space and time. 
A truly deep discovery made by Einstein is that the fundamental degrees of freedom in Nature are essentially geometrical. 
According to General Relativity we may formulate all laws in Nature in terms of differential geometry.  
The successes of General Relativity are many. 
General Relativity is the basis of the cosmological standard model, 
which predicts the birth of the Universe at a finite time in the past 
and describes the evolution of the Universe. 
The discovery of the cosmic microwave background radiation by Penzias and Wilson in 1965 
and the more recent measurements of the anisotropy in this background radiation revealing the density fluctuations of the Universe at early times
probably compromise the best evidence so far for the Big Bang model of the Universe.    
Quantum Mechanics was developed in the first half of the twentieth century and concerns Nature on or below the atomic scale. 
The foundation of Quantum Mechanics was developed by Niels Bohr, Werner Heisenberg, Max Planck, Louis 
De Broglie, Albert Einstein, Erwin Schr\"odinger, Max Born, John von Neumann, Paul Dirac, Wolfgang Pauli and others. 
According to (the Copenhagen interpretation of) Quantum Mechanics 
the fundamental laws governing Nature are probabilistic rather than deterministic as in classical Physics. 
Even though we know the initial state of a physical system 
we cannot predict the outcome of a given experiment with certainty. 
We can only predict the probability of a given outcome. 
The discovery of Quantum Mechanics has had a crucial impact on 
both Chemistry and modern day technology.  
The models of the electron configuration within atoms and molecules obtained from Quantum Mechanics 
explains the structure of the periodic table and explains how individual atoms combine into 
chemicals and molecules. 
The invention of semi-conductor devices such as the transistor has truly revolutionized our modern day society. 
The development of Quantum Mechanics has led to Quantum Field Theory and 
has so far culminated in the formulation 
of the standard model in particle physics, which describes three of the four known interactions in Nature binding matter together.  
The standard model has been verified numerous times in accelerators all around the world. 

Einsteins theory of Gravity is a classical theory. 
Given the energy momentum tensor and a set of suitable boundary conditions we may in principle determine 
the gravitational field from Einsteins field equations. 
Thus, General Relativity cannot be the fundamental theory of Gravity and we need to formulate a quantum theory of Gravity,  
in which we reconcile the probabilistic nature of Quantum Mechanics with the geometrical nature of General Relativity. 
The formulation of a quantum theory of Gravity is probably the first step toward a 
unified theory of all known interactions in Nature, that is a theory of ``everything''. 
Moreover, a quantum theory of Gravity would allow us to study the evolution of the Universe 
all the way back to the moment of big bang. 

In this thesis we will study 2 dimensional euclidean Quantum Gravity, in which 
case General Relativity simplifies due to the topological nature of the Einstein-Hilbert action.
2D euclidean Quantum Gravity serves as a toy model,  
from which we hope to gain some insight, which will help us formulate 
a theory of Quantum Gravity in higher dimensions. 
In particular we will consider non-compact geometries in this thesis, which are clearly of fundamental 
importance to a quantum theory of Gravity. 
In chapter \ref{chapter1} we discuss how Liouville theory emerges in the process of gauge-fixing 2D euclidean Quantum Gravity. 
In chapters \ref{chapterLiouville} and \ref{chapter3} 
we discuss Liouville theory and so-called FZZT and ZZ boundary conditions in Liouville theory. 
In chapter \ref{chapter4} we consider the $(p,q)$ minimal model coupled to 
2D euclidean Quantum Gravity and discuss the spectrum of physical closed string states and the so-called FZZT and ZZ branes. 
In chapter \ref{chaptercyl} we calculate the FZZT-FZZT cylinder amplitude for all pairs of Cardy matter states 
in the $(p,q)$ minimal model coupled to 2D euclidean 
Quantum Gravity and discuss the structure of these cylinder amplitudes. 
In chapter \ref{chapterZZ} we discuss the ZZ branes and propose  
a particular interpretation of the so-called principal ZZ branes. 
In chapter \ref{ZZbranesworldsheet} we study a transition from compact to non-compact worldsheet geometry and show, 
how the boundary conditions obtained in this limit are related to the ZZ branes.  
In the chapters \ref{chapter1}-\ref{chapter4} and the sections \ref{sectionDT} and \ref{sectioncyldis} 
we review the nessecary background material. 
In the chapters \ref{chaptercyl} and \ref{chapterZZ} and in section \ref{worldsheetperspective} we present the results obtained by the author 
of this thesis in collaboration with his academic adviser.  
Finally, I would like to thank my academic adviser Prof. Jan Ambj\o rn for introducing me to the fascinating world of Quantum Gravity 
and for having faith in me.  

\chapter{Gauge-fixing 2D euclidean Quantum Gravity} \label{chapter1}
\section{Riemannian manifolds in two dimensions}
In this thesis we will consider both compact and non-compact two-dimensional connected and oriented Riemannian manifolds. 
We will often refer to these manifolds as world-sheets.
Every oriented two-dimensional real manifold is a complex manifold of complex dimension 1. 
If the manifold is also compact, we refer to the manifold as a Riemann surface.\cite{Nakahara} 

Two manifolds are homeomorphic, if there exists a homeomorphism between the two, 
that is if we can deform one into the other by a continuous transformation. 
Two homeomorphic manifolds are topologically indistinguishable. 
The group of homeomorphisms defines an equivalence relation and thus 
introduces a partition of the set of manifolds into equivalence classes of manifolds with the 
same topology. 
Any two-dimensional compact, connected and orientable surface can be continuously transformed into a sphere to which we have added a 
number of handles and a 
number of holes. The particular numbers of handles and holes added to the sphere depend uniquely on the surface we consider. 
Hence, the topology, that is the equivalence class discussed in the above, 
of a two-dimensional compact, connected and orientable surface is completely characterized 
by the number of handles and the number of connected boundaries of the surface. 
The different equivalence classes are characterized by so-called topological invariants, which are properties of the differentiable manifolds 
invariant under homeomorphisms.        
An important topological invariant of a manifold is the Euler number. 
The Euler number $\chi$ of a two-dimensional compact, connected and orientable surface is given by
\[
\label{euler}
\chi = 2-2h-b
\]
in terms of the number of connected boundaries $b$ and the genus $h$, that is the number of handles of the surface.\cite{Nakahara,pol} 

If two parametrizations of two Riemannian manifolds are related by a coordinate transformation, that is the two manifolds are related by a diffeomorphism, 
then the two manifolds are physically indistinguishable. 
Indeed, the physics cannot depend on our choice of parametrization. 
Thus, we identify Riemannian manifolds up to diffeomorphisms, that is  
the group of diffeomorphisms partitions the set of Riemannian manifolds into physically distinct equivalence classes. 

Let us for simplicity consider a connected, compact and oriented manifold of genus $h$ without boundaries and let $\mathcal{G}_{h}$ be the space of metrics 
on this manifold. A Weyl rescaling of the metric $g$ is defined as
\[
\label{Weyl}
g_{\mu \nu}(x) \to g'_{\mu \nu}(x) = e^{\phi(x)} g_{\mu \nu}(x)
\]
The space of metric modulo Weyl rescalings and diffeomorphisms continuously connected with the identity 
is a finite dimensional space called Teichm\"uller space
\[
\textrm{Teich} \equiv \frac{\mathcal{G}_{h}}{\textrm{Weyl} \times \textrm{Diff}_{0}}
\] 
Teichm\"uller space is $0$-dimensional in the case of the sphere, 
$2$-dimensional in the case of the torus of genus $1$ and $(6-6h)$-dimensional in the case of the torus of genus $h$, 
$h \geq 2$.\footnote{In this thesis we refer to the \textit{real} dimension of a given space.}
There exists so-called large coordinate transformations, which are not continuously connected with the identity. 
The large coordinate transformations and the identity transformation 
form a group known as the modular group
\[
\textrm{Mod} \equiv \frac{\textrm{Diff}}{\textrm{Diff}_{0}}
\]
These large coordinate transformations relate different points $\tau$ in Teichm\"uller space. 
If we identify points in Teichm\"uller space under the modular group we obtain a compact space known as moduli space 
\[
\mathcal{M}_{h} \equiv \frac{\mathcal{G}_{h}}{\textrm{Weyl} \times \textrm{Diff}}
\]
An equivalence class in moduli space is referred to as a conformal structure.
From each conformal structure $\tau$ in moduli space we choose a representative $\hat{g}_{\tau}$ known as the fiducial metric.\cite{Nakahara}

Let us consider the group of compositions of Weyl rescalings and diffeomorphisms in more detail. 
This group is \emph{not} a direct product of Weyl rescalings and diffeomorphisms, 
a fact, which is crucial in the continuum formulation of euclidean 2D Quantum Gravity. 
Indeed, diffeomorphisms exist, under which the metric transforms by a Weyl rescaling as in (\ref{Weyl}). 
These are the conformal transformations and they form a group, the conformal group.\cite{Nakahara}
The conformal group as a abstract group\footnote{Isomorphic groups correspond to the same abstract group.} 
is independent of the particular metric on the manifold. 
Actually, the conformal group is completely determined by the topology of the manifold. 
However, the particular realization of the group off course depends on the particular metric on the manifold.   
It is easily seen from eq. (\ref{Weyl}), that a conformal transformation preserves angles, while changing the scale locally.   

Let us take a closer look at the particular realization of the group of conformal transformations in the case 
when the metric is proportional to the unit metric
\[
g_{\mu \nu}(x)=f(x)\delta_{\mu \nu}.
\]  
Under the infinitesimal coordinate transformation 
\[
x^{\mu} \to x'^{\mu}=x^{\mu}-\xi^{\mu}(x)
\]
the metric transforms as
\[
g_{\mu \nu}(x)=f(x)\delta_{\mu \nu} 
\quad \to \quad 
g'_{\mu \nu}(x')=f(x)\delta_{\mu \nu}+f(x)\left( \prt_{\mu}\xi_{\nu}+\prt_{\nu}\xi_{\mu} \right)
\] 
Under an infinitesimal Weyl rescaling the metric transforms according to
\st
 g_{\mu \nu}(x')&  = & f(x')\delta_{\mu \nu} 
\quad  \to  \nonumber \\ 
g'_{\mu \nu}(x') 
= (1+\epsilon(x'))f(x')\delta_{\mu \nu} & = & 
f(x)\delta_{\mu\nu}+\left( \epsilon(x)f(x)-\xi^{\alpha}\prt_{\alpha}f   \right) \delta_{\mu\nu}
\en
Demanding that the coordinate transformation corresponds to a Weyl rescaling, we obtain the following two equations concerning the generator 
of the diffeomorphism
\st
\prt_{1}\xi_{1} &=& \prt_{2}\xi_{2} \\
\prt_{1}\xi_{2} &=& -\prt_{2}\xi_{1}
\en
which we recognize as the Cauchy-Riemann equations.
Hence, if we introduce complex coordinates
\[
\label{compcoor}
z=x_{1}+ix_{2}, \qquad \bar{z}=x_{1}-ix_{2}
\]
we realize, that an infinitesimal holomorphic coordinate transformation \label{conformaltrans} 
\[\label{conftr}
z \to z'=z-\xi(z)
\]
is a conformal transformation in the flat region of our manifold. 

\section{The partition function in conformal gauge}\label{partfunc}
Let us couple some matter to 2D euclidean gravity.  
This is achieved by expressing the action $S_{M}[X]$ governing the matter fields $X$ and valid in special relativity in a covariant manner, 
\[
S_{M}[X] \to S_{M}[X,g].
\] 
The principle of general covariance assures us, that we have obtained an action for the matter 
fields valid in general relativity.     
Let us for simplicity consider the topology of a compact, connected and oriented manifold of genus $h$ without boundaries.  
For this topology we formally define the euclidean partition function as
\[
\label{part0}
\mathcal{Z}_{h} \equiv \int \frac{\mathcal{D}_{g}g \, \mathcal{D}_{g}X}{\textrm{Vol}(\textrm{Diff})} \, \exp(-S_{G}[g]-S_{M}[X,g])
\]
where $S_{G}[g]$ is the euclidean Einstein-Hilbert action modified by a cosmological term 
\[
\label{graac}
S_{G}[g] = \frac{\lambda}{4 \pi} \int d^{2}x \sqrt{g} R + \mu_{0} \int d^{2}x \sqrt{g}
\]
where $G=\frac{1}{4\pi\lambda}$ is the gravitational constant, $R$ is the Ricci scalar and $\mu_{0}$ is the bare cosmological constant. 
In the partition function (\ref{part0}) we integrate over the space of matter configurations and the space of metrics $\mathcal{G}_{h}$ 
defined on the considered topology.  
Using the Gauss-Bonnet theorem, 
which in the case of a compact Riemannian manifold $\mathcal{S}$ with a boundary $\prt\mathcal{S}$ 
reads \cite{pol}
\[
\label{Gaussbonnet}
\chi = \frac{1}{4\pi} \int_{\mathcal{S}} \sqrt{g} R 
+ \frac{1}{2\pi} \int_{\prt \mathcal{S}} ds \: k,
\]
where $k$ is the geodesic curvature of the boundary $\prt \mathcal{S}$ and $ds$ is the line element, 
we realize from eq. (\ref{euler}), that the euclidean Einstein-Hilbert term in the gravitational action 
only depends on which topology we consider and we may express the gravitational action as
\[
S_{G}[g]=\lambda\chi(h) + \mu_{0} A[g]
\]  
where $\chi(h)$ is the Euler number and $A[g]$ is the area of the Riemannian manifold with metric $g$. 
Hence, as long as we consider a fixed topology we may leave out the euclidean Einstein-Hilbert term in the gravitational action and 
we are left with a quite simple gravitational action in 2D Quantum Gravity depending only on the area of the Riemannian manifold in 
consideration.\footnote{\label{claslim}Notice, pure classical euclidean gravity in 2 dimensions is actually ill-defined, 
since the Riemannian manifold, which minimizes the classical action, has vanishing area.}

As mentioned previously configurations related by coordinate transformations are physically equivalent. 
Thus, both the actions and the measures appearing 
in the partition function should be invariant under diffeomorphisms such that physically equivalent configurations 
give the same contribution to the partition function. 
From eq. (\ref{graac}) and the above discussion it is clear that both the action governing the matter degrees of freedom and the gravitational action are 
invariant under diffeomorphisms. 
In order to define a measure, 
that is an infinitesimal volume element in the corresponding space of configurations, 
we should first define a metric on the space of configurations, 
that is for each configuration we should define 
an inner product on the 
the linear space of infinitesimal deformations of the particular configuration. 
The metric allows us to introduce a notion of angles between deformations 
and length.\footnote{In the case of a finite dimensional space the metric defines the infinitesimal volume element explicitly. 
However, in the present case of an infinite dimensional space of configurations it is hard to construct a volume element explicitly from the metric 
although attempts have been made.\cite{Kawai} 
Fortunately, we only need to determine various Jacobians, 
which we may obtain from a particular Gaussian integral over the tangent space of deformations.\cite{Nakahara}}
If the line element of a given deformation is invariant under diffeomorphisms, then the corresponding measure will also be invariant.\cite{Nakahara} 
In the space of metrics $\mathcal{G}_{h}$ we define the local metric by the line element
\[
\Vert \delta g \Vert^{2} = 
\int d^{2}x \sqrt{g} \left( g^{\mu \alpha}g^{\nu \beta} + u g^{\mu \nu} g^{\alpha \beta}  \right) \delta g_{\mu \nu} \delta g_{\alpha \beta}
\] 
where $u$ is a positive real number.\cite{Distler} This line element is clearly invariant under diffeomorphisms.
Similarly, a suitable metric should be defined on the space of matter configurations such that the matter measure is invariant under diffeomorphisms.

Each physically distinct configuration should only contribute once to the partition function. 
This is formally implemented by dividing the partition function by the volume of the group of diffeomorphisms, $\textrm{Vol}(\textrm{Diff})$. 
However, we wish to fix the gauge explicitly in order to obtain a more well-defined partition function. 
Any given metric $g \in \mathcal{G}_{h}$ may be related to the fiducial metric $\hat{g}(\tau)$ at some particular point $\tau$
in moduli space $\mathcal{M}_{h}$ 
by a Weyl rescaling and a diffeomorphism.
Hence, instead of integrating over the space of metrics $\mathcal{G}_{h}$ in the partition function (\ref{part0}) 
we may integrate over moduli space, the space of Weyl rescalings modulo 
conformal transformations and the space of diffeomorphisms $\zeta$. 
We implement the integration over Weyl rescalings modulo  
the subgroup corresponding to 
conformal transformations of the metric 
by integrating over the entire group of Weyl rescalings and then dividing the partition function by the volume of the conformal group. 
Under this change of variables the measure transforms as 
\[
\mathcal{D}g \quad \to \quad d\tau \: \mathcal{D}\phi \: \mathcal{D}\zeta \,,
\] 
where the Jacobian $J$ may be expressed in terms of Fadeev-Popov ghosts 
as\footnote{Actually, the above expression for the Jacobian is not entirely correct. 
As we will discuss shortly we have not fixed the gauge symmetry completely. 
In order to the fix the gauge symmetry completely we need to consider the partition function with some operators inserted 
and then we may fix the residual symmetry by fixing the positions of some of these operators. 
Moreover, in this case the Jacobian becomes more complicated. 
In addition to the factor $\exp(-S_{gh})$ the integrand in the Jacobian also contains a factor
\[
\frac{1}{4\pi}(b,\prt_{\mu}\hat{g}) 
= 
\frac{1}{4\pi} \int d^{2}x \sqrt{g} b^{ab}\prt_{\mu}\hat{g}_{ab} 
\]
for each metric moduli $\tau^{\mu}$ and a factor $c^{a}(x_{i})$ for each fixed coordinate $x_{i}^{a}$. 
The interested reader is referred to the extensive literature on this subject such as \cite{pol}.} 
\[
J = \int \mathcal{D}b \: \mathcal{D}c \: \exp(-S_{gh}) 
\]  
where the ghost action is given by
\[
\label{ghoac}
S_{gh}= \frac{1}{2\pi} \int d^{2}x \sqrt{g} \, b_{\alpha \beta}g^{\beta \sigma} \nabla_{\sigma} c^{\alpha}\,. 
\]
In the above ghost action $b_{\alpha \beta}$ is a traceless symmetric Grassmann tensor field and $c^{\alpha}$ is a Grassmann vector field. 
Using the invariance of the actions and the measures under diffeomorphisms we may perform the integration over the group of diffeomorphisms 
obtaining the volume of the group of diffeomorphisms, which cancels the same factor in the denominator. 
We are left with the partition function
\[\label{part3}
\mathcal{Z}_{h} = \int d\tau \; \mathcal{D}\!_{\!e^{\phi}\!\hat{g}_{\tau}}\!\phi \; \mathcal{D}\!_{\!e^{\phi}\!\hat{g}_{\tau}}\! X \;
\mathcal{D}\!_{\!e^{\phi}\!\hat{g}_{\tau}}\!b \; \mathcal{D}\!_{\!e^{\phi}\!\hat{g}_{\tau}}\!c \; 
\exp(-\mu_{0} A[e^{\phi}\hat{g}_{\tau}]-S_{M}[X,e^{\phi}\hat{g}_{\tau}]-S_{gh}[b,c,e^{\phi}\hat{g}_{\tau}])
\]
where we have explicitly included the dependence of the measures on the metric.\cite{Moore} 

We define a classical conformal field theory as a field theory defined by an action 
invariant under the composition of any given conformal transformation and the corresponding inverse Weyl 
rescaling leaving the 
background metric\footnote{In this context the background metric refers to the metric of the world-sheet. 
This background metric should not be confused with the background metric of target space in string theory.} 
unaltered.\footnote{Notice, we define a conformal transformation as a diffeomorphism, 
under which the metric transforms by a rescaling in accordance to eq. (\ref{Weyl}). 
Some authors prefer to define a conformal transformation as the composition of a diffeomorphism, 
under which the metric transforms according to (\ref{Weyl}), and the corresponding inverse Weyl rescaling of the metric. 
A given action typically descends from an action invariant under all diffeomorphisms including conformal transformations as defined in this thesis. 
Classical conformal field theories typically distinguish themselves 
from other theories by being invariant under Weyl rescalings of the metric.}
Hence, changing the scale locally maps one solution to a given classical conformal field theory into another solution.
It is easily seen, that the ghost action actually defines a classical conformal field theory. 
This follows from the fact, that the ghost action is invariant under diffeomorphisms and the fact, that the ghost action is invariant under Weyl 
transformations of the metric due to the traceless and symmetric nature of the ghost field $b_{\alpha \beta}$. 
In this thesis we will only consider matter field theories given by an action, which classically defines a conformal field theory, 
that is the matter actions will be invariant under Weyl rescalings of the metric. 
We wish to factor out the dependence of the measures on the Weyl factor in order to make the dependence of the partition function 
on the Liouville field $\phi$ explicit.
In \cite{Moore,pol} it is argued that
\[
\label{matmea}
\mathcal{D}\!_{\!e^{\phi}\!\hat{g}_{\tau}}\! X 
= 
\e^{\frac{C_{M}}{48\pi}S_{L}} \: \mathcal{D}\!_{\!\hat{g}_{\tau}}\! X
\]
where $c_{M}$ is the central charge of the conformal matter theory and 
where the Liouville action $S_{L}$ is given
\[
\label{liac}
S_{L}[\phi,\hat{g}_{\tau}]=\int d^{2}x \sqrt{\hat{g}}
\left(\frac{1}{2}\hat{g}^{\alpha \beta}\prt_{\alpha}\phi\prt_{\beta}\phi+R[\hat{g}]\phi\right)
+ \tilde{\mu} \int d^{2}x \sqrt{\hat{g}} \e^{\phi} 
\]
where $R$ is the Ricci scalar. 
In particular, in \cite{Unified} it is shown that
\[
\label{ghmea}
\mathcal{D}\!_{\!e^{\phi}\!\hat{g}_{\tau}}\!b \; \mathcal{D}\!_{\!e^{\phi}\!\hat{g}_{\tau}}\!c
= 
\e^{-\frac{26}{48\pi}S_{L}[\phi,\hat{g}_{\tau}]} \: \mathcal{D}\!_{\hat{g}_{\tau}}\!b \; \mathcal{D}\!_{\hat{g}_{\tau}}\!c
\]
With regard to the Liouville measure $\mathcal{D}\phi$ the corresponding analysis is much more complicated due to an implicit 
dependence of the Liouville measure on $\phi$.\footnote{The local metric in the space of Liouville configurations $\phi$ is given by the line element 
\[
\Vert \delta \phi \Vert^{2} = \int d^{2}x \sqrt{g} \: \delta \phi \delta \phi = \int d^{2}x \sqrt{\hat{g}} e^{\phi} \:\delta \phi \delta \phi
\]
The implicit dependence of the line element on $\phi$ through the metric $g = e^{\phi}\hat{g}$ complicates calculations.}
Yet, one simply assumes that the Liouville measure transforms in a way similar to (\ref{ghmea}) and (\ref{matmea}) 
and thus arrives at the ansatz \cite{Moore} 
\[
\mathcal{D}\!_{\!e^{\phi}\!\hat{g}_{\tau}}\!\phi \; \mathcal{D}\!_{\!e^{\phi}\!\hat{g}_{\tau}}\! X \;
\mathcal{D}\!_{\!e^{\phi}\!\hat{g}_{\tau}}\!b \; \mathcal{D}\!_{\!e^{\phi}\!\hat{g}_{\tau}}\!c
= 
\exp(-S_{L}[\phi,\hat{g}_{\tau}])
\mathcal{D}\!_{\!\hat{g}_{\tau}}\!\phi \; \mathcal{D}\!_{\!\hat{g}_{\tau}}\! X \;
\mathcal{D}\!_{\!\hat{g}_{\tau}}\!b \; \mathcal{D}\!_{\!\hat{g}_{\tau}}\!c
\]  
where we have rescaled the Liouville field $\phi$ in order to obtain the conventional Liouville action
\[
\label{lioac}
S_{L}[\phi,\hat{g}_{\tau}]=\frac{1}{4 \pi}\int d^{2}x \sqrt{\hat{g}}
\left(\hat{g}^{\alpha \beta}\prt_{\alpha}\phi\prt_{\beta}\phi+QR[\hat{g}]\phi\right)
+ \mu \int d^{2}x \sqrt{\hat{g}} \e^{2b\phi} 
\]
where $Q$ and $b$ are constants to be determined in the following. 
From \cite{Alvarez} we expect the cosmological constant $\mu$ to be divergent in the cutoff introduced in the calculation of the Jacobians.
However, we may absorb $\mu$ into the bare cosmological constant $\mu_{0}$ in the gravitational action (\ref{graac}) obtaining a 
finite renormalized cosmological constant $\mu$, which is a free parameter of the theory.
The final expression for the partition function is given by 
\[
\label{part2}
\mathcal{Z}_{h} = \int d\tau \; \mathcal{D}\!_{\!\hat{g}_{\tau}}\! X \; \mathcal{D}\!_{\!\hat{g}_{\tau}}\!\phi \; 
\mathcal{D}\!_{\!\hat{g}_{\tau}}\!b \; \mathcal{D}\!_{\!\hat{g}_{\tau}}\!c \; 
\exp(-S_{M}[X,\hat{g}_{\tau}]-S_{L}[\phi,\hat{g}_{\tau}]-S_{gh}[b,c,\hat{g}_{\tau}])
\]   
where the Liouville action is normalized according to (\ref{lioac}). 
The development of a partition function formalism for 2D euclidean Quantum Gravity was pioneered by Polyakov in \cite{Polyakov}.

In the above partition function we integrate over the entire group of Weyl rescalings 
instead of integrating over the group of Weyl rescalings modulo the subgroup corresponding 
to conformal transformations of the metric.
Thus, we have not fixed the gauge completely. We are left with a residual gauge symmetry, 
namely the invariance of the partition function under conformal transformations.  
This conformal symmetry is crucial in the continuum formulation of 2D euclidean Quantum Gravity. 
The rescaling of the metric under a conformal transformation may be viewed as a transformation of the Liouville field.   
The fiducial metric does not transform under conformal transformations. 
Both the matter and the ghost sector indeed define 
quantum conformal field theories.\footnote{That is, both the measures and 
the actions are invariant under conformal transformations of the ghost and matter fields.} 
The central charge of the ghost theory is given by
\[\label{cengho}
c_{gh} = -26.
\]
With regard to the Liouville sector we first consider the case $\mu = 0$.
In this case the action (\ref{lioac}) and the Liouville measure define a 
quantum conformal field theory, which is known as the linear dilaton theory, with central charge \cite{pol}
\[
\label{cencharlin}
c_{L} = 1 + 6Q^{2}.
\]
We now turn on the cosmological constant $\mu$. The Liouville sector should remain a quantum conformal field theory 
even for non-zero values of the cosmological constant $\mu$. 
Turning on $\mu$ should correspond to a marginal deformation of the linear dilaton theory to quantum Liouville theory. 
We will study quantum Liouville theory in more detail in the next couple of chapters. However, for the purpose of completing 
the gauge-fixing procedure let us mention here, that the condition of 
marginality implies that 
\[
\label{Q}
Q = b + \frac{1}{b}
\]
This condition and the so-called Seiberg bound eq. (\ref{seibound}) 
determine the value of $b$. 
The marginal deformation of the linear dilaton theory into Liouville theory 
does not alter the value of the central charge $c_{L}$.\cite{Moore} 

In the partition function (\ref{part3})
the Liouville field only enters through the metric $g=\e^{\phi}\hat{g}$. 
Taking the rescaling of $\phi$ into account the partition function (\ref{part2}) should be invariant under the transformation
\[
\label{abs}
\hat{g} \to \e^{2 b \rho}\hat{g}, \qquad \phi \to \phi - \rho, \qquad \mathcal{D}\phi \to \mathcal{D}\left( \phi - \rho \right)
\]
leaving the actual metric $g$ invariant. 
If we redefine the Liouville field as $\phi' = \phi - \rho$, we recognize the invariance of the partition function 
under the above transformation as the condition, 
that the partition function (\ref{part2}) should be invariant under Weyl rescalings of the fiducial metric.
Indeed, the partition function should be independent of our choice of the fiducial metric.  
In \cite{Moore,pol} it is argued that the partition function $\mathcal{Z}_{\textrm{CFT}}$ of a given quantum conformal field theory transforms 
as\footnote{Eqs. (\ref{ghmea}) and (\ref{matmea}) may be considered as special cases of (\ref{partcft})}
\[
\label{partcft}\mathcal{Z}_{\textrm{CFT}}[\e^{\omega}\hat{g}]=\mathcal{Z}_{\textrm{CFT}}[\hat{g}] \: \e^{\frac{c}{48\pi}S_{L}[\omega,\hat{g}]} 
\]
under a Weyl rescaling of the metric $\hat{g}$, where $c$ is the central charge of the 
theory and the Liouville action is normalized according to (\ref{liac}). 
Hence, under a Weyl rescaling of the fiducial metric the partition function (\ref{part2}) transforms as
\st
\mathcal{Z}_{h} & = &\int d\tau \mathcal{Z}_{M}[\hat{g}_{\tau}] \, \mathcal{Z}_{L}[\hat{g}_{\tau}] \, \mathcal{Z}_{gh}[\hat{g}_{\tau}] \\ \nonumber
& \to & 
\int d\tau \mathcal{Z}_{M}[\e^{\omega}\hat{g}_{\tau}] \, \mathcal{Z}_{L}[\e^{\omega}\hat{g}_{\tau}] \, \mathcal{Z}_{gh}[\e^{\omega}\hat{g}_{\tau}] \\ \nonumber
& = & 
\int d\tau \mathcal{Z}_{M}[\hat{g}_{\tau}] \, \mathcal{Z}_{L}[\hat{g}_{\tau}] \, \mathcal{Z}_{gh}[\hat{g}_{\tau}]
\e^{\frac{c_{tot}}{48\pi}S_{L}[\omega,\hat{g}_{\tau}]}
\en 
where $c_{tot}$ is the total central charge.
Demanding that the partition function is invariant under Weyl rescalings of the fiducial metric we obtain the condition\label{totcen}
\[
\label{c0}
c_{tot} = c_{M} + c_{L} + c_{gh} = 0,
\]
which fixes the value of $Q$. 
Given the above discussion we somewhat understand the physical significance of the linear term in the Liouville action. 
This term is needed in order for the coupled theory to be independent of our choice of the fiducial metric.  

The uniformization theorem states, that we may choose a fiducial metric $\hat{g}$ 
from each equivalence class $\tau$ in moduli space proportional to the flat metric $\delta_{\mu \nu}$, that is $\hat{g}_{\mu \nu} = f(x)\delta_{\mu\nu}$, 
given that we parametrize the manifold on an appropriate coordinate region and impose the appropriate periodicity conditions on the coordinate region 
depending on $\tau$.\cite{Nakahara,pol} 
We may choose $f(x)$ independent of $tau$, that is $f(x)$ depends only on the topology of the manifold.  
The conformal structure is encoded in the coordinate region on which we parametrize the manifold and in the periodicity conditions 
imposed on the coordinate region. 
This choice of the fiducial metric is known as the conformal gauge.\cite{pol} 
For a generic topology the factor $f(x)$ do not belong to the group of globally defined Weyl 
rescalings.\footnote{f may for instance be zero at some point in the coordinate region.} 
Hence, for a generic topology 
we may \emph{not} choose the fiducial metric flat everywhere, that is $\hat{g}_{\mu\nu}=\delta_{\mu\nu}$. 
Indeed, this would contradict 
the Gauss-Bonnet theorem (\ref{Gaussbonnet}).  
In the case of the torus of genus $1$ or the cylinder, we may actually choose the fiducial metric flat everywhere.\cite{pol} 
The general approach to dealing with the factor $f(x)$ in the fiducial metric is to 
absorb $f(x)$ into the Liouville field $\phi$ by a transformation similar to (\ref{abs}). 
The new Liouville field inherits the boundary conditions on $f(x)$.  
Instead of integrating over configurations of the Liouville field corresponding to Weyl rescalings of the fiducial metric, 
we are now integrating over Liouville configurations satisfying some explicit boundary condition ensuring 
that $g=e^{\phi}\delta_{\mu\nu}$ belongs to the space of metrics defined on the given topology. 
Due to this approach we end up dealing with the flat metric $\delta_{\mu\nu}$, when imposing the conformal gauge.
In the case of the sphere and the disc we may parametrize them either on the extended complex plane or the upper half plane. 
In both cases the above procedure leads to the boundary condition 
\[
\phi(z,\bar{z}) = -Q \ln(z\bar{z}) +\mathcal{O}(1) \qquad \textrm{for } \vert z \vert \to \infty 
\]    
on the Liouville field.\cite{FZZ}

In complex coordinates (\ref{compcoor}) and introducing complex derivatives
\[
\prt_{z}=\frac{1}{2}(\prt_{1}-i\prt_{2}), \quad \prt_{\bar{z}}=\frac{1}{2}(\prt_{1}+i\prt_{2})
\]
we may express the ghost action (\ref{ghoac}) in unit gauge, $\hat{g}_{\mu\nu}=\delta_{\mu\nu}$, as
\[
S_{gh}=\frac{1}{2\pi} \int d^{2}z 
\left( b \prt_{\bar{z}} c + \bar{b} \prt_{z} \bar{c}\right)
\]
where $b = \frac{1}{2}(b_{11}-ib_{12})$ 
and $c = c^{1}+ic^{2}$. 
We treat $c$, $\bar{c}$, $b$ and $\bar{b}$ as independent fields in the partition function.\cite{pol}

\section{An interpretation in terms of propagating strings} \label{target}
In this thesis our main focus will be on understanding the gauge-fixed theory in terms of quantum geometry.
However, at least for some matter theories we may interpret the gauge-fixed theory as describing strings propagating in some fixed target space. 
This interpretation lies at the core of string theory, on which the literature is quite immense.  
We will only briefly discuss this interpretation. 
Let us for simplicity consider the case, where the matter sector consists of $d$ free Bose fields $X^{\mu}$, which transform as scalars under world sheet 
diffeomorphisms.  
We may view the fields $X^{\mu}$ and the Liouville field $\phi$ as $d+1$ coordinates embedding the world sheet into a $d+1$ dimensional target 
space.\footnote{The Liouville direction always appears from the integration over the conformal anomaly in the gauge-fixing procedure, 
except in the critical case 
when the central charge $c_{M}$ equals $26$.} 
As strings propagate, split and join in the target space they trace out a 2 dimensional surface. 
Each configuration of the fields $X^{\mu}$ and $\phi$ corresponds to a particular two-dimensional surface 
traced out by a set of propagating strings in target space. 
Inserting a gauge invariant operator in the bulk of the world-sheet corresponds to 
adding an incoming or outgoing closed string in some particular state.  
Hence, evaluating a given number of gauge invariant operators 
in the gauge-fixed partition function $\mathcal{Z}_{h}$ corresponds to calculating a particular  
string diagram of a given topology with a given number of incoming and outgoing string states.
The string diagrams are the string theory counterparts of Feynman diagrams in field theory and the nature of the process described by a particular  
string diagram is determined by the particular 
topology of the world-sheet and the particular insertions of gauge-fixed operators into the partition function. 
Now, string theory is a rather vast subject and trying to summarize string theory in a couple of sentences 
will not do justice to the theory. 
In this thesis I will assume, that the reader is familiar with the basics 
of string theory presented in for instance \cite{pol}. 
I will from time to time return to this interpretation and try to understand the results presented in this thesis 
in terms of string theory. 
In this section I will describe the features of the fixed target space background generic to the non-critical string theories, that arise 
when we gauge-fix 2D euclidean Quantum Gravity.  

In string theory we may only calculate the scattering amplitudes of small perturbations of a fixed background. 
The fixed background is given by the target space geometry and the additional target space fields 
and the fixed background determines the conformal non-linear sigma model governing the embedding coordinates living on the world-sheet.
Since the appearance of Liouville theory is generic to 2D euclidean Quantum 
Gravity, 
let us shortly describe the background associated with Liouville theory. 
Firstly, let us consider a configuration of the matter fields $X^{\mu}$ and $\phi$ describing a set of propagating closed strings  
localized in a small region in the $\phi$-direction of target space centered around $\phi_{0}$. 
In this case we may evaluate the linear term in the Liouville action using the Gauss-Bonnet theorem (\ref{Gaussbonnet}) and (\ref{euler})
\st
S_{L} 
& = & 
\frac{1}{4 \pi}\int d^{2}x \sqrt{\hat{g}}
\left(\hat{g}^{\alpha \beta}\prt_{\alpha}\phi\prt_{\beta}\phi+QR[\hat{g}]\phi\right)
+ \mu \int d^{2}x \sqrt{\hat{g}} \e^{2b\phi} \nonumber \\
& \approx &
Q (2-2h) \phi_{0} +
\frac{1}{4 \pi}\int d^{2}x \sqrt{\hat{g}}
\hat{g}^{\alpha \beta}\prt_{\alpha}\phi\prt_{\beta}\phi
+ \mu \int d^{2}x \sqrt{\hat{g}} \e^{2b\phi}
\en
where $h$ is the genus of the closed world-sheet. 
It follows from the previous discussion that two times the number of handles of the closed world-sheet 
equals the number of times a closed string either splits into two intermediate strings or joins with another intermediate closed string 
in the corresponding string diagram. 
Hence, we may view $\exp(Q\phi)$ as the local closed string coupling constant. 
Notice, the closed string coupling constant increases with $\phi$. 
At $\phi \to -\infty$ higher order string diagrams are suppressed, 
while at $\phi \to \infty$ the contributions from higher order diagrams dominate the scattering amplitudes.
  
Secondly, the contributions to the partition function from configurations localized in the large $\phi$-region of target space 
are suppressed by the interaction term in the Liouville action. The interaction term introduces a wall in the Liouville direction 
at $\phi \approx \frac{1}{2b}\ln\mu$ preventing the strings from entering the strong coupling region of target space. 

Different target space backgrounds correspond to different conformal field theories on the world-sheet 
with a total central charge equal to zero, which is one of the defining features of a string theory background. 
However, the particular conformal matter theories we will consider in this thesis do not justify any obvious interpretation 
in terms of strings propagating in some target space. 
The above target space picture of Liouville theory is appropriate in the case of 2D string theory, 
in which the matter consists of a single free Bose field describing a Wick rotated time-direction in target space. 
Yet, the mathematical structure of the conformal field theories we are going to consider in this thesis 
is very much similar to the mathematical structure of the conformal field theories, which we more obviously may associate 
with some target space backgrounds. 
Some authors have actually tried to interpret the results obtained in the models, 
we are going to consider, in terms of the above target space picture of Liouville theory.\cite{KOPSS} 
Furthermore, a new target space interpretation of the models we are going to consider 
has actually been proposed recently.\cite{SS,KOPSS,MMSS} 
We will return to this string theory interpretation later on in the thesis.   

\chapter{Liouville Theory} \label{chapterLiouville}
In the previous chapter Liouville theory appeared in the procedure of the gauge fixing euclidean 2D quantum gravity. 
Liouville theory concerns the Liouville factor relating the physical metric to the fiducial metric, $g=e^{\phi}\hat{g}$,   
and defines a theory of 2D euclidean geometry within a given equivalence class $\tau$ in moduli space. 
The role of the fiducial metric in Liouville theory is to determine, which particular equivalence class $\tau$ in moduli space we consider. 
In this chapter we first study classical Liouville theory in order to obtain some intuition about Liouville theory. 
We then turn our attention to a particular solution to Liouville theory describing a \emph{non-compact} geometry.
Finally, we discuss semi-classical Liouville theory and then quantum Liouville theory.

\section{Classical Liouville Theory}
Let us start out by studying classical Liouville theory. 
If we define the classical Liouville field as
\[
\phi_{cl}=2 b \phi
\]
and define
\[\label{muclas}
\mu_{cl} = 4 \pi b^{2} \mu
\]
from 
(\ref{lioac}) we obtain the action
\[
\label{cllioac}
S_{L}=\frac{1}{16 \pi b^{2}} \int d^{2}x \sqrt{\hat{g}} 
\left( \hat{g}^{\mu \nu}\prt_{\mu}\phi_{cl} \prt_{\nu}\phi_{cl} + 2 R[\hat{g}]\phi_{cl} + 4 \mu_{cl} e^{\phi_{cl}}  \right)
\]
in the limit $b \ll 1$.
Notice, in this limit $b^{2} \sim \hbar$ essentially measures the rigidity of the world sheet geometry toward quantum fluctuations and 
in the strict limit $b \to 0$ only the Liouville 
configuration, that minimizes the above Liouville action, contributes to the Liouville partition function. 
We may define the limit $b \to 0$ as the classical limit of Liouville 
theory.\footnote{This does not contradict our previous statement made in footnote \ref{claslim}  
regarding the non-existence of a classical theory of pure gravity in two dimensions defined by the Einstein-Hilbert action. 
On page \pageref{totcen} we argued, that the total central charge has to vanish. 
In the case of pure 2D euclidean quantum gravity we get from eqs. (\ref{cengho}), (\ref{cencharlin}) and (\ref{Q}), that $b=\sqrt{\frac{2}{3}}$. 
Hence, in the limit $b \to 0$ we do not study pure classical gravity defined by the Einstein-Hilbert action. 
In this limit we study some effective theory of 2D euclidean gravity obtained by integrating out some
exotic matter field theory with central charge $c_{M} \to -\infty$.}   
Using the formula \cite{Moore} 
\[
\label{Ricci}
R[\e^{\phi}\hat{g}] =\e^{-\phi}  \left( R[\hat{g}] - \hat{\nabla}^{2}\phi\right) 
\]
concerning the Ricci scalar $R$ and valid in two dimensions 
the classical equation of motion of the Liouville field is easily obtained from the classical Liouville action (\ref{cllioac})
\[
\label{eomlio}
R[g] = -2 \mu_{cl}.
\]
where $g=e^{\phi_{cl}}\hat{g}$. The classical Liouville equation transforms covariantly under diffeomorphisms. 
Thus, classical Liouville theory defines a consistent theory concerning the metric $g=\e^{\phi}\hat{g}$ 
within a given equivalence class $\tau$ in moduli space. 
The classical Liouville equation (\ref{eomlio}) describes a surface of constant curvature.

Whether the classical Liouville equation has a solution within a given equivalence class $\tau$ depends on, whether the sign of $\mu_{cl}$ 
is consistent with the topology of the Riemannian manifolds belonging to $\tau$. 
The sign of $\mu_{cl}$ must be choosen in accordance with the Gauss-Bonnet theorem (\ref{Gaussbonnet}), 
which relates the Euler number (\ref{euler}) to the curvature.\footnote{In the case of the torus of genus $1$ we have to choose $\mu_{cl}=0$.}
If this is the case, the uniformization theorem states, that 
there exists a Riemannian manifold of constant curvature $- 2 \mu_{cl}$ within any given equivalence 
class $\tau$ in moduli space.\cite{Nakahara}

We may view a conformal transformation of the physical metric $g$ as a transformation of the classical Liouville field 
keeping the fiducial metric $\hat{g}$ fixed. In conformal gauge and applying complex coordinates 
the classical Liouville field transforms according to
\[
\phi_{cl}(z,\bar{z}) \to \phi'_{cl}(z',\bar{z}') = \phi_{cl}(z,\bar{z}) - \ln\left\vert \frac{\prt z'}{\prt z}  \right\vert^{2}
\]
under the conformal transformation $z \to z'(z)$. 
Since the classical Liouville equation (\ref{eomlio}) transforms covariantly under a diffeomorphism, 
a conformal transformation of the Liouville field maps one solution to the classical Liouville equation into another solution. 
Hence, classical Liouville theory is a conformal field theory.\footnote{This 
may also be deduced from the classical Liouville action (\ref{cllioac}). In conformal gauge the classical Liouville action changes 
by a boundary term under an infinitesimal conformal transformation of the classical Liouville field.}

\section{The Lobachevskiy plane} \label{Lobachevskiy}
Let us consider the case $\mu_{cl}>0$ 
and let us introduce the length scale 
\[
R_{0} = \frac{1}{\sqrt{\mu_{cl}}}.
\] 
In the upper half plane\footnote{In this case the upper half plane does not include the real axis.} Liouville theory admits the solution
\[
ds^{2} = \exp(\phi_{cl}(z,\bar{z})) dz d\bar{z} = \frac{R_{0}^{2}}{(\textrm{Im}(z))^{2}}dz d\bar{z}
\]
expressed in complex coordinates.
This solution describes a non-compact geometry with constant negative curvature $-2/R_{0}^{2}$ known as  
the Lobachevskiy plane, which is an euclidean version of AdS$_{2}$.\cite{ZZ} 
The isometry group of the Lobachevskiy plane is the group of M\"obius transformation $\textrm{PSL}(2,\mathbf{R})$
mapping the upper half plane to the upper half plane 
\[
z \to z' = \frac{az+b}{cz+d}, \qquad ad-bc=1, \quad a,b,c,d \in \mathbf{R}.
\]
Since $\textrm{PSL}(2,\mathbf{R})$ is the group of holomorphic bijective maps from the upper half plane to the upper half plane, 
it follows from our discussion on page \pageref{conformaltrans}, that the group of isometries 
coincides with the group of conformal transformations on the upper half plane. 
Thus, the Lobachevskiy plane is maximally symmetric.
Solving the geodesic equation we may easily determine the geodesic distance $D$ between the two point $z_{1}$ and $z_{2}$ 
in the Lobachevskiy plane parametrized in the upper half plane 
\[\label{distanceloba}
\eta = \tanh^{2}\left(\frac{D}{2R_{0}}\right)
\]
where 
\[\label{eta}
\eta = \frac{(z_{1}-z_{2})(\bar{z}_{1}-\bar{z}_{2})}{(z_{1}-\bar{z}_{2})(\bar{z}_{1}-z_{2})}.
\]
Notice, the geodesic distance $D$ diverges as $\eta \to 1$.

Using the transformation
\[\label{uppertodisk}
z \to \omega(z) = \frac{iz+1}{z+i}
\]
we may map the Lobachevskiy plane to the interior of the unit disk
\[
\label{Lobcircle}
ds^{2} = \frac{4R_{0}^{2}}{(1-\omega\bar{\omega})^{2}} d\omega d\bar{\omega}.
\]  
Expressed in these coordinates $\eta$ is given by
\[
\eta = \frac{\vert \omega_{1} - \omega_{2} \vert^{2}}{\vert \omega_{1}\bar{\omega}_{2} - 1 \vert^{2}} 
= 
\frac{\vert \omega_{1} - \omega_{2} \vert^{2}}{\vert \omega_{2} \vert^{2} \vert \omega_{1} - \frac{1}{\bar{\omega}_{2}} \vert^{2}}
\]
Notice, as either $\omega_{1}$ or $\omega_{2}$ approach the boundary of the unit disk, the geodesic distance $D$ diverges 
reflecting the non-compact nature of the Lobachevskiy plane. 
The Lobachevskiy plane do not have a boundary. 
However, the unit circle forms a one-dimensional infinity known as the absolute. 
Let us consider any given point in the Lobachevskiy plane parametrized in the interior of the unit disk. 
Using an isometry we map the given point to the center of the unit disk.
Let us moreover consider the set of points $\mathcal{S}$ within a given geodesic distance $D$ of the marked point. 
From the rotational symmetry of the metric (\ref{Lobcircle}) it is obvious, that the set of points $\mathcal{S}$ is given by 
a disk of radius $r_{b}<1$ within the unit disk and centered at the origin. 
The geodesic distance $D$ from the marked point to the boundary of $\mathcal{S}$ is given in terms of $r_{b}$ by
\[
D = \int ds = \int_{0}^{r_{b}} dr \frac{2R_{0}}{1-r^{2}} = 2R_{0} \tanh^{-1}(r_{b})
\]
The area of $\mathcal{S}$ is given by
\[
A = \int_{\mathcal{S}} d^{2}x \sqrt{g} = \int_{0}^{r_{b}} \int_{0}^{2\pi} dr d\theta \frac{4R_{0}^{2}r}{(1-r^{2})^{2}} 
=  4 \pi R_{0}^{2} \sinh^{2}\left(\frac{D}{2R_{0}}\right)
\]
Finally, the length of the boundary $\prt\mathcal{S}$ is given by
\[
L = \oint_{\prt\mathcal{S}} ds = \int_{0}^{2\pi} d\theta \frac{2R_{0}r_{b}}{1-r_{b}^{2}} 
= 4 \pi R_{0} \sinh\left(\frac{D}{2R_{0}}\right) \cosh\left(\frac{D}{2R_{0}}\right)
\]
For $D \gg R_{0}$ the area of $\mathcal{S}$ and the length of the boundary $\prt\mathcal{S}$ are proportional to each other  
and they both grow exponentially with the geodesic distance $D$. 
This signature of the Lobachevskiy plane will be important later on, when we turn our attention to the 
non-compact geometries coming from dynamical triangulations.

\section{The semi-classical spectrum of Liouville theory.} \label{semiclassical}
As mentioned previously our main focus in this thesis will be on studying 2D euclidean quantum gravity, 
that is the random geometry of the two-dimensional world-sheet.
However, by a closed string state\footnote{Not all closed string states are 
physical gauge-invariant states. We will return to this discussion in section \ref{physoperator}.} 
in a given 2D conformal field theory such as Liouville theory  
we conventionally mean a state of the field quantized on the circle, which 
we parametrize by $\sigma \in [0,2\pi)$. 
In order to obtain the spectrum of closed string states and the Hamiltonian we consider the particular conformal field 
theory on the cylinder with time $t$ propagating along the cylinder.
We may map the infinite cylinder to the sphere by the transformation
\[
\label{cylsphere}
z = \exp(-i\omega), \quad \omega \equiv \sigma + it.
\] 
In the complex $z$-plane time evolves radially outwards.
Under this transformation the circle at $t=-\infty$ is mapped to the origin in the complex $z$-plane. 
Inserting an incoming state on the circle at $t \to -\infty$, 
that is weighing each field configuration in the cylinder partition function 
by a wave functional depending on the field configuration on the circle at $t \to -\infty$, 
corresponds to inserting a vertex operator at the origin in the complex $z$-plane. 
Thus, an incoming state at $-\infty$ is mapped to a local vertex operator under the transformation (\ref{cylsphere}).
This is the famous state-operator correspondence in conventional conformal field theory. 
While the spectrum of closed string states is conventionally introduced on the cylinder, 
the spectrum of vertex operators is conventionally introduced on the sphere.\cite{pol} 
As we will discuss later on in this chapter, 
the conventional state-operator correspondence is not valid in Liouville theory.  

In order to obtain the Hilbert space of closed string states in Liouville theory, 
we need some insight obtained from a semi-classical analysis of the spectrum.
The entropy of surfaces with fixed area $A$ and fixed topology is typically given by a power relation $A^{\gamma}$, where the exponent 
$\gamma$ depends on $b$ and the topology of the surfaces \cite{Moore,ADJ}. 
This implies that the Laplace transform of the partition function 
with fixed area $A$ is convergent only for positive values of the cosmological constant $\mu$.
Hence, we expect the Liouville partition function to be convergent, if and only if the cosmological constant $\mu$ is positive. 
This also follows from the fact, that the Liouville action is unbounded from below if $\mu < 0$.
Due to the fact that $\mu > 0$ in quantum Liouville theory semi-classical Liouville theory obtained in the limit $b \to 0$ 
describes geometries of constant \emph{negative} curvature. 
There exists three families of solutions with constant negative curvature 
to the classical Liouville equation (\ref{eomlio}) on the cylinder, the elliptic, the parabolic and the hyperbolic solutions. 
These three families of solutions are each characterized by their monodromy properties.\cite{Seiberg,Moore}
\begin{figure}[!t]
\begin{center}
\includegraphics{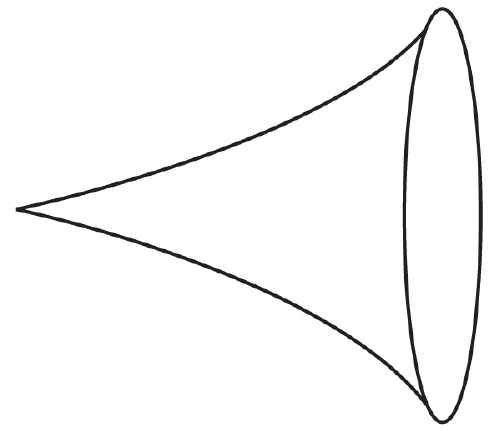}
\end{center}
\caption{The elliptic solutions}
\end{figure}
The elliptic solutions are given by
\st
ds^{2}
&=&
\frac{1}{4\pi b^{2}\mu}\frac{\nu^{2}}{\sinh^{2}(\nu t)}
\left(dt^{2}+d\sigma^{2}\right)
\nonumber \\ \label{ellsol}
&=&
\frac{1}{4\pi b^{2}\mu}\frac{\nu^{2}}{\sinh^{2}(\nu t)} 
d\omega d\bar{\omega}, \quad t > 0, \quad \nu > 0.
\en 
The elliptic solutions have curvature singularities at $t=0$. 
The parabolic solution is obtained from the family of elliptic solutions in the limit $\nu \to 0$ 
and also has a curvature singularity at $t=0$.\cite{Seiberg}
The hyperbolic solutions are given by
\st
ds^{2}
&=&
\frac{1}{4\pi b^{2}\mu}\frac{\epsilon^{2}}{\sin^{2}(\epsilon t)}
\left(dt^{2}+d\sigma^{2}\right) \nonumber \\ \label{hypsol}
&=&
\frac{1}{4\pi b^{2}\mu}\frac{\epsilon^{2}}{\sin^{2}(\epsilon t)} 
d\omega d\bar{\omega}, \quad \frac{n\pi}{\epsilon} < t < \frac{(n+1)\pi}{\epsilon}, \quad \epsilon > 0.
\en
\begin{figure}[b]
\begin{center}
\includegraphics{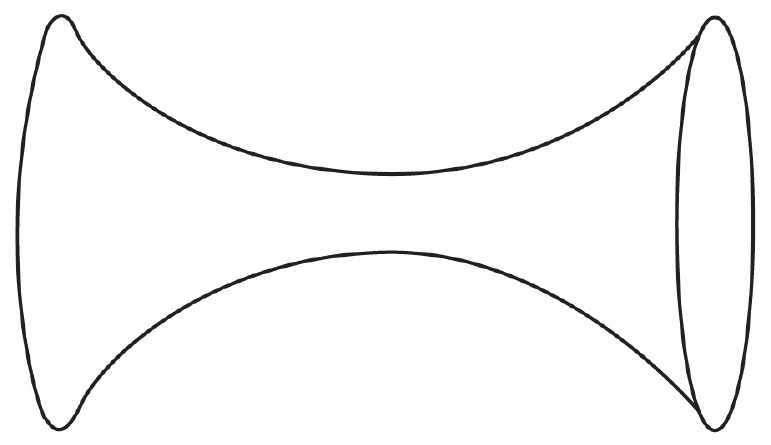}
\end{center}
\caption{The hyperbolic solutions}
\end{figure}
From (\ref{lioac}) we obtain the energy momentum tensor in the semi-classical limit $b \ll 1$ 
\st
T^{\mu\nu} & \equiv & 4\pi\frac{\delta S_{L}}{\delta \hat{g}_{\mu\nu}}\Big\vert_{\hat{g}_{\alpha\beta}=\delta_{\alpha\beta}} 
\nonumber \\
& = & 
\frac{1}{2}\delta^{\mu\nu}
\left( \prt_{\alpha}\phi\prt_{\alpha}\phi + 4\pi\mu\exp(2b\phi)-\frac{2}{b}\prt_{\alpha}\prt_{\alpha}\phi \right)
-\prt^{\mu}\phi\prt^{\nu}\phi+\frac{1}{b}\prt^{\mu}\prt^{\nu}\phi
\label{energyxy}
\en
In order to obtain the above result we have to perform the functional derivative of the Ricci scalar in the Liouville action (\ref{lioac}) 
before setting the fiducial metric equal to the flat metric $\delta_{\alpha\beta}$. This functional derivative is easily obtained from \cite{Weinberg}. 
Applying the equation of motion (\ref{eomlio}) together with eqs. (\ref{Ricci}) and (\ref{muclas}) 
we realize, that the energy momentum tensor is traceless, which is actually 
a defining feature of a conformal field theory. 
Applying the complex coordinates $\omega$ and $\bar{\omega}$ introduced in (\ref{cylsphere}) we get that 
\[
T(\omega)=-(\prt_{\omega}\phi)^{2}+\frac{1}{b}\prt_{\omega}^{2}\phi, \qquad
\bar{T}(\bar{\omega})=-(\prt_{\bar{\omega}}\phi)^{2}+\frac{1}{b}\prt_{\bar{\omega}}^{2}\phi
\]
In section \ref{QLT} we will discuss the energy momentum tensor in quantum Liouville theory. 
It actually turns out, that a constant term appears in the energy momentum tensor on the cylinder 
in quantum Liouville theory due to the Schwarzian derivative appearing 
in the transformation law (\ref{schwarz}) governing the energy momentum tensor under conformal transformations. 
Since we eventually want to compare the energies of the classical solutions to Liouville theory on the cylinder 
with the energies of the eigenstates of the Hamiltonian in quantum Liouville theory we need to modify the above classical 
energy momentum tensor by the same constant term 
\[
\label{energymom}
T(\omega)=-(\prt_{\omega}\phi)^{2}+\frac{1}{b}\prt_{\omega}^{2}\phi + \frac{1}{24}, \qquad
\bar{T}(\bar{\omega})=-(\prt_{\bar{\omega}}\phi)^{2}+\frac{1}{b}\prt_{\bar{\omega}}^{2}\phi + \frac{1}{24} \,.
\]
The Hamilton operator on the cylinder is given by
\[
\label{hamcyl}
H_{c} 
=  
\frac{1}{2\pi} \int_{0}^{2\pi} \!d\sigma \, T_{00}
=
- \frac{1}{2\pi} \int_{0}^{2\pi} d\sigma \left( T(\omega) + T(\bar{\omega}) \right)  
\]
Notice, the addition of the constant term to the energy momentum tensor simply corresponds to changing the zero of the energy scale. 
Applying the above expression for the Hamilton operator on the cylinder  
we may calculate the energies at a given time $t$ of each of the solutions to semi-classical Liouville theory on the cylinder.
In the case of the elliptic surfaces we get
\[
\label{eell}
E_{\textrm{elliptic}}=-\frac{\nu^{2}}{2b^{2}}- \frac{1}{12}.
\]  
The energy of the parabolic solution at a given time $t$ is obtained from the above energy in the limit $\nu \to 0$.
The energies of the hyperbolic solutions at given time $t$ are given by
\[
\label{ehyp}
E_{\textrm{hyperbolic}}=\frac{\epsilon^{2}}{2b^{2}} - \frac{1}{12}.
\]
The energy of each solution is conserved. This is a consequence of the fact, that the Lagrangian  
in Liouville theory do not depend on time explicitly. 
We expect each of these solutions to correspond to some particular state in the semi-classical limit of 
quantum Liouville theory. In the following we will clarify this statement to some extent.  

\section{The mini-superspace approximation.}
Before we turn our attention to quantum Liouville field theory  
it will actually prove fruitful to study quantum Liouville theory in the so-called mini-superspace approximation. 
In this approximation we only consider configurations of the Liouville field independent of the spatial direction $\sigma$ on the cylinder, 
that is $\phi=\phi(t)$. 
We then quantize the Liouville field applying this approximation.
Our expection is that all states of the Liouville field on the circle invariant under rigid rotations will be captured by this approximation.
In order to obtain the Hamilton operator 
we first perform a Wick rotation to 
Minkowski space. The Liouville action in Minkowski space $\hat{g}_{\mu\nu}=\eta_{\mu\nu}$ 
and in the mini-superspace approximation is obtained from (\ref{lioac})
\[
S_{L}^{M} = \int dt L = \frac{1}{2}\int dt 
\left( \dot{\phi}^{2} - 4\pi\mu\exp(2b\phi) \right)
\]  
where $L$ is the Lagrangian, $\dot{\phi}=\prt_{t}\phi$ and where we have performed the integration over $\sigma$.
The canonical momentum associated with the Liouville field in the mini-superspace approximation is given by
\[
p = \frac{\prt L}{\prt(\dot{\phi})} = \dot{\phi}
\]
From the above equations we obtain the Hamiltonian 
\[
\label{hammini}
H = 
p\dot{\phi} - L 
= \frac{1}{2}p^{2} + 2\pi\mu\exp(2b\phi)
\]
So far in this section we have discussed classical Liouville theory in the mini-superspace approximation. 
We now quantize Liouville theory in this approximation. 
Applying the operator identity
\[\label{poperator}
p = -i\frac{\prt}{\prt\phi}
\]
in the Hamiltonian (\ref{hammini}) 
and taking the shift of the zero of the energy scale discussed in the previous section into account we 
obtain\footnote{The reader may have noticed, that the shift in the zero of the energy scale deviates from the shift  
derived in \cite{Moore,Seiberg}, which depends on $Q^{2}$. 
The deviation is a consequence of the fact, that the authors in \cite{Moore,Seiberg} demand, that the Fourier coefficients appearing in the 
Fourier expansion of the energy momentum tensor on the \emph{cylinder} satisfy the Virasoro algebra, while we demand, that the Virasoro generators 
defined in the $z$-plane satisfy the Virasoro algebra. These two condition are not the equivalent due to the Schwarzian derivative appearing in the 
transformation law governing the energy momentum tensor under conformal transformations. 
Moreover, the overall conventions applied in \cite{Moore,Seiberg} are somewhat different from the conventions applied in this thesis.}
\[
\label{hamterm}
H = -\frac{1}{2}\frac{\prt^{2}}{\prt\phi^{2}} + 2\pi\mu\exp(2b\phi) - \frac{1}{12}.
\]
The circumference $l$ of the one-dimensional ``universe'' at fixed time $t$ is given by
\[
\label{length}
l = 2\pi\exp(b\phi).
\] 
Expressing the Hamiltonian (\ref{hamterm}) in terms of $l$ we obtain
\[
H = -\frac{b^{2}}{2}\left(l\frac{\prt}{\prt l}\right)^{2} + \frac{\mu}{2\pi}l^{2} - \frac{1}{12}
\] 
We now impose the boundary condition, that the wave functions fall off at large $l$. 
In the target space picture introduced in section \ref{target} 
this corresponds to imposing the physical condition, that the wave functions fall off behind the Liouville wall. 
The eigenstates of the Hamiltonian in the mini-superspace approximation satisfying this boundary condition are given by
\[
\label{wave}
\Psi_{P}(l) \propto  K_{\frac{i2P}{b}}(\sqrt{\mu}l) 
\]
with energies
\[
\label{emini}
E_{P}= 2 P^{2} - \frac{1}{12}.
\]
In (\ref{wave}) we have absorbed the factor $\frac{1}{b^{2}\pi}$ 
in the cosmological constant appearing in the argument of the modified Bessel function $K_{\frac{i2P}{b}}$. 
The above eigenstates are normalizable, if and only if $P$ is real. 
For $P$ imaginary the eigenstates diverge at $l \to 0$.
Furthermore, $\Psi_{P}(l) \propto \Psi_{-P}(l)$, that is the two wave functions correspond to the same physical state.

\section{Quantum Liouville theory} \label{QLT}
Let us apply the conformal gauge. 
In this gauge the action (\ref{lioac}) in quantum Liouville theory is given by
\[
\label{lioaccom}
S_{L} = 
\frac{1}{2\pi} \int d^{2}z \,\prt_{z}\phi \prt_{\bar{z}}\phi + \frac{\mu}{2} \int d^{2}z \exp(2b\phi). 
\] 
The central charge of Liouville theory is 
\[
c_{L}=1+6Q^{2}.
\]
The primary operators $V_{\alpha}$ in Liouville theory are spinless and are given by
\[
\label{primope}
V_{\alpha}(z,\bar{z})=\,:\!\exp(2\alpha\phi(z,\bar{z}))\!:
\]
with dimension
\[ 
\label{confdim}
\Delta(\alpha)=\alpha(Q-\alpha).
\]
In eq. (\ref{primope}) the dots ``$:$'' denote the operation of normal ordering, 
in which all the divergences are removed, which come from evaluating several factors of $\phi$    
at the same point on the world-sheet, when inserting the Taylor expansion of $V_{\alpha}$ in terms of $\phi$ 
into the partition function.\cite{ZZ} 
In the operator formalism this normal ordering is implemented by placing 
all lowering operators to the right of all raising operators. 
(With regard to lowering and raising operators see the discussion below.) 
We demand, that the Virasoro generators appearing in the Laurent expansion of the energy momentum tensor 
in the complex $z$-plane
\[\label{Virgen}
T(z) = \sum_{m=-\infty}^{\infty} \frac{L_{m}}{z^{m+2}}, 
\quad 
\bar{T}(\bar{z}) = \sum_{n=-\infty}^{\infty} \frac{\bar{L}_{n}}{\bar{z}^{n+2}} 
\]
satisfy the Virasoro algebras
\[\label{viralgebra}
[L_{m},L_{n}] 
= 
(m-n)L_{m+n}+\frac{c}{12}(m^{3}-m)\delta_{m,-n}
\]
and 
\[
[\bar{L}_{m},\bar{L}_{n}] 
= 
(m-n)\bar{L}_{m+n}+\frac{c}{12}(m^{3}-m)\delta_{m,-n},
\]
where $c$ is the central charge.
The energy momentum tensor obtained from (\ref{lioac}) by a calculation identical to (\ref{energyxy}) 
indeed satisfies this condition \cite{FZZ} 
\[\label{energysphere}
T(z)=-(\prt_{z}\phi)^{2}+Q\prt_{z}^{2}\phi,
\quad
\bar{T}(\bar{z})=-(\prt_{\bar{z}}\phi)^{2}+Q\prt_{\bar{z}}^{2}\phi.
\]
Under a conformal transformation $z \to z'=z'(z)$ 
the Liouville field transforms according to \cite{Moore}
\[\label{contransphi}
\phi(z',\bar{z}') = \phi(z,\bar{z}) -\frac{Q}{2} \ln \vert \prt_{z}z' \vert^{2}.
\]
Mapping the $z$-plane to the infinite cylinder with the transformation
\[
\omega = i \, \textrm{Ln}\,(z)
\]
and applying the above transformation law (\ref{contransphi}) and the transformation law 
\[
\label{schwarz}
T(\omega)=(\prt_{\omega}z)^{2}T(z)
+
\frac{c_{L}}{12}\frac{2\,\prt_{\omega}^{3}z\,\prt_{\omega}z-3\,\prt_{\omega}^{2}z\,\prt_{\omega}^{2}z}{2\,\prt_{\omega}z\,\prt_{\omega}z}
\]   
governing the energy momentum tensor under a conformal transformation, we obtain 
the energy momentum tensor on the cylinder
\[
T(\omega)=-(\prt_{\omega}\phi)^{2}+Q\prt_{\omega}^{2}\phi + \frac{1}{24}, \qquad
\bar{T}(\bar{\omega})=-(\prt_{\bar{\omega}}\phi)^{2}+Q\prt_{\bar{\omega}}^{2}\phi + \frac{1}{24}
\]
Notice, this energy momentum tensor indeed reduces to (\ref{energymom}) in the semi-classical limit. 
From eq. (\ref{hamcyl}) and applying eqs. (\ref{schwarz}) and (\ref{Virgen}) we obtain the Hamiltonian on the cylinder 
given in terms of the Virasoro generators in the $z$-plane. 
\[
\label{hamcyl2}
H_{c} = L_{0} + \bar{L}_{0} - \frac{c}{12}
\]

Before we proceed let us shortly consider Liouville theory with $\mu = 0$. 
In this case the Liouville action in conformal gauge reduces to the action of a free massless bose field discussed in \cite{pol}.  
Applying the equation of motion of the Liouville field obtained from (\ref{lioaccom}) with $\mu = 0$
\[
\prt_{\bar{z}}\prt_{z}\phi = 0
\] 
we may expand the Liouville field in the complex $z$-plane and insert $z=\exp(t-i\sigma)$ giving us   
\[
\phi(\sigma,t) = \phi_{0} - i p t   
+ \frac{i}{\sqrt{2}}\sum_{n \neq 0} \frac{1}{n}
\left[ a_{n}\e^{-nt+in\sigma} + \tilde{a}_{n} \e^{-nt-in\sigma}  \right],
\]
where $p$ is the canonical momentum operator introduced in the mini-superspace approximation.
Performing a Wick rotation to Minkowski space and imposing the standard equal time commutator 
\[
[\phi(\tilde{\sigma},t_{M}),\Pi(\sigma,t_{M})] = i 2\pi \delta(\tilde{\sigma}-\sigma)
\]
where the canonical momentum field is given by 
\[
\Pi(\sigma,t_{M}) 
\equiv 
\frac{\prt\phi(\sigma,t_{M})}{\prt t_{M}} 
\]
we obtain the commutation relations
\st
&&[a_{m},a_{n}]  =  [\tilde{a}_{m},\tilde{a}_{n}] = m\delta_{m,-n} \nonumber\\
&&[\phi_{0},p]  =  i
\en   
with all other commutators vanishing. 
Using standard terminology from the discussion of the harmonic oscillator in quantum mechanics, 
we will refer to the ladder operators $a_{m}$ and $\tilde{a}_{n}$ as raising operators for $m,n > 0$ and lowering operators for $m,n < 0$.
Turning on the cosmological constant $\mu$, the ladder operators are no longer constants of the motion. 
On the contrary they depend on time in some complicated way. 

If we cut out a disk in the $z$-plane, insert a vertex operator at the center of the disk and fix the configuration of the Liouville field at 
the boundary of the disk, we may define the wave functional associated with the inserted vertex operator and evaluated at the 
particular configuration of the Liouville field imposed at the boundary as the amplitude obtained by integrating over all configurations of the 
Liouville field on the disk satisfying the particular boundary condition.\cite{pol} 
This is the inverse map in the state-operator correspondance discussed previously. 
Using this map on the entire set of vertex operators in Liouville theory, 
we may define the so-called Hartle-Hawking space of wave functionals in Liouville theory \cite{Seiberg}. 
Applying the Hamiltonian (\ref{hamcyl2}) we may determine the energy of the primary state 
$\vert \alpha \rangle$ associated with a given primary operator $V_{\alpha}$ 
in the Hartle-Hawking space of wave functionals from eq. (\ref{confdim})
\[
\label{eprim}
H_{c} \vert \alpha \rangle  = \left( -\frac{1}{2}(Q-2\alpha)^{2}-\frac{1}{12} \right)  \vert \alpha \rangle
\] 
Comparing the states in the Hartle-Hawking space of wave functionals, the eigenstates of the Hamiltonian in the mini-superspace 
approximation and the solutions to classical Liouville theory on the cylinder a consistent picture arises, 
which sheds light on the Hilbert space and the set of operators in Liouville theory.
Let us start out by discussing the set of eigenfunctions of the Hamiltonian in the mini-superspace approximation. 
The eigenstate $K_{\frac{i2P}{b}}(\sqrt{\mu}l)$ is normalizable, if and only if $P$ is real. 
Applying the target space picture introduced in section \ref{target} we may give a physical interpretation of $P$. 
The strings propagate freely in the region $\phi \to -\infty$ of target space. 
Taking the Liouville wall located at $\phi \approx \frac{1}{2b}\ln\mu$ into account we expect a given closed string eigenstate 
of the mini-superspace Hamiltonian
to be the sum of an incoming wave and an outgoing wave in this region of target space. 
The outgoing wave is caused by the scattering of the incoming wave on the Liouville wall.
Moreover, we expect, that the momenta of the incoming and outgoing wave 
measured by the operator $p$ defined in eq. (\ref{poperator}) coincide up to a sign. 
Indeed, from (\ref{length}) we obtain, that the wave function $K_{\frac{i2P}{b}}(\sqrt{\mu}l)$
behaves as 
\[
K_{\frac{i2P}{b}}(\sqrt{\mu}l) 
\sim 
\frac{1}{\Gamma(1-i2P/b)}\left(\pi\sqrt{\mu}\right)^{-i2P/b}\e^{-i2P\phi} 
- 
\frac{1}{\Gamma(1+i2P/b)}\left(\pi\sqrt{\mu}\right)^{i2P/b}\e^{i2P\phi} 
\]
in the region $\phi \to -\infty$. 
Hence, we may regard $P$ as the momentum carried by the waves in the region $\phi \to -\infty$. 
From now on we will refer to $P$ as the Liouville momentum and label the eigenstate with momentum $P$ by $\vert\frac{Q}{2}+iP\rangle$. 
A given normalizable eigenstate of the Hamiltonian in the mini-superspace approximation  
corresponds to a state in the Hilbert space of quantum Liouville theory invariant under rigid rotations, 
that is a state, where none of the degrees of freedom associated with ladder operators $a_{m}$ and $a_{n}$ are excited. 
\[\label{highestlad}
a_{m} \big\vert Q/2+iP  \big\rangle_{\mathcal{H}} = \tilde{a}_{n} \big\vert Q/2+iP \big\rangle_{\mathcal{H}} = 0, \quad \textrm{for } m,n > 0 \;
\textrm{and } P>0.  
\]
We refer to such a state as a highest weight state. 
We have attached the subscript $\mathcal{H}$ in order to remind ourselves, 
that this state belongs to the Hilbert space of normalizable states.
We expect, that states exist in the Hilbert space, which are not invariant under rigid rotations.
Furthermore, the Hilbert space should define a representation of the 
algebra satisfied by the ladder operators.  
This condition ensures, that the Hilbert space defines a representation of the Virasoro algebra, 
since the Virasoro generators may be expressed in terms of the ladder operators.
Acting on a highest weight state repeatedly with the raising operators $a_{m}$, $m<0$, 
we generate a highest weight representation of the algebra satisfied by the ladder operators $a_{m}$  
known as the Feigin-Fuchs module  
\[\label{feigin}
\mathcal{F}(c_{L},P) 
\equiv
\textrm{Span}\bigg\{ \prod_{i=1}^{n}a_{k_{i}} \big\vert Q/2+iP \big\rangle_{\mathcal{H}}  \bigg\vert n \geq 0, k_{n} \leq k_{n-1} \leq \ldots \leq k_{1}<0\bigg\}.
\]
Similarly, we may generate a highest weight representation $\widetilde{\mathcal{F}}(c_{L},P)$ by acting repeatedly 
with the raising operators $\tilde{a}_{n}$, $n<0$, on the highest weight state with momentum $P$.
Due to the fact that all the raising operators commute with the momentum operator $p$ 
we associate the same Liouville momentum with all the states belonging to a given Feigin-Fuchs module.  
We expect the Hilbert space in quantum Liouville theory to be given by
\[\label{Hilbertlio}
\mathcal{H}_{L} = \int_{0}^{\infty}\! dP \,\mathcal{F}(c_{L},P)\!\otimes\! \widetilde{\mathcal{F}}(c_{L},P) 
\]   
where the integral is a direct integral. 
Notice, the fact, that the state in the mini-superspace approximation with Liouville momentum $-P$ is
identical to the state with Liouville momentum $P$, implies, that we only integrate over the positive real axis. 

As mentioned previously, in conventional conformal field theory 
there is a one-to-one correspondence between states in the Hilbert space 
and vertex operators generating local disturbances on the world-sheet. 
This is not the case in Liouville theory.\cite{Seiberg} 
Given a state in a conformal field theory 
we may impose this state at a given time $t_{1}$ in the complex $z$-plane by removing the disk of radius $\e^{t_{1}}$ 
and imposing the wave functional corresponding to the state 
as a boundary condition on the boundary of the hole.    
Alternatively, we may impose the state at time $t_{1}$ by removing a disk of radius $r = \e^{t_{0}} < \e^{t_{1}}$, imposing the wave functional 
corresponding to the state as a boundary condition on the boundary of the hole and compensating for the time evolution of the state from $t_{0}$ to $t_{1}$ by 
acting on the wave functional with the operator $\exp((t_{1}-t_{0})(L_{0}+\bar{L}_{0}))$, 
where $L_{0}+\bar{L}_{0}$ is the generator of scale transformations in the complex $z$-plane. 
In the limit $r \to 0$ the hole shrinks to a point 
and imposing the wave function as a boundary condition now corresponds to inserting a vertex operator.\cite{pol} 
In the case of Liouville theory the physical circumference $l$ of the hole measured in terms of 
the actual metric, $g = \exp(2b\phi)\hat{g}$, 
also vanishes in this limit as long as we consider wave functionals, which decay sufficiently fast for large values of 
$\phi$. Assuming this is the case imposing the wave functional in the limit $r \to 0$ corresponds to inserting a vertex operator, 
which generates a local disturbance on the world-sheet with metric $g=\e^{2b\phi}\hat{g}$.\footnote{Such a vertex operator is known as a local operator.} 
However, in this case the state corresponding to wave functional 
has to be peaked on small circumferences $l$. 
Otherwise, we are not able to associate a finite vertex operator with the state in the limit, where the hole shrinks to a point.
Hence, the state corresponding to a local vertex operator has to be non-normalizable. 
The normalizable states in the Hilbert space of Liouville theory do not correspond to local operators. 
Now, the eigenstates of the Hamiltonian in the mini-superspace 
approximation with imaginary momentum $P$ are not normalizable due the fact, 
that they grow as $l^{-\frac{2\vert P \vert}{b}}$ for $l \to 0$. 
From eqs. (\ref{emini}) and (\ref{eprim}) we realize, that the energies of these non-normalizable states exactly match the 
energies of the states in the Hartle-Hawking space of wave functionals associated with 
the primary operators $V_{\alpha}$, $\alpha \in \mathbf{R}$. 
Thus, we identify these states\footnote{Notice, the states in the Hartle-Hawking space of wave functionals corresponding to  
primary operators are highest weight states with respect to the Virasoro algebra. (See eq. (\ref{highestvir}).)  
Given the decomposition of the Virasoro generators in terms of the ladder operators in \cite{pol}, we realize, that these states 
in the Hartle-Hawking space of wave functionals are also highest weight states with respect to the ladder operators as defined in eq. (\ref{highestlad}). 
Hence, we expect these states to be invariant under rigid rotations.} 
and we expect, that the corresponding primary operators $V_{\alpha}$, $\alpha \in \mathbf{R}$, 
are local operators.\cite{Seiberg} 
Similarly, we observe from eqs. (\ref{emini}) and (\ref{eprim}), that the energies of the normalizable states in the mini-superspace approximation 
match with the energies of the states in the Hartle-Hawking space of wave functionals associated with 
the primary operators $V_{Q/2+iP}$, $P \in \mathcal{R}$.
Due to this fact we identify these states. 
This implies, that the operators $V_{Q/2+iP}$, $P \in \mathcal{R}$, do not correspond to local operators. 
Instead, they create holes in the world-sheet with metric $g$.
Moreover, the fact, that the state in the mini-superspace approximation with Liouville momentum $-P$ is identical 
to the state with Liouville momentum $P$, translates into the fact,  
that the operator $V_{\alpha}$ should be identified with the operator $V_{Q-\alpha}$. 
We may therefore impose the so-called Seiberg bound
\[\label{seibound}
\alpha \leq \frac{Q}{2}
\]
on the set of local primary operators in Liouville theory and we may impose the bound $P>0$ on the set of non-local primary operators.\cite{Moore}
This implies in particular, that we may choose $b \leq 1$.      
From the above we conclude, that the linear space of local operators in Liouville theory is given by 
\[
\mathcal{A}=\bigg\{\hat{A}_{\alpha}(z,\bar{z})\, \bigg\vert \, \alpha \leq \frac{Q}{2}   \bigg\}
\]
where $\hat{A}_{\alpha}$ is either the primary operator $V_{\alpha}$ or one of its conformal descendants.

The classical solutions to Liouville theory on the cylinder discussed in section \ref{semiclassical} fit nicely into the above picture.
Matching the energies of the classical solutions given by eqs. (\ref{eell}) and (\ref{ehyp}) with the energies (\ref{emini}) 
of the eigenstates of the Hamiltonian in the mini-superspace approximation, 
it seems obvious, that we should identify the elliptic and the  
parabolic solutions with the non-normalizable states in the classical limit and 
the hyperbolic solutions with the normalizable states in the classical limit. 
It is tempting to view the singularity of a given  
elliptic or parabolic solution as the point of insertion of the local operator creating 
the non-normalizable state. 
The hyperbolic solutions on the opposite do not have unique points, which we can associate with the insertion of local operators. 
This is consistent with the fact, that we cannot associate a local operator with a normalizable state in Liouville theory.\cite{Moore,Seiberg}

The above picture raises a question. 
Since the states discussed in the mini-superspace approximation are eigenstates of the Hamiltonian, they 
do not evolve with time. 
The average circumference of the one-dimensional universe obtained from one of these eigenstates is independent of time. 
However, if we consider any of the classical solutions given by (\ref{ellsol}) and (\ref{hypsol}), 
the circumference of the one-dimensional universe do change with time. 
I propose the following solution to this puzzle. 
The point is, that the operator $l$ and the Hamiltonian $H$ 
do \emph{not} commute in the mini-superspace approximation. 
Hence, we cannot determine both the circumference $l$ and the energy $E$ at the same time.
Given that the one-dimensional universe is a eigenstate of the Hamiltonian 
we cannot with certainty locate the one-dimensional universe on either of the figures showing the classical time evolution of the 
one-dimensional universe in the elliptic, parabolic or hyperbolic case. 
In order to observe the classical geometry emerge from one of the eigenstates in the mini-superspace approximation we should 
define a classical probability distributions of the circumference $l$ for each of the classical solutions and then verify, that the probability 
distribution obtained from a given eigenstate in the mini-superspace approximation 
reduces to the probability distribution of the corresponding classical solution in the classical limit.  

In the above we have discussed the Feigin-Fuchs modules. 
Let us now introduce the so-called Verma modules. 
Due to the fact, that the operator $V_{\alpha}$ is a primary operator, 
the corresponding state $\vert \alpha \rangle$ 
in the linear space of Hartle-Hawking wave functionals is a highest weight state with regard to the Virasoro algebra, that is
\[\label{highestvir}
L_{m} \vert \alpha \rangle = \bar{L}_{m} \vert \alpha \rangle = 0, \quad \textrm{for } m>0. 
\]
Acting on the highest weight state $\vert \alpha \rangle$ repeatedly with the Virasoro generators $L_{m}$, $m<0$, we obtain a 
highest weight representation of the Virasoro algebra known as a Verma module
\[
\label{Verma}
\mathcal{V}(c_{L},\Delta(\alpha)) 
\equiv 
\textrm{Span}\bigg\{ \prod_{i=1}^{n}L_{-k_{i}} \vert \alpha \rangle  \bigg\vert n \geq 0, k_{n} \geq k_{n-1} \geq \ldots \geq k_{1}> 0 \bigg\}.
\]
The operator $L_{0}-\Delta(\alpha)$ introduces a grading in the Verma module. 
$\mathcal{V}(c_{L},\Delta(\alpha))$. 
The corresponding degree is known as the level $N$. 
It is easily seen from the Virasoro algebra, 
that the level of the state $\prod_{i=1}^{n}L_{-k_{i}} \vert \alpha \rangle$ is given by 
\[
N=\sum_{i=1}^{n}k_{i}.
\]
Acting repeatedly with the raising operators $a_{m}$, $m<0$, 
on the state $\vert \alpha \rangle$ we may generate a Feigin-Fuchs module as in (\ref{feigin}). 
Since we may express the Virasoro generators $L_{m}$ in terms of the raising and lowering operators  
the Verma module $\mathcal{V}(c_{L},\Delta(\alpha))$ is embedded in this Feigin-Fuchs module. 
In the case when the Verma module defines an irreducible representation of the Virasoro algebra, the Verma module and the Feigin-Fuchs module are 
actually isomorphic.\cite{Lian}

\section{The dual state space }\label{dualspace}
In this section we will introduce the dual space 
of the Hartle-Hawking space of wave functionals. Our discussion is based upon \cite{T2,pol}. 
In order to proceed we first need to discuss the case $\mu = 0$, 
in which Liouville theory reduces to the linear dilaton theory. 
Let us consider the linear dilaton theory with a generic fiducial metric $\hat{g}$. 
In this case the linear dilaton action is given by (\ref{lioac}) with $\mu=0$. 
The normal ordering of a given primary operator in the linear dilaton theory is given by \cite{pol}
\[
V_{\alpha}(x)
= \:
:\!\exp(2\alpha\phi(x))\!: \: = \: 
1+\sum_{n=1}^{\infty} \frac{(2\alpha)^{n}}{n!} \left( \prod_{i=1}^{n} \lim_{x_{i} \to x} \right) \mathcal{F} \prod_{j=1}^{n}\phi(x_{j})
\] 
where
\[
\mathcal{F} \equiv 
\exp\left( \frac{1}{4} \int d^{2}x \: d^{2}x' \, 
\ln \left( d^{2}(x, x')\right) \,
\frac{\delta}{\delta \phi(x)} \frac{\delta}{\delta \phi(x')}  \right) 
\]
where $d(x, x')$ is the geodesic distance between $x$ and $x'$ measured in terms of the fiducial metric $\hat{g}$.
Under the infinitesimal transformation
\[\label{itr}
\phi(x) \to \phi'(x) = \phi(x) + \delta \phi_{0}
\] 
the normal ordered primary operator transforms according to
\st
:\!\exp(2\alpha\phi'(x))\!: 
& = & 
1+\sum_{n=1}^{\infty} \frac{(2\alpha)^{n}}{n!} \left( \prod_{i=1}^{n} \lim_{x_{i} \to x} \right) \mathcal{F}' \prod_{j=1}^{n}\phi'(x_{j})
\nonumber\\
& = &
1+ \sum_{n=1}^{\infty} \frac{(2\alpha)^{n}}{n!} \left( \prod_{i=1}^{n} \lim_{x_{i} \to x} \right) \mathcal{F} 
\left( \prod_{j=1}^{n} \phi(x_{j}) + \delta\phi_{0}\sum_{k=1}^{n} \prod_{j \neq k}^{n} \phi(x_{j}) \right)
\nonumber\\
& = &
:\!\exp(2\alpha\phi(x))\!:(1 + 2\alpha\delta\phi_{0})  
\label{sidst}
\en
Using the invariance of the linear dilaton measure under the transformation (\ref{itr})
and applying the Gauss-Bonnet theorem (\ref{Gaussbonnet}) 
we obtain the following relation concerning the n-point function on the sphere in the linear dilaton theory
\st
0 & = &
\int \mathcal{D}\phi' \exp(-S_{L}(\phi'))\prod_{i=1}^{n} V_{\alpha_{i}}(x_{i})    
\;
- \;
\int \mathcal{D}\phi \exp(-S_{L}(\phi))\prod_{i=1}^{n} V_{\alpha_{i}}(x_{i})     
\nonumber \\
&=&
\int \mathcal{D}\phi \exp(-S_{L}(\phi)-2Q\delta\!\phi_{0})\,\prod_{i=1}^{n}
V_{\alpha_{i}}(x_{i})
(1\!+\!2\alpha_{i}\delta\phi_{0})\,
\nonumber\\
&&-
\int \mathcal{D}\phi \exp(-S_{L}(\phi))\prod_{i=1}^{n} V_{\alpha_{i}}(x_{i})     
\nonumber \\
& = &
2\left(\sum_{i=1}^{n}\alpha_{i}-Q\right)\delta\phi_{0}
\int \mathcal{D}\phi \exp(-S_{L}(\phi))\prod_{i=1}^{n} V_{\alpha_{i}}(x_{i})     
\label{relconsphere}
\en
Thus, the n-point function on the sphere in the linear dilaton theory vanishes, unless
\[
\label{conditionsphere}
\sum_{i=1}^{n}\alpha_{i} = Q.
\]
In the case when this condition is satisfied 
we may actually determine the n-point function through a calculation almost identical to the calculation performed in \cite{pol} 
in case of the free bosonic string. 
In the calculation performed in \cite{pol} one determines the n-point function by expanding the fields in a 
complete set of eigenfunctions of the Laplace operator 
and thereafter performing the Gaussian integrals.       
In order to determine the n-point function in the linear dilaton theory 
we simply modify this calculation by including the linear term in the linear dilaton action in the background field $J$ introduced in 
the calculation performed in \cite{pol}. 
In conformal gauge we obtain the following n-point function on the sphere in the linear dilaton theory
\[\label{npointlindil}
\langle \prod_{i=1}^{n}
V_{\alpha_{i}}(z_{i},\bar{z}_{i}) \rangle_{\mu=0}
=\,
\delta(\sum_{i=1}^{n}\alpha_{i}-Q)\,
\prod_{i < j} \vert z_{i}-z_{j} \vert^{-4\alpha_{i}\alpha_{j}}
\]

Let us return to Liouville theory with $\mu > 0$ in conformal gauge.
As discussed previously, we may generate the state associated 
with a given wave functional in the Hartle-Hawking space of wave functionals 
by inserting the corresponding vertex operator at the origin in the complex $z$-plane. 
In the case of a primary state $\vert \alpha \rangle$ we may express this as
\[\label{correspondence}
\vert \alpha \rangle 
= 
\lim_{z,\bar{z} \to 0} V_{\alpha}(z,\bar{z}) \vert 0 \rangle
\]
where $\vert 0 \rangle$ is the so-called $SL(2,\mathbf{C})$ invariant state. 
The conformal descendants of the primary state $\vert \alpha \rangle$ are obtained 
by acting on $\vert \alpha \rangle$ with the Virasoro generators $L_{m}$ and $\bar{L}_{n}$, $m,n\!<\!0$, as mentioned 
previously in relation with the introduction of the Verma modules. 
Let us consider the two-point function on the sphere in Liouville theory. 
In \cite{francesco} it is shown, that the two point function of two primary operators inserted on the sphere vanishes 
in any given conformal field theory, 
unless the conformal dimensions of the two operators coincide.
In the case where the two primary operators are $V_{\alpha}$ and $V_{Q-\alpha}$, the conformal dimensions coincide  
and we may calculate the two-point function 
by performing a perturbative expansion in the cosmological constant $\mu$. 
From eqs. (\ref{lioaccom}), (\ref{conditionsphere}) and (\ref{npointlindil}) we obtain that
\st
&&\lim_{z,\bar{z} \to \infty} \vert z \vert^{4\Delta(\alpha')}
\langle V_{Q-\alpha'}(z,\bar{z}) V_{\alpha}(0,0) \rangle
\nonumber
\\
& = &
\lim_{z,\bar{z} \to \infty} \vert z \vert^{4\Delta(\alpha')}
\sum_{n=0}^{\infty} \frac{(-\mu)^{n}}{2^{n}n!}
\int \prod_{i=1}^{n} d^{2}\omega_{i}
\langle V_{Q-\alpha'}(z,\bar{z}) \prod_{i=1}^{n} V_{b}(\omega_{i},\bar{\omega}_{i}) V_{\alpha}(0,0) \rangle_{\mu = 0}
\nonumber\\
& = &
\lim_{z,\bar{z} \to \infty} \vert z \vert^{4\Delta(\alpha')}
\langle V_{Q-\alpha'}(z,\bar{z}) V_{\alpha}(0,0) \rangle_{\mu = 0}
\nonumber\\
& = &
\label{twopoint}
\delta(\alpha'-\alpha)
\en  
In the light of the above result 
it seems natural to define the state dual to the primary state $\vert \alpha \rangle$ as
\[\label{dualprimestate}
\langle \alpha \vert \equiv 
\lim_{z,\bar{z} \to \infty} 
\langle 0 \vert 
V_{Q-\alpha}(z,\bar{z})
\vert z \vert^{4\Delta(\alpha)}
\]
and to define the inner product as\footnote{Notice, this 
inner product is not related to the inner product of states in the mini superspace approximation.}    
\[\label{inner}
\langle \alpha \vert \beta \rangle 
\equiv 
\lim_{z,\bar{z} \to \infty} \vert z \vert^{4\Delta(\alpha)}
\langle V_{Q-\alpha}(z,\bar{z}) V_{\beta}(0,0) \rangle
\]
If we restrict our discussion to states satisfying the Seiberg bound (\ref{seibound}) we obtain 
from eq. (\ref{twopoint}) and the above definition of the inner product that
\[\label{normaldual}
\langle \alpha \vert \beta \rangle = \delta(\alpha-\beta)
\]
We define the state dual to a given descendant state as
\st
L_{-m_{1}}\ldots L_{-m_{k}}\bar{L}_{-n_{1}}\ldots \bar{L}_{-n_{l}}\vert \alpha \rangle
&\leftrightarrow&
\langle \alpha \vert 
\bar{L}_{n_{l}}\ldots \bar{L}_{n_{1}}
L_{m_{k}}\ldots L_{m_{1}}
\nonumber\\
&=&
\langle \alpha \vert  
\prod_{j=1}^{l} \, \int_{-\tilde{C}_{j}} \, \frac{d\bar{\omega}_{j}}{2\pi i} \, \bar{\omega}_{j}^{n_{j}+1} \, \bar{T}(\bar{\omega}) 
\,\prod_{i=1}^{k} \,\int_{C_{i}} \,\frac{d\omega_{i}}{2\pi i} \,\omega_{i}^{m_{i}+1} \,T(\omega) 
\nonumber\\
\label{dualdescendant}
\en
where $C_{i}$ and $\tilde{C}_{j}$ are circles 
of radii $r_{i}$ and $\tilde{r}_{j}$ respectively satisfying  $r_{1} < r_{2}< \ldots < r_{k}$ and $\tilde{r}_{1} < \tilde{r}_{2}< \ldots < \tilde{r}_{l}$. 
The contours $C_{i}$ and $\tilde{C}_{j}$ encircle the origin in the complex $z$-plane counterclockwise.  
Hence,
\[
L_{-m}^{\dag} = L_{m} \quad \textrm{and} \quad \bar{L}_{-m}^{\dag} = \bar{L}_{m}.
\]
The definition of the inner product given in (\ref{inner}) generalizes in an obvious way to include descendant states. 
Applying the transformation law 
\[\label{conftransprim}
V_{\alpha}(z',\bar{z}') = \vert \prt_{z}z'  \vert^{-2\Delta(\alpha)} V_{\alpha}(z,\bar{z})
\]
governing the spinless primary operator $V_{\alpha}$ under the conformal transformation $z \to z'(z)$ and 
eqs. (\ref{dualprimestate}) and (\ref{correspondence}) 
it is easily seen, that the conformal transformation 
\[\label{inverse}
z \to z' = \frac{1}{z}
\]
maps
\[\label{dualtodual}
\langle \alpha \vert 
\to 
\vert Q-\alpha \rangle.
\] 
Applying eq. (\ref{schwarz}) we obtain, 
that the conformal transformation (\ref{inverse}) maps 
\[\label{LtoL}
L_{m} = \int_{C} \frac{d\omega}{2\pi i} \omega^{m+1} T(\omega)
\; \to \; 
L_{-m} = \int_{\tilde{C}} \frac{d\omega}{2\pi i} \omega^{-m+1} T(\omega)
\]
where $C$ is a circle of radius $r$ encircling the origin in the complex $z$-plane counterclockwise, while 
$\tilde{C}$ is a circle of radius $\frac{1}{r}$ encircling the origin counterclockwise.
It follows from the two above equations, that 
\[
\langle \alpha \vert 
\prod_{j=1}^{l}\bar{L}_{n_{j}}
\prod_{i=1}^{k}L_{m_{i}}  
\,L_{0} 
=
\langle \alpha \vert 
\prod_{j=1}^{l}\bar{L}_{n_{j}}
\prod_{i=1}^{k}L_{m_{i}}
\left( \sum_{i=1}^{k}m_{j} + \Delta_{\alpha} \right).
\]
from which we obtain 
\st
&&
\left(
\sum_{i=1}^{k}m_{j} + \Delta_{\alpha}
\right)
\langle \tilde{\alpha} \vert 
\prod_{j=1}^{\tilde{l}}\bar{L}_{\tilde{n}_{j}}
\prod_{i=1}^{\tilde{k}}L_{\tilde{m}_{i}} 
\prod_{i=1}^{k}L_{-m_{i}} 
\prod_{j=1}^{l}\bar{L}_{-n_{j}} 
\vert \alpha \rangle 
\nonumber\\
&=&
\langle \tilde{\alpha} \vert 
\prod_{j=1}^{\tilde{l}}\bar{L}_{\tilde{n}_{j}}
\prod_{i=1}^{\tilde{k}}L_{\tilde{m}_{i}} 
\,L_{0}\,
\prod_{i=1}^{k}L_{-m_{i}} 
\prod_{j=1}^{l}\bar{L}_{-n_{j}} 
\vert\alpha \rangle 
\nonumber\\
& = &
\left(
\sum_{i=1}^{\tilde{k}}\tilde{m}_{j} + \Delta_{\tilde{\alpha}}
\right)
\langle \tilde{\alpha} \vert 
\prod_{j=1}^{\tilde{l}}\bar{L}_{\tilde{n}_{j}}
\prod_{i=1}^{\tilde{k}}L_{\tilde{m}_{i}} 
\prod_{i=1}^{k}L_{-m_{i}} 
\prod_{j=1}^{l}\bar{L}_{-n_{j}} 
\vert\alpha\rangle 
\label{sidsteafsnit}
\en
From the fact, that the two-point function of two descendant operators is proportional to the two-point function 
of the two corresponding primary operators, and from eq. (\ref{normaldual}), we conclude, 
that the inner product of two given states vanishes, unless 
they belong to the same representation of the two copies of the Virasoro algebra. 
Furthermore, we conclude from the above equation 
and the corresponding equation obtained from considering $\bar{L}_{0}$ instead of $L_{0}$, 
that the inner product of two given states belonging 
to same representation of the two copies of the Virasoro algebra vanishes, unless the two states belong to the same levels $N$ 
and $\bar{N}$. 
The introduction of the dual space of the Hartle-Hawking space of wave functionals 
allows us to introduce the concept of an orthonormal basis for the Hartle-Hawking space of wave functionals. 

\section{The structure of the Verma modules} \label{sectionVerma}
A generic Verma module is irreducible. 
However, for particular values of $\alpha$ the corresponding Verma module actually defines a reducible representation of the Virasoro algebra.  
The structure of these reducible Verma modules in terms of irreducible representations of the Virasoro algebra 
will play an important role, when we later on in this thesis turn our attention to the non-compact geometries 
in 2D euclidean Quantum Gravity. 
Let us therefore discuss the structure of these Verma modules in more detail. 
A given highest weight representation $\mathcal{V}(c_{L},\Delta(\alpha))$ is reducible, if and only if it contains another highest weight 
representation $\mathcal{V}(c_{L},\Delta(\alpha'))$, $\alpha' \neq \alpha$.
We denote this by
\[
\mathcal{V}(c_{L},\Delta(\alpha')) \to \mathcal{V}(c_{L},\Delta(\alpha)) 
\quad
\Leftrightarrow 
\quad
\mathcal{V}(c_{L},\Delta(\alpha')) \subset \mathcal{V}(c_{L},\Delta(\alpha)).
\]
The highest weight state $\vert \alpha' \rangle$ associated with 
the Verma module $\mathcal{V}(c_{L},\Delta(\alpha'))$ embedded in $\mathcal{V}(c_{L},\Delta(\alpha))$ 
is actually a null state, that is the state $\vert \alpha' \rangle$ is orthogonal to all states including 
itself. 
This is easily derived from the Virasoro algebra using the fact, that $\vert \alpha' \rangle$ is both a highest weight state and 
a descendant state of the highest weight state $\vert \alpha \rangle$.\cite{francesco} 
This property is inherited by all descendant states of $\vert \alpha' \rangle$, that is  
all states in $\mathcal{V}(c_{L},\Delta(\alpha'))$ are actually null states. 
We refer to a state, which is both a highest weight state and a null state, as a 
singular state.
In order to determine, whether or not a given Verma module is reducible, that is whether or not it contains null states, 
we consider the so-called Kac determinant. 
The set of states 
\[
\label{VermaN}
\bigg\{
\prod_{i=1}^{n} L_{-k_{i}} \vert \alpha \rangle \bigg\vert k_{n} \geq k_{n-1} \geq \ldots \geq k_{1}> 0, \; \sum_{i=1}^{n}k_{i} = N
\bigg\}
\]
defines a basis for the linear subspace of $\mathcal{V}(c_{L},\Delta(\alpha))$ at level $N$. 
The Kac determinant at level $N$ 
is the determinant of the Gram matrix of inner product of all states belonging to the above basis. 
If the Verma module $\mathcal{V}(c_{L},\Delta(\alpha))$ contains a null state at level $N$, then the Kac determinant vanishes as level $N$ 
due to the definition of a null state discussed in the above. 
Thus, we may determine whether or not a given Verma module is reducible by 
studying the Kac determinant. 
We may even determine 
the structure of a given reducible Verma module in terms of irreducible representations of the Virasoro algebra from the Kac determinant.
Let us describe the results of the analysis of the Kac determinant.  
A given Verma module $\mathcal{V}(c_{L},\Delta(\alpha))$ is reducible if and only if
\[\label{acon}
2\alpha=\frac{1}{b}(1-\tilde{r})+b(1-\tilde{s}) 
\equiv 
2 \alpha_{\tilde{r},\tilde{s}}
\]
that is
\[
\label{redu}
\Delta(\alpha) = \frac{1}{4}
\left(
Q^{2}-\left(\frac{\tilde{r}}{b}+\tilde{s}b\right)^{2}
\right)
\, ,
\]
where $\tilde{r}$ and $\tilde{s}$ are positive integers.\cite{SS} 
The structure of a given reducible Verma module actually depends on whether $b^{2}$ is rational or irrational.     
Let us start out by considering the generic case $b^{2}$ irrational.
In this case the Verma module $\mathcal{V}(c_{L},\Delta(\alpha))$ with $\alpha$ given by eq. (\ref{acon}) 
only contains one singular state at level $\tilde{r}\tilde{s}$ \cite{SS} 
and we may define an irreducible representation of the Virasoro algebra by
\[
\label{irr}
L(c_{L},\Delta(\alpha))
\equiv
\frac{\mathcal{V}(c_{L},\Delta(\alpha))}{\mathcal{V}(c_{L},\Delta(\alpha'))}
\]
where $\Delta(\alpha') = \Delta(\alpha)+\tilde{r}\tilde{s}$. 

Let us turn our attention to the case $b^{2}$ rational, that is
\[
b^{2}=\frac{p}{q} 
\] 
where $p<q$ and $\textrm{gcd}(p,q)=1$.\footnote{In this thesis we will not consider the case $b^{2}=1$.}
In this case we may express the condition (\ref{redu}) as
\[
\Delta(\alpha) = \frac{(p+q)^{2}-(\tilde{r}q+\tilde{s}p)^{2}}{4pq}.
\] 
The Verma module $\mathcal{V}(c_{L},\Delta(\alpha))$ is completely characterized by the conformal dimension 
$\Delta(\alpha)$ of the highest weight state from which it is   
constructed. 
Thus, a given reducible Verma module is uniquely labelled by the positive integer 
\[
n=\tilde{r}q+\tilde{s}p.
\] 
It is easily seen, that we may express this positive integer uniquely as 
\[
\label{labelling}
n = tpq + rq+sp \equiv n(t,r,s),
\] 
where 
\[\label{labellinglimit}
r,s,t \in \mathbf{Z}, \quad t \geq 0, \quad  1 \leq r \leq p \quad \textrm{and} \quad 1 \leq s \leq q.  
\]
Different sets of labels $(\tilde{r},\tilde{s})$ may actually correspond to the same value of $n$ in the case $b^{2}$ rational. 
The labelling of the reducible Verma modules by $(\tilde{r},\tilde{s})$ is not convenient in this case. 
Instead, we label a given reducible Verma module by the integers $r,s$ and $t$ introduced in the above equation.\cite{SS}

A given Verma module with $r=p$ or $s=q$ may contain several singular states. 
However, in both these cases the structure of the Verma module is given by \cite{SS,francesco}
\[
\label{structure2}
\mathcal{V}(c_{L},\Delta(\alpha)) \leftarrow \mathcal{V}(c_{L},\Delta(\alpha')) \leftarrow \mathcal{V}(c_{L},\Delta(\alpha'')) \leftarrow 
\ldots \leftarrow \mathcal{V}(c_{L},\Delta(\gamma))
\]  
in terms of embedded highest weight representations,
where $\Delta(\alpha') = \Delta(\alpha)+(t+1)ps $ in the case $r=p$ and $\Delta(\alpha') = \Delta(\alpha) + (t+1)qr$ in the case $s=q$. 
We may define an irreducible representation of the Virasoro algebra as in eq. (\ref{irr}). 

The structures of the reducible Verma modules with $r < p$ and $s < q$ are much more complicated   
and it is convenient to introduce an additional way of labelling these reducible Verma modules.
In the case $r < p$ and $s < q$ we can express the positive integer $n$ given in (\ref{labelling}) uniquely as
\[
n 
= 
\vert 2kpq + mq \pm np \vert  
\]
where 
\[
k\in\mathbf{Z}, \; 1 \leq m \leq p-1, \; 1 \leq n \leq q-1 \; \textrm{and} \; mq-np>0. 
\]
Hence, we may uniquely label a reducible Verma module with $r < p$ and $s < q$ by the set of integers $m,n$ and $k$. 
We define
\[\label{Acondim}
A_{m,n}(k) =  \frac{(p+q)^{2}-(2kpq + mq +np)^{2}}{4pq}
\]
and
\[\label{Bcondim}
B_{m,n}(k) =  \frac{(p+q)^{2}-(2kpq + mq -np)^{2}}{4pq}.
\] 
From the Kac determinant one may derive the so-called reflection property of reducible Verma modules \cite{Lian}
\[\label{reflection}
\mathcal{V}(c_{L},\Delta(\alpha')) \rightarrow \mathcal{V}(c_{L},\Delta(\alpha))
\quad
\Leftrightarrow
\quad
\mathcal{V}(26 - c_{L}, 1 - \Delta(\alpha')) \leftarrow  \mathcal{V}(26 - c_{L},1 - \Delta(\alpha))
\]
This property relates the structure of a given reducible Verma module in Liouville theory 
with the structure of a corresponding reducible Verma module in the so-called $(p,q)$ minimal model. 
The structure of the reducible Verma modules appearing in the minimal models have been studied in detail.\cite{francesco,Lian}
Using the reflection property we obtain the following diagram illustrating the structure of the reducible Verma modules labelled by $m,n$ and $k$  
in Liouville theory \cite{Lian}
\\ 
\\
\[
\label{structure1}
\begin{array}{cccccccc}
\mathcal{V}\left(c_{L},B_{m,n}(0)\right)  & \!\LARGE{\rightarrow}\! & 
\mathcal{V}\left(c_{L},A_{m,n}(-1)\right) & \!\LARGE{\rightarrow}\! &
\mathcal{V}\left(c_{L},B_{m,n}(-1)\right) & \!\LARGE{\rightarrow}\! &
\mathcal{V}\left(c_{L},A_{m,n}(-2)\right) & \dots \\
&\!\LARGE{\searrow}\!&&\!\darrow\!&&\!\darrow\! \\
&&
\mathcal{V}\left(c_{L},A_{m,n}(0)\right) & \!\LARGE{\rightarrow}\! & 
\mathcal{V}\left(c_{L},B_{m,n}(1)\right) & \!\LARGE{\rightarrow}\! &
\mathcal{V}\left(c_{L},A_{m,n}(1)\right) & \dots
\\
&&&&&&&
\end{array}
\]
Given a reducible Verma module $\mathcal{V}(c_{L},\Delta(\alpha))$ labelled by $m,n$ and $k$ in Liouville theory 
we simply locate the Verma module in the above diagram and read of the structure of the given Verma module in terms 
of the embedded highest weight representations. 
We may define an irreducible representation $L(c_{L},\Delta(\alpha))$ of the Virasoro algebra  
as the given Verma module $\mathcal{V}(c_{L},\Delta(\alpha))$ modulo the sum of the 
representations nested in $\mathcal{V}(c_{L},\Delta(\alpha))$.    
From the diagram (\ref{structure1}) we may define the distance $d$ between two Verma modules in the diagram 
as the number of arrows along the shortest path from one Verma module to the other. For instance, $d(B_{m,n}(0),A_{m,n}(-2))=3$.  

The structure of a given reducible Verma module $\mathcal{V}(c_{L},\Delta(\alpha))$ 
is encoded in the so-called Virasoro character of the corresponding irreducible 
representation $L(c_{L},\Delta(\alpha))$ 
constructed as described in the above. 
The Virasoro character of a given representation $\mathcal{R}$ of the Virasoro algebra is defined as
\[\label{defcharac}
\chi_{[\mathcal{R}]}(q) 
\equiv
\textrm{Tr}_{\mathcal{R}} \: q^{L_{0}-c/24} 
\] 
where $c$ is the central charge and $\vert q \vert < 1$.
We may easily calculate the Virasoro character of a given Verma module by 
determining the dimension of the subspace at level $N$ from (\ref{VermaN}) and inserting the result  
in the Virasoro character expressed as a sum over levels.
\st
\chi_{[\mathcal{V}(c_{L},\Delta(\alpha))]}(q)
&=&
\textrm{Tr}_{\mathcal{V}(c_{L},\Delta(\alpha))} \: q^{L_{0}-c_{L}/24} \nonumber\\
& = &
q^{-(Q/2-\alpha)^{2}-1/24}\:\textrm{Tr}_{\mathcal{V}(c_{L},\Delta(\alpha))} \: q^{L_{0}-\Delta(\alpha)} \nonumber \\
& = &
q^{-(Q/2-\alpha)^{2}-1/24} \sum_{N=0}^{\infty} p(N) q^{N} \nonumber\\
& = & 
\frac{q^{-(Q/2-\alpha)^{2}}}{\eta(q)}
\en       
where $p(N)$ is the number of decompositions of the positive integer $N$ into sums of positive integers without regard to order. 
Moreover, the Dedekind eta function is given by
\[
\eta(q) = q^{1/24}\prod_{n=1}^{\infty} (1-q^{n}).
\]
In the case of a Verma module associated with a normalizable primary operator $V_{Q/2+iP}$ 
we may express the above result as 
\[\label{virchaP}
\chi_{[\mathcal{V}(c_{L},\Delta(Q/2+iP))]}(q)
= 
\frac{1}{\eta(q)}q^{P^{2}}
\equiv 
\chi_{P}(q)\, .
\]
Applying this result we may obtain the Virasoro character of the irreducible representation $L(c_{L},\Delta(\alpha))$ 
defined in eq. (\ref{irr})
\[\label{virchab3irrational}
\chi_{[L(c_{L},\Delta(\alpha))]}(q)
=
\chi_{[\mathcal{V}(c_{L},\Delta(\alpha))]}(q) - \chi_{[\mathcal{V}(c_{L},\Delta(\alpha'))]}(q)
\]
In the case where the structure of the reducible Verma module is given by a sub-diagram of (\ref{structure1}) 
the Virasoro character is more complicated. The key observation, which makes the calculation of the Virasoro character possible in this case, 
is, that the intersection of two Verma modules appearing in the same column in (\ref{structure1}) is exactly given by the sum 
of the two Verma modules appearing in the following column on the left hand side. 
Due to this fact we may calculate the Virasoro character of the irreducible representation $L(c_{L},\Delta(\alpha))$ obtained from a given 
Verma module $\mathcal{V}(c_{L},\Delta(\alpha))$ appearing in the diagram (\ref{structure1}) as 
\[\label{inout}
\chi_{[L(c_{L},\Delta(\alpha))]}(q)
=
\sum_{i} (-1)^{d(\Delta(\alpha),\Delta(\alpha_{i}))}\chi_{[\mathcal{V}(c_{L},\Delta(\alpha_{i}))]}(q)
\]  
where we sum over all Verma modules $\mathcal{V}(c_{L},\Delta(\alpha_{i}))$ embedded in $\mathcal{V}(c_{L},\Delta(\alpha))$ 
including $\mathcal{V}(c_{L},\Delta(\alpha))$ and 
where the oscillating sign $(-1)^{d(\Delta(\alpha),\Delta(\alpha_{i}))}$ enforces the successive addition-subtraction of Virasoro characters 
as we move to the left column upon column in the diagram (\ref{structure1}).
Applying the labelling introduced in (\ref{labelling}) we may express the Virasoro character $\chi_{[t,r,s]}(q)$ of the irreducible 
representation obtained from a given reducible Verma module labelled by $r,s$ and $t$ as \cite{SS}
\[\label{virchab2rational}
\chi_{[t,r,s]}(q)
=
\frac{1}{\eta(q)} \sum_{j=0}^{t} 
\left(q^{-n(t-2j,r,s)^{2}/4pq} - q^{-n(t-2j,r,-s)^{2}/4pq } \right)
\]
where $n(t,r,s)$ is defined in eq. (\ref{labelling}).

As observed in the above the structure of Liouville theory changes drastically going from $b^{2}$ irrational to $b^{2}$ rational. 
This behaviour is intrinsic to Liouville theory and affects many important quantities in Liouville theory.
Later on we will consider Liouville theory coupled to the $(p,q)$-minimal models in order to study 2D euclidean Quantum Gravity 
in the case, when the matter sector consists of a $(p,q)$-minimal model. 
In this case the condition (\ref{c0}) implies, that $b^{2}$ is rational.
The author of this thesis is not aware of any conformal matter theory, which leads to $b^{2}$ being irrational. 
Many results in pure Liouville theory 
such as the three point function and the generic bulk-boundary structure constants 
are only known in the case $b^{2}$ irrational. 
This is off course unfortunate, since the main application of Liouville theory is 
2D euclidean Quantum Gravity.

\section{The operator product expansion}\label{sectionope}
In a given conformal field theory we may expand the product of two nearby operators in a series of single operators. 
This is known as the operator product expansion (OPE). 
In the case of two spinless primary operators the operator product expansion is 
given by\cite{francesco} 
\[\label{ope}
V_{\alpha}(z_{1},\bar{z}_{1}) V_{\sigma}(z_{2},\bar{z}_{2})
=
\sum_{\gamma}C_{\alpha\sigma}^{\gamma} \: 
\vert z_{12} \vert^{2(\Delta(\gamma) - \Delta(\alpha)-\Delta(\sigma))} \: 
\Psi_{\gamma}^{\alpha\sigma}(z_{12},\bar{z}_{12}\vert z_{2}, \bar{z}_{2})
\] 
where the operator $\Psi_{\gamma}^{\alpha\sigma}$  
is given by
\[\label{psi}
\Psi_{\gamma}^{\alpha\sigma}(z_{12},\bar{z}_{12}\vert z_{2}, \bar{z}_{2})
=
\sum_{\{k ,\bar{k}\}} 
\beta_{\alpha\sigma}^{\gamma,\{k\}}\bar{\beta}_{\alpha\sigma}^{\gamma,\{\bar{k}\}} \,
z_{12}^{K}\,\bar{z}_{12}^{\bar{K}} \,
\prod_{i}L_{-k_{i}} \prod_{j}\bar{L}_{-\bar{k}_{j}} V_{\gamma}(z_{2},\bar{z}_{2})
\]
where we sum over all finite sets of positive integers $\{k_{i}\}$ and $\{\bar{k_{j}}\}$ ordered 
such that $k_{i} \leq k_{i+1}$ and $\bar{k}_{j} \leq \bar{k}_{j+1}$.
Moreover, we define
\[
K=\sum_{i}k_{i} \quad \textrm{and} \quad \bar{K}=\sum_{j}\bar{k}_{j}
\]
and we apply the definitions
\[\label{vironoperator}
L_{-m}\:\hat{A}(z_{0},\bar{z}_{0}) \equiv \oint_{C_{z_{0}}} \frac{dz}{2\pi i} \frac{1}{(z-z_{0})^{m-1}}T(z)\hat{A}(z_{0},\bar{z}_{0}),
\]
and
\[
\bar{L}_{-n}\:\hat{A}(z_{0},\bar{z}_{0}) 
\equiv 
- \oint_{C_{z_{0}}} \frac{d\bar{z}}{2\pi i} \frac{1}{(\bar{z}-\bar{z}_{0})^{n-1}}\bar{T}(\bar{z})\hat{A}(z_{0},\bar{z}_{0})  
\]
where $\hat{A}$ is a given vertex operator and where $C_{z_{0}}$ is a closed contour enclosing $z_{0}$ counterclockwise in the complex $z$-plane.
Finally, the constants $\beta_{\alpha\sigma}^{\gamma,\{k\}}$ and $\bar{\beta}_{\alpha\sigma}^{\gamma,\{\bar{k}\}}$ are determined completely by the 
conformal symmetry of the theory.\cite{francesco} 
In the case of Liouville theory the sum over conformal families appearing in the OPE is actually an integral. 
If we consider the OPE of two local operators in Liouville theory we only integrate over the conformal families of local operators.  

In the next chapter we will apply the OPE (\ref{ope}) in the case $V_{\sigma}=V_{-b/2}$. 
Let us therefore discuss the OPE in this case.
In order to proceed we start out by considering the Verma module generated 
from the highest weight state $\vert -\frac{b}{2}  \rangle$ belonging to the 
Hartle-Hawking space of wave-functionals. From (\ref{acon}) and the following discussion we obtain, 
that this Verma module is reducible and that there exists a singular state $\vert \chi\rangle$ belonging to this Verma module at level 2, 
that is  
\[\label{singular}
L_{m} \vert \chi \rangle = 0, \quad \textrm{for }m>0, 
\]
where
\[
\vert \chi \rangle 
\propto
\left( L_{-1}^{2} + \beta  L_{-2} \right)   
\vert\! -\!b/2  \rangle.
\]   
Applying the fact, that $\vert -b/2 \rangle$ is a highest weight state, and the Virasoro algebra (\ref{viralgebra}), we obtain that  
\[
L_{1} \vert \chi \rangle
= 
\left( [L_{1},L_{-1}^{2}] + \beta [L_{1},L_{-2}] \right) 
\vert \!-\!b/2  \rangle
 = 
\left( 4\Delta(-b/2) + 2 + 3\beta \right) L_{-1} \vert\! -\!b/2  \rangle 
\]
Since $L_{-1} \vert -\frac{b}{2}  \rangle$ is not a null state, it follows from 
eq. (\ref{singular}) and the above equation that\footnote{Acting on 
$\vert \chi \rangle$ with any other Virasoro generator $L_{m}$, $m>0$, 
does not lead to any additional constraints.}
\[
4\Delta(-b/2) + 2 + 3\beta = 0 
\quad \Rightarrow \quad
\beta = b^{2}.
\]
As discussed in section \ref{dualspace} the highest weight state $\vert -\frac{b}{2}  \rangle$ is given by
\[
\vert \!-\!b/2 \rangle
= 
\lim_{z,\bar{z} \to 0}V_{-b/2}(z,\bar{z}) \vert 0 \rangle
\]
where $\vert 0 \rangle$ is the $SL(2,\mathbf{C})$-invariant state in Liouville theory. 
From the above discussion we obtain the following condition on the vertex operator $V_{-b/2}$ corresponding to 
the highest weight state $\vert -\frac{b}{2}  \rangle$
\[\label{conb2}
\left( L_{-1}^{2} + b^{2}  L_{-2} \right) V_{-b/2}(z,\bar{z}) = 0,
\]
where the action of the Virasoro generators on the primary state $\vert -b/2 \rangle$ is defined in eq. (\ref{vironoperator}). 
Let us consider the n-point function on the sphere. 
The above condition implies that
\[
\bigg\langle
\left[\left(L_{-1}^{2}+b^{2}L_{-2}\right)V_{-b/2}(z_{1},\bar{z}_{1})\right]\prod_{i=2}^{n}V_{\alpha_{i}}(z_{i},\bar{z}_{i})
\bigg\rangle 
= 
0.
\]
From eq. (\ref{vironoperator}) and the standard OPE of the energy momentum tensor 
and a given vertex operator given in \cite{pol}, we obtain 
\st
\bigg\langle
\left[L_{-1}^{2} V_{-b/2}(z_{1},\bar{z}_{1})\right] 
\prod_{i=2}^{n}V_{\alpha_{i}}(z_{i},\bar{z}_{i})
\bigg\rangle 
& = & 
\oint_{C_{z_{1}}} \frac{d\omega}{2\pi i}
\bigg\langle
T(\omega)
\left[ L_{-1} V_{-b/2}(z_{1},\bar{z}_{1})\right] 
\prod_{i=2}^{n}V_{\alpha_{i}}(z_{i},\bar{z}_{i})
\bigg\rangle
\nonumber\\
& = &
\oint_{C_{z_{1}}} \frac{d\omega}{2\pi i} \frac{1}{\omega-z_{1}}
\bigg\langle
\prt_{z_{1}}
\left[ L_{-1} V_{-b/2}(z_{1},\bar{z}_{1})\right] 
\prod_{i=2}^{n}V_{\alpha_{i}}(z_{i},\bar{z}_{i})
\bigg\rangle
\nonumber\\
& = &
\prt_{z_{1}}
\bigg\langle
\left[ L_{-1} V_{-b/2}(z_{1},\bar{z}_{1})\right] 
\prod_{i=2}^{n}V_{\alpha_{i}}(z_{i},\bar{z}_{i})
\bigg\rangle
\nonumber\\
& = &
\prt_{z_{1}}
\oint_{C_{z_{1}}} \frac{d\omega}{2\pi i} 
\bigg\langle
T(\omega)
V_{-b/2}(z_{1},\bar{z}_{1}) 
\prod_{i=2}^{n}V_{\alpha_{i}}(z_{i},\bar{z}_{i})
\bigg\rangle
\nonumber\\
& = &
\prt_{z_{1}}
\oint_{C_{z_{1}}} \frac{d\omega}{2\pi i} \frac{1}{\omega-z_{1}} 
\bigg\langle
\prt_{z_{1}} V_{-b/2}(z_{1},\bar{z}_{1}) 
\prod_{i=2}^{n}V_{\alpha_{i}}(z_{i},\bar{z}_{i})
\bigg\rangle
\nonumber\\
& = &
\prt_{z_{1}}^{2}
\bigg\langle 
V_{-b/2}(z_{1},\bar{z}_{1})
\prod_{i=2}^{n}V_{\alpha_{i}}(z_{i},\bar{z}_{i})
\bigg\rangle
\label{denforste}
\en
where $C_{z_{1}}$ is a small closed contour encircling the $z_{1}$ in the complex $z$-plane counterclockwise.
From eq. (\ref{vironoperator}), Cauchy's theorem and the OPE of the energy momentum tensor 
and a given primary vertex operator given in \cite{pol}, we obtain 
\st
&&
\bigg\langle 
[L_{-2}V_{-b/2}(z_{1},\bar{z}_{1})]
\prod_{i=2}^{n} V_{\alpha_{i}}(z_{i},\bar{z}_{i})
\bigg\rangle
\nonumber\\
&=&
\int_{C_{z_{1}}} \frac{d\omega}{2\pi i} \frac{1}{\omega-z_{1}}
\bigg\langle 
T(\omega)V_{-b/2}(z_{1},\bar{z}_{1})
\prod_{i=2}^{n}V_{\alpha_{i}}(z_{i},\bar{z}_{i})
\bigg\rangle
\nonumber\\
&=&
-\sum_{j=2}^{n}
\int_{C_{z_{i}}} 
\frac{d\omega}{2\pi i} \frac{1}{\omega-z_{1}}
\bigg\langle 
T(\omega) V_{\alpha_{j}}(z_{j},\bar{z}_{j})
V_{-b/2}(z_{1},\bar{z}_{1})
\prod_{i \neq j}V_{\alpha_{i}}(z_{i},\bar{z}_{i}) 
\bigg\rangle
\nonumber\\
& = &
-\sum_{j=2}^{n}
\int_{C_{z_{i}}} 
\frac{d\omega}{2\pi i} \left\{\frac{1}{z_{j}-z_{1}}-\frac{1}{(z_{j}-z_{1})^{2}}(\omega-z_{j})+\ldots\right\}
\nonumber\\
&&
\phantom{mapmapmap}
\times
\left\{
\frac{\Delta(\alpha)}{(\omega-z_{j})^{2}}+\frac{1}{\omega-z_{j}}\prt_{z_{j}} + \ldots
\right\}
\bigg\langle 
V_{\alpha_{j}}(z_{j},\bar{z}_{j})
V_{-b/2}(z_{1},\bar{z}_{1})
\prod_{i \neq j}V_{\alpha_{i}}(z_{i},\bar{z}_{i}) 
\bigg\rangle
\nonumber\\
& = &
\sum_{j=1}^{n}\left(
\frac{\Delta(\alpha)}{(z_{j}-z_{1})^{2}}-\frac{1}{z_{j}-z_{1}}\prt_{z_{j}}
\right)
\bigg\langle 
V_{-b/2}(z_{1},\bar{z}_{1})
\prod_{i=2}^{n} V_{\alpha_{i}}(z_{i},\bar{z}_{i})
\bigg\rangle
\label{denanden}
\en
Hence, we obtain the equation 
\[
\left( 
\frac{1}{b^{2}}\prt_{z_{1}}^{2}
+
\sum_{j=1}^{n}\left(
\frac{\Delta(\alpha)}{(z_{j}-z_{1})^{2}}-\frac{1}{z_{j}-z_{1}}\prt_{z_{j}}
\right)
\right)
\bigg\langle 
V_{-b/2}(z_{1},\bar{z}_{1})
\prod_{i=2}^{n} V_{\alpha_{i}}(z_{i},\bar{z}_{i})
\bigg\rangle 
= 
0.
\]
Let us insert the OPE (\ref{ope}) of the two primary operators $V_{-b/2}$ and $V_{\alpha}$ into the above equation.
Evaluating the contribution from each conformal family to lowest order in $z_{12}$ we obtain the condition, 
that $C_{-\frac{b}{2}\,\alpha_{2}}^{\gamma} \neq 0$, only if 
\st
&
\frac{1}{b^{2}}
\bigg\{
\Delta(\gamma)-\Delta(-b/2)-\Delta(\alpha_{2})
\bigg\}
\bigg\{
\Delta(\gamma)-\Delta(-b/2)-\Delta(\alpha_{2})-1
\bigg\}
\phantom{mapmapmapmap}&
\nonumber\\
&
\phantom{mapmapmapmapmapmap}
+
\Delta(\alpha_{2})
-
\bigg\{
\Delta(\gamma)-\Delta(-b/2)-\Delta(\alpha_{2})
\bigg\}=0&
\nonumber\\
&&
\en
We solve the above equation in terms of $\Delta(\gamma)$ by first solving the equation in terms of 
$\Delta(\gamma)-\Delta(-b/2)-\Delta(\alpha_{2})$. From this calculation we obtain the condition, 
that the only conformal families, which appear in the OPE of $V_{-b/2}$ and $V_{\alpha_{2}}$, 
are the conformal families corresponding to the primary operators $V_{\alpha_{2}-b/2}$ and $V_{\alpha_{2}+b/2}$.
In the following we omit the subscript in $\alpha_{2}$ for convenience.
Applying the OPE (\ref{ope}), eqs. (\ref{dualprimestate}) and (\ref{normaldual}) 
and the condition discussed in the paragraph following eq. (\ref{sidsteafsnit}), we realize, that we may 
express the structure constants appearing in the OPE of the primary operators $V_{-b/2}$ and $V_{\alpha}$ as 
\[\label{opestructurecon}
C_{-\frac{b}{2}\,\alpha}^{\alpha'-b/2}
=
\lim_{z,\bar{z}\to\infty}
\vert z \vert^{-4\Delta(\alpha'-b/2)}
\bigg\langle
V_{\alpha}(0) V_{-b/2}(1)V_{Q-\alpha'+b/2}(z,\bar{z})
\bigg\rangle
\]
Performing a perturbative expansion in $\mu$ and applying the condition (\ref{conditionsphere}) and eq. (\ref{npointlindil}) 
we obtain that 
\st
C_{-\frac{b}{2}\,\alpha}^{\alpha'-b/2}
& = &
\lim_{z,\bar{z}\to\infty}
\vert z \vert^{-4\Delta(\alpha'-b/2)}
\bigg\langle
V_{\alpha}(0) V_{-b/2}(1)V_{Q-\alpha'+b/2}(z,\bar{z})
\bigg\rangle
\nonumber\\
& = & 
\lim_{z,\bar{z}\to\infty}
\vert z \vert^{-4\Delta(\alpha'-b/2)}
\sum_{n=0}^{\infty} \frac{(-\mu)^{n}}{2^{n}n!} 
\nonumber\\
&&
\phantom{map}
\times
\int \prod_{i=1}^{n} d^{2}\omega_{i}
\bigg\langle
V_{\alpha}(0) V_{-b/2}(1)\prod_{i=1}^{n}V_{b}(\omega_{i},\bar{\omega}_{i})V_{Q-\alpha'+b/2}(z,\bar{z})
\bigg\rangle_{\mu=0}
\nonumber\\
& = & 
\lim_{z,\bar{z}\to\infty}
\vert z \vert^{-4\Delta(\alpha'-b/2)}
\bigg\langle
V_{\alpha}(0) V_{-b/2}(1)V_{Q-\alpha'+b/2}(z,\bar{z})
\bigg\rangle_{\mu=0}
\nonumber\\
& = & 
\delta(\alpha'-\alpha)
\label{bulkstructure1}
\en
In the above calculation we assume, that $\alpha'$ is in a small neighborhood of $\alpha$. 
We apply this assumption in the third line, where we pick out the only term in the perturbative expansion in $\mu$, 
which is different from zero. 
Hence, the final expression is only valid for $\alpha'$ belonging to a small neighborhood of $\alpha$. 
Setting $\alpha' = \tilde{\alpha} + b$ in eq. (\ref{opestructurecon}) we obtain
\st
C_{-\frac{b}{2}\,\alpha}^{\tilde{\alpha}+b/2}
& = &
\lim_{z,\bar{z}\to\infty}
\vert z \vert^{-4\Delta(\tilde{\alpha}+b/2)}
\bigg\langle
V_{\alpha}(0) V_{-b/2}(1)V_{Q-\tilde{\alpha}-b/2}(z,\bar{z})
\bigg\rangle
\nonumber\\
& = & 
\lim_{z,\bar{z}\to\infty}
\vert z \vert^{-4\Delta(\tilde{\alpha}+b/2)}
\sum_{n=0}^{\infty} \frac{(-\mu)^{n}}{2^{n}n!} 
\nonumber\\
&&
\phantom{map}
\times
\int \prod_{i=1}^{n} d^{2}\omega_{i}
\bigg\langle
V_{\alpha}(0) V_{-b/2}(1)\prod_{i=1}^{n}V_{b}(\omega_{i},\bar{\omega}_{i})V_{Q-\tilde{\alpha}-b/2}(z,\bar{z})
\bigg\rangle_{\mu=0}
\nonumber\\
& = & 
-\frac{\mu}{2}
\delta(\tilde{\alpha}-\alpha)
\int d^{2}\omega
\,\vert \omega \vert^{-4\alpha b}\,\vert 1-\omega \vert^{2b^{2}}
\nonumber\\
& = &
-\pi\mu
\frac{\gamma(2\alpha b-1-b^{2})}{\gamma(2\alpha b)\gamma(-b^{2})}
\,\delta(\tilde{\alpha}-\alpha)
\nonumber\\
& \equiv & 
C_{-}
\,\delta(\tilde{\alpha}-\alpha)
\label{bulkstructure2}
\en
where
\[
\gamma(x) 
\equiv 
\frac{\Gamma(x)}{\Gamma(1-x)}
\]
and where the integral appearing in the second last expression is evaluated in appendix B in \cite{DF}.
The above calculation is only valid for $\tilde{\alpha}$ belonging to a small neighborhood of $\alpha$. 
As in the previous calculation we have applied this assumption in the third line, where we pick out the only term different from zero 
in the perturbative expansion in $\mu$. 
Notice however, in the two above calculation we have determined the structure constant $C_{-\frac{b}{2}\,\alpha_{2}}^{\gamma}$  
precisely in the two cases $\gamma=\alpha_{2}-b/2$ and $\gamma=\alpha_{2}+b/2$, where it is different from zero. 

\chapter{Boundary conditions in Liouville theory} \label{chapter3}
\section{The boundary state formalism} \label{boundstates}
So far we have discussed conformal field theory defined on a closed world-sheet. 
Let us now discuss conformal field theory in conformal gauge defined on a world-sheet with the topology of the disk. 
The following discussion is not specific to Liouville theory. 
The discussion concerns a generic 2D conformal field theory. 
According to the Riemann mapping theorem we may map 
any given simply connected proper subset of the complex plane to the upper half plane by a conformal transformation. 
Hence, it is sufficient to consider a generic conformal field theory defined in the upper half plane.
In this case we may impose a boundary condition on the real axis. 
In the following we want to deduce a necessary condition ensuring 
that the boundary condition does not break conformal invariance.  
Now, any given infinitesimal conformal transformation 
\[\label{contradisk}
z \to z' = z + \epsilon(z)\, ,
\] 
which maps the upper half plane to the upper half plane, satisfies the condition
\[\label{conboundary}
\epsilon(x) = \bar{\epsilon}(x), \quad \textrm{for } x \in \mathbf{R}
\]
ensuring, that the real axis is mapped to the real axis.
A quantum conformal field theory is defined by the condition, 
that the product of the weight $\exp(-S[\phi])$, where $S$ is the action, and the measure $\mathcal{D}\phi$ is invariant 
under a given conformal transformation, that is 
\[\label{defcontheo}
\e^{-S[\phi']}\mathcal{D}\phi'
=
\e^{-S[\phi]}\mathcal{D}\phi \,.
\] 
Let us assume, that the field $\phi$  
transforms according to
\[\label{conphi}
\phi(z,\bar{z}) \to 
\phi'(z,\bar{z}) = \phi(z,\bar{z}) + \delta\phi(z,\bar{z}) 
\]
under a given infinitesimal conformal transformation. 
Moreover, let us consider a slight modification of the above transformation of the field $\phi$
\[\label{almostcon}
\phi(z,\bar{z}) \to 
\phi'_{D}(z,\bar{z}) = \phi(z,\bar{z}) + \rho(z,\bar{z})\delta\phi(z,\bar{z}) \, ,
\]
where $\rho(z,\bar{z})=1$ inside the region $D$ and $\rho(z,\bar{z})=0$ outside $D$, and 
let us consider the case, 
where the region $D$ lies entirely in the bulk of upper half plane. 
Since the field $\phi$ close to the boundary is invariant under this transformation, 
the arguments given in section 2.3 in \cite{pol} in the case of a closed world-sheet are also valid in our case. 
From these arguments we obtain   
\st
\mathcal{D}\phi_{D}' \exp(-S[\phi'_{D}])
& = &
\mathcal{D}\phi \exp(-S[\phi])
\left[
1+\frac{i}{2\pi} \int d^{2}x \,j^{a}(x) \,\prt_{a} \rho(x) 
\right]
\nonumber\\
& = &
\mathcal{D}\phi \exp(-S[\phi])
\left[
1+\frac{1}{2\pi i} \int_{D} d^{2}x \, \prt_{a} j^{a}(x)  
\right]
\nonumber\\
& = &
\mathcal{D}\phi \exp(-S[\phi])
\left[
1+\frac{1}{2\pi i}\oint_{\prt D} \left( dz \, \epsilon(z) \,T(z) -d\bar{z} \, \bar{\epsilon}(\bar{z}) \, \bar{T}(\bar{z})      \right)
\right]
\nonumber\\
\label{almostdefcontheo}
\en
where we have used the divergence theorem and the fact, 
that the conserved current associated with the conformal transformation (\ref{contradisk}) is given by \cite{pol}
\[
j(z) = i\epsilon(z)T(z), \quad \tilde{j}(\bar{z}) = i \bar{\epsilon}(\bar{z}) \bar{T}(\bar{z}) 
\] 
in the case of a closed world-sheet. 
Let us gradually approach the limit, in which the region $D$ becomes the entire upper half plane. 
In this limit the transformation (\ref{almostcon}) reduces to the transformation law (\ref{conphi}) governing $\phi$ under 
the given conformal transformation. 
From (\ref{defcontheo}) and (\ref{almostdefcontheo}) we obtain the following necessary condition in the limit, where $D$ becomes 
the entire upper half plane, 
\[
\frac{1}{2\pi i} \int_{-\infty}^{\infty} dx \:\epsilon(x) \left(T(x) - \bar{T}(x)\right) = 0\, , 
\]
ensuring that, the the imposed boundary condition does not break conformal invariance defined by eq. (\ref{defcontheo}). 
This condition is satisfied, if
\[\label{Tcondition}
T(x) = \bar{T}(x) \quad \textrm{for }x \in \mathbf{R}.  
\]
Integrating over the configuration space of the field $\phi$ the above equation turns into 
a condition concerning the energy momentum tensor viewed as an operator. 
Notice, this condition implies, 
that we may extend the definition of the holomorphic energy momentum tensor $T(z)$ to the entire complex plane.
\[\label{doubling}
T(z) \equiv \bar{T}(z) \quad \textrm{for }  \textrm{Im}(z) < 0.
\]
Performing a Laurent expansion of the holomorphic energy momentum tensor $T(z)$ we obtain one set of Virasoro generators satisfying the Virasoro algebra. 
The anti-holomorphic energy momentum tensor $\bar{T}(\bar{z})$ does not give rise to an additional 
independent set of Virasoro generators due to the relation (\ref{doubling}). 

Applying the transformation  
\[
\omega = i\, \textrm{Ln} \, (z) + \pi
\]
we may map the upper half plane to the infinite strip parametrized by $\omega = \sigma + it$, 
where $\sigma \in [0,\pi]$ and $t \in \mathbf{R}$.
As in the case of the infinite cylinder discussed in section \ref{semiclassical}, 
we may view time evolving along the infinite strip 
and introduce the concept of an open string state quantising the field on the finite line element 
$\sigma \in [0,\pi]$.\footnote{A given open string state is not necessarily a physical gauge-invariant state.}  
This is the so-called open string picture.  
Since we only have one set of independent Virasoro generators, 
the spectrum of open string states only transforms under one copy of the Virasoro algebra as opposed to the spectrum of closed string states, 
which transforms under two independent copies of the Virasoro algebra. 
In general we do not have to impose the same boundary condition on each of the two sides of the infinite strip.  
The spectrum of open string states depends crucially on the boundary conditions imposed on the opposite sides of the infinite strip. 
Applying the transformation law (\ref{schwarz}) the condition (\ref{Tcondition}) becomes
\[
T_{\sigma t}(\sigma,t) = 0, \quad \textrm{for } \sigma=0,\pi
\]
on the infinite strip. 
From this condition we realize, 
that conformal invariance of a given boundary condition is equivalent to momentum conservation, that is no momentum 
flows across the boundaries of the infinite strip. 

Due to the fact, that the signature of the world-sheet is euclidean, 
we may view $\sigma$ as the time direction and $t$ as the spatial direction instead of considering time evolving along the infinite strip. 
Moreover, let us impose the periodicity condition
\[\label{perio}
t \sim t + 2\pi.
\]  
transforming the infinite strip into a finite cylinder of circumference $2\pi$.
In this so-called closed string picture the boundary conditions become boundary states $\vert B \rangle$, 
which we may expand in terms of states belonging to the Hilbert space of closed string states. 
Applying the map
\[
z' = \exp(-i\omega'), \quad \omega' = t + i\sigma
\]
we may map the cylinder to the annulus in the complex $z'$-plane with time evolving radially outwards. 
Applying the transformation law (\ref{schwarz}) we may translate the condition (\ref{Tcondition}) into a condition 
satisfied by the energy momentum tensor on both boundaries of the annulus in the complex $z'$-plane
\[
z'^{2} T(z') = \bar{z}'^{2} \bar{T}(\bar{z}')\, .  
\]  
Expanding the energy momentum tensor in terms of the Virasoro generators we obtain the 
following condition satisfied by the boundary state $\vert B \rangle$ on the inner most boundary of the annulus
\[\label{Ishibashicon}
\left( L_{m} - \bar{L}_{-m} \right) \vert B \rangle = 0.
\] 
We may express the Hilbert space of closed string states as a direct sum over 
tensor products $L(c,\Delta) \otimes \tilde{L}(c,\tilde{\Delta})$ of irreducible representations 
of the Virasoro algebra.   
(See for instance eq. (\ref{Hilbertlio})). 
Moreover, we may express the boundary state $\vert B \rangle$ as a sum over 
contributions from each of these tensor products of irreducible representation of the Virasoro algebra.
The above linear equation (\ref{Ishibashicon}) imposes an independent condition on the contribution  
from each of these tensor products.  
Given the tensor product $L(c,\Delta) \otimes \tilde{L}(c,\bar{\tilde{\Delta}})$ 
we may construct a solution to the above equation (\ref{Ishibashicon}) belonging to this tensor product, 
if and only if $\Delta = \tilde{\Delta}$, that is if and only if the left and the right representations of the Virasoro algebra are isomorphic. 
If this is the case the solution is unique and is given by \cite{Ishibashi,Zuber,OA}
\[\label{Ishibashistate}
\vert\vert \Delta \rangle \rangle 
=
\sum_{n} \vert \Delta, n \rangle \otimes \vert \Delta, n \rangle   
\]
where
\[
\bigg\{
\vert \Delta, n \rangle
\bigg\}_{n}
\]
is an orthonormal basis of the irreducible representation 
$L(c,\Delta)$.\footnote{In the case of a non-unitary representation we cannot choose an orthonormal basis 
due to the existence of negative norm states. 
In this case we normalize the vectors in the above basis such, 
that the norm of any given vector belonging to the basis is either $1$ or $-1$. 
In the following we also refer to such a basis as an orthonormal basis.} 
In the case of Liouville theory 
the concept of orthonormality is defined  
with respect to the chiral part of inner product introduced in section \ref{dualspace}.  
In the case of a generic conformal field theory we may introduce a similar inner product. 
Furthermore, we may choose an orthonormal basis of eigenstates of $L_{0}$ and $\bar{L}_{0}$.   
The state defined in the above equation  
is known as an Ishibashi state. 
We conclude, that we may express a given boundary state as 
\[\label{genboundstate}
\vert B \rangle 
=
\sum_{\Delta} \Psi_{\!\Delta} \vert\vert \Delta \rangle\rangle
\]
where $\Psi_{\!\Delta}$ is known as the wave function of the boundary state.

As shown in appendix \ref{apptwopoint} in the case of Liouville theory 
the one-point function in the upper half plane 
of a given spinless primary operator $\mathcal{O}_{\!\Delta}$ of conformal dimension $\Delta$ is given by 
\[
\langle
\mathcal{O}_{\!\Delta}(z,\bar{z}) 
\rangle
= \frac{U_{\!\Delta}}{\vert z-\bar{z} \vert^{2\Delta}} \,.
\]
Applying the transformation (\ref{uppertodisk}) mapping the upper half plane to the unit disk and the transformation law (\ref{conftransprim}) 
we obtain the one-point function on the disk
\[
\langle
\mathcal{O}_{\!\Delta}(\omega,\bar{\omega}) 
\rangle 
= 
\frac{U_{\!\Delta}}{\vert 1-\omega\bar{\omega} \vert^{2\Delta}} \, .
\]
Evaluating the given primary operator at the center of the unit disk we obtain from eqs. (\ref{Ishibashistate}) and (\ref{genboundstate}) 
\[\label{onepointboundarywave}
U_{\!\Delta} 
= 
\langle
\mathcal{O}_{\!\Delta}(0) 
\rangle 
= 
\sum_{\Delta'} \Psi^{c}_{\!\Delta'} \langle\langle \Delta' \vert\vert \Delta \rangle
=
\Psi^{c}_{\!\Delta}\, ,
\]
where $\Psi^{c}$ is the wave function conjugate to $\Psi$ and where $\vert \Delta \rangle$ is the highest weight state 
corresponding to the given primary operator 
$\mathcal{O}_{\!\Delta}$.\footnote{Notice, we assume, that the primary operators are normalized such that 
$\langle \Delta \vert \Delta' \rangle = \delta_{\Delta,\Delta'}$. This in indeed the case with regard to the $(p,q)$ minimal model, 
which we will consider later on in this thesis. 
In Liouville theory the sum appearing in (\ref{genboundstate}) is actually an integral and the primary states are delta-function normalized. 
(See eq. (\ref{normaldual})). 
With this in mind the above discussion also applies to boundary states in Liouville theory.} 
Hence, the wave function of a given boundary state is determined by the one-point function. 
In the case of Liouville theory we use the notation 
\[\label{Liouvilleconwave}
U(\alpha) 
\equiv U_{\alpha(Q-\alpha)} \qquad \textrm{and} \qquad  
\Psi(P) \equiv \Psi^{c}_{Q^{2}/2+P^{2}} \,.
\] 

Let us return to the infinite strip.  
However, let us consider the generic case, in which the width of the strip is $\pi\tau$.
In the open string picture the Hamilton operator is given by \cite{Cardy}
\[\label{openham}
H_{o} = \frac{1}{2\pi} \int_{0}^{\pi\tau} d\sigma \, T_{tt}(\sigma,t) = \frac{1}{\tau}\left( L_{0} - \frac{c}{24} \right)\, .
\]
In deriving the last identity we have applied the map 
\[
z=-\exp\left(-\frac{i}{\tau}\omega\right), \quad \omega= \sigma + it
\]
mapping the infinite strip to the upper half plane and eqs. (\ref{schwarz}), (\ref{doubling}) and (\ref{Virgen}). 
Let us once again impose the periodicity condition (\ref{perio}) transforming the infinite strip into a cylinder of circumference $2\pi$ 
and length $\pi\tau$. 
Due to the periodic boundary condition the cylinder amplitude $\mathcal{Z}$ evaluated in the open string picture is given by
\st
\mathcal{Z} 
= 
\textrm{Tr}\exp\left(-2\pi H_{o}\right)
& = & \textrm{Tr}\exp\left(-\frac{2\pi}{\tau}\left(L_{0} - \frac{c}{24}\right)\right) 
\nonumber\\
&=&
\sum_{\Delta} n(\Delta)\; \chi_{[L(c,\Delta)]}(\tilde{q}), \qquad \tilde{q}=\e^{-\frac{2\pi}{\tau}}
\label{Cardycon}
\en 
where we take the trace over the entire spectrum of open string states, that couple to the two boundary conditions imposed on the opposite sides of the cylinder 
and where we partition the spectrum of open string states into irreducible representation of the Virasoro algebra and 
apply definition (\ref{defcharac}).  
$n(\Delta)$ denotes the number of times the irreducible representation $L(c,\Delta)$ appears in the spectrum of open string states, that flow 
in between the two boundaries in the open string channel. 
The set of possible boundary states is constrained by the condition, 
that $n(\Delta)$ has to be a non-negative integer for all $\Delta$ for any given 
wave-function.\footnote{More precisely, this is true with regard to the discrete part of the spectrum of open 
states flowing in the open string channel.} 
This condition is known as the Cardy condition. 

The above discussion concerns 2D conformal field theory in general. 
Let us now return to Liouville theory. 
The primary boundary operators in Liouville theory are given by \cite{FZZ}
\[
B_{\beta}(x) = :\!\exp(\beta\phi(x))\!:
\]
with conformal dimension
\[
\Delta(\beta) 
= 
\beta(Q-\beta).
\]
In \cite{Moore} it is argued, that we should identify $B_{Q-\beta} \simeq B_{\beta}$. 
Hence, we may also impose the Seiberg bound (\ref{seibound}) with regard to boundary operators. 
The conformal descendants of a given primary boundary operator are obtained by acting on the primary operator 
with the Virasoro generators $L_{m}$, $m<0$.
As in the case of bulk operators there exists a state-operator correspondence between open string states and boundary vertex operators.  
As mentioned previously we do not necessarily have to impose the same boundary condition along an entire connected boundary. 
However, at any given point, where the boundary condition changes, we have to insert a boundary operator, 
which couple between the two boundary conditions.
Given the state-operator correspondence the spectrum of boundary operators, which couple between two given boundary conditions, is determined 
by the spectrum of open string states, that flow in between the two boundary condition in the open string channel. 
If we choose the two boundary condition to be the same on both sides of the strip, we may determine 
the spectrum of boundary operators associated with a given boundary condition.

\section{The FZZT boundary state}
In chapter 1 we gauge-fixed 2D euclidean Quantum Gravity coupled to a given conformal field theory  
in the case, where we only include closed geometries with a given fixed topology in the statistical ensemble of surfaces. 
Let us now consider 2D euclidean Quantum Gravity coupled to a given conformal field theory in the case, 
where we fix the topology of the world-sheet $M$ to be the topology of the disk.
In this case the gravitational action (\ref{graac}) is modified by a boundary term
\[\label{graacboundary}
S_{G}[g] = \mu_{0} \int_{M} d^{2}x \sqrt{g} + \mu_{B_{0}} \int_{\prt M} ds \,,
\]     
where $ds$ is the infinitesimal line-element. 
Notice, the boundary term in the above gravitational action is proportional to the length $l$ of the boundary 
and the constant of proportionality $\mu_{B_{0}}$ is known as the bare boundary cosmological constant.
In \cite{Yamaguchi} the gauge-fixing procedure of 2D euclidean quantum gravity is performed in this case.
The result of this procedure is similar to the result obtained in section \ref{partfunc}. 
However, in the Liouville action a boundary term appears in addition to the bulk action (\ref{lioac}) 
\st
S_{L}[\phi,\hat{g}] 
& = &
\frac{1}{4 \pi}\int_{M} d^{2}x \sqrt{\hat{g}}
\left(\hat{g}^{\alpha \beta}\prt_{\alpha}\phi\prt_{\beta}\phi+QR[\hat{g}]\phi\right)
+ \mu \int_{M} d^{2}x \sqrt{\hat{g}} \e^{2b\phi} \nonumber\\
& + &
\frac{Q}{2\pi} \int_{\prt M} d\hat{s} \, k[\hat{g}] \, \phi + \mu_{B} \int_{\prt M} d\hat{s} \e^{b\phi} \,,
\label{lioacboundary}
\en
where $d\hat{s}$ is the infinitesimal line element and $k[\hat{g}]$ is 
the geodesic curvature both defined in terms of the fiducial metric $\hat{g}$. 
The coefficient of the boundary term linear in $\phi$ is determined by considering the case $\mu = \mu_{B} = 0$ and demanding, 
that the gauge-fixed theory is independent of our choice of fiducial metric $\hat{g}$. 
The constant appearing in the exponent of the vertex operator in the nonlinear boundary interaction term 
is determined by the condition, that the conformal symmetry is maintained 
even for non-zero values of the boundary cosmological constant, that is 
turning on the boundary cosmological constant $\mu_{B}$ has to correspond 
to a marginal deformation of Liouville theory. 
Notice, given the fact, that the physical metric is given by $g = \e^{2b\phi}\hat{g}$, 
the value $b$ obtained by imposing this condition of marginality 
is consistent with the interpretation, 
that the nonlinear boundary term measures the length of the boundary. 
From \cite{Alvarez} we expect, that in the process of gauge-fixing this particular boundary interaction term  
appears in the action with a coefficient divergent in the cut-off introduced in the gauge-fixing procedure. 
In the above Liouville action we have absorbed this divergent coefficient 
into the bare boundary cosmological constant $\mu_{B_{0}}$ 
giving us a finite renormalized boundary cosmological constant $\mu_{B}$, which is a free parameter of the theory.     

The boundary condition imposed on the Liouville field 
by the above boundary action is easily determined by evaluating the functional derivative 
of the Liouville action (\ref{lioacboundary}) with respect to the Liouville field and setting it to zero. 
\[\label{bcon}
\prt_{n}\phi + Q k[\hat{g}] + 2 \pi b \mu_{B}\e^{b\phi} = 0
\]  
In this equation the normal derivative $\prt_{n}\phi$ appears from 
the functional derivative 
of the kinetic term in the bulk Liouville action. 
In the semi-classical limit $b \ll 1$ this boundary condition corresponds to a generalized Neumann boundary condition.

Let us parametrize Liouville theory in the upper half plane and 
let us consider Liouville theory in the conformal gauge. 
Due to the origin of the boundary action given in (\ref{lioacboundary}) it is clear, 
that this boundary condition does not break conformal invariance.  
However, given our discussion in section \ref{boundstates} we should verify, that the energy momentum tensor satisfies 
the condition (\ref{Tcondition}), when we impose the boundary condition given in eq. (\ref{lioacboundary}). 
The authors in \cite{Yamaguchi} merely verify, that the condition (\ref{Tcondition}) is satisfied in the case $\mu_{B}=0$. 
In the generic case $\mu_{B} \neq 0$ we may argue as follows. 
Firstly, the boundary Liouville field transforms according to
\[
\phi(x',0) = \phi(x,0) - Q \ln \vert\prt_{x}x'\vert 
\]  
under a given conformal transformation $z\to z'(z)$ mapping the upper half plane to the upper half 
plane.\footnote{Notice, 
this transformation law indeed reduces to the classical transformation law in the semi-classical limit, where $Q \approx 1/b$.} 
Secondly, applying this transformation law, 
the transformation law governing a primary boundary operator under a given conformal transformation and the fact, that 
the conformal dimension of the primary boundary operator $\e^{b\phi}$ is 1, we 
obtain\footnote{Notice, in the following calculation the manipulations are performed on the operators inserted in the partition function.}
\st
\prt_{x} \e^{b\phi(x,0)}
& \equiv &
\lim_{x' \to x} 
\frac{\e^{b\phi(x',0)}-\e^{b\phi(x,0)}}{x'-x}
\nonumber\\
& = &
\prt_{x}\phi
\lim_{x' \to x} 
\frac{\e^{b\phi(x',0)}-\e^{b\phi(x,0)}}{\phi(x',0)-\phi(x,0)}
\nonumber\\
& = &
-\frac{1}{Q}\e^{b\phi(x,0)} \prt_{x} \phi  
\lim_{x' \to x} 
\frac{\prt_{x'}x-1}{\ln \prt_{x}x'}
\nonumber \\
& = &
\frac{1}{Q}
\e^{b\phi(x,0)}
\prt_{x} \phi\,, 
\en
where we have applied an infinitesimal conformal transformation $z \to z' = z + \epsilon(z)$ 
mapping $x$ to $x'$ on the real axis.
Inserting the boundary condition (\ref{bcon}) into the energy momentum tensor 
(\ref{energysphere}) evaluated on the real axis 
we finally obtain
\[
T(x)-\bar{T}(x) 
= 
i\left(\prt_{x}\phi\prt_{y}\phi-Q\prt_{x}\prt_{y}\phi \right)
=
-i2\pi b \mu_{B}\left(\e^{b\phi}\prt_{x}\phi-Q\prt_{x}\e^{b\phi}\right)
= 0\,. 
\] 

Let us determine the boundary state corresponding to the boundary condition given in (\ref{lioacboundary}) 
using the so-called bootstrap techniques. 
These techniques are based upon and utilize the conformal symmetry of the field theory to full extent 
and from a theoretical point of view these techniques are interesting, since they provide a powerful tool to obtain 
exact results in conformal field theory. 
The beauty of these techniques becomes evident, when one compares the complexity of and the limited information obtained from 
a standard perturbative calculation in terms of Feynman diagrams with the abundance of information obtained from a 
exact calculation based upon these bootstrap techniques.  
We have actually already applied some of these bootstrap techniques in our discussion of the OPE in section \ref{sectionope}. 
In addition to these bootstrap techniques we will need to know some particular structure constants in Liouville theory, 
which are not fixed by the conformal symmetry, in order to 
set up an exact equation determining the wave function of the boundary state. 
We have already determined two of these structure constants appearing in the OPE in section \ref{sectionope}. 
The additional structure constant needed in the calculation of the boundary state will be determined by a method similar to the 
``perturbative'' method applied in section \ref{sectionope}.  
The boundary state corresponding to the boundary condition given in eq. (\ref{lioacboundary}) 
was first determined Fateev, Zamolodchikov and Zamolodchikov in \cite{FZZ} by the above method,  
which relies on some tricks developed by Teschner \cite{T1,T2}. 
We therefore refer to this boundary state as the FZZT boundary state.    

The basic idea for determining the boundary state is to consider the two-point function 
of the operator $V_{-b/2}$ and a generic primary operator $V_{\alpha}$ in the upper half plane. 
It follows from eq. (\ref{conb2}), that this two-point function satisfies the condition 
\[\label{contwopoint}
0 
= 
\bigg\langle
\left[\left( L_{-1}^{2} + b^{2}  L_{-2} \right) V_{-b/2}(z_{1},\bar{z}_{1})\right] 
V_{\alpha}(z_{2},\bar{z}_{2})
\bigg\rangle
\]
By a calculation identical to (\ref{denforste}) we obtain
\[\label{contopnkt}
\bigg\langle
\left[ L_{-1}^{2} V_{-b/2}(z_{1},\bar{z}_{1})\right] 
V_{\alpha}(z_{2},\bar{z}_{2})
\bigg\rangle 
= 
\prt_{z_{1}}^{2} 
\bigg\langle
V_{-b/2}(z_{1},\bar{z}_{1}) 
V_{\alpha}(z_{2},\bar{z}_{2})
\bigg\rangle 
\]
From eq. (\ref{vironoperator}) 
we obtain 
\[
\bigg\langle 
\left[L_{-2}V_{-b/2}(z_{1},\bar{z}_{1})V_{\alpha}(z_{2},\bar{z}_{2})\right]
\bigg\rangle
= 
\oint_{C_{z_{1}}} \frac{d\omega}{2\pi i} 
\frac{1}{\omega-z_{1}}
\bigg\langle 
T(\omega) V_{-b/2}(z_{1},\bar{z}_{1})V_{\alpha}(z_{2},\bar{z}_{2})
\bigg\rangle
\nonumber
\]
\[
= 
\!\!\int_{-\infty}^{\infty} \frac{dx}{2\pi i} 
\frac{1}{\!x\!-\!z_{1}}
\bigg\langle 
T(x) V_{-b/2}(z_{1},\bar{z}_{1})V_{\alpha}(z_{2},\bar{z}_{2})
\bigg\rangle
-
\oint_{C_{z_{2}}} \frac{d\omega}{2\pi i} 
\frac{1}{\!\omega\!-\!z_{1}}
\bigg\langle 
T(\omega) V_{-b/2}(z_{1},\bar{z}_{1})V_{\alpha}(z_{2},\bar{z}_{2})
\bigg\rangle  
\]
where we have used Cauchy's theorem and the fact, that there is no pole at infinity.
Applying the condition (\ref{Tcondition}) and Cauchy's theorem we may transform the integral along the real axis into a path integral 
in the complex $\bar{z}$-plane. 
\[
\phantom{map}
\bigg\langle 
\left[L_{-2}V_{-b/2}(z_{1},\bar{z}_{1})V_{\alpha}(z_{2},\bar{z}_{2})\right]
\bigg\rangle =
\phantom{-
\oint_{C_{\bar{z}_{1}}} \frac{d\bar{\omega}}{2\pi i} 
\frac{1}{\!\bar{\omega}\!-\!z_{1}}
\bigg\langle 
\bar{T}(\bar{\omega}) V_{-b/2}(z_{1},\bar{z}_{1})V_{\alpha}(z_{2},\bar{z}_{2})
\bigg\rangle}
\nonumber
\]
\[
-
\oint_{C_{\bar{z}_{1}}} \frac{d\bar{\omega}}{2\pi i} 
\frac{1}{\!\bar{\omega}\!-\!z_{1}}
\bigg\langle 
\bar{T}(\bar{\omega}) V_{-b/2}(z_{1},\bar{z}_{1})V_{\alpha}(z_{2},\bar{z}_{2})
\bigg\rangle\phantom{mapmapmapmapmap}
\nonumber
\]
\[ 
-
\oint_{C_{\bar{z}_{2}}} \frac{d\bar{\omega}}{2\pi i} 
\frac{1}{\!\bar{\omega}\!-\!z_{1}}
\bigg\langle 
\bar{T}(\bar{\omega}) V_{-b/2}(z_{1},\bar{z}_{1})V_{\alpha}(z_{2},\bar{z}_{2})
\bigg\rangle
\nonumber
\]
\[\phantom{mapmapmapmapmapmap}
-
\oint_{C_{z_{2}}} \frac{d\omega}{2\pi i} 
\frac{1}{\!\omega\!-\!z_{1}}
\bigg\langle 
T(\omega) V_{-b/2}(z_{1},\bar{z}_{1})V_{\alpha}(z_{2},\bar{z}_{2})
\bigg\rangle
\label{int}
\]
These three integrals may be determined in the same way as we determined the integrals in eq. (\ref{denanden}). 
Inserting the result of this calculation and eq. (\ref{contopnkt}) into eq. (\ref{contwopoint}) we obtain the partial differential equation  
\[
\left\{
\frac{1}{b^{2}}\prt_{z_{1}}^{2} + 
\frac{\Delta(-b/2)}{(\bar{z}_{1}-z_{1})^{2}}-\frac{1}{\bar{z}_{1}-z_{1}}\prt_{\bar{z}_{1}} +
\frac{\Delta(\alpha)}{(z_{2}-z_{1})^{2}}-\frac{1}{z_{2}-z_{1}}\prt_{z_{2}} 
\right.
\phantom{\frac{\Delta(\alpha)}{(\bar{z}_{2}-z_{1})^{2}}-\frac{1}{\bar{z}_{2}-z_{1}}\prt_{\bar{z}_{2}}mapmap}
\nonumber
\]
\[\label{partdiff}
\phantom{mapmapmapmapmap}
\left.+
\frac{\Delta(\alpha)}{(\bar{z}_{2}-z_{1})^{2}}-\frac{1}{\bar{z}_{2}-z_{1}}\prt_{\bar{z}_{2}} 
\right\}
\bigg\langle 
V_{-b/2}(z_{1},\bar{z}_{1})V_{\alpha}(z_{2},\bar{z}_{2})
\bigg\rangle = 0 
\]
In appendix \ref{apptwopoint} we show, that the two point function in the upper half plane is given by
\[\label{t1}
\bigg\langle 
V_{-b/2}(z_{1},\bar{z}_{1})V_{\alpha}(z_{2},\bar{z}_{2})
\bigg\rangle 
=
\frac{1}{\vert z_{2}-\bar{z}_{2} \vert^{2(\Delta(\alpha)-\Delta(-b/2))} \vert z_{1}-\bar{z}_{2} \vert^{4\Delta(-b/2)}} \mathcal{F}(\eta)
\]
where
\[\label{F}
\mathcal{F}(\eta)
=
\frac{2^{2(\Delta(\alpha)+\Delta(-b/2))}}{(1+\sqrt{\eta})^{4\Delta(-b/2)}}
\bigg\langle 
V_{-b/2}(it,-it)V_{\alpha}(i,-i)
\bigg\rangle 
\]
and
\[\label{teta}
t = \frac{1-\sqrt{\eta}}{1+\sqrt{\eta}}
\]
Furthermore, $\eta$ is given by eq. (\ref{eta}).
Inserting this expression for the two-point function into 
the partial differential equation (\ref{partdiff}) we obtain a second order differential equation 
governing $\mathcal{F}(\eta)$.
\[
\left\{
\frac{1}{b^{2}}\frac{d^{2}}{d\eta^{2}} +\left(\frac{1}{1-\eta}-\frac{1}{\eta}\right)\frac{d}{d\eta} +\frac{\Delta(\alpha)}{\eta^{2}} +
\Delta(\!-\!b/2)\frac{2-\eta}{\eta(1-\eta)^{2}}
\right\}
\mathcal{F}(\eta) 
= 
0
\]
The two linearly independent solutions to this equation known as conformal blocks is given by \cite{T2}
\[
\mathcal{F}_{1}(\eta) = 
\eta^{\alpha b}(1-\eta)^{-\frac{b^{2}}{2}}F(2\alpha b-1-2b^{2},-b^{2},2\alpha b - b^{2};\eta)
\]
and
\[
\mathcal{F}_{2}(\eta) = 
\eta^{1+b^{2}-\alpha b}(1-\eta)^{-\frac{b^{2}}{2}}F(-b^{2},1-2\alpha b,2-2\alpha b + b^{2};\eta)
\]
where $F(a,b,c;\eta)=\phantom{}_{2}F_{1}(a,b,c;\eta)$ is the hypergeometric function.  
$\mathcal{F}$ is given by a linear combination of these two solutions and 
let us determine the coefficients in this linear combination by inserting the OPE discussed 
in section \ref{sectionope} into eq. (\ref{F}) and evaluating the contributions 
from each of the two conformal families appearing in the OPE to lowest order in $\eta$.
From eqs. (\ref{ope}), (\ref{bulkstructure1}), (\ref{bulkstructure2}), (\ref{teta}), (\ref{psi}) 
and (\ref{onepointupperappendix}) we obtain 
\[
\mathcal{F}(\eta) 
= 
\frac{2^{2(\Delta(\alpha)+\Delta(-b/2))}}{(1+\sqrt{\eta})^{4\Delta(-b/2)}}
\bigg\langle 
V_{-b/2}(it,-it)V_{\alpha}(i,-i)
\bigg\rangle 
\nonumber
\]
\st
& = & 
\frac{2^{2(\Delta(\alpha)+\Delta(-b/2))}}{(1+\sqrt{\eta})^{4\Delta(-b/2)}}
\int \! d\sigma \: C_{-\frac{b}{2} \, \alpha}^{\sigma}\; 
\vert t\!-\!1 \vert^{2(\Delta(\sigma)-\Delta(\alpha)-\Delta(-b/2))}\:
\bigg\langle\! 
\Psi_{\sigma}^{-\frac{b}{2}\,\alpha}(it-i,-it+i\vert i, -i)
\!\bigg\rangle 
\nonumber\\
& = &
2^{2\Delta(\alpha-b/2)}\:
\langle
V_{\alpha-b/2}(i,-i)
\rangle\:
\eta^{\Delta(\alpha-b/2)-\Delta(\alpha)-\Delta(-b/2)}
(1 + \sum_{i=1}^{\infty} a_{i}\eta^{i} )
\nonumber\\
& + &
C_{-}\:
2^{2\Delta(\alpha+b/2)}\:
\langle
V_{\alpha+b/2}(i,-i)
\rangle\:
\eta^{\Delta(\alpha+b/2)-\Delta(\alpha)-\Delta(-b/2)}
(1 + \sum_{j=1}^{\infty} b_{j} \eta^{j})
\nonumber\\
& = &
U(\alpha-b/2)\,
\eta^{\alpha b}
(1 + \sum_{i=1}^{\infty} a_{i}\eta^{i} )
+
C_{-}
U(\alpha+b/2)\,
\eta^{1+b^{2}-\alpha b}
(1 + \sum_{j=1}^{\infty} b_{j}\eta^{j} ) 
\en
where $a_{i}$ and $b_{j}$ are some undetermined constants. 
Comparing this result with the expansions of $\mathcal{F}_{1}$ and $\mathcal{F}_{2}$ to lowest order in $\eta$ 
we obtain the coefficients appearing in $\mathcal{F}$ expressed as a linear combination of $\mathcal{F}_{1}$ and $\mathcal{F}_{2}$.
\[
\mathcal{F}(\eta) 
= 
U(\alpha-b/2)
\mathcal{F}_{1}(\eta) 
+
C_{-}
U(\alpha+b/2)
\mathcal{F}_{2}(\eta) 
\] 
Applying eqs. (15.3.6) and (15.3.3) in \cite{Ab} we may express 
\st
\mathcal{F}(\eta) 
& = &  
A\, 
\eta^{1-\alpha b + b^{2}}
\,(1-\eta)^{-b^{2}/2}
\,F(1-2\alpha b,-b^{2},-2b^{2};1-\eta)
\nonumber\\
&+& 
B \,
\eta^{\alpha b} 
\,(1-\eta)^{1+\frac{3}{2}b^{2}}
\,F(2\alpha b,1+b^{2},2+2b^{2};1-\eta)
\label{F2}
\en
where
\[
A 
= 
U(\alpha-\frac{b}{2})
\frac{\Gamma(2\alpha b - b^{2})\Gamma(1+2b^{2})}{\Gamma(1+b^{2})\Gamma(2\alpha b)}
+ 
C_{-}
U(\alpha+\frac{b}{2})
\frac{\Gamma(2-2\alpha b + b^{2})\Gamma(1+2b^{2})}{\Gamma(2-2\alpha b+2b^{2})\Gamma(1+b^{2})}
\]
and 
\[\label{Bcons}
B
= 
U(\alpha-\frac{b}{2})
\frac{\Gamma(2\alpha b - b^{2})\Gamma(-1-2b^{2})}{\Gamma(-b^{2})\Gamma(2\alpha b-1-2b^{2})}
+ 
C_{-}
U(\alpha+\frac{b}{2})
\frac{\Gamma(2-2\alpha b + b^{2})\Gamma(-1-2b^{2})}{\Gamma(1-2\alpha b)\Gamma(-b^{2})}
\]

So far the imposed boundary condition has not entered into the calculation explicitly. 
The above expression (\ref{F2}) is valid independent of the conformal invariant boundary condition imposed on the real axis. 
In order to determine the FZZT boundary state we have to consider some quantity, 
which depends explicitly on the boundary condition. 
Let us therefore consider the bulk-boundary operator expansion. 
In \cite{CL} it is argued, that given a boundary condition, which does not break conformal invariance, 
we may expand a given bulk operator in terms of boundary operators 
in the limit, where the bulk operator approaches the boundary. 
This bulk-boundary expansion is given by  
\[\label{bulkboundary}
V_{\alpha}(z,\bar{z}) 
= 
\int_{-\infty}^{Q/2} d\beta \, R_{\alpha}^{\beta} \,\vert z-\bar{z} \vert^{\Delta(\beta)-2\Delta(-b/2)} 
\sum_{\{k\}}\kappa_{\alpha}^{\beta \{k\}} \, \vert z-\bar{z}\vert^{K}\, \prod_{i}L_{-k_{i}}B_{\beta}(x) \, ,
\]
where we integrate over all conformal families of boundary operators   
and where we sum over all finite sets of positive integers ordered such that $k_{i} \leq k_{i+1}$. 
Moreover, $R_{\alpha}^{\beta}$ is known as the bulk-boundary structure constant, $x=\frac{1}{2}(z-\bar{z})$ 
and the remaining constants $\kappa_{\alpha}^{\beta \{k\}}$ are fixed completely by the conformal symmetry. 
The bulk-boundary structure constant is normalized such that
\[
\kappa_{\alpha}^{\beta \{\}} = 1\,,
\]
where $\{\}$ denotes the empty set. Finally, 
\[
K = \sum_{i}k_{i}
\,.
\] 
Let us consider the expansion of the primary operator $V_{-b/2}$ in terms of boundary operators. 
This expansion has to be consistent with eq. (\ref{conb2}).
Inserting the bulk-boundary operator expansion into eq. (\ref{partdiff}), 
which is equivalent to eq. (\ref{conb2}),  
and evaluating the contribution from each conformal family to lowest order in $z_{1}-\bar{z}_{1}$, 
we obtain the condition, that the bulk-boundary structure constant $R_{-b/2}^{\beta} \neq 0$, only if 
\[
\bigg\{
\Delta(\beta)-2\Delta(-b/2)
\bigg\}
\bigg\{
\Delta(\beta)-2\Delta(-b/2)-1
\bigg\}
+b^{2}\Delta(-b/2) 
- 
b^{2}\bigg\{
\Delta(\beta)-2\Delta(-b/2)
\bigg\}
=
0
\] 
In order to solve this equation in terms of $\Delta(\beta)$ we first determine $\Delta(\beta)-2\Delta(-b/2)$ and then determine $\Delta(\beta)$. 
From this calculation and the Seiberg bound we obtain, 
that the bulk-boundary structure constant $R_{-b/2}^{\beta} \neq 0$, only if 
\[
\beta = 0 
\quad \textrm{or} \quad   
\beta = -b\,. 
\]
Hence, only two conformal families of boundary operators appear in the bulk-boundary expansion of $V_{-b/2}$.  
One of these families is the family descending from the identity operator and we conclude, 
that we may express the bulk-boundary structure constant as 
\[\label{bu}
R_{-b/2}^{\beta} 
= 
\tilde{R}_{-b/2}^{0} \, \delta(\beta) 
+ 
\tilde{R}_{-b/2}^{-b} \, \delta(\beta + b).
\]

Let us determine the structure constant $\tilde{R}_{-b/2}^{0}$ in the case, where the boundary condition is given by 
(\ref{lioacboundary}). 
As in the case of the structure constants determined in section \ref{sectionope} we first need to consider Liouville theory with $\mu=\mu_{B}=0$, 
that is the limit, in which Liouville theory reduces to the linear dilaton theory. 
The correlation function of $m$ bulk operators and $n$ boundary operators 
in the linear dilaton theory is given by \cite{FZZ}\footnote{By an argument similar to (\ref{relconsphere}) we may show, 
that this correlation function vanishes, unless 
\[
2\sum_{i=1}^{m}\alpha_{i}+\sum_{j=1}^{n}\beta_{j} 
= 
Q
\]} \newpage 
\[
\bigg\langle
\prod_{i=1}^{m} V_{\alpha_{i}}(z_{i},\bar{z}_{i})
\prod_{j=1}^{n} B_{\beta_{j}}(x_{j})
\bigg\rangle_{\mu=\mu_{B}=0}
\nonumber
\]
\[
= 
\delta(2\sum_{i=1}^{m}\alpha_{i}+\sum_{j=1}^{n}\beta_{j}-Q) \:
\frac{\displaystyle\prod_{i=1}^{m}\vert z_{i} - \bar{z}_{i} \vert^{-2\alpha_{i}^{2}} \prod_{i,j} \vert z_{i} - x_{j} \vert^{-4\alpha_{i}\beta_{j}}}
{\displaystyle\prod_{i>j}^{n} \vert x_{i} - x_{j} \vert^{2\beta_{i}\beta_{j}}\prod_{i>j}^{m} \vert (z_{i} - z_{j}) ( z_{i} - \bar{z}_{j}) \vert^{4\alpha_{i}\alpha_{j}}}
\label{bulkboundarycorr} 
\]
Let us return to Liouville theory with $\mu > 0$ and $\mu_{B} \neq 0$. 
In the upper half plane the two-point function of two primary boundary operators vanishes, 
unless the conformal dimensions of the two operators coincide. 
This is a consequence of the conformal symmetry. 
Performing a perturbative expansion in $\mu$ and $\mu_{B}$ 
we obtain from the above equation the following two-point function valid for $\beta,\beta' \leq Q/2$.
\[
\bigg\langle
B_{\beta'}(0) B_{Q-\beta}(x)
\bigg\rangle 
\]
\[ 
=  
\sum_{m=0}^{\infty} 
\frac{(-\mu)^{m}}{2^{m}m!} 
\sum_{n=0}^{\infty}
\frac{(-\mu_{B})^{n}}{n!} 
\int \prod_{i=1}^{m} d^{2}\omega_{i} 
\int \prod_{j=1}^{n} dx_{j} 
\bigg\langle
\prod_{i=1}^{m} V_{b}(\omega_{i},\bar{\omega}_{i})
B_{\beta'}(0) 
\prod_{j=1}^{n} B_{b}(x_{j})
B_{Q-\beta}(x)
\bigg\rangle_{0}
\,  
\nonumber
\] 
\[
\phantom{mapmapmapmapm}
= 
\bigg\langle
B_{\beta'}(0) B_{Q-\beta}(x)
\bigg\rangle_{\mu=\mu_{B}=0}
= \;
\delta(\beta'-\beta) \frac{1}{\vert x \vert^{2\Delta(\beta)}}
\phantom{mapmapmapmapmapmapmapmapmapmapmapmapma} 
\]
In the above calculation we have assumed, that $\beta'$ belongs to a small interval around $\beta$. 
We apply this assumption, when we identify the term in the perturbative expansion different from zero. 
However, it is sufficient to determine two-point function for $\beta'$ belonging to a small interval around $\beta$, 
since the two-point function vanishes for all $\beta \neq \beta'$ assuming $\beta,\beta' \leq Q/2$.    
From the bulk-boundary expansion of $V_{-b/2}$ given in eq. (\ref{bulkboundary}) and the above equation we obtain, that   
\st
&&
\lim_{x \to \infty} 
\vert x \vert^{2\Delta(\beta)}
\bigg\langle
V_{-b/2}(i,-i)B_{Q-\beta}(x)
\bigg\rangle
\nonumber\\
& = & 
\lim_{x \to \infty} 
\vert x \vert^{2\Delta(\beta)} 
\int_{-\infty}^{Q/2} d\beta' \; R_{-b/2}^{\beta'} \; 2^{\Delta(\beta')-2\Delta(-b/2)}
\bigg\{\bigg\langle
B_{\beta'}(0)B_{Q-\beta}(x)
\bigg\rangle+\ldots \bigg\}
\nonumber\\
& = &
2^{\Delta(\beta)-2\Delta(-b/2)}
R_{-b/2}^{\beta} 
\en
valid for $\beta \leq Q/2$.
In the above calculation the $\ldots$ refers to the contribution from descendant operators. 
The two point function involving a descendant operator is proportional 
to the corresponding two point function involving the corresponding primary operator. 
This follows from the integral representation of the Virasoro generators given in eq. (\ref{vironoperator})  
and from the OPE of the energy momentum tensor and a given primary boundary operator. 
However, the two-point function involving a descendant operator decays faster in the limit $x \to \infty$ 
than the corresponding two-point function involving 
the corresponding primary operator. 
Hence, in the limit $x \to \infty$   
we project out the contribution from descendant operators in the above calculation. 
In order to determine the bulk-boundary structure constant $R_{-b/2}^{0}$ we consider the above equation for $\beta$ belonging to a small 
interval around zero and perform a perturbative expansion in $\mu$ and $\mu_{B}$. 
From eq. (\ref{bulkboundarycorr}) we obtain 
\st
R_{-b/2}^{\beta} 
& = &
2^{2\Delta(-b/2)-\Delta(\beta)} 
\lim_{x \to \infty} 
\vert x \vert^{2\Delta(\beta)}
\bigg\langle
V_{-b/2}(i,-i) B_{Q-\beta}(x)
\bigg\rangle
\nonumber\\
& = &
2^{2\Delta(-b/2)-\Delta(\beta)}  
\lim_{x \to \infty} 
\vert x \vert^{2\Delta(\beta)}
(-\mu_{B})
\int_{-\infty}^{\infty} d x'
\bigg\langle
V_{-b/2}(i,-i) B_{b}(x') B_{Q-\beta}(x)
\bigg\rangle_{\mu=\mu_{B}=0}
\nonumber\\
& = &
-2^{-1-2b^{2}} \mu_{B} 
\,\delta(\beta)
\int_{-\infty}^{\infty} d x' 
\left(
x'^{2}+1
\right)^{b^{2}}
\nonumber\\
& = &
-2^{-1-2b^{2}} \mu_{B} 
\,\delta(\beta)
\int_{0}^{\infty} du \, u^{-1/2}(u+1)^{b^{2}}
\nonumber\\
& = &
-2^{-1-2b^{2}} \mu_{B} 
\,\delta(\beta)
B(1/2,-1/2-b^{2})
\nonumber\\
& = &
-2\mu_{B}\pi\frac{\Gamma(-1-2b^{2})}{\Gamma^{2}(-b^{2})} 
\,\delta(\beta) 
\nonumber\\
& = &
\tilde{R}_{-b/2}^{0} 
\,\delta(\beta)\,, 
\en
where we consider the integral as a formal representation of the beta function $B$. See eq. (6.2.1) in \cite{Ab}. 
Moreover, we have used eqs. (6.2.2), (6.1.18), (6.1.15) and (6.1.8) in \cite{Ab}. 
Let us insert the expansion of $V_{-b/2}$ in terms of boundary operators given by eq. (\ref{bulkboundary}) into eq. (\ref{F}). 
Applying eqs. (\ref{teta}), (\ref{bu}) and (\ref{onepointupperappendix}) and performing an expansion in $1-\eta$ we obtain 
\st
\mathcal{F}(\eta) 
& = &
\frac{2^{2(\Delta(\alpha)+\Delta(-b/2))}}{(1+\sqrt{\eta})^{4\Delta(-b/2)}}
\bigg\langle 
V_{-b/2}(it,-it)V_{\alpha}(i,-i)
\bigg\rangle 
\nonumber\\
& = &
\frac{2^{2\Delta(\alpha)}}{(1+\sqrt{\eta})^{4\Delta(-b/2)}}
\int_{-\infty}^{Q/2} d\beta \, R_{-b/2}^{\beta} \, 
2^{\Delta(\beta)} \,
t^{\Delta(\beta)- 2\Delta(-b/2)} 
\nonumber\\
&&
\phantom{mapmapmapmapmapmap}
\times\sum_{\{k\}} \kappa_{-b/2}^{\beta \{k\}}\, 2^{K} \, t^{K} 
\bigg\langle
\left[\prod_{i}L_{-k_{i}} B_{\beta}(0) \right]
V_{\alpha}(i,-i)
\bigg\rangle
\nonumber\\
& = &
\tilde{R}_{-b/2}^{0}
\,U(\alpha)\,(1-\eta)^{1+\frac{3}{2}b^{2}}
\left(
1+\sum_{k=1}^{\infty} 
\tilde{a}_{k}(1-\eta)^{k}
\right)
\nonumber\\
&+&
\tilde{R}_{-b/2}^{-b}\, 
2^{2\Delta(\alpha)-\Delta(-b)}
\,\bigg\langle
B_{-b}(0) V_{\alpha}(i,-i)
\bigg\rangle\,
(1-\eta)^{-\frac{1}{2}b^{2}}
\left(
1+\sum_{l=1}^{\infty} 
\tilde{b}_{l}(1-\eta)^{l}
\right)
\nonumber\\
\en
Comparing this equation with eq. (\ref{F2}) we identify the first term in eq. (\ref{F2}) as coming from the fusion 
of $V_{-b/2}$ to the boundary operator $B_{-b}$, while the second term comes from the fusion of $V_{-b/2}$ to the identity operator on the boundary.   
Moreover,  
from this comparison we obtain the equation 
\st
&&
\phantom{mapmapmapmapmapmap}
B = 
\tilde{R}_{-b/2}^{0}
\,U(\alpha)
\nonumber\\
&
\Updownarrow
&
\nonumber\\
&&
-\frac{2\pi\mu_{B}}{\Gamma(-b^{2})} U(\alpha) 
= 
U(\alpha-b/2)\frac{\Gamma(2\alpha b - b^{2})}{\Gamma(2\alpha b -1 - 2b^{2})} 
- 
\pi \mu \frac{\Gamma(2\alpha b -1 -b^{2})}{\gamma(-b^{2})\Gamma(2\alpha b)}
\label{onepointeq}
\en
where the constant $B$ is defined in eq. (\ref{Bcons}). 
We have obtained the above exact equation  
by exploiting the reducible structure 
of the Verma module associated with the primary operator $V_{-b/2}$. 
However, instead of considering the two-point function involving $V_{-b/2}$ we might as well 
consider the two-point function involving the primary operator $V_{-\frac{1}{2b}}$. 
The Verma module associated with the primary operator $V_{-\frac{1}{2b}}$ also has a singular state at level 2. 
In this case we would obtain an exact equation concerning the one-point function on the disk related 
to the above equation by the duality transformation \cite{FZZ} 
\[\label{duality1}
b \to 
\frac{1}{b} 
\]
and
\[\label{duality2}
\mu \to 
\tilde{\mu} = \frac{1}{\pi\gamma(1/b^{2})} \left( \pi \mu \gamma(b^{2}) \right)^{1/b^{2}}\,.
\] 
The solution to eq. (\ref{onepointeq}) and the dual equation is given by \cite{FZZ}
\[
U(\alpha) 
= 
\frac{2}{b} 
\left(
\pi\mu\gamma(b^{2})
\right)^{(Q-2\alpha)/2b} \Gamma(2b\alpha-b^{2}) \Gamma(\frac{2\alpha}{b}-\frac{1}{b^{2}}-1)\cosh\left((2\alpha-Q)\pi \sigma\right)
\] 
where the parameter $\sigma$ is related to the boundary cosmological constant by 
\[\label{cosmolcon}
\cosh^{2}(\pi b \sigma) 
= 
\frac{\mu_{B}^{2}}{\mu} \sin(\pi b^{2})
\]
Applying eq. (\ref{onepointboundarywave}) we obtain the FZZT boundary wave function 
\[\label{FZZTboundfunc}
\Psi_{\sigma}(P) 
=
-\frac{1}{2^{5/4}\pi}
\left(
\pi \mu \gamma(b^{2})
\right)^{-iP/b} 
\Gamma(1+2ibP) 
\Gamma(1+2iP/b) 
\frac{\cos(2 \pi \sigma P)}{iP}
\] 
where we have introduced an overall factor $-\frac{1}{2^{5/4}\pi}$. 
We will comment on this particular normalization of the FZZT wave function at the end of the next section.

\section{The ZZ boundary states} \label{sectionZZ}

For each given solution to classical Liouville theory with negative curvature we expect, that there exists 
at least one corresponding solution to quantum Liouville theory, which reduces to the 
given classical solution in the classical limit $b \to 0$. 
In section \ref{Lobachevskiy} we discussed 
a particular non-compact solution to classical Liouville theory with constant negative curvature, the Lobachevskiy plane. 
In \cite{ZZ} Zamolodchikov and Zamolodchikov quantize the Lobachevskiy plane.   
Let us now discuss their derivation.  
By the quantum Lobachevskiy plane we mean a random geometry, 
whose properties on average are similar to the properties of the classical Lobachevskiy plane.
Specifically, Zamolodchikov and Zamolodchikov assume, 
that the geodesic distance between two operators inserted on the quantum Lobachevskiy plane diverges in the limit $\eta \to 1$ 
as in the case of the classical Lobachevskiy plane. (See eqs. (\ref{distanceloba}) and (\ref{eta}).)
Moreover, they assume, that the two-point function factorizes into a product of 
one-point functions in the limit, where the geodesic distance between the two operators approaches infinity. 
This assumption seems reasonable at least in a unitary theory, where the correlation functions decay at large distance.  
By a derivation identical to the derivation given in appendix \ref{apptwopoint} in the case of the two-point function 
we may show, that the product of two one-point functions in the upper half plane is given by 
\[
\langle
V_{\alpha'}(z_{1},\bar{z}_{1}) 
\rangle
\langle
V_{\alpha}(z_{2},\bar{z}_{2}) 
\rangle
\nonumber
\]
\[ 
=
\frac{2^{2(\Delta(\alpha)+\Delta(\alpha'))}}
{\vert z_{2}-\bar{z}_{2} \vert^{2(\Delta(\alpha)-\Delta(\alpha'))} \vert z_{1}-\bar{z}_{2} \vert^{4\Delta(\alpha')}(1+\sqrt{\eta})^{4\Delta(\alpha')}}
\bigg\langle 
V_{\alpha'}(it,-it)
\bigg\rangle
\bigg\langle
V_{\alpha}(i,-i)
\bigg\rangle 
\]
where $t$ is defined in eq. (\ref{teta}). 
In the case where $\alpha'=-b/2$ and in the limit $\eta \to 1$, 
where the geodesic distance between the two operators diverges and $V_{-b/2}$ approaches the absolute, 
we obtain the following leading order behavior of the product of the two one-point functions 
from eqs. (\ref{onepointupperappendix}) and (\ref{teta}).   
\[
\langle
V_{\alpha'}(z_{1},\bar{z}_{1}) 
\rangle
\langle
V_{\alpha}(z_{2},\bar{z}_{2}) 
\rangle
\approx 
\frac{U(-b/2)U(\alpha)}{\vert z_{2}-\bar{z}_{2} \vert^{2(\Delta(\alpha)-\Delta(-b/2))} \vert z_{1}-\bar{z}_{2} \vert^{4\Delta(-b/2)}}
(1-\eta)^{1+\frac{3}{2}b^{2}}
\]
The above leading order behaviour is identical to the leading order behaviour of the term in the two-point function (\ref{t1}) 
coming from the second conformal block in eq. (\ref{F2}), 
which describes the fusion of the operator $V_{-b/2}$ to the boundary identity operator in the limit, 
where $V_{-b/2}$ approaches the absolute. 
Moreover, we obtain from eqs. (\ref{t1}) and (\ref{F2}), that 
the fusion of the operator $V_{-b/2}$ to the boundary operator $B_{-b}$ in the limit, where $V_{-b/2}$ approaches the absolute, 
is responsible for a large distance correlation between the two operators. 
This large distance correlation between the two operators 
is only present, if the bulk-boundary structure constant $\tilde{R}_{-b/2}^{-b} \neq 0$. 
Zamolodchikov and Zamolodchikov now assume, that the second conformal block in eq. (\ref{F2}) 
indeed captures the decay of the two point function into a product of one-point functions. 
Equating the leading order term of the second conformal block  
with the corresponding leading order term in the above product of two one-point functions 
Zamolodchikov and Zamolodchikov obtain the equation
\st
&&
\phantom{mapmapmapmapmapmap}
U(-b/2)U(\alpha) = B
\nonumber\\
&\Updownarrow
&
\nonumber\\
&&
\frac{\Gamma(-b^{2})U(\alpha)U(-b/2)}{\Gamma(-1-2b^{2})\Gamma(2\alpha b -b^{2})}
=
\frac{U(\alpha-b/2)}{\Gamma(2\alpha b -1-2b^{2})} 
- 
\pi \mu \frac{U(\alpha+b/2)\Gamma(1+b^{2})}{\Gamma(-b^{2})\Gamma(2\alpha b)(2\alpha b -1-b^{2})}
\nonumber\\
\en   
where $B$ is defined in eq. (\ref{Bcons}). 
If we consider the primary operator $V_{-\frac{1}{2b}}$ instead of the primary operator $V_{-b/2}$, we obtain a 
dual equation through a similar argument. 
This dual equation is related to the above equation by the duality transformation 
given by eqs. (\ref{duality1}) and (\ref{duality2}). 
Zamolodchikov and Zamolodchikov argue, 
that at least for incommensurable values of $b$ and $1/b$ 
the complete set of solutions to these two equations is given by a family of solutions parametrized by two positive integers $m$ and $n$.
\[\label{onepointmn}
U_{m,n}(\alpha) = 
\frac{\sin(\pi b^{-1} Q)\sin(\pi mb^{-1}(2\alpha-Q))}{\sin(\pi mb^{-1}Q)\sin(\pi b^{-1}(2\alpha-Q))}
\frac{\sin(\pi b Q)\sin(\pi nb(2\alpha-Q))}{\sin(\pi nbQ)\sin(\pi b(2\alpha-Q))}
U_{1,1}(\alpha)
\] 
where the basic (1,1) one-point function is given by 
\[\label{onepoint11}
U_{1,1}(\alpha) 
= 
\frac{\{\pi\mu\gamma(b^{2})\}^{-\alpha/b}\Gamma(Qb)\Gamma(Q/b)Q}{\Gamma(b(Q-2\alpha))\Gamma(b^{-1}(Q-2\alpha))(Q-2\alpha)}
\]
Given the derivation of this family of one-point functions we ought to verify, whether each of these solutions really 
describe a quantum Lobachevskiy plane. 
In \cite{ZZ} Zamolodchikov and Zamolodchikov show the following. 
Let us define the quantum Liouville field as 
\[
\phi 
=
\frac{1}{2}\prt_{\alpha}V_{\alpha}\bigg\vert_{\alpha=0} 
= \,
\frac{1}{2}\prt_{\alpha} :\! e^{2\alpha\phi}\!:\bigg\vert_{\alpha=0}
\]
From the three above equations we obtain that 
\st
\nabla^{2} \langle \phi(z,\bar{z}) \rangle 
& = &
\nabla^{2} \left( 
\frac{1}{2}\prt_{\alpha}\langle V_{\alpha}(z,\bar{z}) \rangle \bigg\vert_{\alpha=0}
\right)
=
\nabla^{2}\left( -Q \ln \vert z-\bar{z} \vert +\frac{1}{2}\prt_{\alpha}U_{m,n}(\alpha)\bigg\vert_{\alpha=0}  \right)
\nonumber\\
& = &
\frac{4Q}{\vert z - \bar{z} \vert^{2}}
=
4\pi\mu b \,\langle V_{b}(z,\bar{z})  \rangle 
\en
independent of which solution labelled by $(m,n)$ we consider. 
This is precisely the classical Liouville equation (\ref{eomlio}) in conformal gauge 
satisfied by the classical Lobachevskiy plane.\footnote{Applying (\ref{Ricci}) 
we may express the classical Liouville equation (\ref{eomlio}) in conformal gauge as 
\[
\nabla^{2} \phi 
= 
4\pi\mu b e^{2b\phi}.
\]} 

Let us discuss the classical limit $b \to 0$ of the above two parameter family of one-point functions. 
For $m>1$ the one-point function $U_{m,n}(\alpha)$ does not approach a well-defined function in the limit $b \to 0$.  
Thus, it seems hard to associate a classical geometry with the one-point function $U_{m,n}(\alpha)$ in the classical limit for $m>1$. 
However, the one-point function $U_{1,n}(\alpha)$ is well-behaved in the classical limit $b \to 0$. 
We may view the insertion of a single vertex operator $V_{\alpha}(\omega,\bar{\omega})$ 
into the partition function as the introduction of an external current into the Liouville action 
\[
S_{L}[\phi] \to 
S_{L}[\phi] 
- 
\int d^{2}\omega' J(\omega',\bar{\omega}') \phi(\omega',\bar{\omega}') \,,
\]
where 
\[
J(\omega',\bar{\omega}') 
= 
2 \alpha \delta^{2}(\omega'-\omega) \,.
\] 
In the language of standard quantum field theory we may interpret  
\st
-W[J] &\equiv& 
\ln 
\langle 
V_{\alpha}(\omega,\bar{\omega})
\rangle
= 
\ln \left( \frac{U_{1,n}(\alpha)}{\vert 1- \omega \bar{\omega} \vert^{2\Delta{\alpha}}}  \right)
\nonumber\\
& = &
\sum_{k=1}^{\infty} 
\frac{(2\alpha)^{k}}{k!} 
G_{k}(b)
\en
as the generating function of all connected Feynman diagrams.\footnote{Notice, 
in the above we have parametrized Liouville theory on the disk for convenience.} 
Applying the definition of the generating function of all connected Feynman diagrams we obtain  
\st
-W[J] 
& = &
\sum_{k=0}^{\infty} \frac{1}{k!} 
\int d^{2}\omega_{1} \ldots \int d^{2}\omega_{k}
J(\omega_{1},\bar{\omega}_{1}) \ldots J(\omega_{k},\bar{\omega}_{k}) 
\tilde{G}_{k}(\omega_{1},\bar{\omega}_{1},\ldots,\omega_{k},\bar{\omega}_{k}) 
\nonumber\\
& = & 
\sum_{k=0}^{\infty} \frac{(2\alpha)^{k}}{k!} \tilde{G}_{k}(\omega,\bar{\omega},\ldots,\omega,\bar{\omega}) 
\en
where $\tilde{G}_{k}$ is the sum over all connected Feynman diagrams with $k$ external legs 
and where we have inserted the explicit expression for the external current. 
Comparing the two above equations we realize, that we may interpret $G_{k}(b)$ as the sum over all 
connected Feynman diagrams with $k$ external legs all ending at $\omega$. 
Given the one-point function $U_{1,n}(\alpha)$ we may determine the asymptotic expansion of $G_{k}(b)$ in $b$ order by order. 
In the case of $G_{1}(b)$ the first term depending on $n$ appears at third order.
In the case of $G_{2}(b)$ the first term depending on $n$ appears at second order.
Finally, the lowest order term in $G_{3}(b)$ do not depend on $n$. 
Expanding the Liouville action (\ref{lioac}) around the solution (\ref{Lobcircle}) describing the classical Lobachevskiy plane 
we may perform a standard perturbative calculation of $G_{k}(b)$ in terms of Feynman diagrams. 
This is done in \cite{ZZ} by Zamolodchikov and Zamolodchikov with regard to $G_{1}(b)$, $G_{2}(b)$ and $G_{3}(b)$. 
Firstly, they show, that the perturbative calculation in terms of Feynman diagrams of $G_{1}(b)$ to second order, 
of $G_{2}(b)$ to first order and of $G_{3}(b)$ to lowest order agrees with the asymptotic expansion of $G_{1}(b)$, $G_{2}(b)$ and $G_{3}(b)$  
independent of $n$. 
Secondly, they calculate $G_{2}(b)$ to second order and show, that the perturbative calculation in terms of Feynman diagrams  
match the asymptotic expansion obtained from the one-point function $U_{1,n}(\alpha)$, only if $n=1$. 
From this they conclude, that the basic $(1,1)$ solution corresponds to the "natural" quantization of the Lobachevskiy plane.      
This raises the question, how we should interpret the remaining members of the two-parameter family of solutions.  
One of the three main results obtained by the author of this thesis in collaboration with his academic adviser 
and presented in this thesis is a possible interpretation of the remaining members of the two parameter family of solutions. This 
interpretation is based upon and is in agreement with calculation performed both on the disk and the cylinder. 
We will return to this interpretation later on in this thesis.  

Even though the Lobachevskiy plane is a non-compact geometry we may, as we have done in eq. (\ref{Lobcircle}), 
parametrize the Lobachevskiy plane in the interior of a compact region of the complex plane, for instance the unit disk.      
In order to solve the Liouville equations in the interior of the unit disk 
we have to impose a boundary condition satisfied by the Liouville field as we approach the boundary. 
With time evolving radially outwards on the disk a given boundary condition in quantum Liouville theory 
corresponds to a boundary state imposed at the boundary of the unit disk as discussed previously in section \ref{boundstates}. 
In the case of the Lobachevskiy plane this out-vacuum state is imposed at the absolute of the Lobachevskiy plane at infinite. 
Zamolodchikov and Zamolodchikov assume, that the boundary conditions corresponding to the two parameter family of one-point 
functions $U_{m,n}(\alpha)$ do not break conformal invariance, that is the corresponding boundary states may be expressed as 
linear combinations of Ishibashi states. 
Applying eq. (\ref{onepointboundarywave}) we may define the corresponding two parameter family of boundary wave functions 
as
\[\label{ZZstatemn} 
\Psi_{m,n}(P) 
\equiv 
\frac{\sinh(2\pi mP/b)\sinh(2\pi nbP)}{\sinh(2\pi P/b)\sinh(2\pi bP)}
\Psi_{1,1}(P) 
\]
where 
\[\label{ZZstate11}
\Psi_{1,1}(P) 
= 
\frac{2^{3/4} 2 i \pi P}{\Gamma(1-2ibP)\Gamma(1-2iP/b)}\left\{ \pi\mu\gamma(b^{2}) \right\}^{-iP/b}
\]
Strictly speaking, we have omitted an overall constant factor independent of $P$ in the above definition compared with the expression 
for the wave function given in eq. (\ref{onepointboundarywave}). 
We will motivate this particular normalization in a short while. 
The boundary states corresponding to the above boundary wave functions are known as ZZ-boundary states 
named after Zamolodchikov and Zamolodchikov. 
Notice the somewhat peculiar fact, that we may express a given ZZ-boundary state as the difference between two 
FZZT-boundary states
\[\label{FZZTZZb2irr}
\Psi_{m,n}(P) 
= 
\Psi_{\sigma(m,n)}(P) - \Psi_{\sigma(m,-n)}(P)\, ,
\]
where 
\[\label{sigmamn}
\sigma(m,n) = i \left( \frac{m}{b} + nb \right)\, .
\] 
This fact was first noticed by Martinec in \cite{Martinec}. 
We will discuss this fact later on in this thesis.

Let us determine the spectrum of open string states flowing in the open string channel between two given ZZ boundary conditions 
labelled by $(r,s)$ and $(m,n)$ respectively. 
In order to determine this spectrum we consider the cylinder amplitude with circumference $2\pi$ and length $\pi\tau$ 
and impose a ZZ boundary state on each boundary. 
Applying the definition of an Ishibashi state given in eq. (\ref{Ishibashistate}) 
and applying eqs. (\ref{hamcyl2}) and (\ref{defcharac})
we obtain\footnote{In Liouville theory we denote the Ishibashi state associated with the irreducible representation  
$\mathcal{F}(c_{L},P)\otimes\tilde{\mathcal{F}}(c_{L},P)$ of the left and right copies 
of the Virasoro algebra by $\vert\vert Q/2 + i P \rangle\rangle$. This notation differs 
from the notation applied in section \ref{boundstates}.}    
\st
&&
\langle\langle Q/2 + iP' \vert\vert \e^{-\pi\tau H_{c}} \vert\vert Q/2 + iP \rangle\rangle
\nonumber\\
& = &
\sum_{k,l} 
\langle Q/2+i P', k \vert \otimes \langle Q/2+i P', k \vert 
\e^{-\pi\tau (L_{0}+\bar{L}_{0}-c_{L}/12)} 
\vert Q/2 + i P, l \rangle \otimes \vert Q/2 + i P, l \rangle
\nonumber\\
& = &
\delta(P'-P) 
\sum_{k} \langle Q/2+ i P, k \vert 
\e^{-2 \pi\tau (L_{0} - c_{L}/24)} 
\vert Q/2 + i P \rangle
\nonumber\\
& = &
\delta(P'-P)
\chi_{P}(q_{c})   
\label{IshibashiIshibashi}
\en
where 
\[\label{qclosed}
q_{c} = \e^{-2\pi\tau}.
\]
Applying eq. (\ref{virchaP}), performing a standard Gaussian integration and applying the formula \cite{pol}
\[
\eta(-1/\tau) 
= 
(-i\tau)^{1/2}\eta(\tau)\, ,
\]
where $\eta(\tau)$ is the Dedekind eta function, we obtain the following identity
\[\label{intform}
\sqrt{2} 
\int_{-\infty}^{\infty}  dP' 
\chi_{P'}(q_{c}) \e^{4i\pi PP'} 
= 
\frac{1}{\eta(\tilde{q})}\tilde{q}^{P^{2}} \,,
\]   
where 
\[
\tilde{q} = \e^{-2\pi/\tau}\, .
\]
Applying the relation
\[\label{sinsin}
\sinh(kx) \sinh(k'x) 
= 
\sinh(x) 
\sideset{}{'}\sum_{l = \vert k'-k \vert + 1}^{k'+k-1} 
\sinh(lx)
\]
where we have introduced a prime in the summation symbol in order to denote, 
that the summation runs in steps of two, and applying 
eqs. (\ref{ZZstatemn}), (\ref{ZZstate11}), (\ref{IshibashiIshibashi}) and (\ref{intform})  
we obtain
\st
\mathcal{Z}_{L} 
& = & 
\int_{0}^{\infty} d P \int_{0}^{\infty} d P' 
\Psi_{r,s}(P')\Psi^{\dag}_{m,n}(P)
\langle\langle Q/2 + i P' \vert\vert e^{-\pi\tau H_{c}} \vert\vert Q/2 +i P \rangle\rangle
\nonumber\\
& = &
2\sqrt{2} \int_{0}^{\infty} d P
\frac{\sinh(2\pi r P/b)\sinh(2\pi m P/b)}{\sinh(2\pi P/b)}  
\frac{\sinh(2\pi s P b)\sinh(2\pi n P b)}{\sinh(2\pi P b)}
\chi_{P}(q_{c}) 
\nonumber\\
& = &
\sideset{}{'}\sum_{k = \vert r - m \vert + 1}^{r+m-1} 
\:\sideset{}{'}\sum_{l = \vert s - n \vert + 1}^{s+n-1} 
\frac{1}{\eta(\tilde{q})} 
\left\{
\tilde{q}^{-\frac{1}{4}\left(\frac{k}{b} + l b\right)^{2}}
-
\tilde{q}^{-\frac{1}{4}\left(\frac{k}{b} - l b\right)^{2}}
\right\}
\en
In the case of $b^{2}$ irrational it follows from the discussion regarding reducible Verma modules in section \ref{QLT} 
that we may express the above Liouville cylinder amplitude in the open string channel as
\[\label{fusionb2irr}
\mathcal{Z}_{L}((r,s),(m,n)) 
= 
\sideset{}{'}\sum_{k = \vert r - m \vert + 1}^{r+m-1} 
\:\sideset{}{'}\sum_{l = \vert s - n \vert + 1}^{s+n-1} 
\chi_{[L(c_{L},\Delta(\alpha_{k,l}))]}(\tilde{q})
\]
where $\alpha_{k,l}$ is defined in eq. (\ref{acon}). 
The normalizations of the ZZ wave functions are chosen 
such that the overall factor in front the sum over Virasoro characters is one. 
Comparing the above equation with eq. (\ref{Cardycon}) we realize, 
that for $b^{2}$ irrational  
the spectrum $\mathcal{S}$ of open string states flowing in the open string channel in between the two ZZ boundary conditions  
is given by the direct sum
\[
\mathcal{S}(r,s;m,n)
=
\sideset{}{'}\bigoplus_{k = \vert r - m \vert + 1}^{r+m-1} 
\:\sideset{}{'}\bigoplus_{l = \vert s - n \vert + 1}^{s+n-1} 
L(c_{L},\Delta(\alpha_{k,l}))
\]
where the primes refer to the fact, that the summations run in steps of two.    
For $b^{2}$ irrational the fusion\footnote{The fusion rules usually pertain to only the left 
or the right part of the theory. 
Given two chiral operators belonging to the left part of the theory the fusion rules state, which conformal 
families of chiral operators appear in the OPE or fusion of the two operators. 
However, in the case of a conformal field theory, in which only spinless operators exist, 
we might as well apply the term fusion in relation with operators belonging to the tensor product 
of both the left and right part of the theory.}  
of the two primary operators 
$V_{\alpha_{r,s}}$ and $V_{\alpha_{m,n}}$ is given by \cite{francesco}
\[
V_{\alpha_{r,s}} 
\times 
V_{\alpha_{m,n}} 
= 
\sideset{}{'}\sum_{k = \vert r - m \vert + 1}^{r+m-1} 
\:\sideset{}{'}\sum_{l = \vert s - n \vert + 1}^{s+n-1} 
[V_{\alpha_{k,l}}] \,,
\]
where $[V_{\alpha_{k,l}}]$ refers to the conformal family of operators descending from  
the primary operator $V_{\alpha_{k,l}}$. 
Comparing the two above equation we realize, that 
for $b^{2}$ irrational there is the following one-to-one correspondence between the ZZ boundary states  
and the degenerate primary operators $V_{\alpha_{r,s}}$: The spectrum of open string states flowing in the open string channel between 
two ZZ-boundary states labelled by $(r,s)$ and $(m,n)$ is given by the fusion rule associated with the corresponding 
operators $V_{\alpha_{r,s}}$ and $V_{\alpha_{m,n}}$.  
In \cite{ZZ} Zamolodchikov and Zamolodchikov show, that we may associate the primary operator $V_{Q/2+i\sigma/2}$ with the 
FZZT boundary state given by the wave function $\Psi_{\sigma}(P)$ in a similar way.  
In particular by a calculation similar to calculation performed in the above we may show, that the Liouville 
cylinder amplitude with a FZZT boundary condition imposed on the one boundary and the $(m,n)$ ZZ boundary condition 
imposed on the other boundary is given by
\st
\mathcal{Z}_{L}(\sigma;m,n) 
& = &
\int_{0}^{\infty} dP \,\Psi_{m,n}(P) \,\Psi^{\dag}_{\sigma}(P) \,\chi_{P}(q_{c})
\nonumber\\
& = & 
\sideset{}{'}\sum_{k = -(m-1)}^{m-1} 
\:\sideset{}{'}\sum_{l = -(n-1)}^{n-1} 
\chi_{\frac{\sigma + i(k/b+lb)}{2}}(\tilde{q})
\label{lioFZZT-ZZ}
\en
in the open string channel. 
The normalization of the FZZT wave function given in eq. (\ref{FZZTboundfunc}) is exactly chosen such, 
that the overall coefficient in front of the sum over the Virasoro characters in the above equation is equal to one.  

\chapter{The (p,q) minimal model coupled to Quantum Gravity} \label{chapter4}
\section{The (p,q) minimal model} \label{sectionpq}
The set of minimal models consists of a family of two-dimensional conformal field theories labelled 
by two relatively prime positive integers $(p,q)$ satisfying $p<q$. 
The central charge of the $(p,q)$ minimal model is given by 
\[
c_{p,q} = 1 - 6\frac{(p-q)^{2}}{pq}\, .
\]  
The term minimal refers to the fact, that a given minimal model only consists of a finite number of irreducible 
representations of the two copies of the Virasoro algebra. 
Due to the simplicity of the minimal models, the $(p,q)$ minimal model coupled to Liouville theory 
defines a laboratory well suited for the study of matter coupled to Quantum Gravity in two dimensions. 
The literature on the minimal models is quite extensive. 
This short introduction to the minimal models is mainly based upon \cite{francesco}.  
In the so-called diagonal $(p,q)$ minimal model, which we will consider in this thesis, 
the primary operators are spinless and are given by
\[\label{minprimoperator}
\mathcal{O}_{r,s}\, ,\quad \textrm{where} \quad 1 \leq r \leq p-1, \quad 1 \leq s \leq q-1 \quad \textrm{and} \quad rq-sp > 0.
\]     
with conformal dimensions 
\[\label{condimmin}
\Delta_{r,s} 
= 
\frac{(rq-sp)^{2}-(p-q)^{2}}{4pq}\, .
\] 
The $(p,q)$ minimal model is unitary, if and only if $q=p+1$. 
By unitary we mean a conformal field theory, which do not contain any negative norm states.   
Applying the state-operator correspondence defined in eq. (\ref{correspondence}) 
we may associate a highest weight state with each of the primary operators. 
Acting on a given highest weight state repeatedly with the Virasoro generators $L_{m}$, $m<0$, 
we generate a Verma module as defined in (\ref{Verma}) in the case of Liouville theory. 
The structure of a given Verma module in terms of irreducible representations of 
the Virasoro algebra may be determined from the Kac determinant discussed in section \ref{sectionVerma}. 
The structure of the Verma module $\mathcal{V}\left(c_{p,q},\Delta_{r,s}\right)$ is given by \cite{Lian}
\\ 
\\
\[
\label{structuremin1}
\begin{array}{cccccccc}
\mathcal{V}\left(c_{p,q},\Delta_{r,s}\right)  & \!\LARGE{\leftarrow}\! & 
\mathcal{V}\left(c_{p,q},\tilde{A}_{r,s}(-1)\right) & \!\LARGE{\leftarrow}\! &
\mathcal{V}\left(c_{p,q},\tilde{B}_{r,s}(-1)\right) & \!\LARGE{\leftarrow}\! &
\mathcal{V}\left(c_{p,q},\tilde{A}_{r,s}(-2)\right) & \dots \\
&\!\LARGE{\nwarrow}\!&&\!\ddarrow\!&&\!\ddarrow\! \\
&&
\mathcal{V}\left(c_{p,q},\tilde{A}_{r,s}(0)\right) & \!\LARGE{\leftarrow}\! & 
\mathcal{V}\left(c_{p,q},\tilde{B}_{r,s}(1)\right) & \!\LARGE{\leftarrow}\! &
\mathcal{V}\left(c_{p,q},\tilde{A}_{r,s}(1)\right) & \dots
\\
&&&&&&&
\end{array}
\]
where 
\[
\tilde{A}_{r,s}(k) =  \frac{(2kpq + rq +sp)^{2}-(p-q)^{2}}{4pq} = 1 - A_{r,s}(k)
\]
and 
\[
\tilde{B}_{r,s}(k) =  \frac{(2kpq + rq -sp)^{2}-(p-q)^{2}}{4pq} = 1 - B_{r,s}(k) \, ,
\]
where $A_{r,s}(k)$ and  $B_{r,s}(k)$ are defined in eqs. (\ref{Acondim}) and (\ref{Bcondim}).
Notice, this diagram is related to the corresponding diagram (\ref{structure1}) in Liouville theory with $m=r$ and $n=s$  
by the previously discussed reflection property. 
We define the irreducible representation $L(c_{p,q},\Delta_{r,s})$ associated with 
the reducible Verma module $\mathcal{V}(c_{p,q},\Delta_{r,s})$ 
as in Liouville theory   
\[
L(c_{p,q},\Delta_{r,s}) 
\equiv 
\frac{\mathcal{V}(c_{p,q},\Delta_{r,s})}{\mathcal{V}(c_{p,q},\tilde{A}_{r,s}(0)) + \mathcal{V}(c_{p,q},\tilde{A}_{r,s}(-1))}
\, .
\]
As in the case of Liouville theory we determine the Virasoro character of the irreducible representation 
$L(c_{p,q},\Delta_{r,s})$ from the above diagram. (See eq. (\ref{inout}))
\st
\chi_{[L(c_{p,q},\Delta_{r,s})]}(q_{c}) 
& = & 
\sum_{i} (-1)^{d(\Delta_{r,s},\Delta_{i})}
\chi_{[\mathcal{V}(c_{p,q},\Delta_{i})]}
\nonumber\\ 
& = &
\frac{1}{\eta(q_{c})}
\sum_{n \in \mathbf{Z}} 
\left\{
q_{c}^{\frac{(2pqn+rq-sp)^{2}}{4pq}}
-
q_{c}^{\frac{(2pqn+rq+sp)^{2}}{4pq}}
\right\}
\nonumber\\
& \equiv &
\chi_{r,s}(q_{c})
\label{virchamin}
\en
where we sum over all Verma modules $\mathcal{V}(c_{p,q},\Delta_{i})$ embedded in $\mathcal{V}(c_{p,q},\Delta_{r,s})$ 
including $\mathcal{V}(c_{p,q},\Delta_{r,s})$. 
The Hilbert space of the diagonal $(p,q)$ minimal model is given by 
\[
\mathcal{H}_{p,q} 
= 
\bigoplus_{r,s} L(c_{p,q},\Delta_{r,s}) \otimes \tilde{L}(c_{p,q},\Delta_{r,s})
\]
where we sum over the set of labels $(r,s)$ given in eq. (\ref{minprimoperator}). 

The structure of the Verma modules  
is essential to the $(p,q)$ minimal model.
Due to the fact, that there exists two unrelated singular states in the Verma module 
$\mathcal{V}(c_{p,q},\Delta_{r,s})$ at level $rs$ and level $(p-r)(q-s)$ respectively,  
the corresponding primary operator $\mathcal{O}_{r,s}$ satisfies two independent constraints similar to the constraint (\ref{conb2}). 
This follows from an argument similar to the argument given in section \ref{sectionope} with regard to 
the primary operator $V_{-b/2}$ in Liouville theory. 
These constraints, the corresponding anti-holomorphic constraints and the constraint,  
that only spinless primary operators appear in the OPE of two spinless primary operators, 
imply, that the set of conformal families descending from the primary operators 
given in (\ref{minprimoperator}) is closed under the OPE, which is off course a requirement in order for the model to be well-defined. 
The fusion of two primary operators in the diagonal $(p,q)$ minimal model is given by \cite{francesco}
\[
\mathcal{O}_{r,s} 
\times 
\mathcal{O}_{m,n} 
= 
\sideset{}{'}\sum_{k = \vert r - m \vert + 1}^{\textrm{min}(r+m-1,2p-1-r-m)}
\:\sideset{}{'}\sum_{l = \vert s - n \vert + 1}^{\textrm{min}(s+n-1,2q-1-s-n)}
[\mathcal{O}_{k,l}]
\label{fusionmin}
\]
The term diagonal applied in the above discussion refers to the fact, that 
the Hilbert space is the direct sum over direct products of copies of the same representation, 
which is equivalent to the statement, that all primary operators are spinless.
However, minimal models exist, which are not diagonal. 
The three-state Potts model at the critical point is an example of a non-diagonal model.\cite{francesco}  
The operator content of a given conformal field theory is constrained by the fact, 
that the genus $1$ torus amplitude must be invariant under the modular group $PSL(2,\mathbf{Z})$. 
This condition follows from the fact, that any given torus may be 
defined as the upper half plane modulo some lattice spanned by two linearly independent vectors. 
These two linearly independent vectors enter the torus amplitude through the modular parameter $\tau$. 
However, there exists a whole family of pairs of linearly independent vectors each spanning the same lattice 
and the torus amplitude should not depend on, which pair of vectors we choose to define the lattice by. 
Going from one pair of vectors spanning the lattice to another pair of vectors, 
the modular parameter $\tau$ transforms under the action of the modular group $PSL(2,\mathbf{Z})$.     
Hence, the torus amplitude must be invariant under the group of modular transformations. 
Any given torus amplitude may be obtained by imposing some   
periodic boundary condition on the cylinder of finite length. 
The torus amplitude may therefore be expressed as the trace over the Hilbert space of closed string states. 
The condition, that the torus amplitude has to be invariant under the group of modular transformations, 
imposes a constraint on the Hilbert space and the corresponding operator content of the conformal field theory.  
Both the diagonal minimal models and the three-state Potts model at the critical point 
satisfy the above modular constraint. 

Let us consider the so-called Cardy boundary states in the $(p,q)$ minimal model introduced in \cite{Cardy2,Zuber}. 
These boundary states are defined by two conditions.  
Firstly, the boundary states should not be break conformal invariance. 
This condition implies, that the boundary states may be 
expanded in terms of the Ishibashi states defined in eq. (\ref{Ishibashistate}). 
Secondly, the family of Cardy states should satisfy the Cardy condition discussed in section \ref{boundstates}. 
For each primary operator $\mathcal{O}_{r,s}$ we obtain a Cardy state
\[\label{Cardydef}
\vert r,s \rangle_{_{C}} 
= 
\begin{array}{c}
\displaystyle\sum_{m=1}^{p-1} \,\,\, 
\sum_{n=1}^{q-1}
\\
\textrm{\scriptsize{$mq\!-\!np>0$}}
\end{array} 
\frac{S_{(r,s);(m,n)}}{\sqrt{S_{(1,1);(m,n)}}}
\vert\vert \Delta_{m,n} \rangle\rangle
\]
where we sum over all conformal families in the diagonal $(p,q)$ minimal model 
and where $S_{(r,s);(m,n)}$ is the so-called modular S-matrix \cite{francesco}
\[\label{mods}
S_{(r,s);(m,n)} 
= 
2\sqrt{\frac{2}{pq}}(-1)^{1+sm+rn}
\sin\left(\frac{\pi q rm}{p}\right)
\sin\left(\frac{\pi p sn}{q}\right)
\]
The correspondence between the Cardy state labelled by $(r,s)$ and the primary operator $\mathcal{O}_{r,s}$ is similar 
to the correspondence between the ZZ-boundary state labelled by $(m,n)$ and the primary operator $V_{\alpha_{m,n}}$ in Liouville theory 
in the case $b^{2}$ irrational. 
The spectrum of open string states flowing in the open string channel between the $(r,s)$ Cardy state 
and the $(m,n)$ Cardy state is determined by the fusion rule associated with the primary operators $\mathcal{O}_{r,s}$ 
and $\mathcal{O}_{m,n}$. 
Hence, in the open string channel the cylinder amplitude of length $\pi\tau$ 
with the $(r,s)$ Cardy boundary condition imposed 
on one boundary and the $(m,n)$ Cardy boundary condition imposed on the other boundary is given by 
\[
\mathcal{Z}_{M}((r,s);(m,n)) 
= 
\sideset{}{'}\sum_{k = \vert r - m \vert + 1}^{\textrm{min}(r+m-1,2p-1-r-m)}
\:\sideset{}{'}\sum_{l = \vert s - n \vert + 1}^{\textrm{min}(s+n-1,2q-1-s-n)}
\chi_{k,l}(\tilde{q})
\] 
where $\chi_{k,l}(\tilde{q})$ denotes the Virasoro character $\chi_{[L(c_{p,q},\Delta_{k,l})]}(\tilde{q})$. 

Finally, let us shortly touch on the physics described by the minimal models. 
Let us for simplicity consider the unitary $(3,4)$ minimal model. 
The $(3,4)$ minimal model describes the Ising model in the thermodynamic limit 
at the critical temperature $T_{c}$, 
which separates the ordered ferromagnetic phase from the unordered symmetric phase.  
More generally, the $(3,4)$ minimal model describes the universality class of models, to which the Ising model at the critical point belongs.
Due to the fact, that the correlation length $\xi$ diverges at the critical point $T_{c}$ of the Ising model, 
the theory no longer contains any length scale nor mass scale $ m \sim 1/\xi$ at the critical point 
and the theory becomes scale-invariant, conformal.  
Moreover,  
we are able to define a continuum model describing the Ising model at the critical point. 
Comparing critical exponents we may identify this continuum model as the $(3,4)$ minimal model.    
The three primary operators existing in the $(3,4)$ minimal model, $\mathcal{O}_{1,1}$, $\mathcal{O}_{2,1}$ and $\mathcal{O}_{2,2}$, 
correspond to the identity operator, the energy-density operator 
and the spin-density operator respectively in the Ising model.\cite{BPZ} 
The two Cardy boundary states $\vert 1,1 \rangle_{_{C}}$ and $\vert 2,1 \rangle_{_{C}}$ correspond 
to the two boundary conditions fixing the spins on the boundary to be either up or down respectively. 
The third Cardy state $\vert 2,2 \rangle_{_{C}}$ corresponds to the free boundary condition, 
where the spins on the boundary are free to fluctuate.\cite{Cardy2} 
The remaining $(p,q)$ minimal models also describe some lattice models at criticality.\cite{francesco}    
However, in this thesis we will not put to much emphasis 
on the particular realizations of the minimal models in terms of lattice models.  
Our main focus will be on studying the effect of matter on geometry in 2D euclidean Quantum Gravity. 
We consider the minimal models coupled to 2D euclidean Quantum Gravity as toy models,  
since the simplicity of the minimal models enables us to perform actual calculations.   

\section{The (p,q) minimal model coupled to Quantum Gravity} \label{physoperator}
Let us couple the $(p,q)$ minimal model to 2D euclidean Quantum Gravity 
and let us discuss the space of \emph{physical} closed string states. 
Performing the gauge-fixing procedure discussed in section \ref{partfunc} we obtain  
the tensor product of the $(p,q)$ minimal model, Liouville theory 
and the ghost theory, where the central charge of Liouville theory is fixed by the condition  
(\ref{c0}), that is    
\[
c_{L} 
= 
1 + 6\left( \sqrt{\frac{p}{q}}+\sqrt{\frac{q}{p}} \right)^{2}\, .
\]    
The parameter $b$ in Liouville theory is fixed by eqs. (\ref{cencharlin}), (\ref{Q}) 
and the Seiberg bound (\ref{seibound}), 
\[\label{ekstra}
b = \sqrt{\frac{p}{q}}\, .
\]

Any given physical amplitude should not depend on our choice of gauge. 
Considering a small variation of the fiducial metric $\hat{g}_{ab}$, we may show, that 
this condition of gauge-independence is equivalent to the condition   
\[
\langle \alpha \vert T^{ab} \vert \beta \rangle = 0
\] 
for all physical states $\vert\alpha\rangle$ and $\vert\beta\rangle$, where $T_{ab}$ is the total energy momentum tensor.\cite{pol} 
Expanding the energy momentum tensor in terms of the Virasoro generators we obtain the condition,  
that any matrix element obtained by inserting a Virasoro generator in between two physical states must vanish. 
We may express this condition as a constraint on the space of physical closed string states in the so-called BRST formalism. 
Actually, the two independent copies of the Virasoro algebra give rise to two independent constraints on the space of physical closed string states. 
Let us for simplicity only consider the constraint imposed by the left copy of the Virasoro 
algebra.\footnote{We may perform a similar discussion with regard to the constraint imposed by the right copy of the Virasoro algebra.}   
We may express this constraint in terms of the BRST charge\footnote{Actually, 
we only consider the term in the BRST charge, which acts on the left part of theory. The complete BRST charge consists of the above term 
and a corresponding term acting on the right part of the theory.} \cite{Lian,pol} 
\[
Q 
= 
\sum_{n=-\infty}^{\infty} c_{n} L_{-n}  
+ 
\sum_{m,n=-\infty}^{\infty} \frac{(m-n)}{2} :\!c_{m}c_{n}b_{-m-n}\!: - c_{0}
\]
where the operators $b_{k}$ and $c_{l}$ are the conventional ghost operators defined in \cite{pol} and 
where the dots ``:'' refer to the operation of creation-annihilation normal ordering, in which we place all lowering operators to the right of all 
raising operators.\footnote{Notice, we add a minus sign each time we change the order of two neighboring ghost operators.}   
Furthermore, the Virasoro generators $L_{m}$ appearing in the above expression for the BRST charge are given by the sums of the corresponding 
Virasoro generators in the matter sector and in the Liouville sector. 
As discussed in various textbooks on 2D Quantum Gravity or string theory such as \cite{pol,Moore} 
the condition of gauge independence may be expressed partially\footnote{We obtain 
a similar constraint with regard to the right part of the complete BRST charge, which acts on the right part of the theory.}  
as the constraint, that any given physical state $\vert \alpha \rangle$ must belong to the kernel of $Q$, that is  
\[\label{Q0}
Q \vert \alpha \rangle = 0
\,.
\]   
A state annihilated by $Q$ is known as a closed state. 
Due to the fact, that the BRST charge is nilpotent  
\[
Q^{2} = 0\, ,
\]
the image of $Q$ is actually a subspace of the kernel of $Q$. 
A state belonging to the image of $Q$ is known as an exact state.
Because $Q$ is self-adjoint  
\[
Q^{\dag} = Q \,,
\]
any given exact state is actually orthogonal to all closed states including itself. 
Thus, two given closed states, which differ by an exact state, have the same inner product with all closed states 
and are therefore physically equivalent. 
We identify two given physical states, if they differ by an exact states. 
In addition to the condition (\ref{Q0}) any given physical state $\vert \alpha \rangle$ must satisfy 
the condition 
\[\label{constraintb}
b_{0}\vert \alpha \rangle 
= 
0, \qquad
\bar{b}_{0}\vert \alpha \rangle 
= 0\,.
\] 
The origin of this condition is essentially kinematic.\cite{pol} 

Applying the anti-commutator relations concerning the ghost operators given in \cite{pol}   
we may easily show, that the commutator of the BRST charge 
and the so-called ghost number operator $N_{g}$  
defined in \cite{pol} is given by 
\[
[N_{g},Q] = Q\,.   
\]      
Hence, the BRST charge raises  
the ghost number by one. 
Let us define 
\[
L_{0}^{\textrm{tot}} = L_{0}^{M} + L_{0}^{L} + L_{0}^{\textrm{Ghost}}\,.
\] 
Due to the fact, that
\[\label{Qbb}
\{Q,b_{0}\} = L_{0}^{\textrm{tot}}, 
\]
which follows from the anti-commutator relations concerning the ghost operators given in \cite{pol}, 
it follows from eqs. (\ref{Q0}) and (\ref{constraintb}), 
that any given physical state is annihilated by $L_{0}^{\textrm{tot}}$. 
Hence, in our search for physical states we may restrict our attention to the subspace 
annihilated by both $b_{0}$  
and $L_{0}$.     
Let us consider  
\[
C_{n}(\Delta_{r,s},P) \equiv
\nonumber
\]
\[
\bigg\{ \vert \sigma \rangle \in  
\mathcal{M}(\Delta_{r,s},P)
\bigg\vert \: 
b_{0} \vert \sigma \rangle = 
L_{0}^{\textrm{tot}} \vert \sigma \rangle = 0, \: 
N_{g} \vert \sigma \rangle = n \vert \sigma \rangle 
\bigg\} 
\]
where 
\[
\mathcal{M}(\Delta_{r,s},P)
\equiv  
L(c_{p,q},\Delta_{r,s})  
\otimes
\mathcal{F}(c_{L},P)  
\otimes \mathcal{H}^{L}_{\textrm{Ghost}}\,, 
\]
where $\mathcal{H}^{L}_{\textrm{Ghost}}$  
is the left part of the Hilbert space of the ghost theory.
It follows from eq. (\ref{Qbb}), that the linear space annihilated by 
$b_{0}$  
and $L_{0}$  
is closed under $Q$. 
Due to this fact we obtain the cochain complex 
\[
\ldots 
-\!\!\!\longrightarrow \!\!\!\!\!\!\!\!\!\!\phantom{}^{Q}\,\,\,\,\,\,\,\,
C_{n-1}(\Delta_{r,s},P) \,\, 
-\!\!\!\longrightarrow 
\!\!\!\!\!\!\!\!\!\!\phantom{}^{Q}\,\,\,\,\,\,\,\,
C_{n}(\Delta_{r,s},P) \,\,
-\!\!\!\longrightarrow 
\!\!\!\!\!\!\!\!\!\!\phantom{}^{Q}\,\,\,\,\,\,\,\,
C_{n+1}(\Delta_{r,s},P) \,\,
-\!\!\!\longrightarrow  
\!\!\!\!\!\!\!\!\!\!\phantom{}^{Q}\,\,\,\,\,\,\,\,
\ldots
\] 
Let us define the cohomology groups  
\[
\mathcal{H}_{n}\left(\Delta_{r,s},P\right) 
\equiv
\frac{\textrm{Ker}\left( Q : C_{n}(\Delta_{r,s},P) \longmapsto C_{n+1}(\Delta_{r,s},P)    \right)}
{\textrm{Im}\left( Q : C_{n-1}(\Delta_{r,s},P) \longmapsto C_{n}(\Delta_{r,s},P)    \right)}
\]
and  
\[
\mathcal{H}_{\ast}\left(\Delta_{r,s},P\right) 
\equiv 
\bigoplus_{n} 
\mathcal{H}_{n}\left(\Delta_{r,s},P\right) \,.
\]
Given the above discussion, it seems natural to identify $\mathcal{H}_{\ast}\left(\Delta_{r,s},P\right)$ 
as the left part of the linear space of physically distinct closed string states belonging to the tensor product of the irreducible matter representation 
$L(c_{p,q},\Delta_{r,s}) \otimes \tilde{L}(c_{p,q},\Delta_{r,s})$ 
and the irreducible Liouville representation $\mathcal{F}(c_{L},P) \otimes \tilde{\mathcal{F}}(c_{L},P)$.  
The entire linear space is given by the tensor product 
$\mathcal{H}_{\ast}\left(\Delta_{r,s},P\right)\otimes\tilde{\mathcal{H}}_{\ast}\left(\Delta_{r,s},P\right)$, where 
$\tilde{\mathcal{H}_{\ast}}\left(\Delta_{r,s},P\right)$ denotes the corresponding right part.

The above BRST cohomology groups were determined by Lian and Zuckerman in \cite{Lian}. 
The proof presented in \cite{Lian} is quite complicated and 
employs the rather advanced technology of spectral sequences.  
Moreover, the formula given in \cite{Lian} for  
the non-trivial BRST invariant states  
is quite complex and involves the so-called Felder resolution, 
which was originally introduced in the construction of the minimal models in the 
Coulomb-Gas formalism.\cite{Felder} 
In order not to go to far astray we will only discuss the above BRST cohomology groups to the extent necessary 
for our further discussion in this thesis.   
Lian and Zuckerman show, that   
\[\label{hom1}
\mathcal{H}_{\ast}\left(\Delta_{r,s},P\right) 
\neq 
0\,,
\quad \textrm{if and only if} \quad 
P = \pm i \frac{2kpq + rq \pm sp}{2\sqrt{pq}}\,,\; k\in\mathbf{Z}. 
\]
Hence, $\mathcal{H}_{\ast}\left(\Delta_{r,s},P\right)$ is non-trivial, 
if and only if the Verma module $\mathcal{V}(c_{L},\Delta(Q/2+i P))$ belongs the diagram (\ref{structure1}) with $m=r$ and $n=s$ related to 
the diagram (\ref{structuremin1}) by the reflection principle (\ref{reflection}),  
that is if and only if 
\[
\mathcal{V}(c_{p,q},1-\Delta(Q/2+iP)) \to 
\mathcal{V}(c_{p,q},\Delta_{r,s}) \, .
\]
Notice, the Liouville momenta $P$, at which non-trivial cohomologies exist, all lie on the imaginary axis 
and are located symmetric around the origin. 
Assuming that the above condition is satisfied, Lian and Zuckerman prove, that
\[
\mathcal{H}_{n}\left(\Delta_{r,s},P\right) 
\cong 
\delta_{n+1/2,\textrm{sign}(iP) d(\Delta(Q/2+iP),1-\Delta_{r,s})} \:
\mathbf{C}\,,
\]
where the distance $d(\Delta(Q/2+iP),1-\Delta_{r,s})$ is defined in the paragraph following the diagram given in (\ref{structure1}). 
Thus, in the case when the condition (\ref{hom1}) is 
satisfied, the BRST cohomology $\mathcal{H}_{\ast}\left(\Delta_{r,s},P\right)$ is one-dimensional and 
we may determine the ghost number of the non-trivial physical state 
spanning the BRST cohomology from the above 
theorem.\footnote{Notice, Lian and Zuckerman consider $C_{n}(\Delta_{r,s},P)$ as a linear space over the field $\mathbf{C}$.} 

We may construct a single physical closed string state for each non-trivial cohomology 
group $\mathcal{H}_{\ast}\left(\Delta_{r,s},P\right)$ determined by Lian and Zuckerman. 
This follows from the fact, that both the $(p,q)$ minimal model and Liouville theory are diagonal, and from the fact, 
that the BRST cohomology $\mathcal{H}_{\ast}\left(\Delta_{r,s},P\right)$ is one-dimensional. 
This physical closed string state is given by the tensor product of the non-trivial state spanning the BRST cohomology 
$\mathcal{H}_{\ast}\left(\Delta_{r,s},P\right)$ and the corresponding non-trivial state spanning the BRST cohomology 
$\tilde{\mathcal{H}}_{\ast}\left(\Delta_{r,s},P\right)$. 
Since the ghost numbers $N_{g}$ and $\tilde{N}_{g}$ associated with this physical closed string state obviously coincide, 
for simplicity we will refer to the two ghost numbers as one in the following. 
In section \ref{QLT} we argued, that the conformal family of states descending from the primary state corresponding 
to the primary operator $V_{\alpha}$ should 
be identified with the conformal family of states descending from the primary state corresponding 
to the primary operator $V_{Q-\alpha}$. 
In the $(p,q)$ minimal model coupled to euclidean Quantum Gravity this observation turns into the statement, 
that we should identify any given physical closed string state 
with Liouville momentum $P=i t$, $t>0$, with the corresponding physical closed string state with Liouville momentum $P=-it$. 
Hence, we may impose the bound, 
that we only consider physical closed string states with Liouville momentum on the upper imaginary axis. 
Given the state-operator correspondence discussed previously 
we may define a physical observable corresponding to each physical closed string state.
We may view the result obtained by Lian and Zuckerman as a classification 
of the physical observables in the $(p,q)$ minimal model coupled to 2D euclidean Quantum Gravity. 
Due to the non-tensor transformation laws given in \cite{pol} and governing the ghost number currents under a given conformal transformation 
the ghost number associated with the physical observable corresponding to a given physical state with ghost number $n$ is actually 
given by $n+3/2$.\cite{pol} 
Imposing the condition,  
that we only consider physical closed string states with Liouville momentum $P$ belonging to the upper imaginary axis, 
is equivalent to imposing the Seiberg bound on the set of physical observables. 
The ghost number of a given physical observable satisfying the Seiberg bound is either $1$, $0$ or $-n$, $n \in \mathbf{Z}$. 
Now, the explicit construction of a generic physical observable is quite complicated. 
(See \cite{Lian} with regard to the so-called zig-zag method.) 
However, in this thesis we will only need the fact, 
that we may express the physical observable $\hat{\mathcal{A}}^{r,s}_{P}$ 
spanning the one-dimensional linear space 
$\mathcal{H}_{n-3/2}\left(\Delta_{r,s},P\right)\otimes\tilde{\mathcal{H}}_{n-3/2}\left(\Delta_{r,s},P\right)$ as \cite{SS}
\[\label{obs}
\hat{\mathcal{A}}^{r,s}_{P} 
= 
\mathcal{L}_{r,s}^{P} \mathcal{O}_{r,s} V_{Q/2+iP}\,, 
\]    
where $\mathcal{L}_{r,s}^{P}$ refers to a certain combination of Virasoro raising operators and ghost operators depending on $r$, $s$ and $P$.  
Moreover, we may actually construct the observables $\mathcal{T}_{r,s}$ with ghost number $1$ explicitly. 
In this case the zig-zag method becomes trivial. 
\[
\mathcal{T}_{r,s} 
= 
c \bar{c} 
\mathcal{O}_{r,s} V_{\beta_{r,s}} 
\]
where 
\[\label{KPZeq}
2\beta_{r,s} = \frac{p+q-rq+sp}{\sqrt{pq}} \,.
\]
We refer to these observables as tachyon operators. 
The above equation for $\beta_{r,s}$ determining the gravitational dressing of a given matter operator 
was first determined independently by Knizhnik, Polyakov and Zamolodchikov in the light cone gauge in \cite{KPZ} 
and by David in conformal gauge in \cite{David}. 
This equation has been instrumental to the development of euclidean 2D Quantum Gravity. 
The tachyon operator $\mathcal{T}_{r,s}$ should actually be identified with the physical observable  
\[\label{tachyon}
\mathcal{T}_{r,s} =
\int d^{2}z \, \mathcal{O}_{r,s}(z,\bar{z}) V_{\beta_{r,s}}(z,\bar{z})\,. 
\]
The integration in the above expression for the tachyon operator $\mathcal{T}_{r,s}$ 
is replaced by the factor $c\bar{c}$, 
if we fix the position of the vertex operator appearing in the integrand applying and fixing some of the remaining conformal symmetry.\cite{pol} 
It follows from the above expression, that the tachyon operator $\mathcal{T}_{1,1}$ measures the area of the world-sheet. 

To each of the physical observables in the BRST formalism we may associate a gravitational scaling dimension $\rho$ defined 
by \cite{Moore}
\st
\mathcal{F}^{r,s}_{P}(A,h) 
& = & 
\frac{1}{\mathcal{Z}(A)} 
\int \mathcal{D}\phi \,\mathcal{D}b \, \mathcal{D}c  
\,\e^{-S_{L}-S_{gh}} \, \delta\!\left( \int d^{2}x \sqrt{g} e^{2b\phi} - A \right)
\mathcal{L}_{r,s}^{P} \,\langle \mathcal{O}_{r,s} \rangle \, V_{Q/2+iP}
\nonumber\\
& \sim & 
A^{1 - \rho}
\en
where $\mathcal{F}^{r,s}_{P}(A,h)$ denotes the normalized expectation value of the observable $\hat{\mathcal{A}}^{r,s}_{P}$  
evaluated on closed Riemann surface on genus $h$ with fixed area $A$. 
Under the rigid translation  
\[\label{rig}
\phi \to \phi' = \phi +  \frac{1}{2b}\ln(A)
\]
the different factors in the above integrand transform according to 
\[
\delta\!\left( \int d^{2}x \sqrt{g} e^{2b\phi'} - A \right) 
= 
\frac{1}{A}\:\delta\!\left( \int d^{2}x \sqrt{g} e^{2b\phi} - 1 \right) \,,
\]
\[
V'_{Q/2+iP}
=\;
:\!e^{(Q+2iP)\phi'}\!\!: 
\;= \; 
A^{\frac{Q+2iP}{2b}}
\;:\!e^{(Q+2iP)\phi}\!\!:
\]
and 
\[
S_{L}[\phi'] = \frac{1}{4\pi} \int d^{2}x \sqrt{\hat{g}} 
\left( \hat{g}^{ab}\prt_{a}\phi' \prt_{b}\phi' + QR[\hat{g}]\phi' \right) 
= A^{-\frac{Q\chi(h)}{2b}} S_{L}[\phi]\,,
\]
where we have applied eq. (\ref{sidst}) in deriving the two first equations. 
Using the fact, that the Liouville measure is invariant under the rigid translation (\ref{rig}) we obtain 
\[
\mathcal{F}^{r,s}_{P}(A,h) 
= 
\frac{A^{\frac{Q+2iP}{2b}-\frac{Q\chi(h)}{2b}-1}}{A^{-\frac{Q\chi(h)}{2b}-1}} \mathcal{F}^{r,s}_{P}(1,h) 
= 
A^{\frac{Q+2iP}{2b}} \mathcal{F}^{r,s}_{P}(1,h)
\]
from which we obtain the gravitational scaling dimension 
\[\label{gradim}
1-\rho = \frac{p+q+2i\sqrt{pq}P}{2p}\,. 
\]
Notice, the scaling dimension of the observable $\hat{\mathcal{A}}^{r,s}_{P}$ is determined by the Liouville momentum $P$. 
From eq. (\ref{hom1}) we see, that the set of gravitational scaling dimensions 
associated with the set of physical observables with ghost number $1$, $0$ or $-k$, $k \in \mathbf{N}$, 
is given by   
\[
1 - \rho = \frac{p+q-n}{2p}\,,\qquad n\geq1\,,\quad p \ndiv n\,,\quad q \ndiv n\,. 
\]
In the matrix model approach to 2D euclidean Quantum Gravity we may construct a set of scaling operators. 
The gravitational scaling dimensions associated with these operators are given by \cite{Moore}
\[
1 - \rho = \frac{p+q-n}{2p}\,,\qquad n\geq1\,,\quad p \ndiv n\,.
\]
Notice, the condition $q \ndiv n$ has been lifted. 
The nature of the additional physical operators appearing in the matrix model formalism compared with the BRST formalism 
is still an open question.

\section{The FZZT and ZZ-branes} \label{sectionTheFZZTandZZbranes}
Seiberg and Shih define a FZZT 
brane\footnote{In this thesis 
we refer to a conformal invariant boundary condition in a conformal field theory with vanishing 
central charge as a brane.} 
as the tensor product of a Cardy state and a FZZT boundary state \cite{SS}
\[
\vert \sigma; k,l \rangle
= 
\begin{array}{c}
\displaystyle\sum_{m=1}^{p-1} \,\,\, 
\sum_{n=1}^{q-1}
\\
\textrm{\scriptsize{$mq\!-\!np>0$}}
\end{array} 
\int_{0}^{\infty} dP
\frac{S_{(k,l);(m,n)}}{\sqrt{S_{(1,1);(m,n)}}}
\Psi_{\sigma}^{\dag}(P)
\vert\vert \Delta_{m,n} \rangle\rangle
\otimes
\vert\vert Q/2 + iP \rangle\rangle 
\]
where the modular S-matrix is defined in (\ref{mods}) 
and the FZZT wave function $\Psi_{\sigma}(P)$ is given by (\ref{FZZTboundfunc}). 
Due to the fact that we may express any given Virasoro raising operator appearing in a correlator  
as a linear differential operator acting on the position dependent part of the correlator as in 
eqs. (\ref{denforste}) and (\ref{denanden}), 
we may determine the one-point function evaluated on the disk from eqs. (\ref{obs}), (\ref{onepointboundarywave}), (\ref{mods}) 
and (\ref{FZZTboundfunc})  
up to a constant independent of the boundary 
condition\footnote{The boundary state $\phantom{}_{_{C}}\!\langle k,l \vert$ conjugate to the Cardy state $\vert k,l \rangle_{_{C}}$ 
is defined in \cite{Zuber} as 
\[
\phantom{}_{_{C}}\!\langle r,s \vert
= 
\begin{array}{c}
\displaystyle\sum_{m=1}^{p-1} \,\,\, 
\sum_{n=1}^{q-1}
\\
\textrm{\scriptsize{$mq-np>0$}}
\end{array} 
\frac{S_{(r,s),;(m,n)}}{\sqrt{S_{(1,1);(m,n)}}}
\langle\langle \Delta_{m,n} \vert\vert
\]} 
\st
\langle \sigma; k,l \vert \hat{\mathcal{A}}^{r,s}_{P} \rangle
& = &
\tilde{\Omega}_{r,s}^{P} \frac{S_{(k,l);(r,s)}}{\sqrt{S_{(1,1);(r,s)}}}  \Psi_{\sigma}(P)  
\nonumber\\
& = &
\grave{\Omega}_{r,s}^{P}(-1)^{ks+lr}\cosh\left(\frac{\pi(2 k p q + r q \pm s p)\sigma}{\sqrt{pq}} \right) 
\sin\left( \frac{\pi q k r}{p}  \right)
\sin\left( \frac{\pi p l s}{q}  \right)
\nonumber\\
\label{onepointFZZT}
\en 
where $\grave{\Omega}_{r,s}^{P}$ and $\tilde{\Omega}_{r,s}^{P}$ are constants independent of $\sigma$, $k$ and $l$ labelling the FZZT brane 
and where we used the fact that 
\[
P = \pm i \frac{2kpq + rq \pm sp}{2\sqrt{pq}} \, ,
\]
which follows from eq. (\ref{hom1}).
Notice, the one-point function is invariant under each of the transformations 
\[
\sigma \to -\sigma\,, \qquad 
\sigma \to \sigma \pm 2i\sqrt{pq} \, .
\]  
Assuming that a given boundary state is completely characterized by the set of physical gauge invariant 
one-point functions evaluated on the disk,  
Seiberg and Shih view this invariance as evidence, 
that FZZT branes related by the above transformations are identical modulo BRST exact states. 
In order to identify the physically distinct FZZT branes they introduce a new label $z$ instead of $\sigma$ 
invariant under the above symmetries    
\[\label{z}
z = \cosh\left(\frac{\pi\sigma}{\sqrt{pq}}\right)\,.  
\]   

Before we define a ZZ-brane in the $(p,q)$ minimal model coupled to 2D euclidean Quantum Gravity 
let us shortly discuss the definition of a ZZ boundary condition in Liouville theory in the case $b^{2}$ rational. 
In the following discussion we will apply the notation concerning the reducible Verma modules 
introduced in eqs. (\ref{labelling}) and (\ref{labellinglimit}).  
In section \ref{sectionZZ} we showed, that in the case $b^{2}$ irrational 
there is a one-to-one correspondence between the degenerate primary operators in Liouville theory associated with the reducible Verma modules  
and the ZZ boundary conditions. 
However, as we discussed in section \ref{sectionVerma} the structure of a generic reducible Verma module changes, 
when we go from $b^{2}$ irrational to $b^{2}$ rational. 
Eq. (\ref{fusionb2irr}) is not valid in the case $b^{2}$ rational, since the Virasoro character 
of the irreducible representation obtained from a given generic reducible Verma module 
is more complicated in the case $b^{2}$ rational than in the case $b^{2}$ irrational. 
(Compare eqs. (\ref{virchab3irrational}) and (\ref{virchab2rational}).) 
Furthermore, the fusion rules concerning the degenerate primary operators in the case $b^{2}$ rational differ 
from the corresponding fusion rules valid in the case $b^{2}$ irrational. 
Thus, the one-to-one correspondence between the degenerate primary operators and the ZZ boundary states 
defined in eqs. (\ref{ZZstatemn}) and (\ref{ZZstate11})  
is not valid in the case $b^{2}$ rational.  
Seiberg and Shih suggest, that we alter the definition of the ZZ boundary states   
in the case $b^{2}$ rational in order to restore 
this one-to-one correspondence.\cite{SS} 
Let us shortly discuss their definition of the ZZ boundary states in Liouville theory.      
Since we label a given reducible Verma module by $(t,r,s)$ defined in eq. (\ref{labelling}) 
in the case $b^{2}$ rational, they argue, that we should label the ZZ boundary states 
in Liouville theory likewise.
Now, it follows from eq. (\ref{lioFZZT-ZZ}), which is valid for $b^{2}$ rational, 
that we may associate the identity operator with the basic $(1,1)$ or rather $(0,1,1)$ ZZ boundary state  
even in the case $b^{2}$ rational. 
Moreover, Seiberg and Shih argue, that we should 
define the ZZ boundary state labelled by $(t,r,s)$ 
as the following sum over FZZT-boundary states
\[\label{ZZboundb2ra}
\vert t,r,s \rangle
\equiv 
\sum_{j=0}^{t} 
\left( 
\vert \sigma = i\frac{n(t-2j,r,s)}{\sqrt{pq}} \rangle 
-
\vert \sigma = i\frac{n(t-2j,r,-s)}{\sqrt{pq}} \rangle 
\right) \,,
\]  
where $n(t,r,s)$ is defined in eq. (\ref{labelling}).
This definition ensures, 
that the spectrum of open string states flowing in the open string channel 
between the $(0,1,1)$ ZZ-brane and $(t,r,s)$ ZZ-brane is indeed given by the fusion of the identity operator 
with the degenerate primary operator associated with reducible Verma module labelled by $(t,r,s)$.
This follows from eqs. (\ref{lioFZZT-ZZ}), (\ref{virchaP}) and (\ref{virchab2rational}). 
However, given the definition by Seiberg and Shih of the ZZ boundary states in Liouville theory  
in the case $b^{2}$ rational it would be nice to verify explicitly, 
that the spectrum of states flowing in the open string channel 
between two generic ZZ boundary conditions is indeed given 
by the direct sum of the conformal families appearing in the fusion of the two corresponding degenerate primary operators.  

Seiberg and Shih define a ZZ brane as the tensor product of a Cardy matter state and a ZZ boundary state. 
Due to the fact, that we 
identify FZZT branes labelled by the same $z$ and with the same Cardy indices  
we may express a given ZZ brane as 
\st
&&
\vert (k,l)_{_{C}}; (t,r,s)_{_{ZZ}} \rangle
\nonumber\\
& \equiv & 
\vert k,l \rangle_{_{C}} \otimes 
\left\{
\sum_{j=0}^{t} 
\left( 
\vert z = \cos(\frac{\pi n(t-2j,r,s)}{pq} )  \rangle 
-
\vert z = \cos(\frac{\pi n(t-2j,r,-s)}{pq} ) \rangle 
\right)
\right\}
\nonumber\\
& = &
(t+1)
\vert k,l \rangle_{_{C}} \otimes 
\left\{ 
\vert z = (-1)^{t}\cos(\frac{\pi (r q + s p)}{p q} )  \rangle 
-
\vert z = (-1)^{t}\cos(\frac{\pi (r q - s p)}{p q} ) \rangle 
\right\}
\nonumber\\
& = &
\left\{ \begin{array}{ll} 
+(t+1) \vert (k,l)_{_{C}}; (0,r,s)_{_{ZZ}} \rangle &   \quad \textrm{$t$ even} \\ & \\
-(t+1) \vert (k,l)_{_{C}}; (0,r,q-s)_{_{ZZ}} \rangle & \quad \textrm{$t$ odd}
\end{array} \right.
\label{principal1}
\en
Furthermore, it follows from the above calculation that
\[
\vert (k,l)_{_{C}}; (t,p-r,q-s)_{_{ZZ}} \rangle 
=
\vert (k,l)_{_{C}}; (t,r,s)_{_{ZZ}} \rangle  
\label{principal2}
\]
and 
\[
\vert (k,l)_{_{C}};(t,r,s)_{_{ZZ}} \rangle = 0 \quad \textrm{for $r=p$ or $s=q$.}
\label{principal3}
\]
These three equations should be considered true modulo BRST exact states. 
We will refer to the ZZ boundary states in Liouville theory labelled by $(0,r,s)$, where $rq-sp>0$, 
as principal ZZ boundary states. 
With regard to these principal ZZ boundary states we leave out the redundant label $t$. 
Moreover, we will refer to a given principal ZZ boundary state tensored 
with \emph{any} given Cardy matter state as a principal ZZ brane. 
This terminology differs somewhat from the terminology of Seiberg and Shih. 
We will motivate this terminology shortly.    
Thus, we realize from the three above equations, that the linear space of ZZ branes is smaller than one might naively think. 
Any given ZZ brane may be expressed as a linear combination of the principal ZZ branes.
In the case of a principal ZZ boundary state the definition (\ref{ZZboundb2ra}) reduces 
to eq. (\ref{FZZTZZb2irr}).\footnote{Going from $b^{2}$ irrational 
to $b^{2}$ rational the structure of a reducible Verma module associated with a given principal ZZ boundary state 
does actually not change. Even in the case $b^{2}$ rational the reducible Verma module   
only contains one singular state.  
These Verma modules are located in the second column counted 
from the left hand side in the diagram (\ref{structure1}).} 
Hence, the definition of a ZZ boundary state given in eqs. (\ref{ZZstatemn}) and (\ref{ZZstate11}) is valid with regard to the 
principal ZZ boundary states. 
It follows from eq. (\ref{cosmolcon}), that the cosmological constants associated 
with the FZZT boundary states appearing 
on the left hand side of eq. (\ref{FZZTZZb2irr}) actually coincide. 
Hence, even though the Lobachevskiy plane is non-compact, 
we may actually associate a cosmological constant with each of the principal ZZ boundary state. 
The cosmological constant associated with the principal ZZ boundary state labelled by $(r,s)$ is given by   
\[\label{cosmologicalZZ}
\mu_{B}(r,s) = (-1)^{r}\cos(\pi s b^{2})\sqrt{\mu}\,,
\]  
where we have absorbed the factor $\pi \gamma(b^{2})$ into $\mu$ and 
the factor $\sqrt{\pi \gamma(b^{2})\sin(\pi b^{2})}$ into $\mu_{B}$.\footnote{In 
the remaining part of this thesis we apply this normalization of $\mu$ and $\mu_{B}$.} 
We will comment on this surprising fact later on in this thesis.

Differentiating the disk amplitude $\mathcal{Z}$ once with respect to the bulk cosmological constant corresponds to 
inserting the area operator $\mathcal{T}_{1,1}$ into the disk partition function.  
From the first identity in (\ref{onepointFZZT}) and eqs. (\ref{mods}), (\ref{FZZTboundfunc}) and (\ref{KPZeq}) we obtain 
\[
-\prt_{\mu} \mathcal{Z}
=
\langle \sigma; k,l \vert \mathcal{T}_{1,1} \rangle 
\propto 
(-1)^{k+l} 
\sin\left(\frac{\pi q k}{p}\right)
\sin\left(\frac{\pi p l }{q}\right)
(\sqrt{\mu})^{\frac{q-p}{p}}\cosh\left( \frac{\pi(q-p)\sigma}{\sqrt{pq}} \right) \,,
\]
where the omitted dimensionless constant of proportionality is independent $\mu$, $\mu_{B}$ and the Cardy indices $k$ and $l$. 
Integrating this amplitude with respect to the bulk cosmological constant 
and thereafter differentiating the amplitude with respect to the boundary cosmological constant 
we obtain the disk amplitude $W$ with one marked point on the boundary 
\[\label{diskamplitude}
W(\mu_{B},\mu)
= 
(\sqrt{\mu})^{\frac{q}{p}} \cosh\left(\frac{\pi\sigma}{b}\right)\,, 
\]
where we have absorbed a constant factor independent of $\mu$ and $\mu_{B}$ into $W$. 
Introducing the dimensionless variables
\[
t = \frac{\mu_{B}}{\sqrt{\mu}}\,,\qquad
y = \frac{W(\mu_{B},\mu)}{(\sqrt{\mu})^{q/p}}
\]
and applying the expression for the boundary cosmological constant in terms of $\sigma$  
we may express eq. (\ref{diskamplitude}) as 
\[\label{Chebyshev}
T_{p}(y)-T_{q}(t) = 0 \,,
\]
where $T_{k}(\cos\theta)=\cos(k\theta)$ is the Chebyshev polynomial of the first kind. 
This polynomial equation defines a Riemann surface 
\[\label{mpq}
\mathcal{M}_{p,q} 
= 
\bigg\{
(t,y) \in \mathbf{C}^{2} \bigg\vert T_{p}(y)-T_{q}(t) = 0
\bigg\} 
\] 
which we may interpret as the moduli space of FZZT-branes with Cardy indices $(k,l)$. 
Each point on this Riemann surface corresponds to a particular FZZT-brane with Cardy indices $(k,l)$. 
Notice, our discussion of the moduli space of FZZT-branes differs somewhat from the discussion by Seiberg and Shih in the sense, that we introduce a Riemann surface for each set of Cardy indices $(k,l)$. 
We will comment on this shortly. 
Due to fact that 
\[
t = T_{p}(z) \,,\quad
y = T_{q}(z)
\]
we may consider $z$ introduced in eq. (\ref{z}) as an uniformizing parameter of $\mathcal{M}_{p,q}$ covering the Riemann surface once. 
Since $z \in \mathbf{C}$  
the topology of the Riemann surface $\mathcal{M}_{p,q}$ appears to be the topology of the sphere. 
However, this is not correct.
There exists a finite number of 
pairs $(z_{r,s}^{+},z_{r,s}^{-})$ of distinct values of $z$, where both values $z_{r,s}^{+}$ and $z_{r,s}^{-}$
belonging to a given pair correspond to the same point $(t_{r,s},y_{r,s})$ 
on the Riemann surface $\mathcal{M}_{p,q}$. 
These pairs are given by 
\[
(z_{r,s}^{+},z_{r,s}^{-}) = (\cos\left(\frac{\pi(rq+sp)}{pq}\right),\cos\left(\frac{\pi(rq-sp)}{pq}\right))\,,
\]
where 
\[
1 \leq r \leq p-1, \; 1 \leq s \leq q-1, \; \textrm{and} \; rq-ps>0\,,
\] 
and the corresponding points on the Riemann surface $\mathcal{M}_{p,q}$ are given by  
\[
(t_{r,s},y_{r,s}) = \left( (-1)^{r}\cos\left( \frac{\pi s p}{q}  \right),(-1)^{s}\cos\left( \frac{\pi r q}{p} \right) \right) \,.
\] 
We may think of each of these singularities 
on the Riemann surface $\mathcal{M}_{p,q}$ as a pinched A-cycle of a higher genus surface.\cite{SS} 

The above discussion is based on the truly exciting work \cite{SS} of Seiberg and Shih. 
However, we have to some degree modified the discussion appearing in \cite{SS}. 
Let us shortly discuss the points of deviation. 
It follows from eq. (\ref{onepointFZZT}), that the one-point function on the disk satisfies 
\[
\langle \sigma; k,l \vert \hat{\mathcal{A}}^{r,s}_{P} \rangle 
= 
\sideset{}{'}\sum_{m = -(k-1)}^{k-1}     
\;\sideset{}{'}\sum_{n = -(l-1)}^{l-1}    
\langle \sigma + i\frac{mq+np}{\sqrt{pq}}; 1,1 \vert \hat{\mathcal{A}}^{r,s}_{P} \rangle
\]
regardless of which physical gauge invariant operator $\hat{\mathcal{A}}^{r,s}_{P}$ we insert.  
Seiberg and Shih regard this equation as evidence, that the following identity regarding the FZZT-branes is valid 
modulo BRST exact states 
\[\label{SSproposal}
\vert \sigma; k,l \rangle 
= 
\sideset{}{'}\sum_{m = -(k-1)}^{k-1}        
\;\sideset{}{'}\sum_{n = -(l-1)}^{l-1}        
\vert \sigma + i\frac{mq+np}{\sqrt{pq}}; 1,1 \rangle\,. 
\]
From this equation they argue, that we only need to consider FZZT-branes and ZZ-branes with Cardy matter indices $(1,1)$.\footnote{Recall, 
that Seiberg and Shih define the ZZ boundary states in terms of the FZZT boundary states.} Thus, they 
define a principal ZZ brane as the tensor product of a principal ZZ boundary state and the $(1,1)$ Cardy matter state. 
In the above discussion we view the Riemann surface $\mathcal{M}_{p,q}$ as the moduli space of FZZT branes with fixed Cardy matter indices $(k,l)$. 
In light of the above equation 
Seiberg and Shih consider the Riemann surface $\mathcal{M}_{p,q}$ as the moduli space of \emph{all} FZZT branes 
regardless of the values of the Cardy indices. 
Moreover, they actually view the Riemann surface $\mathcal{M}_{p,q}$ as the target space in $(p,q)$ 
minimal string theory.\cite{MMSS}  
However, 
in the following chapter we will show, 
that the above equation (\ref{SSproposal}) is not consistent with the generic cylinder amplitude in the $(p,q)$ minimal model 
coupled to 2D euclidean Quantum Gravity. 
It remains to be seen, if the above target space interpretation  
can be reconciled with this fact.  

\chapter{The generic FZZT-FZZT cylinder amplitude} \label{chaptercyl}
In this chapter we calculate the FZZT-FZZT cylinder amplitudes  
in the $(p,q)$ minimal model coupled to 2D euclidean Quantum Gravity 
for all pairs of Cardy matter states imposed on the two boundaries.  
This calculation is based upon the work of the author of this thesis 
done in collaboration with his academic adviser and is published in \cite{AG2,AG4} in the case of the $(2,2m-1)$ minimal model 
coupled to 2D euclidean Quantum Gravity. 
The calculation of the cylinder amplitude for a generic pair of Cardy matter states imposed on the two boundaries in the $(p,q)$ minimal model coupled to 2D euclidean Quantum Gravity, $p > 2$, will be submitted for publication in the near future. 
That said, the specific FZZT-FZZT cylinder amplitude in the unitary $(p,p+1)$ minimal model coupled to 2D euclidean Quantum Gravity with the Cardy matter states 
$\vert r, 1 \rangle$ and $\vert s, 1 \rangle$ imposed on the two boundaries of the cylinder has previously been determined in \cite{KOPSS,Martinec}. 
The cylinder amplitude obtained here generalizes the results 
obtained in \cite{KOPSS,Martinec}.     
As we will see in the following discussion the structure of the FZZT-FZZT cylinder amplitude becomes clear once 
we consider the cylinder amplitude in a generic $(p,q)$ minimal model coupled to 2D euclidean Quantum Gravity.

In the closed string channel the FZZT-FZZT cylinder amplitude is given by 
\[\label{cyldef}
\mathcal{Z}((r,s)_{_{C}},\sigma_{1};(k,l)_{_{C}},\sigma_{2}) 
= 
\int_{0}^{\infty} d\tau \, \mathcal{Z}_{M}((r,s);(k,l)) \, \mathcal{Z}_{L}(\sigma_{1};\sigma_{2}) \, \mathcal{Z}_{G}\,, 
\]
where $\tau$ is the single real moduli of the cylinder.  
\begin{figure}
\begin{center}
\includegraphics[width=0.5\textwidth]{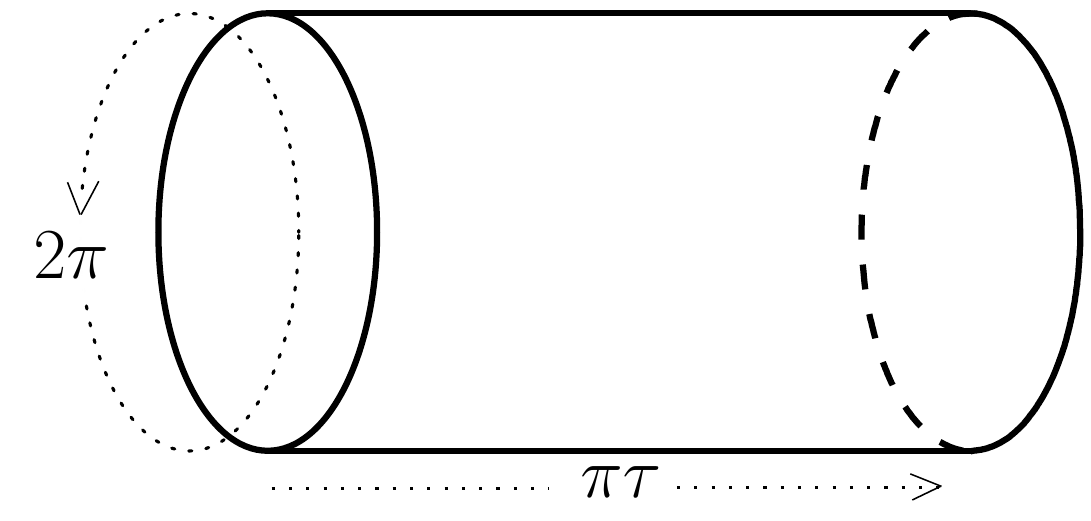}
\end{center}
\caption{Finite cylinder worldsheet of length $\pi\tau$, where $\tau$ is the single real moduli of the cylinder.}
\end{figure}
The matter cylinder amplitude is given by 
\st
\mathcal{Z}_{M}((r,s)_{_{C}};(k,l)_{_{C}}) 
& = & 
\phantom{}_{_{C}}\langle r,s \vert e^{-\pi\tau H_{c}^{M}} \vert k,l \rangle_{_{C}}
\nonumber\\
& = &
\begin{array}{c}
\displaystyle\sum_{m=1}^{p-1} \,\,\, 
\sum_{n=1}^{q-1}
\\
\textrm{\scriptsize{$mq\!-\!np>0$}}
\end{array} 
\frac{S_{(r,s);(m,n)}S_{(k,l);(m,n)}}{S_{(1,1);(m,n)}} 
\textrm{Tr}_{L(c_{p,q},\Delta_{m,n})} \,
e^{-2\pi\tau (L^{M}_{0}-c_{p,q}/24)} 
\nonumber\\
& = &
\sum_{m,n}
\frac{S_{(r,s);(m,n)}S_{(k,l);(m,n)}}{S_{(1,1);(m,n)}} 
\chi_{m,n}(q_{c})\,,  \qquad q_{c} = e^{-2\pi\tau}
\label{Zm}
\en
where the cylinder Hamiltonian in the closed string channel is given by eq. (\ref{hamcyl2}) 
and where we have applied the definition (\ref{Cardydef}) for the Cardy matter states 
and the definition (\ref{Ishibashistate}) for the Ishibashi states. 
The Virasoro character $\chi_{m,n}(q_{c})$ is determined in eq. (\ref{virchamin}). 
The Liouville cylinder amplitude is given by 
\st
\mathcal{Z}_{L}(\sigma_{1};\sigma_{2}) 
& = & 
\phantom{}_{_{FZZT}}\langle \sigma_{1} \vert e^{-\pi\tau H_{c}^{L}} \vert \sigma_{2} \rangle_{_{FZZT}}
\nonumber\\
& = &
\int_{0}^{\infty} d P 
\,\Psi_{\sigma_{1}}(P) 
\,\bar{\Psi}_{\sigma_{2}}(P) 
\,\textrm{Tr}_{\mathcal{F}(c_{L},P)} \,
e^{-2\pi\tau (L^{L}_{0}-c_{L}/24)} 
\nonumber\\
& = &
\int_{0}^{\infty} d P 
\,\Psi_{\sigma_{1}}(P) 
\,\bar{\Psi}_{\sigma_{2}}(P) 
\,
\chi_{P}(q_{c})\,,
\label{Zl}
\en
where we have expanded both the FZZT boundary states in Ishibashi states according 
to eqs. (\ref{genboundstate}) and (\ref{Liouvilleconwave}) and inserted the definition (\ref{Ishibashistate}) for the Ishibashi states. 
The Virasoro character $\chi_{P}(q_{c})$ is defined in eq. (\ref{virchaP}). 
Finally, in the closed string channel the ghost cylinder amplitude is given 
by\footnote{The single real moduli $\tau$ of the cylinder gives rise to the insertion of the ghost operator 
$b_{0}$ into the ghost cylinder amplitude. The single conformal Killing vector of the cylinder 
generating rigid rotations parallel to the boundaries  
gives rise to the insertion of the ghost operator $c_{0}$ into the ghost cylinder amplitude. \cite{pol}} \cite{pol}
\[
\mathcal{Z}_{G} 
= 
\langle B \vert c_{0} b_{0} e^{-\pi\tau H_{c}^{gh}} \vert B \rangle\,, 
\]
where the ghost boundary state is defined by
\[\label{boundghost}
\vert B \rangle 
= 
(c_{0}+\bar{c}_{0})
\exp\left( -\sum_{n=1}^{\infty}(b_{-n}\bar{c}_{-n}+\bar{b}_{-n}c_{-n}) \right)
\vert \downarrow, \downarrow \rangle\,, 
\] 
where $\vert \downarrow,\downarrow \rangle$ is the ghost vacuum defined in \cite{pol}. 
Applying the anti-commutator relations 
concerning the ghost operators given in \cite{pol} we obtain that  
\[\label{inghost}
\langle B \vert c_{0}b_{0} (c_{0}+\bar{c}_{0})
\prod_{i=1}^{m}(-b_{-k_{i}}\bar{c}_{-k_{i}})
\prod_{j=1}^{n}(-\bar{b}_{-\tilde{k}_{j}}c_{-\tilde{k}_{j}}) 
\vert \downarrow,\downarrow \rangle 
= 
(-1)^{m+n}\,.
\]
From the expressions for the Virasoro generators $L_{0}^{gh}$ and $\bar{L}_{0}^{gh}$ in terms of ghost operators given in \cite{pol}, 
the anti-commutator relations concerning the ghost operators and 
the above relation we may easily show that   
\[\label{Zg}
\mathcal{Z}_{G} = 
\eta(q_{c})^{2}\,,
\]
where $\eta$ is the Dedekind eta function. 
Inserting eqs. (\ref{Zm}), (\ref{Zl}) and (\ref{Zg}) into eq. (\ref{cyldef}) we obtain  
\begin{samepage}
\[
\mathcal{Z}((r,s)_{_{C}},\sigma_{1};(k,l)_{_{C}},\sigma_{2}) 
\nonumber
\]
\st
& = &  
\sum_{m,n}
\frac{S_{(r,s);(m,n)}S_{(k,l);(m,n)}}{S_{(1,1);(m,n)}} 
\int_{0}^{\infty}\!\! d P 
\,\Psi_{\sigma_{1}}(P) 
\,\bar{\Psi}_{\sigma_{2}}(P) 
\int_{0}^{\infty} \!\!d\tau
\,\chi_{m,n}(q_{c})  
\,\chi_{P}(q_{c})
\,\eta(q_{c})^{2}
\nonumber\\
\label{truecyl}
\en
\end{samepage}We will determine the cylinder amplitude by transforming the integration over the Liouville momentum $P$ 
into a sum over residues in the complex $P$-plane. 
The analytical structure of the integrand viewed as a function of $P$ is therefore crucial to our derivation of the cylinder amplitude. 
Let us start out by performing the integral our the moduli $\tau$. 
This integral is obtained in eq. (\ref{remark2}) in appendix \ref{app2}. 
\st
&&
\phantom{mapmapmapmapmap}
\int_{0}^{\infty} \!\!d\tau
\,\chi_{m,n}(q_{c})  
\,\chi_{P}(q_{c})
\,\eta(q_{c})^{2} 
\nonumber\\
& = & 
\frac{\sinh\left(\frac{2 \pi P}{\sqrt{pq}}\right)}{2\sqrt{pq}P}
\left\{ 
\frac{1}{\cosh\left(\frac{2 \pi P}{\sqrt{pq}}\right) - \cos\left(\frac{\pi(mq-np)}{pq} \right)}
- 
\frac{1}{\cosh\left(\frac{2 \pi P}{\sqrt{pq}}\right) - \cos\left(\frac{\pi(mq+np)}{pq} \right)}
\right\}
\nonumber\\
& \equiv &
h_{mn}(P)\,.
\label{remark}
\en
This result is quite remarkable. 
The poles in the above expression are located at 
\[
P  =  \pm i \frac{2tpq + mq \pm np}{2\sqrt{pq}}\,,\; t\in\mathbf{Z}
\]
which are exactly the points in the complex $P$-plane 
at which the cohomology group $\mathcal{H}_{\ast}\left(\Delta_{m,n},P\right)$ is non-vanishing, that is 
the above expression has a pole at each   
value of the complex Liouville momentum $P$ associated with a physical closed string state 
with matter content belonging to 
the irreducibel matter representation $L(c_{p,q},\Delta_{m,n}) \otimes \tilde{L}(c_{p,q},\Delta_{m,n})$. 
If we take the summation in eq. (\ref{truecyl}) 
over all the different irreducible matter representations 
$L(c_{p,q},\Delta_{m,n}) \otimes \tilde{L}(c_{p,q},\Delta_{m,n})$ into account 
we realize, that the integrand in eq. (\ref{truecyl}) has a simple pole at each point in the complex Liouville momentum plane 
associated with a physical closed string state.    
Let us define 
\st
f\left(i\frac{2\pi P}{\sqrt{pq}}\right) 
& \equiv & 
\frac{p q P}{\sqrt{2}\sinh\left(\frac{2\pi P}{\sqrt{pq}}\right)} 
\begin{array}{c}
\displaystyle\sum_{m=1}^{p-1} \,\,\, 
\sum_{n=1}^{q-1}
\\
\textrm{\scriptsize{$mq\!-\!np>0$}}
\end{array} 
\!\!\frac{S_{(r,s);(m,n)}S_{(k,l);(m,n)}}{S_{(1,1);(m,n)}} 
\,h_{m n}(P)
\nonumber\\
& = &
\sum_{m,n} 
(-1)^{1+m(s+l+1)+n(r+k+1)} 
\frac{ \sin\left(\frac{\pi q r m}{p}\right) \sin\left(\frac{\pi q k m}{p}\right) }{\sin\left(\frac{\pi q m}{p}\right)}
\frac{ \sin\left(\frac{\pi p s n}{q}\right) \sin\left(\frac{\pi p l n}{q}\right) }{\sin\left(\frac{\pi p n}{q}\right)}
\nonumber\\
&&
\times 
\left\{ 
\frac{1}{\cosh\left(\frac{2 \pi P}{\sqrt{pq}}\right) - \cos\left(\frac{\pi(mq-np)}{pq} \right)}
- 
\frac{1}{\cosh\left(\frac{2 \pi P}{\sqrt{pq}}\right) - \cos\left(\frac{\pi(mq+np)}{pq} \right)}
\right\}\,,
\nonumber\\
\label{ffunc}
\en
where we have applied the definition of the modular S-matrix given in eq. (\ref{mods}). 
It follows from eqs. (\ref{defsetA}) and (\ref{setA}) derived in appendix \ref{app2}, that 
we may express the set of points in the complex $P$-plane, 
at which $f$ has a pole, as     
\st
\mathcal{A}
& = &
\bigg\{
\pm i \frac{2tpq + mq \pm np}{2\sqrt{pq}}\bigg\vert\; 
t\in\mathbf{Z}\,,\; 
1 \leq m \leq p-1\,,\;
1 \leq n \leq q-1\,,\;
mq-np>0 
\bigg\}
\nonumber\\
& = & 
\bigg\{
\frac{it}{2\sqrt{pq}}\bigg\vert 
t\in\mathbf{Z}\,,\;
p \ndiv t \,,\;
q \ndiv t \bigg\}\,,
\label{paraphyspoles}
\en 
where the symbol $\ndiv$ means "does not divide". 

Quite surprisingly, the integrand in expression (\ref{truecyl}) for the cylinder amplitude contains more 
poles than the poles associated with physical closed string states. 
From eq. (\ref{FZZTboundfunc}) and from eq. (6.1.31) in \cite{Ab} we obtain 
\[\label{prodFZZTwavefunc}
\Psi_{\sigma_{1}}(P)\bar{\Psi}_{\sigma_{2}}(P) 
= 
\frac{1}{\sqrt{2}}\frac{\cos(2\pi P \sigma_{1})\cos(2\pi P \sigma_{2})}{\sinh(\frac{2\pi P}{b})\sinh(2\pi Pb)}\, 
\]
which has poles at 
\[
P = \frac{it}{2\sqrt{pq}}\,,\qquad t\in \mathbf{Z}\,,\quad t \;\textrm{mod } p = 0 \quad \textrm{or} \quad t \;\textrm{mod } q = 0\,.  
\] 

All the poles in the integrand in eq. (\ref{truecyl}) are simple poles except the pole at zero, which is a second order pole. 
In order to regularize this second order pole at zero and make the integral in eq. (\ref{truecyl}) convergent, 
we introduce the factor  
\[\label{regularization}
\frac{P^{2}}{(P+i\epsilon)(P-i\epsilon)}
\]
into the integrand, 
which splits the second order pole at zero into pole at $i\epsilon$ and a pole at $-i\epsilon$. 
In the following calculation we will assume that $ 0 < \sigma_{2} < \sigma_{1}$. 
The cylinder amplitude for generic values of $\sigma_{1}$ and $\sigma_{2}$ 
is obtained by analytical continuation of the result obtained under the assumption $ 0 < \sigma_{2} < \sigma_{1}$. 
Let us split the FZZT wave function $\Psi_{\sigma_{1}}(P)$ given by eq. (\ref{FZZTboundfunc}) into two terms 
\[
\Psi_{\sigma_{1}}(P) 
= 
\frac{1}{2}
\left\{
\Psi^{+}_{\sigma_{1}}(P) 
+ 
\Psi^{-}_{\sigma_{1}}(P)
\right\}\,,
\]
where 
\[
\Psi^{\pm}_{\sigma_{1}} 
\equiv 
-\frac{1}{2^{5/4}\pi}
\left(
\pi \mu \gamma(b^{2})
\right)^{-iP/b} 
\Gamma(1+2ibP) 
\Gamma(1+2iP/b) 
\frac{e^{\pm i 2 \pi \sigma P}}{iP}\,.
\]
Corresponding to eq. (\ref{prodFZZTwavefunc}) we obtain 
\[\label{boundprod}
\Psi^{\pm}_{\sigma_{1}}(P)\bar{\Psi}_{\sigma_{2}}(P) 
= 
\frac{1}{\sqrt{2}}\frac{e^{\pm i 2\pi P \sigma_{1}}\cos(2\pi P \sigma_{2})}{\sinh(\frac{2\pi P}{b})\sinh(2\pi Pb)}
\,.
\]
Due to the fact, that the integrand in (\ref{truecyl}) is an even function in $P$ we may express the cylinder amplitude as 
an integral along the entire real axis in the complex $P$-plane.  
Splitting $\Psi_{\sigma_{1}}(P)$ into the two above terms 
and applying the assumption $0 < \sigma_{2} < \sigma_{1}$ 
we may divide the expression (\ref{truecyl}) for the cylinder amplitude 
into two terms, one integral, which we close in the upper half plane, 
and one integral, which we close in the lower half plane. 
\st
\mathcal{Z}((r,s)_{_{C}},\sigma_{1};(k,l)_{_{C}},\sigma_{2}) 
\!\!
& = & 
\!\!
\frac{1}{4}
\left\{
\sum_{m,n}
\frac{S_{(r,s);(m,n)}S_{(k,l);(m,n)}}{S_{(1,1);(m,n)}} 
\oint_{\gamma_{+}}\!\!\! d P 
\frac{\Psi^{+}_{\sigma_{1}}(P) 
\,\bar{\Psi}_{\sigma_{2}}(P) 
\, h_{mn}(P)\,P^{2}}{(P+i\epsilon)(P-i\epsilon)}
\right.
\nonumber\\
&&\!\!
\left. +  
\sum_{m,n}
\frac{S_{(r,s);(m,n)}S_{(k,l);(m,n)}}{S_{(1,1);(m,n)}} 
\oint_{\gamma_{-}}\!\!\! d P 
\frac{\Psi^{-}_{\sigma_{1}}(P) 
\,\bar{\Psi}_{\sigma_{2}}(P) 
\, h_{mn}(P)\,P^{2}}{(P+i\epsilon)(P-i\epsilon)}
\right\}
\nonumber\\
\label{cylint}
\en
where the closed contour $\gamma_{+}$ encircle all the poles in the upper half plane counterclockwise 
and where the closed contour $\gamma_{-}$ encircle all the poles in the lower half plane clockwise.  
Applying Cauchy's theorem we may express the cylinder amplitude as a sum over residues. 
Taking the opposite directions of the two contours $\gamma_{+}$ and $\gamma_{-}$ into account 
it follows from eqs. (\ref{remark}) and (\ref{boundprod}) 
that the contribution to the cylinder amplitude from a given pole at $-P$ equals the contribution to the cylinder amplitude 
from the corresponding pole at $P$. 
Hence, we may express the cylinder amplitude entirely as a sum over residues in the upper half plane of the complex $P$-plane. 
Let us parametrize the set of points in the upper half plane at which $h_{mn}(P)$ has a pole as 
\[
P^{+}_{m,n,t,\delta} = i \frac{2pqt + mq + \delta\,np}{2\sqrt{pq}}\,,\quad t \in \mathbf{N}_{0}\,,\quad \delta \in \{+,-\} 
\] 
and 
\[
P^{-}_{m,n,t,\delta} = i \frac{2pqt - mq - \delta\,np}{2\sqrt{pq}}\,,\quad t \in \mathbf{N}\,,\quad \delta \in \{+,-\} \,.
\]
The residue at a given pole in the function $h_{mn}(P)$ is easily determined from eq. (\ref{remark}) 
and from Cauchy's theorem we obtain 
\st
&&
\mathcal{Z}((r,s)_{_{C}},\sigma_{1};(k,l)_{_{C}},\sigma_{2}) 
\nonumber\\
& = & 
\begin{array}{c}
\displaystyle\sum_{m=1}^{p-1} \,\,\, 
\sum_{n=1}^{q-1}
\\
\textrm{\scriptsize{$mq\!-\!np>0$}}
\end{array} 
\left\{
\sum_{t=0}^{\infty} \sum_{\delta \in \{+,-\}} 
\frac{\delta}{4 i P^{+}_{m,n,t,\delta}} 
\left[
\frac{S_{(r,s);(m,n)}S_{(k,l);(m,n)}}{S_{(1,1);(m,n)}} \Psi^{+}_{\sigma_{1}}(P^{+}_{m,n,t,\delta})\bar{\Psi}(P^{+}_{m,n,t,\delta})
\right]\right.
\nonumber\\
&&
\phantom{ma}
\left.+
\sum_{t=1}^{\infty} \sum_{\delta \in \{+,-\}} 
\frac{\delta}{4 i P^{-}_{m,n,t,\delta}} 
\left[
\frac{S_{(r,s);(m,n)}S_{(k,l);(m,n)}}{S_{(1,1);(m,n)}} 
\Psi^{+}_{\sigma_{1}}
(P^{-}_{m,n,t,\delta})\bar{\Psi}(P^{-}_{m,n,t,\delta})
\right]\right\}
\nonumber\\
&&
\phantom{mapm}
+ 
\frac{1}{pq}\sum_{j=1, \,\,p \ndiv j}^{\infty}(-1)^{j}
\frac{e^{-\frac{\pi j \sigma_{1}}{b}}\cosh(\frac{\pi j \sigma_{2}}{b})\sin(\frac{\pi j}{p})f(\frac{\pi j}{p})}
{j\sin(\frac{\pi j}{b^{2}})}
\nonumber\\
&&
\phantom{mapmapm}
+ 
\frac{1}{pq}\sum_{j=1,\,\,q \ndiv j}^{\infty}(-1)^{j}
\frac{e^{-\pi j \sigma_{1}b}\cosh(\pi j \sigma_{2}b)\sin(\frac{\pi j}{q})f(\frac{\pi j}{q})}
{j\sin(\pi j b^{2})}
\nonumber\\
&&
\phantom{mmapmapmap}
+ 
\frac{1}{(p q)^{2}}\sum_{j=1}^{\infty}\frac{(-1)^{j(p+q+1)}}{j}
e^{-\pi j \sqrt{p q} \sigma_{1}}\cosh(\pi j \sqrt{p q} \sigma_{2})f(\pi j)
\nonumber\\
&&
\phantom{mmapmapmapmap}
+
\frac{f(0)}{4(p q)^{3/2}}\left(\frac{1}{\epsilon}-2\pi\sigma_{1}\right)\,,
\label{cylsumres}
\en
where the even function $f$ is defined in eq. (\ref{ffunc}). 
The last line in the above expression contains the divergent part 
and the finite part in the cut off $\epsilon$ of the residue at $P=i\epsilon$. 
In the second last line we sum over the residues at $P=\frac{ik}{2\sqrt{pq}}$, $k\;\textrm{mod } pq = 0$. 
In the third line from the bottom we sum over the residues at $P=\frac{ik}{2\sqrt{pq}}$, 
$k\;\textrm{mod } p = 0$, $q \ndiv k$, 
and in the fourth line from the bottom we sum over the residues at $P=\frac{ik}{2\sqrt{pq}}$, 
$k\;\textrm{mod } q = 0$, $p \ndiv k$.     
All the residues in the second line to the fourth line counted from the bottom  
are associated with poles appearing in the product of the two FZZT wave functions given by eq. (\ref{boundprod}). 
Finally, in the first two lines  
we sum over all the residues,  
which we may associate with physical closed string states as discussed previously. 
This sum is particularly interesting and let us denote it by $\mathcal{S}$. 
Inside the curly bracket we sum over all the residues associated with the closed string states 
with a matter content belonging to 
the irreducible matter representation $L(c_{p,q},\Delta_{m,n}) \otimes \tilde{L}(c_{p,q},\Delta_{m,n})$. 
Outside the curly bracket we sum over all the irreducible representations in the $(p,q)$ minimal model.   
From the expression (\ref{onepointFZZT}) for the one-point function evaluated on the disk we realize, 
that  
the residue associated with a given physical closed string state $\vert \hat{\mathcal{A}}_{r,s,p} \rangle$ 
is essentially given by the product of the two disk amplitudes 
$\langle \sigma_{1}; r,s \vert \hat{\mathcal{A}}_{r,s,p} \rangle$ 
and      
$\langle \hat{\mathcal{A}}_{r,s,p} \vert \sigma_{2}; k,l \rangle$. 
This is true up to a constant independent of the boundary conditions. 
\emph{Thus, we may interpret a given residue in the sum $S$ as the amplitude of the corresponding 
closed string state propagating between the two branes.} 
The observant reader may object, that the boundary wave function $\Psi^{+}_{\sigma_{1}}(P)$ is 
not really the FZZT wave function $\Psi_{\sigma_{1}}(P)$. 
However, we may decompose the boundary wave function $\Psi^{+}_{\sigma_{1}}(P)$ 
into an even part in $\sigma_{1}$ equal to the FZZT wave function $\Psi_{\sigma_{1}}(P)$ and an odd part and 
we may \emph{formally} divide the sum $S$ into a sum involving the even part of $\Psi^{+}_{\sigma_{1}}(P)$ and a sum involving the 
odd part of $\Psi^{+}_{\sigma_{1}}(P)$. 
We know from eqs. (\ref{truecyl}) and (\ref{FZZTboundfunc}), that 
the cylinder amplitude is an even function in $\sigma_{1}$. 
As we will see in a moment the even nature of the cylinder amplitude is restored by the finite part of the residue at $P=i\epsilon$. 
Thus, all the odd terms effectively cancel and the significant part of the boundary wave function 
$\Psi^{+}_{\sigma_{1}}(P)$ is the even part, which is identical to the FZZT boundary wave function $\Psi_{\sigma_{1}}(P)$. 
This argument is clearly of a formal nature. 
If we change $\Psi^{+}_{\sigma_{1}}(P) \to \Psi_{\sigma_{1}}(P)$ in the sum $S$ we obtain a divergent sum. 
The sum obtained by inserting a complete set of physical closed string states 
into the cylinder amplitude is strictly speaking ill-defined. 
\emph{Hence, we may view the sum $S$ as a convergent realization of the formal sum obtained by inserting a complete set of 
physical closed string states into the cylinder amplitude.}  
In light of the above discussion we understand why the contribution to the cylinder amplitude 
from a given pole located at $-P$ is equal to the contribution to the cylinder amplitude 
from the corresponding pole located at $P$. 
With regard to the poles associated with physical closed string states this is a 
manifestation of the fact, that we have to identify the physical closed string state with Liouville momentum $-P$ 
with the physical closed string state with Liouville momentum $P$.  
\begin{figure}[t]
\begin{center}
\includegraphics[width=0.7\textwidth]{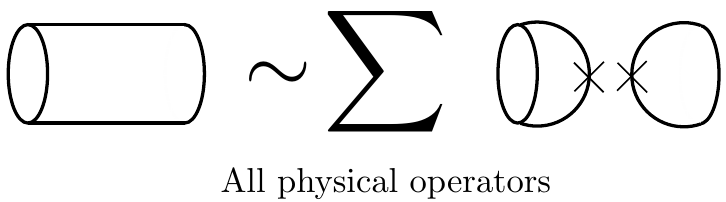}
\end{center}
\caption{The decomposition of the sum $S$ in terms of solutions to the Wheeler-DeWitt equation.} 
\label{figurdecomp}
\end{figure}

The structure of cylinder amplitude is similar to the structure of 
the classical Greens function encountered in the discussion of the Wick rotated harmonic oscillator. 
Let us consider the restriction of the operator 
\[
\hat{D} = -\frac{d^{2}}{dt^{2}} + \omega^{2} 
\]  
to the linear space of functions satisfying the boundary condition  
\[
\psi(t) \to 0 \quad \textrm{for} \quad t \to \pm \infty\,.   
\]
(More precisely, we consider $\hat{D}$ on $L^{2}(\mathbf{R})$). 
With respect to the usual inner product  
the operator $\hat{D}$ is Hermitian 
due to the above boundary condition.
In this linear space of functions there does not exist any solution to the homogeneous equation 
\[\label{homeq}
\hat{D} \psi(t) = 0\,.
\] 
The operator $\hat{D}$ is therefore invertible and we may try to determine the inverse operator 
given in terms of the Greens function $G(t,s)$ defined by  
\[
\hat{D}G(t,s) = \left[-\frac{d^{2}}{dt^{2}} + \omega^{2}\right]G(t,s) = \delta(t-s)\,.
\] 
Given a complete set of eigenfunction $\psi_{n}(t)$ with eigenvalues $\lambda_{n}$   
of the Hermitian operator $\hat{D}$ we may express the Greens function as 
\[\label{invharham}
G(t,s) 
= \sum_{m,n} \langle t \vert n \rangle \langle n \vert \frac{1}{\hat{D}} \vert m \rangle \langle m \vert s \rangle
= \sum_{n} \frac{\bar{\psi}_{n}(s)\psi_{n}(t)}{\lambda_{n}}\,. 
\]
In the case of the Hermitian operator $\hat{D}$ the properly normalized eigenfunctions are given by 
$\psi_{p} = \frac{1}{\sqrt{2\pi}}e^{i p t}$ with eigenvalues $p^{2}+\omega^{2}$. 
These eigenfunction do strictly speaking not belong to $L^{2}(\mathbf{R})$. 
However, they remain bounded as $t \to \pm\infty$ and we may consider them as generalized eigenfunctions. 
From the above expression we obtain 
\[
G(t,s) = \int_{-\infty}^{\infty} \frac{dp}{2\pi} \frac{e^{ip(t-s)}}{p^{2}+\omega^{2}}\,,
\]
which we recognize as the ordinary Fourier transform. 
In the case $t>s$ we perform the above integration by closing the integration contour in the upper half plane. 
In this case the Greens function is given by $2\pi i$ times the residue of the single pole in the upper half plane 
located at $p=i\omega$. 
\[\label{greens}
G(t,s) = 
\frac{e^{-\omega(t-s)}}{2\omega} 
= 
\frac{e^{-\omega \vert t-s\vert}}{2\omega}\,,
\]
where the final expression is valid even for $t<s$. 
Thus, the Greens function is essentially given by the 
product of the two \emph{non-normalizable} solutions $e^{-\omega t}$ and $e^{\omega s}$ to the homogeneous equation (\ref{homeq}).   

With regard to the cylinder amplitude we start out in eq. (\ref{cyldef}) by consider the restriction of the Hermitian operator 
$e^{-\pi\tau(L^{tot}_{0}+\bar{L}^{tot}_{0}) }$ to the Hilbert space of normalizable states  
\[
\mathcal{H}_{tot} 
= 
\mathcal{H}_{p,q} 
\otimes 
\mathcal{H}_{L} 
\otimes 
\mathcal{H}_{ghost}\,.
\]
This is true in the sense, that we may expand the boundary states in terms of normalizable states belonging to this Hilbert space. 
This is obtained by expanding the matter and the Liouville boundary states in Ishibashi states, 
which we furthermore expand as in eq. (\ref{Ishibashistate}), 
and by expanding the exponential operator appearing in the definition (\ref{boundghost}) of the 
ghost boundary state.\footnote{Notice, in addition to the operator $e^{-\pi\tau(L^{tot}_{0}+\bar{L}^{tot}_{0}) }$ the ghost operator 
$c_{0}b_{0}$ appears in the cylinder amplitude. 
The importance of this insertion becomes apparent, when we consider the inner product of the ghost states. 
The two ghost boundary states imposed on the opposite sides of the cylinder each contribute 
with a factor $(c_{0}+\bar{c}_{0})$ in the cylinder amplitude and we may express 
\[
(c_{0}+\bar{c}_{0})c_{0}b_{0}(c_{0}+\bar{c}_{0}) = \bar{c}_{0}c_{0}\,. 
\] 
Due to the fact, that  
\[
\langle \downarrow,\downarrow  \vert \downarrow,\downarrow \rangle = 0\,,
\]
while 
\[
\langle \downarrow,\downarrow \vert \bar{c}_{0}c_{0}  \vert \downarrow,\downarrow \rangle 
= 
\langle \downarrow,\downarrow  \vert \uparrow,\uparrow \rangle = 1\,.
\]
the operator $\bar{c}_{0}c_{0}$ appearing in the cylinder amplitude ensures, 
that the ghost cylinder amplitude is non-vanishing.} 
Notice, this is an expansion in terms of eigenstates of the cylinder Hamiltonian $L_{0}^{tot} + \bar{L}_{0}^{tot}$.
The eigenvalues of the cylinder Hamiltonian $L^{tot}_{0}+\bar{L}^{tot}_{0}$ restricted to the above Hilbert space are all 
strictly positive.\footnote{It follows from B\`ezout's theorem, that there exists $r \in \{1,\dots,p-1\}$ and $s \in \{1,\dots,q-1\}$ satisfying  
\[
rq-sp = \textrm{gcd}(p,q) = 1\,.
\]
Moreover, it follows from the Virasoro algebra (\ref{viralgebra}), 
that the conformal dimension of a given descendant state is larger than the conformal dimension of the corresponding primary state. 
From B\`ezout's theorem, eqs. (\ref{condimmin}) and (\ref{confdim}) 
and from the anti-commutator relations concerning the ghost operators given in \cite{pol} we obtain the following lower bound 
on the set of eigenvalues $\lambda$ of the cylinder Hamiltonian restricted to the Hilbert space $\mathcal{H}_{tot}$      
\[ 
\lambda 
> 
2\left(
\Delta_{r,s} + \Delta(Q/2) - 1
\right)  
= 
\frac{1}{2pq}\,.  
\]
}    
This implies that the integration over $\tau$ is convergent and performing this integration we obtain 
\[
e^{-\pi\tau(L^{tot}_{0}+\bar{L}^{tot}_{0}) } \;
\longrightarrow \;
\frac{1}{\pi}\frac{1}{L^{tot}_{0}+\bar{L}^{tot}_{0}}\,.
\] 
Expanding the boundary states in normalizable eigenstates of the cylinder Hamiltonian     
we obtain an expression similar to 
eq. (\ref{invharham}). 
Due to the analytic structure of the cylinder amplitude  
we may close the integration contour in the complex $P$-plane and by Cauchy's theorem we may 
express the cylinder amplitude as a sum over residues. 
At least with respect to the residues associated with physical closed string states and up to a constant 
independent of the boundary conditions 
we may express a given residue as the product of two disk amplitudes 
$\langle \sigma_{1}; r,s \vert \hat{\mathcal{A}}_{r,s,p} \rangle$ 
and      
$\langle \hat{\mathcal{A}}_{r,s,p} \vert \sigma_{2}; k,l \rangle$, 
where the \emph{non-normalizable} physical closed string state 
$\vert \hat{\mathcal{A}}_{r,s,p} \rangle$ satisfies the homogeneous Wheeler-DeWitt equation 
\[
\left(
L^{tot}_{0}+\bar{L}^{tot}_{0}
\right) 
\vert \hat{\mathcal{A}}_{r,s,p} \rangle = 0\,.
\]  
Thus, we obtain an expression for the cylinder amplitude quite similar to eq. (\ref{greens}). 

Let us discuss the relation (\ref{SSproposal}) proposed by Seiberg and Shih in \cite{SS} and 
let us start out by considering the sum $S$ over residues associated with physical closed string states. 
We will consider the consistency of the relation (\ref{SSproposal}) with the full cylinder amplitude later on in this chapter.   
Instead of struggling with the expression for the sum $S$ given in eq. (\ref{cylsumres}) 
we return to the expression for cylinder amplitude given in eq. (\ref{cylint}). 
Inserting the expression obtained in (\ref{boundprod}) we may express the cylinder amplitude as 
\[\label{truecyl2}
\mathcal{Z}((r,s)_{_{C}},\sigma_{1};(k,l)_{_{C}},\sigma_{2}) 
=
\frac{1}{2pq}
\oint_{\gamma_{+}}\!\! d P \, 
\frac{\exp(i 2 \pi P \sigma_{1})\cos(2 \pi P \sigma_{2})\sinh\left(\frac{2 \pi P}{\sqrt{pq}}\right)f\left(\frac{i 2 \pi P}{\sqrt{pq}}\right)P}
{\sinh\left(2 \pi P\sqrt{\frac{p}{q}}\right)\sinh\left(2 \pi P\sqrt{\frac{q}{p}}\right)(P+i\epsilon)(P-i\epsilon)}\,, 
\]   
where the function $f$ is defined in eq. (\ref{ffunc}). 
Applying the parametrization of the set of poles associated with physical closed string states 
given in eq. (\ref{paraphyspoles}) we may express the sum $S$ as 
\newpage
\[
S((r,s)_{_{C}},\sigma_{1};(k,l)_{_{C}},\sigma_{2}) 
\nonumber
\]
\st
& = & 
\frac{2\pi}{i\sqrt{pq}}\!\!\!
\sum_{\begin{array}{c}
t=1 \\
\phantom{}^{p \,\ndiv \,t\,,\,q \,\ndiv \,t}
\end{array}}^{\infty}\!\!\!
\frac{\exp\left(-\frac{\pi t \sigma_{1}}{\sqrt{pq}}\right)\cosh\left(\frac{\pi t \sigma_{2}}{\sqrt{pq}}\right)
\sin\left(\frac{\pi t}{pq}\right)\textrm{Res}\left(f\left(\frac{i 2 \pi P}{\sqrt{pq}}\right);\frac{i t}{2\sqrt{pq}}\right)}
{t\sin\left(\frac{\pi t}{p}\right)\sin\left(\frac{\pi t}{q}\right)}
\nonumber\\
& = &
-\!\!\!
\sum_{\begin{array}{c}
t=1 \\
\phantom{}^{p \,\ndiv \,t\,,\,q \,\ndiv \,t}
\end{array}}^{\infty}\!\!\!
\frac{1}{t}\exp\left(-\frac{\pi t \sigma_{1}}{\sqrt{pq}}\right)\cosh\left(\frac{\pi t \sigma_{2}}{\sqrt{pq}}\right)
\frac{\sin\left(\frac{\pi r t}{p}\right)\sin\left(\frac{\pi k t}{p}\right)}{\sin^{2}\left(\frac{\pi t}{p}\right)}
\frac{\sin\left(\frac{\pi s t}{q}\right)\sin\left(\frac{\pi l t}{q}\right)}
{\sin^{2}\left(\frac{\pi t}{q}\right)}\,,
\nonumber\\
\label{sss}
\en
where we have inserted the expression for the residue of $f$ obtained in (\ref{resf}).\footnote{Notice, 
in eq. (\ref{resf}) we determine
\[
\textrm{Res}\left(f(z);\frac{\pi t}{pq}\right) 
= 
\frac{i2\pi}{\sqrt{pq}}
\textrm{Res}\left(f\left(\frac{i 2 \pi P}{\sqrt{pq}}\right);-\frac{i t}{2\sqrt{pq}}\right)\,.
\]} 
Applying the identity 
\[\label{sincos}
\frac{\sin(\pi k x)}{\sin(\pi x)} 
= 
\sideset{}{'}\sum_{\tilde{k} = -(k-1)}^{k-1}   
\cos(\pi \tilde{k} x),
\]
where the prime in the summation symbol denotes, that the summation runs in steps of two, 
and the fact that the values, which we sum over in this identity, are located symmetric 
around zero, we obtain 
\[
S((r,s)_{_{C}},\sigma_{1};(k,l)_{_{C}},\sigma_{2}) 
\]  
\[
= 
-
\sideset{}{'}\sum_{\tilde{r} = 1-r}^{r-1}
\;\sideset{}{'}\sum_{\tilde{s} = 1-s}^{s-1}
\;\sideset{}{'}\sum_{\tilde{k} = 1-k}^{k-1}
\;\sideset{}{'}\sum_{\tilde{l} = 1-l}^{l-1}
\sum_{\begin{array}{c}
t=1 \\
\phantom{}^{p \,\ndiv \,t\,,\,q \,\ndiv \,t}
\end{array}}^{\infty}\!\!\!
\left\{
\cos\left(\frac{\pi\tilde{r}t}{p}\right)
\cos\left(\frac{\pi\tilde{s}t}{q}\right)
\cos\left(\frac{\pi\tilde{k}t}{p}\right)
\cos\left(\frac{\pi\tilde{l}t}{q}\right)\right.\phantom{ma}
\nonumber
\]
\[
\left.
\phantom{\cos\left(\frac{\pi\tilde{l}t}{p}\right)}\phantom{mapma}
\times\frac{1}{t}\exp\left(-\frac{\pi t \sigma_{1}}{\sqrt{pq}}\right)\cosh\left(\frac{\pi t \sigma_{2}}{\sqrt{pq}}\right)\right\}
\nonumber
\]
\[
= 
-
\sideset{}{'}\sum_{\tilde{r} = 1-r}^{r-1}
\;\sideset{}{'}\sum_{\tilde{s} = 1-s}^{s-1}
\;\sideset{}{'}\sum_{\tilde{k} = 1-k}^{k-1}
\;\sideset{}{'}\sum_{\tilde{l} = 1-l}^{l-1}
\sum_{\begin{array}{c}
t=1 \\
\phantom{}^{p \,\ndiv \,t\,,\,q \,\ndiv \,t}
\end{array}}^{\infty}\!\!\!
\frac{1}{t}e^{-\frac{\pi t}{\sqrt{pq}} \left(\sigma_{1}+i\left(\frac{\tilde{r}}{b}+\tilde{s}b\right)\right)}
\cosh\!\left(\frac{\pi t}{\sqrt{pq}} \!\left\{\sigma_{2}+i\left(\frac{\tilde{k}}{b}+\tilde{l}b\right)\!\right\}\!\right)\!
\nonumber
\]
\[
= 
\sideset{}{'}\sum_{\tilde{r} = 1-r}^{r-1}
\;\sideset{}{'}\sum_{\tilde{s} = 1-s}^{s-1}
\;\sideset{}{'}\sum_{\tilde{k} = 1-k}^{k-1}
\;\sideset{}{'}\sum_{\tilde{l} = 1-l}^{l-1}
S\left((1,1)_{_{C}},\sigma_{1}+i\left(\frac{\tilde{r}}{b}+\tilde{s}b\right);(1,1)_{_{C}},\sigma_{2}+i\left(\frac{\tilde{k}}{b}+\tilde{l}b\right)\right) 
\phantom{mapmapma}
\label{SSconsistent}
\]
From the above calculation we realize, that the sum over the residues associated with physical closed string states is consistent with the 
relation (\ref{SSproposal}) proposed by Seiberg and Shih for all values of the Cardy matter indices. 
This is not surprising, since we may decompose each residue associated with a physical closed string state 
into the product of two disk amplitudes and since the relation proposed by Seiberg and Shih is valid on the disk. 

After having discussed the sum $S$ over residues associated with physical closed string states, 
let us now briefly comment on the poles in the cylinder amplitude, which we cannot associate with any physical closed string states in the BRST formalism. 
From eq. (\ref{gradim}) we may associate a gravitational scaling dimension 
to each of the poles in the integrand in (\ref{truecyl2}).  
Quite remarkable, the scaling dimensions associated to the poles located at 
\[
P = \frac{it}{2\sqrt{pq}}\,,\quad t \geq 1\,, \quad  t \;\textrm{mod } q = 0\,,\quad p \ndiv t  
\]
precisely match the gravitational scaling dimensions associated with the physical operators present in the matrix model approach 
to the $(p,q)$ minimal model coupled to 2D euclidean Quantum Gravity 
but not present in BRST formalism discussed in section \ref{physoperator}.  
In this sense we may associate these poles with the additional physical operators appearing in the matrix model approach. 
These poles appear in the product of the two FZZT wavefunctions given by eq. (\ref{prodFZZTwavefunc}). 
The sum over the residues corresponding to these poles is given by the fourth line from the bottom in eq. (\ref{cylsumres}).

Let us determine the FZZT-FZZT cylinder amplitude for all pairs of Cardy matter states. 
It follows from eq. (\ref{cylsumres}) that the cylinder amplitude depends explicitly on  
the function $f$ defined in eq. (\ref{ffunc}) and not only on the residues of $f$. 
For generic values of the Cardy indices $r$, $s$, $k$ and $l$ 
we are not able to express $f$ in any simple way. 
However, in the special case $k=l=1$ and $rq-sp>0$ and in the special case $k=l=1$ and $rq-sp<0$ 
we are actually able to determine a simpler expression for the function $f$ than the expression given 
in eq. (\ref{ffunc}).\footnote{Strictly speaking, the cylinder amplitude is only defined for $rq-sp>0$, 
since all Cardy matter states $\vert r,s \rangle_{_{C}}$ satisfy this condition. 
However, we choose to \emph{define} the matter cylinder amplitude by eq. (\ref{Zm}) in the case $rq-sp<0$. 
Applying this definition we may define the FZZT-FZZT cylinder amplitude by eq. (\ref{cyldef}) even in the case $rq-sp<0$.
This will prove convenient in our further discussion.} 
In appendix \ref{app2} we show that 
\[
f\left(\frac{i 2 \pi P}{\sqrt{pq}}\right) 
= 
pq\frac{\sinh\left(2 \pi (p-r) P \sqrt{\frac{q}{p}}\right)\sinh\left(2 \pi s  P \sqrt{\frac{p}{q}}\right)}
{\sinh\left( 2 \pi \sqrt{pq} P \right)\sinh\left(\frac{2 \pi P}{\sqrt{pq}}\right)}\,,
\]
valid for $k=l=1$ and $rq-sp>0$, and 
\[
f\left(\frac{i 2 \pi P}{\sqrt{pq}}\right) 
= 
pq\frac{\sinh\left(2 \pi r P \sqrt{\frac{q}{p}}\right)\sinh\left(2 \pi (q-s)  P \sqrt{\frac{p}{q}}\right)}
{\sinh\left( 2 \pi \sqrt{pq} P \right)\sinh\left(\frac{2 \pi P}{\sqrt{pq}}\right)}\,,
\]
valid for $k=l=1$ and $rq-sp<0$.
Let us begin by considering the case $k=l=1$ and $rq-sp>0$.
Using the above expression for $f$ and applying the identity (\ref{sincos}) 
we may easily determine the residue at $P=\frac{it}{2\sqrt{pq}}$ of the integrand $I$ in eq. (\ref{truecyl2}). 
\[
2 \pi i\,\textrm{Res}\left( I ; \left(\frac{it}{2\sqrt{p q}}\right)\right) 
\nonumber
\]
\[
= 
\pi i
\lim_{P \to \frac{it}{2\sqrt{p q}}}
\left(P - \frac{it}{2\sqrt{p q}}\right)
\frac{\sinh\left(2 \pi (p-r) P\sqrt{\frac{q}{p}}\right)}{\sinh\left(2 \pi P\sqrt{\frac{q}{p}}\right)}
\frac{\sinh\left(2 \pi s P\sqrt{\frac{p}{q}}\right)}{\sinh\left(2 \pi P\sqrt{\frac{p}{q}}\right)}
\frac{e^{i 2 \pi P \sigma_{1}}\cos(2 \pi P \sigma_{2})}
{\sinh\left(2\pi P \sqrt{p q}\right)P}
\phantom{ma}
\nonumber\]
\[
= 
\frac{2 \pi\sqrt{p q}}{t} 
\sideset{}{'}\sum_{\tilde{r}=1+r-p}^{p-r-1}
\;\sideset{}{'}\sum_{\tilde{s}=1-s}^{s-1}
\cos\left(\frac{\pi \tilde{r} t}{p}\right) \cos\left(\frac{\pi \tilde{s} t}{q}\right)
e^{-\frac{\pi t \sigma_{1}}{\sqrt{p q}}} \cosh\left(\frac{\pi t \sigma_{2}}{\sqrt{p q}}\right) 
\lim_{P \to \frac{it}{2\sqrt{p q}}} 
\frac{\left(P - \frac{it}{2\sqrt{p q}}\right)}{\sinh\left(2\pi P \sqrt{p q}\right)}
\nonumber
\]
\[ 
= 
\frac{(-1)^{t}}{t} 
\sideset{}{'}\sum_{\tilde{r}=1+r-p}^{p-r-1}
\;\sideset{}{'}\sum_{\tilde{s}=1-s}^{s-1}
\cos\left(\frac{\pi \tilde{r} t}{p}\right) \cos\left(\frac{\pi \tilde{s} t}{q}\right)
e^{-\frac{\pi t \sigma_{1}}{\sqrt{p q}}} \cosh\left(\frac{\pi t \sigma_{2}}{\sqrt{p q}}\right)
\phantom{mapmapmapmapmap}
\]    
Expanding the trigonometric and hyperbolic functions appearing in the above residue in terms of exponential functions 
we may express the cylinder amplitude as       
\st
&&
\phantom{mapmapmapmapmap}
\mathcal{Z}((r,s)_{_{C}},\sigma_{1};(1,1)_{_{C}},\sigma_{2}) 
\nonumber\\
& = & 
\frac{(p-r)s}{4\sqrt{p q}}\left(\frac{1}{\epsilon}-2\pi\sigma_{1}\right) 
\nonumber\\
&&
+  
\sideset{}{'}\sum_{\tilde{r}=1+r-p}^{p-r-1}
\;\sideset{}{'}\sum_{\tilde{s}=1-s}^{s-1}
\sum_{t=1}^{\infty} 
\frac{(-1)^{t}}{t} 
\cos\left(\frac{\pi \tilde{r} t}{p}\right) \cos\left(\frac{\pi \tilde{s} t}{q}\right)
e^{-\frac{\pi t \sigma_{1}}{\sqrt{p q}}} \cosh\left(\frac{\pi t \sigma_{2}}{\sqrt{p q}}\right)
\phantom{mapmap}
\nonumber\\
& = &
\frac{(p-r)s}{4\sqrt{p q}}\left(\frac{1}{\epsilon}-2\pi\sigma_{1}\right)
\nonumber\\
&&
-\frac{1}{2} 
\sideset{}{'}\sum_{\tilde{r}=1+r-p}^{p-r-1}
\;\sideset{}{'}\sum_{\tilde{s}=1-s}^{s-1} 
\ln\left[
\left\{
1+\exp\left(-\frac{\pi \sigma_{1}}{\sqrt{pq}}\right)
\exp\left(\frac{\pi \sigma_{2}}{\sqrt{pq}}+i\frac{\pi(\tilde{r}q-\tilde{s}p)}{pq}\right)
\right\}
\right.
\nonumber\\
&&
\phantom{mapmapmapmapamp}
\left. \times
\left\{
1+\exp\left(-\frac{\pi \sigma_{1}}{\sqrt{pq}}\right)
\exp\left(-\frac{\pi \sigma_{2}}{\sqrt{pq}}-i\frac{\pi(\tilde{r}q-\tilde{s}p)}{pq}\right)
\right\}
\right] 
\nonumber\\
& = &
-\frac{1}{4}
\sideset{}{'}\sum_{\tilde{r}=1+r-p}^{p-r-1}
\;\sideset{}{'}\sum_{\tilde{s}=1-s}^{s-1} 
\ln\left[
z_{1}^{2}+z_{2}^{2}+2z_{1}z_{2}
\cos\left(\frac{\pi(\tilde{r}q-\tilde{s}p)}{pq}\right)
-
\sin^{2}\left(\frac{\pi(\tilde{r}q-\tilde{s}p)}{pq}\right)
\right]
\nonumber\\
&&
+
\frac{(p-r)s}{4\sqrt{p q}\epsilon} 
-\frac{(p-r)s}{2}\ln 2\,,
\label{speFZZTcyl}
\en
where we have introduced the uniformization parameter (\ref{z}) in the final expression. 
Let us now consider the FZZT-FZZT cylinder amplitude in the special case $k=l=1$ and $rq-sp<0$. 
By a calculation almost identical to the calculation performed in the above we obtain  
\st
&&
\phantom{mapmapmapmapmap}
\mathcal{Z}((r,s)_{_{C}},\sigma_{1};(1,1)_{_{C}},\sigma_{2}) 
\nonumber\\
& = &
-\frac{1}{4}
\sideset{}{'}\sum_{\tilde{r}=1-r}^{r-1}
\;\sideset{}{'}\sum_{\tilde{s}=1+s-q}^{q-s-1}
\ln\left[
z_{1}^{2}+z_{2}^{2}+2z_{1}z_{2}
\cos\left(\frac{\pi(\tilde{r}q-\tilde{s}p)}{pq}\right)
-
\sin^{2}\left(\frac{\pi(\tilde{r}q-\tilde{s}p)}{pq}\right)
\right]
\nonumber\\
&&
+
\frac{r(q-s)}{4\sqrt{p q}\epsilon} 
-\frac{r(q-s)}{2}\ln 2\,.
\label{speFZZTcyl2}
\en
Thus, we have determined the FZZT-FZZT cylinder amplitude in the $(p,q)$ minimal model coupled to 2D euclidean Quantum Gravity 
in the special case, when one of the Cardy matter states is the $(1,1)$ Cardy matter state. 
Let us now determine the cylinder amplitude for generic values of the Cardy matter indices. 
Given the cylinder amplitudes (\ref{speFZZTcyl}) and (\ref{speFZZTcyl2}) this is actually easily done due to the fact, that 
we may express any given matter cylinder amplitude $\mathcal{Z}_{M}((r,s);(k,l))$ in terms of matter 
cylinder amplitudes with the $(1,1)$ Cardy matter state imposed on one of the boundaries. 
Applying the fact, that the spectrum of open string states, which flow between two given Cardy matter boundary conditions, is determined from  
the fusion algebra (\ref{fusionmin}) of the corresponding primary operators, 
we may express a given matter cylinder amplitude in the open string channel 
as\footnote{Strictly speaking, we ought to prove the identity 
\[
\mathcal{Z}_{M}((i,j)_{_{C}};(1,1)_{_{C}}) = \chi_{i,j}(\tilde{q}) 
\]
in the case $iq-jp<0$, in which the matter cylinder amplitude is defined by eq. (\ref{Zm}). 
This is easily done. From eq. (\ref{mods}) and eq. (10.133) in \cite{francesco} we obtain 
\st
\mathcal{Z}_{M}((i,j)_{_{C}};(1,1)_{_{C}}) 
& = & 
\begin{array}{c}
\displaystyle\sum_{m=1}^{p-1} \,\,\, 
\sum_{n=1}^{q-1}
\\
\textrm{\scriptsize{$mq\!-\!np>0$}}
\end{array} 
S_{(i,j);(m,n)} \chi_{m,n}(q_{c}) 
\nonumber\\
& = &
\begin{array}{c}
\displaystyle\sum_{m=1}^{p-1} \,\,\, 
\sum_{n=1}^{q-1}
\\
\textrm{\scriptsize{$mq\!-\!np>0$}}
\end{array} 
S_{(p-i,q-j);(m,n)} \chi_{m,n}(q_{c})
\nonumber\\
& = &
\chi_{p-i,q-j}(\tilde{q}) 
= 
\chi_{i,j}(\tilde{q})\,, 
\en
where we have applied the symmetry of the Kac table.} 
\st
\mathcal{Z}_{M}((r,s)_{_{C}};(k,l)_{_{C}}) 
& = & 
\sideset{}{'}\sum_{i = \vert r - k \vert + 1}^{\textrm{min}(r+k-1,2p-1-r-k)}
\;\sideset{}{'}\sum_{j = \vert s - l \vert + 1}^{\textrm{min}(s+l-1,2q-1-s-l)}
\chi_{i,j}(\tilde{q}) 
\nonumber\\
& = &
\sideset{}{'}\sum_{i = \vert r - k \vert + 1}^{\textrm{min}(r+k-1,2p-1-r-k)}
\;\sideset{}{'}\sum_{j = \vert s - l \vert + 1}^{\textrm{min}(s+l-1,2q-1-s-l)}
\mathcal{Z}_{M}((i,j)_{_{C}};(1,1)_{_{C}})\,. 
\nonumber\\
\label{step1}
\en  
Inserting this result into the initial expression (\ref{cyldef}) for the FZZT-FZZT cylinder amplitude we obtain 
\[
\mathcal{Z}((r,s)_{_{C}},\sigma_{1};(k,l)_{_{C}},\sigma_{2}) 
= 
\sideset{}{'}\sum_{i = \vert r - k \vert + 1}^{\textrm{min}(r+k-1,2p-1-r-k)}
\;\sideset{}{'}\sum_{j = \vert s - l \vert + 1}^{\textrm{min}(s+l-1,2q-1-s-l)}
\mathcal{Z}((i,j)_{_{C}},\sigma_{1};(1,1)_{_{C}},\sigma_{2})\,. 
\] 
Given eqs. (\ref{speFZZTcyl}) and (\ref{speFZZTcyl2}) we have actually determined 
the FZZT-FZZT cylinder amplitude for all values of the Cardy indices.  

Let us future reference explicitly write down the FZZT-FZZT cylinder amplitudes in the $(2,2m-1)$ minimal model coupled to 2D euclidean Quantum Gravity.
In this case the first Cardy matter label is always equal to $1$. 
\\ 
\\ 
\underline{For $s+l\leq m$:}
\st
&&
\phantom{mapmapmapmapmap}
\mathcal{Z}((1,s)_{_{C}},\sigma_{1};(1,l)_{_{C}},\sigma_{2}) 
\nonumber\\
&=&
-
\frac{1}{4}\:
\sideset{}{'}\sum_{i= \vert s-l \vert + 1}^{s+l-1}
\;\sideset{}{'}\sum_{j=1-i}^{i-1} 
\ln\left[
z_{1}^{2}+z_{2}^{2}+2z_{1}z_{2}
\cos\left(\frac{\pi j}{2m-1}\right)
-
\sin^{2}\left(\frac{\pi j}{2m-1}\right)
\right]
\nonumber\\
&&
+
\frac{sl}{4\sqrt{2}\sqrt{2m-1}\epsilon} 
-\frac{sl}{2}\ln 2
\label{cyl1}
\en
\\
\\
\begin{samepage}
\underline{For $s+l> m$, $l+s-m$ even:}
\\
\\
\[
\mathcal{Z}((1,s)_{_{C}},\sigma_{1};(1,l)_{_{C}},\sigma_{2}) 
\nonumber
\]
\[
=
-
\frac{1}{4} \:
\sideset{}{'}\sum_{i = \vert s-l \vert +1}^{m-1}
\;\sideset{}{'}\sum_{j = 1-i}^{i-1} 
\ln\left[
z_{1}^{2}+z_{2}^{2}+2z_{1}z_{2}
\cos\left(\frac{\pi j}{2m-1}\right)
-
\sin^{2}\left(\frac{\pi j}{2m-1}\right)
\right]
\nonumber
\]
\[
\phantom{map}
-
\frac{1}{4} \:
\sideset{}{'}\sum_{i = m+1}^{s+l-1}
\;\sideset{}{'}\sum_{j = 2+i-2m }^{2m-i-2}
\ln\left[
z_{1}^{2}+z_{2}^{2}+2z_{1}z_{2}
\cos\left(\frac{\pi j}{2m-1}\right)
-
\sin^{2}\left(\frac{\pi j}{2m-1}\right)
\right]
\nonumber
\]
\[
+
\frac{2sl-(s+l-m)(s+l-m+1)}{8\sqrt{2}\sqrt{2m-1}\epsilon} 
-\frac{2sl-(s+l-m)(s+l-m+1)}{4}\ln 2
\label{cyl2}
\]
\end{samepage}
\\
\\
\begin{samepage}
\underline{For $s+l> m$, $l+s-m$ odd:}
\\
\\
\[
\mathcal{Z}((1,s)_{_{C}},\sigma_{1};(1,l)_{_{C}},\sigma_{2}) 
\nonumber
\]
\[
=
-
\frac{1}{4} \:
\sideset{}{'}\sum_{i = \vert s-l \vert +1}^{m-2}
\;\sideset{}{'}\sum_{j = 1-i}^{i-1} 
\ln\left[
z_{1}^{2}+z_{2}^{2}+2z_{1}z_{2}
\cos\left(\frac{\pi j}{2m-1}\right)
-
\sin^{2}\left(\frac{\pi j}{2m-1}\right)
\right]
\nonumber
\]
\[
\phantom{m}
-
\frac{1}{4} \:
\sideset{}{'}\sum_{i = m}^{s+l-1}
\;\sideset{}{'}\sum_{j = 2+i-2m}^{2m-i-2}
\ln\left[
z_{1}^{2}+z_{2}^{2}+2z_{1}z_{2}
\cos\left(\frac{\pi j}{2m-1}\right)
-
\sin^{2}\left(\frac{\pi j}{2m-1}\right)
\right]
\nonumber
\]
\[
+
\frac{2sl-(s+l-m)(s+l-m+1)}{8\sqrt{2}\sqrt{2m-1}\epsilon} 
-\frac{2sl-(s+l-m)(s+l-m+1)}{4}\ln 2
\phantom{map}
\label{cyl3}
\]
\end{samepage}

Let us return to the discussion concerning the relation (\ref{SSproposal}) 
proposed by Seiberg and Shih and let us for simplicity consider the FZZT-FZZT 
cylinder amplitude in the $(2,2m-1)$ minimal model coupled to 2D euclidean Quantum Gravity. 
In this case the uniformization parameter $z$ defined in eq. (\ref{z}) is related 
to the boundary cosmological constant given in eq. (\ref{cosmolcon}) by 
\[
z = \frac{1}{\sqrt{2}}\sqrt{ \frac{\mu_{B}}{\sqrt{\mu}} + 1 } 
\]
Let us for simplicity consider the cylinder amplitude 
in the case, where the values of the two boundary cosmological constants coincide, 
that is $\mu_{B_{1}}=\mu_{B_{2}}=\mu_{B}$. 
In this case we may easily show, that 
\st
&&
\ln\left[
z_{1}^{2}+z_{2}^{2}+2z_{1}z_{2}
\cos\left(\frac{\pi j}{2m-1}\right)
-
\sin^{2}\left(\frac{\pi j}{2m-1}\right)
\right]
\nonumber\\
& = & 
\ln\left[
\frac{2}{\sqrt{\mu}}
\cos^{2}\!\left(
\frac{\pi j}{2(2m-1)}
\right)
\left(
\mu_{B} + \sqrt{\mu} \cos\left(
\frac{\pi j}{2m-1}
\right)
\right)
\right]\,.
\en
With regard to the summation index $j$ appearing in the expressions (\ref{cyl1}), (\ref{cyl2}) and (\ref{cyl3}) 
the following estimate is valid  
\[
\vert j \vert \leq m-2 < \frac{2m-1}{2}\,.
\]
From this estimate, the above identity and the cylinder amplitudes given by eqs.  
(\ref{cyl1}), (\ref{cyl2}) and (\ref{cyl3}) we see, that the generic cylinder amplitude 
is finite in the semi-classical region $\mu_{B_{1}}=\mu_{B_{2}}=\mu_{B}>0$ and $\mu > 0$ 
in the $(2,2m-1)$ minimal model coupled to 2D euclidean Quantum Gravity. 
Now, let us apply the relation (\ref{SSproposal})  
in the generic cylinder amplitude in the $(2,2m-1)$ minimal model coupled to 2D euclidean Quantum Gravity 
and let us again consider the case $\mu_{B_{1}}=\mu_{B_{2}}=\mu_{B}$.   
\st
&&
\mathcal{Z}((1,s)_{_{C}},\sigma_{1};(1,l)_{_{C}},\sigma_{1}) 
\nonumber\\
& = &
\sideset{}{'}\sum_{\tilde{s}=1-s}^{s-1}
\;\sideset{}{'}\sum_{\tilde{l}=1-l}^{l-1} 
\mathcal{Z}\left( (1,1)_{_{C}}, \sigma_{1}+i\tilde{s}\sqrt{\frac{2}{2m-1}};
(1,1)_{_{C}}, \sigma_{1}+i\tilde{l}\sqrt{\frac{2}{2m-1}}\right) 
\nonumber\\
& = & 
-
\frac{1}{4}
\sideset{}{'}\sum_{\tilde{s}=1-s}^{s-1}
\;\sideset{}{'}\sum_{\tilde{l}=1-l}^{l-1} 
\ln\left[
\frac{2}{\sqrt{\mu}}
\cos^{2}\!\left(
\frac{\pi(\tilde{s}+\tilde{l})}{2(2m-1)}
\right)
\left(
\mu_{B} + \sqrt{\mu} \cos\left(
\frac{\pi(\tilde{s}+\tilde{l})}{2m-1}
\right)
\right)
\right]
\nonumber\\
&&
+
\frac{sl}{4\sqrt{2}\sqrt{2m-1}\epsilon} 
-\frac{sl}{2}\ln 2
\en
For $s+l > m+2$ and in the semi-classical region $\mu_{B_{1}}=\mu_{B_{2}}=\mu_{B}>0$ 
this amplitude actually diverges at a positive value of the boundary cosmological constant $\mu$. 
This contradicts our previous observation and  
we conclude, that the relation (\ref{SSproposal}) is not consistent with the cylinder amplitude.  
The difference between the correct cylinder amplitude and the amplitude obtained from the relation (\ref{SSproposal}) 
is not an analytical function due to the fact, 
that the correct cylinder amplitude is finite for $\mu_{B_{1}}=\mu_{B_{2}}=\mu_{B}>0$ and $\mu > 0$, 
while the amplitude obtained from the relation (\ref{SSproposal}) diverges at a positive value of $\mu$, if $s+l>m+2$. 
Therefore, we cannot attribute the difference between the two amplitudes to geometries with vanishing area.   
We conclude from our previous discussion, 
that the relation (\ref{SSproposal}) is not valid with regard to the cylinder amplitude 
due to the contribution from the poles, which we cannot associate with physical closed string states in the BRST formalism.     

\chapter{The nature of ZZ-branes} \label{chapterZZ}
As discussed in section \ref{sectionTheFZZTandZZbranes} 
Seiberg and Shih argue in \cite{SS}, that any ZZ boundary state tensored 
with a given Cardy matter state $\vert i,j \rangle_{_{C}}$ in the $(p,q)$ minimal model 
coupled to 2D euclidean Quantum Gravity may be expressed in terms of principal 
ZZ-boundary states tensored with the given Cardy matter state $\vert i,j \rangle_{_{C}}$ modulo BRST exact states. 
(See eqs. (\ref{principal1}), (\ref{principal2}) and (\ref{principal3})).   
The argument presented in \cite{SS} for this identity relies on the fact, that any given ZZ-boundary state in Liouville theory 
may be written as a linear combination of FZZT-boundary states, and on the observation, that the parameter $\sigma$ labeling the different 
FZZT-boundary states only enters the bulk one-point function evaluated on the disk through the uniformization parameter $z$ 
defined in eq. (\ref{z}). A similar observation is valid with regard to the FZZT-FZZT cylinder amplitudes in the $(p,q)$ minimal model 
coupled to 2D euclidean Quantum Gravity. The generic cylinder amplitude derived in the previous chapter may be expressed entirely in terms of the 
uniformization parameters $z_{1}$ and $z_{2}$ associated with the FZZT boundary conditions imposed on the two 
boundaries of the cylinder and the different Cardy matter indices 
labelling the Cardy matter states imposed on the two boundaries of the cylinder. 
Thus, the FZZT-FZZT cylinder amplitudes  
obtained in the previous chapter provide further evidence for the identity proposed by Seiberg and Shih concerning ZZ-branes. 
In section \ref{sectionZZ} we saw, that the random geometry associated with the basic $(1,1)$ ZZ boundary state 
corresponds to the ``natural'' quantization of the Lobachevskiy plane obtained from standard loop perturbation theory.   
However, we did not address the main question raised by Zamolodchikov and Zamolodchikov in \cite{ZZ} 
concerning the nature of the remaining $(m,n)$ ZZ boundary states in Liouville theory.       
Zamolodchikov and Zamolodchikov wrote: ``The most intriguing point is the nature of the ``excited'' vacua...A meaning of these quantum excitations of 
the (physically infinite faraway) absolute remains to be comprehended.'' 
In this chapter we will present a physical interpretation of the principal ZZ boundary states  
consistent with and motivated by the one-point function 
evaluated on the quantum Lobachevskiy plane, the FZZT-ZZ cylinder amplitudes and ZZ-ZZ cylinder amplitudes.
Due to the identities (\ref{principal1}), (\ref{principal2}) and (\ref{principal3}) a physical interpretation of the principal ZZ boundary states is sufficient. 
This interpretation of the principal ZZ boundary states is developed by the author of this thesis in collaboration with his academic adviser and is published in \cite{AG3}. 
Our interpretation concerns the principal ZZ boundary states in the $(p,q)$ minimal model coupled to 2D euclidean Quantum Gravity. 
However, we expect a similar interpretation to be valid in more general conformal field theories coupled to 2D euclidean Quantum Gravity. 

In order to motivate our interpretation we will start out by providing evidence for the following 
identity concerning the principal ZZ-branes\footnote{The possibility of such an identity was first noticed in \cite{KOPSS}.} 
\[\label{id1}
\vert m,n \rangle_{_{C}} \otimes \vert 1,1 \rangle_{_{ZZ}} 
= 
\vert 1,1 \rangle_{_{C}} \otimes \vert m,n \rangle_{_{ZZ}}\,.
\]   
This identity should be understood in the following way: The expectation value of any given 
physical gauge-invariant observable is the same, 
whether we consider the left hand side or the right hand side of the above equation.  
Notice, the above relation is only possible due to the fact, 
that the set of indices labelling the different Cardy matter states 
\[
\bigg\{
(m,n) \bigg\vert\, 1 \leq m \leq p-1\,,\;1 \leq n \leq q-1\,,\;mq-np>0
\bigg\}
\]
is the same as the set of indices labelling 
the different principal ZZ-boundary states.  
Moreover, we will provide evidence for the following generalization of the above identity concerning the principal ZZ-branes 
\[\label{id2}
\vert r,s \rangle_{_{C}} \otimes \vert m,n \rangle_{_{ZZ}} 
= 
\left(
\sideset{}{'}\sum_{k = \vert r - m \vert + 1}^{\textrm{min}(r+m-1,2p-1-r-m)}
\:\sideset{}{'}\sum_{l = \vert s - n \vert + 1}^{\textrm{min}(s+n-1,2q-1-s-n)}
\vert k,l \rangle_{_{C}}
\right)
\otimes \vert 1,1 \rangle_{_{ZZ}}\,,
\]
where the primes in the summation symbols denote, that the summations run in steps of two. 
Notice, this summation is exactly the same, which appears in the fusion of the primary matter operators $\mathcal{O}_{r,s}$ 
and $\mathcal{O}_{k,l}$ in the $(p,q)$ minimal model given by eq. (\ref{fusionmin}). 
In the final section of this chapter we will introduce and justify our interpretation concerning 
the principal ZZ boundary states. 
This interpretation will explain the physics captured by eqs. (\ref{id1}) and (\ref{id2}). 

\section{The one-point function on the pseudo-sphere} 
Let us begin by showing, that the relations (\ref{id1}) and (\ref{id2}) are valid with respect to the bulk 
one-point function evaluated on the quantum Lobachevskiy plane. 
How can this be true? 
Let us for simplicity focus on the relation (\ref{id1}). 
Recall the definition of the Cardy matter states given in eqs. (\ref{Cardydef}) and (\ref{mods})
\[
\vert r,s \rangle_{_{C}} 
= 
\begin{array}{c}
\displaystyle\sum_{m=1}^{p-1} \,\,\, 
\sum_{n=1}^{q-1}
\\
\textrm{\scriptsize{$mq\!-\!np>0$}}
\end{array} 
\frac{S_{(r,s),;(m,n)}}{\sqrt{S_{(1,1);(m,n)}}}
\vert\vert \Delta_{m,n} \rangle\rangle\,, 
\]  
where 
\[
S_{(r,s);(m,n)} 
= 
2\sqrt{\frac{2}{pq}}(-1)^{1+sm+rn}
\sin\left(\frac{\pi q rm}{p}\right)
\sin\left(\frac{\pi p sn}{q}\right)
\]
and recall the definition of the principal ZZ-boundary states given in eqs. (\ref{ZZstatemn}) and (\ref{ZZstate11})  
\[
\vert m,n \rangle_{_{ZZ}} 
= 
\int_{0}^{\infty} dP 
\frac{\sinh(2\pi mP/b)\sinh(2\pi nbP)}{\sinh(2\pi P/b)\sinh(2\pi bP)}
\Psi^{\dag}_{1,1}(P) 
\vert\vert Q/2+iP \rangle\rangle\,,
\]
where 
\[
\Psi_{1,1}(P) 
= 
\frac{2^{3/4} 2 i \pi P}{\Gamma(1-2ibP)\Gamma(1-2iP/b)}\left\{ \pi\mu\gamma(b^{2}) \right\}^{-iP/b}\,.
\]
Up to a factor independent of the boundary condition imposed at the absolute of the Lobachevskiy plane  
the one-point function of the physical observable $\hat{\mathcal{A}}^{k,l}_{P}$ 
is given by the product of the imposed Cardy matter wave function  
evaluated at $k,l$ and the imposed principal ZZ wave function evaluated at $P$. 
This follows from an argument similar to the argument presented in connection with eq. (\ref{onepointFZZT}). 
The validity of the identity (\ref{id1}) 
with regard to physical observables evaluated on the quantum Lobachevskiy plane relies on the fact, 
that the Liouville momentum $P$ of a given physical observable depends crucially on the matter 
content of the given observable. 
If the matter content of a given observable belongs to the conformal family of operators 
descending from the primary matter operator $\mathcal{O}_{k,l}$, then the Liouville momentum 
of the observable is given by 
\[\label{liomom}
P = \pm i \frac{2tpq + kq \pm lp}{2\sqrt{pq}}\,,\quad t\in\mathbf{Z}. 
\]
It follows from this fact that going from the right hand side to the left side of eq. (\ref{id1}) the factor 
$\sinh(2\pi mP/b)\sinh(2\pi nbP)$ appearing in the one point function obtained 
from the right hand side becomes the factor 
$\sin\left(\frac{\pi q km}{p}\right)\sin\left(\frac{\pi p ln}{q}\right)$ appearing in the one-point function 
obtained from the left hand side and we see, that the two expressions match.       

Let us show in more detail, how this simple mechanism explains the more general 
relation (\ref{id2}) with respect to bulk observables evaluated on the quantum Lobachevskiy plane. 
From the above discussion we obtain 
\st
&&
\bigg(
\phantom{}_{_{C}}\langle r,s \vert \otimes \phantom{}_{_{ZZ}}\langle m,n \vert 
\bigg)
\vert \mathcal{A}_{k,l,P} \rangle
\nonumber\\
& = & 
\tilde{\Omega}_{k,l}^{P}
\frac{S_{(r,s);(k,l)}}{\sqrt{S_{(1,1);(k,l)}}} 
\Psi_{m,n}(P)
\nonumber\\
& = &
\Omega_{k,l}^{P} 
(-1)^{r l+ks} 
\sin\left(\frac{\pi k r q}{p}\right)
\sin\left(\frac{\pi l s p}{q}\right)
\sinh\left(\frac{2\pi m P}{b}\right)
\sinh\left(2\pi n P b\right)
\Psi_{1,1}(P)
\nonumber\\
& = & 
\Omega_{k,l}^{P} 
(-1)^{r l+ks + 1} 
\sin\left(\frac{\pi k r q}{p}\right)
\sin\left(\frac{\pi l s p}{q}\right)
\nonumber\\
&&
\phantom{map}
\times
\sin\left(\frac{2\pi m}{b}\frac{2tpq + kq \pm lp}{2\sqrt{pq}}\right)
\sin\left(2\pi n b\frac{2tpq + kq \pm lp}{2\sqrt{pq}}\right)
\Psi_{1,1}(P)
\nonumber\\
& = &
\pm
\,\Omega_{k,l}^{P} 
(-1)^{(r+m)l+k(n+s) + 1} 
\sin\left(\frac{\pi k r q}{p}\right)
\sin\left(\frac{\pi k m q}{p}\right)
\sin\left(\frac{\pi l s p}{q}\right)
\sin\left(\frac{\pi l n p}{q}\right)
\Psi_{1,1}(P)
\nonumber\\
& = & 
\pm
\,\Omega_{k,l}^{P} 
(-1)^{(r+m)l+k(n+s) + 1} 
\sin\left(\!\frac{\pi k q}{p}\!\right)
\sin\left(\!\frac{\pi l p}{q}\!\right)
\nonumber\\
&&
\phantom{map}
\times
\sideset{}{'}\sum_{\tilde{r} = \vert r-m \vert + 1}^{r + m - 1}
\:\sideset{}{'}\sum_{\tilde{s} = \vert s-n \vert + 1}^{s + n - 1}
\!\!\sin\left(\!\frac{\pi k \tilde{r} q}{p}\!\right)
\sin\left(\!\frac{\pi l \tilde{s} p}{q}\!\right)\Psi_{1,1}(P)\,,
\nonumber\\
\label{inter}
\en
where we have applied the identity (\ref{sinsin}) in deriving the last expression 
and where $\tilde{\Omega}_{k,l}^{P}$ and $\Omega_{k,l}^{P}$ do not depend on the specific boundary condition 
imposed at the absolute of the Lobachevskiy plane.  
Applying the identities 
\[
\sideset{}{'}\sum_{\tilde{r} = 2p+1-m-r}^{r + m - 1}
\sin\left(\frac{\pi k \tilde{r} q}{p}\right) 
= 
0
\label{nul1}
\]
valid for $r+m>p$ and 
\[
\sideset{}{'}\sum_{\tilde{s} = 2q+1-n-s}^{s + n - 1}
\sin\left(\frac{\pi l \tilde{s} p}{q}\right) 
= 
0
\label{nul2}
\]
valid for $s+n>q$ 
we may express the above one-point function as 
\[
\bigg(
\phantom{}_{_{C}}\langle r,s \vert \otimes \phantom{}_{_{ZZ}}\langle m,n \vert 
\bigg)
\vert \mathcal{A}^{k,l}_{P} \rangle
\nonumber
\]
\[
\phantom{map}
=  
\pm
\,\Omega_{k,l}^{P} 
(-1)^{(r+m)l+k(n+s)+1} 
\sin\left(\!\frac{\pi k q}{p}\!\right)
\sin\left(\!\frac{\pi l p}{q}\!\right)
\Psi_{1,1}(P)
\phantom{mapmapmapmapmapmapmapmapmap}
\nonumber
\]
\[
\phantom{mapmap}
\sideset{}{'}\sum_{\tilde{r} = \vert r - m \vert + 1}^{\textrm{min}(r+m-1,2p-1-r-m)}
\:\sideset{}{'}\sum_{\tilde{s} = \vert s - n \vert + 1}^{\textrm{min}(s+n-1,2q-1-s-n)}
\!\!\sin\left(\!\frac{\pi k \tilde{r} q}{p}\!\right)
\sin\left(\!\frac{\pi l \tilde{s} p}{q}\!\right)
\phantom{mapmapmap}
\nonumber
\]
\begin{samepage}
\[
= 
\pm
\,\Omega_{k,l}^{P} 
\sin\left(\!\frac{\pi k q}{p}\!\right)
\sin\left(\!\frac{\pi l p}{q}\!\right)
\Psi_{1,1}(P)
\phantom{mapmapmapmapmapmapmapmapmapma}
\nonumber
\]
\[
\phantom{mapm}
\sideset{}{'}\sum_{\tilde{r} = \vert r - m \vert + 1}^{\textrm{min}(r+m-1,2p-1-r-m)}
\;\sideset{}{'}\sum_{\tilde{s} = \vert s - n \vert + 1}^{\textrm{min}(s+n-1,2q-1-s-n)}
(-1)^{(\tilde{r}+1)l+k(\tilde{s}+1)+1} 
\sin\left(\!\frac{\pi k \tilde{r} q}{p}\!\right)
\sin\left(\!\frac{\pi l \tilde{s} p}{q}\!\right)
\nonumber
\]
\end{samepage}
\[
= 
\sideset{}{'}\sum_{\tilde{r} = \vert r - m \vert + 1}^{\textrm{min}(r+m-1,2p-1-r-m)}
\;\sideset{}{'}\sum_{\tilde{s} = \vert s - n \vert + 1}^{\textrm{min}(s+n-1,2q-1-s-n)}
\bigg(
\phantom{}_{_{C}}\langle \tilde{r},\tilde{s} \vert \otimes \phantom{}_{_{ZZ}}\langle 1,1 \vert 
\bigg)
\vert \mathcal{A}^{k,l}_{P} \rangle
\phantom{mapm}
\]
where the last identity is obtained by comparing the second last expression with the intermediate result obtained in the second last line  
in (\ref{inter}).
From the above calculation we conclude, 
that the relations (\ref{id1}) and (\ref{id2}) are valid with respect to physical bulk observables evaluated on the quantum Lobachevskiy plane.

\section{The FZZT-ZZ and ZZ-ZZ cylinder amplitudes} 
Let us start out by considering the non-compact 
FZZT-ZZ cylinder amplitude with a principal ZZ-brane imposed on one of the boundaries.    
Due to the fact, that we may express any given principal ZZ-boundary state in Liouville theory as the difference 
between two FZZT boundary states as in eq. (\ref{FZZTZZb2irr}) we may easily obtain the FZZT-ZZ cylinder amplitude 
from the FZZT-FZZT cylinder amplitude derived in chapter \ref{chaptercyl}. 
It follows from eqs. (\ref{FZZTZZb2irr}) and (\ref{sigmamn}) 
that imposing the principal $(m,n)$ ZZ-boundary state instead of a FZZT boundary state on one of the boundaries of cylinder amounts to 
changing\footnote{Since the final expression for FZZT-FZZT cylinder amplitude is symmetric in $\sigma_{1}$ and $\sigma_{2}$, 
we are free to choose on which boundary we impose a ZZ-boundary state instead of the FZZT boundary state. 
For convenience we choose the boundary labelled by 2.} 
\st
\cos(2 \pi P \sigma_{2}) 
& \to & 
\cos\left(2 \pi P i\left(\frac{m}{b} + nb\right)\right) - \cos\left(2 \pi P i\left(\frac{m}{b} - nb\right)\right) 
\nonumber\\ 
& = & 
2\sinh\left(\frac{2\pi P m}{b}\right)\sinh\left(2\pi P n b\right)
\en   
in the integrand in the expression (\ref{truecyl2}) for the cylinder amplitude, that is 
\newpage 
\[
\mathcal{Z}((r,s)_{_{C}},\sigma_{1};(k,l)_{_{C}},(m,n)_{_{ZZ}}) 
\nonumber
\]
\[
=
\frac{1}{pq}
\oint_{\gamma_{+}}\!\! d P \, 
\frac{\exp(i 2 \pi P \sigma_{1})\sinh\left(\frac{2\pi P m}{b}\right)\sinh\left(2\pi P n b\right)
\sinh\left(\frac{2 \pi P}{\sqrt{pq}}\right)f\left(\frac{i 2 \pi P}{\sqrt{pq}}\right)}
{\sinh\left(\frac{2 \pi P}{b}\right)\sinh\left(2 \pi P b\right)P}
\label{truecylzz}
\]
where the function $f$ is defined in eq. (\ref{ffunc}).  
Notice, the factor $\sinh\left(\frac{2\pi P m}{b}\right)\sinh\left(2\pi P n b\right)$ 
exactly cancels all the poles, which we cannot associate with physical closed string states in the BRST formalism, 
that is the poles coming from $\sinh\left(\frac{2\pi P}{b}\right)\sinh\left(2\pi P b\right)$ in the denominator. 
Only the poles, which we can associate with physical 
closed string states in the BRST formalism, contribute to the non-compact FZZT-ZZ cylinder amplitude. 
Thus, the nature of the poles, which we cannot associate with physical closed string states in the BRST formalism, must be   
closely related to the compact nature of the FZZT-FZZT cylinder amplitude. 
In the above expression for the FZZT-ZZ cylinder amplitude  
we have not included the regularization introduced in eq. (\ref{regularization}), since there is no need 
for this regularization in the FZZT-ZZ cylinder amplitude. 
Notice, in place of $f$  
in the above expression for the FZZT-ZZ cylinder amplitude 
we may insert any function, which is equal to $f$  
up to an entire function of $P$.   
The set of poles in the integrand and the values of the corresponding residues are invariant under this operation. 
This follows from the fact, that 
the function multiplying $f$  
in the integrand is an entire 
function of $P$, and from the fact, that $\gamma_{+}$ is a closed contour.                 
Inserting $h_{1}$  
defined in eq. (\ref{hfunc}) in place of $f$ we obtain 
\[
\mathcal{Z}((r,s)_{_{C}},\sigma_{1};(k,l)_{_{C}},(m,n)_{_{ZZ}}) 
\nonumber
\]
\[
= 
\oint_{\gamma_{+}}\!\!\!\! d P \, 
\frac{e^{i 2 \pi P \sigma_{1}}
\sinh(\frac{2\pi P (p - r)}{b})
\sinh\left(2\pi P s b\right)}
{\sinh\left(2 \pi P \sqrt{pq}\right)P}
\frac{\sinh\left(\frac{2\pi P m}{b}\right)\sinh\left(\frac{2\pi P k}{b}\right)}{\sinh^{2}\left(\frac{2 \pi P}{b}\right)}
\frac{\sinh\left(2\pi P n b\right)\sinh\left(2\pi P l b\right)}{\sinh^{2}\left(2 \pi P b\right)} 
\nonumber
\]
\[
=
-
\oint_{\gamma_{+}}\!\!\!\! d P \, 
\frac{e^{i 2 \pi P \sigma_{1}}
\sinh(\frac{2\pi P r}{b})
\sinh\left(2\pi P s b\right)}
{\tanh\left(2 \pi P \sqrt{pq}\right)P}
\frac{\sinh\left(\frac{2\pi P m}{b}\right)\sinh\left(\frac{2\pi P k}{b}\right)}{\sinh^{2}\left(\frac{2 \pi P}{b}\right)}
\frac{\sinh\left(2\pi P n b\right)\sinh\left(2\pi P l b\right)}{\sinh^{2}\left(2 \pi P b\right)} 
\label{cylinterpretation}
\]
where we have subtracted an entire function of $P$ from the integrand in obtaining the last expression.  
Applying the identity (\ref{sinsin}) we may express the FZZT-ZZ cylinder amplitude as 
\[
\mathcal{Z}((r,s)_{_{C}},\sigma_{1};(k,l)_{_{C}},(m,n)_{_{ZZ}}) 
\nonumber
\] 
\[
= - \!\!\!
\sideset{}{'}\sum_{\tilde{k} = \vert k - m \vert + 1}^{k+m-1}
\;\sideset{}{'}\sum_{\tilde{l} = \vert l - n \vert + 1}^{l+n-1}
\oint_{\gamma_{+}}\!\! d P \, 
\frac{e^{i 2 \pi P \sigma_{1}}
\sinh\left(\frac{2\pi P r}{b}\right)
\sinh\left(2\pi P s b\right)}
{\tanh\left(2 \pi P \sqrt{pq}\right)P}
\frac{\sinh\left(\frac{2\pi P \tilde{k}}{b}\right)}{\sinh\left(\frac{2 \pi P}{b}\right)}
\frac{\sinh\left(2\pi P \tilde{l} b\right)}{\sinh\left(2 \pi P b\right)} 
\]
Using the following identities derived from eq. (\ref{sinsin})
\[
\sideset{}{'}\sum_{\tilde{k} = 2p+1-k-m}^{k+m-1} 
\sinh\left(
\frac{2 \pi P \tilde{k}}{b}
\right)
= 
\frac{\sinh\left(2 \pi P \sqrt{pq} \right) \sinh\left( \frac{2 \pi P (k+m-p)}{b}\right)}{\sinh\left( \frac{2 \pi P}{b}\right)}
\] 
valid for $k+m > p$  
and 
\[
\sideset{}{'}\sum_{\tilde{l} = 2q+1-l-n}^{l+n-1} 
\sinh\left(
2 \pi P \tilde{l} b
\right)
= 
\frac{\sinh\left(2 \pi P \sqrt{pq} \right) \sinh\left( 2 \pi P (l+n-q)b\right)}{\sinh\left( 2 \pi P b\right)}
\label{cy1}
\]
valid for $l+n>q$, we may express the FZZT-ZZ cylinder amplitude as 
\begin{samepage}
\[
\mathcal{Z}((r,s)_{_{C}},\sigma_{1};(k,l)_{_{C}},(m,n)_{_{ZZ}}) 
\nonumber
\] 
\[
= -\!\!\!
\sideset{}{'}\sum_{\tilde{k} = \vert k - m \vert + 1}^{\textrm{min}\{k+m-1;2p-1-k-m\}}
\;\sideset{}{'}\sum_{\tilde{l} = \vert l - n \vert + 1}^{\textrm{min}\{l+n-1;2q-1-l-n\}}
\oint_{\gamma_{+}}\!\! d P \, 
\left\{
\frac{e^{i 2 \pi P \sigma_{1}}
\sinh\left(\frac{2\pi P r}{b}\right)
\sinh\left(2\pi P s b\right)}
{\tanh\left(2 \pi P \sqrt{pq}\right)P}
\right.
\nonumber
\]
\[
\phantom{mapmapmapmapmapmapmapmapmapmapmapmapmap}
\left.
\times
\frac{\sinh\left(\frac{2\pi P \tilde{k}}{b}\right)}{\sinh\left(\frac{2 \pi P}{b}\right)}
\frac{\sinh\left(2\pi P \tilde{l} b\right)}{\sinh\left(2 \pi P b\right)} 
\right\}
\label{cy2}
\]
\end{samepage}The term subtracted from the integrand going from eq. (\ref{cy1}) to eq. (\ref{cy2}) is an entire function of $P$. 
Comparing this expression for the cylinder amplitude with the expression (\ref{cylinterpretation}) we realize, that 
\[
\mathcal{Z}((r,s)_{_{C}},\sigma_{1};(k,l)_{_{C}},(m,n)_{_{ZZ}}) 
\nonumber
\]
\[
= 
\sideset{}{'}\sum_{\tilde{k} = \vert k - m \vert + 1}^{\textrm{min}\{k+m-1;2p-1-k-m\}}
\;\sideset{}{'}\sum_{\tilde{l} = \vert l - n \vert + 1}^{\textrm{min}\{l+n-1;2q-1-l-n\}}
\mathcal{Z}((r,s)_{_{C}},\sigma_{1};(\tilde{k},\tilde{l})_{_{C}},(1,1)_{_{ZZ}})\,. 
\label{FZZTZZconsistent}
\]
Hence, the FZZT-ZZ cylinder amplitude in the $(p,q)$ minimal model coupled to 2D euclidean Quantum Gravity 
is consistent with the relations (\ref{id1}) and (\ref{id2}). Not only is this true with respect 
to the final expression for the FZZT-ZZ cylinder amplitude. 
We may express both the left hand side and the right hand side of the above identity 
as a sum over residues and each residue on the left hand side is actually equal to the corresponding residue on the right hand side.  
This follows from the fact, that going from eq. (\ref{cylinterpretation}) to eq. (\ref{cy2}) 
we have simply subtracted an entire function of $P$ from the integrand.\footnote{It 
is tempting to think, that this entire function corresponds to the contribution to the integrand from  
an exact state with respect to the BRST charge $Q$.}     
This is important due to the fact, that we may interpret a given residue contributing to the FZZT-ZZ cylinder amplitude 
as the amplitude of the corresponding physical closed string state propagating between the FZZT brane and the ZZ brane as 
explained in chapter \ref{chaptercyl}. Recall, all the residue contributing to the FZZT-ZZ cylinder amplitude may be associated 
with physical closed string states.\footnote{The above identity comes as no surprise due to the fact, that the relation (\ref{id2}) 
is valid with respect to physical observables evaluated on the Lobachevskiy plane, 
and due to the fact, that the decomposition illustrated in figure \ref{figurdecomp} 
is valid with regard to the FZZT-ZZ cylinder amplitude 
since we may associate a closed string state with any given residue contributing to the FZZT-ZZ cylinder amplitude.} 

For completeness let us determine the  
FZZT-ZZ cylinder amplitude from eq. (\ref{cylinterpretation}). 
From Cauchy's theorem and eq. (\ref{sinsin})  
we obtain 
\[
\mathcal{Z}((r,s)_{_{C}},\sigma_{1};(k,l)_{_{C}},(m,n)_{_{ZZ}}) 
\nonumber
\]
\[
=  
2
\sum_{t=1,\, p\,\ndiv\, t,\, q\,\ndiv\, t}^{\infty} 
\frac{1}{t}
e^{-\frac{\pi t \sigma_{1}}{\sqrt{pq}}}\sin\left(\frac{\pi r t}{p}\right)\sin\left(\frac{\pi s t}{q}\right)
\frac{\sin\left(\frac{\pi k t}{p}\right)\sin\left(\frac{\pi m t}{p}\right)}{\sin^{2}\left(\frac{\pi t}{p}\right)}
\frac{\sin\left(\frac{\pi l t}{q}\right)\sin\left(\frac{\pi n t}{q}\right)}{\sin^{2}\left(\frac{\pi t}{q}\right)}
\phantom{mapm}
\nonumber
\]
\[
= 
2
\!\!\!\!\!\!\!\!\!
\sideset{}{'}\sum_{\tilde{k} = \vert k - m \vert + 1}^{\textrm{min}\{k+m-1;2p-1-k-m\}}
\;\sideset{}{'}\sum_{\tilde{l} = \vert l - n \vert + 1}^{\textrm{min}\{l+n-1;2q-1-l-n\}}
\!\!\!\!\!\!\!\!\!
\sum_{t=1,\, p\,\ndiv\, t,\, q\,\ndiv\, t}^{\infty} 
\frac{1}{t}
e^{-\frac{\pi t \sigma_{1}}{\sqrt{pq}}}
\frac{\sin\left(\!\frac{\pi r t}{p}\!\right)\sin\left(\!\frac{\pi \tilde{k} t}{p}\!\right)}{\sin\left(\!\frac{\pi t}{p}\!\right)}
\frac{\sin\left(\!\frac{\pi s t}{q}\!\right)\sin\left(\!\frac{\pi \tilde{l} t}{q}\!\right)}{\sin\left(\!\frac{\pi t}{q}\!\right)}
\nonumber
\]
\st 
& = &  
2
\!\!\!\!\!\!\!\!\!
\sideset{}{'}\sum_{\tilde{k} = \vert k - m \vert + 1}^{\textrm{min}\{k+m-1;2p-1-k-m\}}
\;\sideset{}{'}\sum_{\tilde{l} = \vert l - n \vert + 1}^{\textrm{min}\{l+n-1;2q-1-l-n\}}
\nonumber\\
&&
\phantom{mapmap}
\sideset{}{'}\sum_{\tilde{r} = \vert \tilde{k} - r \vert + 1}^{\textrm{min}\{\tilde{k}+r-1;2p-1-\tilde{k}-r\}}
\;\sideset{}{'}\sum_{\tilde{s} = \vert \tilde{l} - s \vert + 1}^{\textrm{min}\{\tilde{l}+s-1;2q-1-\tilde{l}-s\}}
\!\!
\sum_{t=1}^{\infty} 
\frac{1}{t}
e^{-\frac{\pi t \sigma_{1}}{\sqrt{pq}}}
\sin\left(\!\frac{\pi \tilde{r} t}{p}\!\right)
\sin\left(\!\frac{\pi \tilde{s} t}{q}\!\right)
\nonumber\\
& = & 
\frac{1}{2} 
\sideset{}{'}\sum_{\tilde{k} = \vert k - m \vert + 1}^{\textrm{min}\{k+m-1;2p-1-k-m\}}
\;\sideset{}{'}\sum_{\tilde{l} = \vert l - n \vert + 1}^{\textrm{min}\{l+n-1;2q-1-l-n\}}
\nonumber\\
&&
\phantom{mapmap}
\sideset{}{'}\sum_{\tilde{r} = \vert \tilde{k} - r \vert + 1}^{\textrm{min}\{\tilde{k}+r-1;2p-1-\tilde{k}-r\}}
\;\sideset{}{'}\sum_{\tilde{s} = \vert \tilde{l} - s \vert + 1}^{\textrm{min}\{\tilde{l}+s-1;2q-1-\tilde{l}-s\}}
\ln\left[
\frac{z_{1}-\cos\left(\frac{\pi(\tilde{r}q+\tilde{s}p)}{pq}\right)}
{z_{1}-\cos\left(\frac{\pi(\tilde{r}q-\tilde{s}p)}{pq}\right)}
\right]
\nonumber\\
\label{FZZTZZcyl}
\en
where we have expanded the trigonometric functions in terms of exponential functions in order to perform the summation over residues.  
Furthermore, in the final expression we have introduced the uniformization parameter (\ref{z}). 

Finally, let us consider the ZZ-ZZ cylinder amplitude. 
Applying eqs. (\ref{FZZTZZb2irr}) and (\ref{FZZTZZconsistent}) we may easily show, that the ZZ-ZZ cylinder amplitude 
is consistent with eqs. (\ref{id1}) and (\ref{id2}).
\[
\mathcal{Z}((r,s)_{_{C}},(\tilde{m},\tilde{n})_{_{ZZ}};(k,l)_{_{C}},(m,n)_{_{ZZ}})
\nonumber
\]
\[
= 
\mathcal{Z}\left(\!(r,s)_{_{C}},\sigma = i\left\{\!\frac{\tilde{m}}{b}+\tilde{n} b\!\right\};(k,l)_{_{C}},(m,n)_{_{ZZ}}\!\right) 
- 
\mathcal{Z}\left(\!(r,s)_{_{C}},\sigma = i\left\{\!\frac{\tilde{m}}{b}-\tilde{n} b\!\right\};(k,l)_{_{C}},(m,n)_{_{ZZ}}\!\right)  
\nonumber
\]
\[
= 
\!\!\!\!\!\!\!
\sideset{}{'}\sum_{\tilde{k} = \vert k - m \vert + 1}^{\textrm{min}\{k+m-1;2p-1-k-m\}}
\;\sideset{}{'}\sum_{\tilde{l} = \vert l - n \vert + 1}^{\textrm{min}\{l+n-1;2q-1-l-n\}}
\left[
\mathcal{Z}\left((r,s)_{_{C}},\sigma = i\left\{\frac{\tilde{m}}{b}+\tilde{n} b\right\};(\tilde{k},\tilde{l})_{_{C}},(1,1)_{_{ZZ}}\right) 
\right.
\phantom{map}
\nonumber
\]
\[
\phantom{mapmapmapmapmapmapmmapmapmap}
\left.
- 
\mathcal{Z}\left((r,s)_{_{C}},\sigma = i\left\{\frac{\tilde{m}}{b}-\tilde{n} b\right\};(\tilde{k},\tilde{l})_{_{C}},(1,1)_{_{ZZ}}\right)  
\right]
\nonumber
\]
\[
=  
\!\!\!\!\!\!\!
\sideset{}{'}\sum_{\tilde{k} = \vert k - m \vert + 1}^{\textrm{min}\{k+m-1;2p-1-k-m\}}
\;\sideset{}{'}\sum_{\tilde{l} = \vert l - n \vert + 1}^{\textrm{min}\{l+n-1;2q-1-l-n\}}
\left[
\mathcal{Z}\left((r,s)_{_{C}},(\tilde{m},\tilde{n})_{_{ZZ}};(\tilde{k},\tilde{l})_{_{C}},\sigma = i\left\{\frac{1}{b}+ b\right\}\right) 
\right.
\phantom{map}
\nonumber
\]
\[\!
\phantom{mapmapmapmapmapmapmmapmapmapm}
\left.
- 
\mathcal{Z}\left((r,s)_{_{C}},(\tilde{m},\tilde{n})_{_{ZZ}};(\tilde{k},\tilde{l})_{_{C}},\sigma = i\left\{\frac{1}{b}- b\right\}\right)  
\right]
\nonumber
\]
\[
= 
\!\!\!\!\!\!\!\!\!
\sideset{}{'}\sum_{\tilde{k} = \vert k - m \vert + 1}^{\textrm{min}\{k+m-1;2p-1-k-m\}}
\;\sideset{}{'}\sum_{\tilde{l} = \vert l - n \vert + 1}^{\textrm{min}\{l+n-1;2q-1-l-n\}}
\phantom{mapmapmapmapmapmapmapmapmapmapmapm}
\nonumber
\]
\[
\phantom{mapm}
\sideset{}{'}\sum_{\tilde{r} = \vert r - \tilde{m} \vert + 1}^{\textrm{min}\{r+\tilde{m}-1;2p-1-r-\tilde{m}\}}
\;\sideset{}{'}\sum_{\tilde{s} = \vert s - \tilde{n} \vert + 1}^{\textrm{min}\{s+\tilde{n}-1;2q-1-s-\tilde{n}\}}
\mathcal{Z}((\tilde{r},\tilde{s})_{_{C}},(1,1)_{_{ZZ}};(\tilde{k},\tilde{l})_{_{C}},(1,1)_{_{ZZ}})
\] 

\section{The nature of ZZ-branes}
The nature of the different principal ZZ boundary states becomes apparent, 
when we consider the expression (\ref{cylinterpretation}) for the 
FZZT-ZZ cylinder amplitude. 
In this expression we have integrated out all the matter and the ghost degrees of freedom.  
Notice, if we impose the basic $(1,1)$ Cardy matter state on both boundaries the contribution from the matter cylinder amplitude 
to the FZZT-ZZ cylinder amplitude is trivial.  
The only state, which flows in the open string channel, is the state 
corresponding to the identity operator and the matter cylinder amplitude is given by    
\[
\mathcal{Z}_{M}((1,1)_{_{C}};(1,1)_{_{C}}) 
= 
\textrm{Tr}_{L[c_{p,q},\Delta_{1,1}]} e^{-\frac{2\pi}{\tau}(L_{0}-c_{p,q}/24)} 
= \tilde{q}^{-c_{p,q}/24}\,,
\] 
where $\tilde{q}=e^{-\frac{2\pi}{\tau}}$ and where the Hamiltonian in the open string picture is derived in eq. (\ref{openham}). 
The factor $\tilde{q}^{-c_{p,q}/24}$ simply cancels the corresponding factors coming from the Liouville cylinder amplitude 
and the ghost cylinder amplitude due to the fact, that $c_{tot}=0$. 
Imposing the $(r,s)$ and the $(k,l)$ Cardy matter states on the two boundaries instead 
of the basic $(1,1)$ Cardy matter state the integrand in expression (\ref{cylinterpretation}) 
is modified by the factor 
\[
\frac{\sinh\left(\frac{2\pi P r}{b}\right)\sinh\left(2\pi P s b\right)}{\sinh\left(\frac{2\pi P}{b}\right)\sinh\left(2\pi P b\right)}
\frac{\sinh\left(\frac{2\pi P k}{b}\right)\sinh\left(2\pi P l b\right)}{\sinh\left(\frac{2\pi P}{b}\right)\sinh\left(2\pi P b\right)}\,.
\] 
Thus, we may view 
\[
\frac{\sinh\left(\frac{2\pi P k}{b}\right)\sinh\left(2\pi P l b\right)}{\sinh\left(\frac{2\pi P}{b}\right)\sinh\left(2\pi P b\right)}
\]
as a \emph{dressing factor} resulting from the integration over the matter degress of freedom on the ZZ-brane. 
This is exactly the factor, which distinguish the $(k,l)$ ZZ boundary state defined in eq. (\ref{ZZstatemn}) 
from the basic $(1,1)$ ZZ boundary state. 
If we start out by imposing the basic $(1,1)$ ZZ boundary state and $(k,l)$ Cardy matter state 
on one of the boundaries of the cylinder and then integrate out the matter degrees of freedom 
the basic $(1,1)$ ZZ boundary state turns into the $(k,l)$ ZZ boundary state
\[
\Psi_{1,1}(P) \to 
\frac{\sinh\left(2\pi P k / b\right)\sinh\left(2\pi P l b\right)}{\sinh\left(2\pi P /b\right)\sinh\left(2\pi P b\right)}
\Psi_{1,1}(P) 
= 
\Psi_{k,l}(P)\,.
\]
\emph{Thus, we realize, that we may view the $(k,l)$ ZZ boundary condition, $(k,l) \neq (1,1)$, as an effective boundary condition 
in Liouville theory obtained by integrating out the matter degrees of freedom.} 
Moreover, we realize, that we may either impose the $(k,l)$ Cardy matter condition at the absolute by imposing the $(k,l)$ Cardy matter state tensored 
with the basic $(1,1)$ ZZ boundary state 
or by imposing basic $(1,1)$ Cardy matter state tensored with the $(k,l)$ ZZ boundary state. 
This is the physics captured by the relation (\ref{id1}). 
The above interpretation of the principal ZZ boundary states 
is also valid with regard to the one-point function evaluated on the Lobachevskiy plane.  
In this case we do not have to integrate over any moduli $\tau$ and up to a factor independent of the boundary condition 
the one-point function factorizes into a matter part and a Liouville part. 
Imposing the $(k,l)$ Cardy matter state at the absolute, inserting the physical observable $\mathcal{A}^{r,s}_{P}$ 
and integrating over the matter degrees of freedom 
we obtain the matter part of the one-point function, 
which differs from the corresponding matter part 
of the one-point function obtained by imposing the basic $(1,1)$ Cardy matter state at the absolute 
by the factor    
\[
\frac{S_{(k,l);(r,s)}}{S_{(1,1);(r,s)}}\,,
\]
where $S$ is the modular S-matrix defined in eq. (\ref{mods}). 
Similarly, imposing the $(k,l)$ ZZ boundary state at the absolute and 
integrating over the configuration space of the Liouville field 
we obtain the matter part of the one-point function, which differs from the 
corresponding matter part of the one-point function obtained by imposing the basic $(1,1)$ ZZ boundary state 
at the absolute by the factor 
\[
\frac{\sinh\left(\frac{2\pi P k}{b}\right)\sinh\left(2\pi P l b\right)}{\sinh\left(\frac{2\pi P}{b}\right)\sinh\left(2\pi P b\right)}\,.
\]
Due to the special form of the modular S-matrix $S_{(r,s,);(k,l)}$ given in eq. (\ref{mods}) 
and due to the particular Liouville momentum associated with the given physical observable $\mathcal{A}^{r,s}_{P}$  the two factors 
coincide.\footnote{In light of the above interpretation of the principal ZZ boundary conditions in Liouville theory as effective boundary conditions 
it is tempting to view the matter part $\sqrt{S_{(1,1);(r,s)}}$ of the one-point function 
$\left(\phantom{}_{_{C}}\langle 1,1 \vert \otimes \phantom{}_{_{ZZ}}\langle k,l \vert \right) \vert \hat{\mathcal{A}}^{r,s}_{P} \rangle$ 
as a dressing factor of an effective operator in Liouville theory corresponding to $\hat{\mathcal{A}}^{r,s}_{P}$.}  

Even though we interpret the $(m,n)$ ZZ boundary condition, $(m,n) \neq (1,1)$, as 
an effective ZZ boundary condition obtained by integrating out the matter degrees of freedom, 
nothing from a conformal field theory point of view 
prevents us from imposing both the $(k,l)$ Cardy matter state, $(k,l) \neq (1,1)$, 
and the $(m,n)$ ZZ boundary state at the absolute. 
This somewhat unnatural boundary condition is allowed due to the fact, that 
the effective $(m,n)$ ZZ boundary condition is conformally invariant. 
The above interpretation is incomplete without some explanation of the physical meaning of this boundary condition.      
The nature of this boundary condition is captured by the relation (\ref{id2}). 
Recall that there is a one-to-one correspondance between the Cardy matter conditions 
and the conformal families of operators in the $(p,q)$ minimal model. 
Imposing the above somewhat unnatural boundary condition is equivalent to imposing the basic $(1,1)$ ZZ boundary condition at the absolute    
and imposing the matter boundary condition associated with the sum over conformal families appearing in the 
fusion of the two primary matter operators associated with the $(k,l)$ and the $(r,s)$ Cardy matter states
\[
\mathcal{O}_{k,l} 
\times 
\mathcal{O}_{m,n} 
= 
\sideset{}{'}\sum_{\tilde{k} = \vert k - m \vert + 1}^{\textrm{min}(k+m-1,2p-1-k-m)}
\:\sideset{}{'}\sum_{\tilde{l} = \vert l - n \vert + 1}^{\textrm{min}(l+n-1,2q-1-l-n)}
[\mathcal{O}_{\tilde{k},\tilde{l}}]\,.
\]
Notice, due to the fact that the fusion coefficients are positive integers, 
the corresponding matter boundary condition satisfies the Cardy condition discussed in section \ref{boundstates}. 
The introduction of the somewhat unnatural boundary condition discussed in the above is actually redundant, 
since we may express any ZZ-brane in terms of the basic ZZ-branes
\[
\bigg\{
\vert k,l \rangle_{_{C}} \otimes \vert 1,1 \rangle_{_{ZZ}} 
\,\bigg\vert\, 
1 \leq k \leq p-1\,,\;1 \leq l \leq q-1\,,\;kq-lp>0 
\bigg\}\,.
\]
This set of ZZ-branes compromises the set of independent boundary conditions, which we may impose 
at the absolute at infinity on the quantum Lobachevskiy plane. 
In this sense there only exists one ZZ boundary condition in pure Liouville theory, the basic $(1,1)$ ZZ boundary condition. 
All the other principal ZZ-boundary conditions are effective boundary conditions 
obtained by integrating out the matter degrees of freedom. 

How come the effective ZZ boundary conditions appear in the analysis of Zamolodchikov and Zamolodchikov and not only 
the basic $(1,1)$ ZZ boundary condition? 
The main point is that Zamolodchikov and Zamolodchikov consider Liouville theory parametrized on the disk, 
in which case there are no moduli $\tau$. 
This implies that the theory factorizes into a matter part, a Liouville part and a ghost part. 
Integrating out the matter and the ghost parts we end up with Liouville theory with the central charge 
\[
c_{L} = 26 - c_{p,q}
\]     
as an effective theory of the $(p,q)$ minimal model coupled to 2D euclidean Quantum Gravity. 
In this effective theory both the basic and the matter dressed ZZ boundary conditions appear.   

\chapter{ZZ branes from a worldsheet perspective} \label{ZZbranesworldsheet}
So far we have discussed 2D euclidean Quantum Gravity defined in terms of Liouville theory. 
In this chapter we will discuss a rather different approach to 2D euclidean Quantum Gravity, dynamical triangulations. 
For a thorough introduction to dynamical triangulations see \cite{ADJ},  
a classic on Quantum Geometry. 
In chapter \ref{chapterZZ} we studied the non-compact geometries in the $(p,q)$ minimal model coupled to 2D euclidean Quantum Gravity 
defined in terms of the ZZ branes. 
Due to the simplicity and beauty of dynamical triangulations we are able to calculate the cylinder amplitude  
with the exit loop in a fixed distance from the entrance loop, 
an amplitude, which is not obtainable from the continuum approach to  
2D euclidean Quantum Gravity in terms of Liouville theory. 
This amplitude will provide us with a whole new approach to obtaining non-compact geometries in the $(2,2m-1)$ minimal model coupled to 2D euclidean Quantum Gravity 
and will allow us to study the transition from compact to non-compact geometry. 
The cylinder amplitude with the exit loop in a fixed distance from the entrance loop  
was first derived in the seminal paper \cite{KKMW} by Kawai, Kawamoto, Mogami and Watabiki in the case of pure 
2D euclidean Quantum Gravity. In \cite{GK} their calculation was generalized to the so-called conformal background of 
the $(2,2m-1)$ minimal model coupled to 2D euclidean Quantum Gravity by Gubser and Klebanov. 

\section{Dynamical triangulations} \label{sectionDT}
Instead of considering different parametrizations of 2D euclidean geometries and worrying about gauge fixing etc., in dynamical triangulations  
we construct the 2D euclidean geometries explicitly by gluing equilateral polygons together.  
We may view the link length $a$ as a diffeomorphism invariant cutoff of distances. 
The statistical ensemble of triangulations obtained in this way defines a grid in the space of diffeomorphism classes of metrics. 
It is assumed that this grid becomes dense in the limit $a \to 0$ and that we may determine   
the partition function, in which we sum over all diffeomorphism classes of 2D euclidean geometries, by studying this limit. 
For simplicity we consider geometries of a fixed topology. 
In the following we will consider the class of so-called unrestricted triangulations. 
An unrestricted triangulation is defined as a complex obtained by gluing a collection of equilateral polygons  
and double links together. 
We may connect the polygons and the double links either at the vertices or along the links. 
A given complex remains the same if we attach a double link to a link. 
As a case study in dynamical triangulations let us take a closer look at the disk amplitude with one marked link on the boundary 
obtained from gluing triangles and double links together. 
Let us define the generating function of unrestricted triangulations of the disk as 
\st
w(g,z) 
& \equiv & 
\sum_{k=0}^{\infty}\sum_{l=0}^{\infty} w_{k,l}g^{k}z^{-(l+1)}
\nonumber\\
& = &
\sum_{k=0}^{\infty}\sum_{l=0}^{\infty} w_{k,l}e^{-\sigma k}e^{-\lambda(l+1)}
\label{regpartfunc}
\en 
where 
\[
g = e^{-\sigma}\,,\qquad z = e^{\lambda}\,   
\] 
and where $w_{k,l}$ is the number of unrestricted triangulations of the disk 
consisting of $k$ triangles and with $l$ links on the boundary of which one is marked. 
In the class of unrestricted triangulations we include the complex consisting of a single point. 
This implies that $w_{0,0}=1$.  
Since the area $A$ and the length $L$ of the boundary of a given triangulation consisting of $k$ triangles 
and with $l$ links on the boundary are roughly given by 
\[
A = k a^{2}\,,\qquad 
L = l a
\] 
we may view 
\[\label{bareL}
\mu_{0} = \frac{\sigma}{a^{2}}  
\]
as the bare bulk cosmological constant and 
\[\label{bareZ}
\mu_{B_{0}} = \frac{\lambda}{a}
\]
as the bare boundary cosmological constant. 
Hence, in eq. (\ref{regpartfunc}) each triangulation is weighed in accordance with the gravitational action (\ref{graacboundary}) 
and we may view the generating function $w(g,z)$ as the regularized analogue to the continuum 
disk partition function. 
In this way Quantum Gravity or rather Quantum Geometry is turned into a combinatorial problem of counting 
the number of non-isomorphic triangulations consisting of a given number of triangles 
and with a given number of links on the boundary. 
The generating function $w(g,z)$ was first determined by Tutte in 1962 in \cite{Tutte}   
\[
w(g,z) 
= 
\frac{1}{2} 
\left( 
z-gz^{2}+(gz-c_{2})\sqrt{(z-c_{+})(z-c_{-})}
\right)\,,  
\label{tut}
\]
where $c_{+}$, $c_{-}$ and $c_{2}$ are functions of 
$g$ analytical in a neighbourhood of zero. 
From the condition 
\[
w(g,z) = 1/z + \mathcal{O}(1/z^{2})
\] 
we obtain the three equations 
\[
\frac{1}{2} g (c_{+}+c_{-}) + c_{2} = 1\,,
\]
\[
(c_{+}+c_{-})c_{2} = \frac{1}{4}g(c_{+}-c_{-})^{2}
\]
and 
\[
(c_{+}-c_{-})^{2}(2c_{2}- g(c_{+} +c_{-})) = 32
\]
from which we may determine $c_{+}$, $c_{-}$ and $c_{2}$ in terms of $g$ 
\newpage
\[\label{c+} 
c_{+}(g) 
= 
\frac{1}{\sqrt{3}g}\cos\left(\frac{\theta}{3}+\frac{2\pi}{3}\right)
+
\frac{1}{2g}
+
\frac{1}{\sqrt{2}g}\sqrt{1-\frac{4}{3}\cos^{2}\left(\frac{\theta}{3}+\frac{2\pi}{3}\right)}\,, 
\] 
\[\label{c-} 
c_{-}(g) 
= 
\frac{1}{\sqrt{3}g}\cos\left(\frac{\theta}{3}+\frac{2\pi}{3}\right)
+
\frac{1}{2g}
-
\frac{1}{\sqrt{2}g}\sqrt{1-\frac{4}{3}\cos^{2}\left(\frac{\theta}{3}+\frac{2\pi}{3}\right)}\,\,
\]
and 
\[\label{c2} 
c_{2}(g) 
= 
\frac{1}{2}-\frac{1}{\sqrt{3}}\cos\left(\frac{\theta}{3}+\frac{2\pi}{3}\right)
\]
where 
\[\label{theta} 
\theta = \arccos(3^{3/2}4g^{2})\,.
\] 
In deriving these expressions we have imposed the boundary conditions 
\[
c_{2}(0) = 1\,,\quad 
c_{+}(0) = 2\,,\quad
c_{-}(0) = -2 
\]
obtained from the generating function of rooted branched polymers.\cite{ADJ}  
It follows from the above expressions, that the generating function $w(g,z)$ defined by the expansion in eq. (\ref{regpartfunc}) 
has a finite radius of convergence with respect to $g$
\[
g_{0} = e^{-\sigma_{0}} = \frac{1}{3^{3/4}2}\,
\]
and is convergent only if $\vert z \vert > c_{+}$ in the case $0 \leq g <g_{0}$.
Off course we may define $w(g,z)$ by analytic continuation outside the region of convergence. 
However,  
the interpretation of $w(g,z)$ in terms of unrestricted triangulations is only valid inside the region of convergence.  

The asymptotic behaviour of $w_{k,l}$ for large $k$ and bounded $l$ is given by \cite{ADJ}
\[\label{assydyn}
w_{k,l} 
\sim  
e^{\sigma_{0}k} 
k^{\beta}\,,
\]
where $\beta$ is a critical exponent.   
In order to obtain the continuum disk partition function with fixed area and fixed length of the boundary 
we must approach the continuum limit $a \to 0$ keeping both the area of the discretized worldsheet $A=ka^{2}$ 
and the length of the discretized boundary $L=la$ fixed. 
Hence, we are interested in $w_{k,l}$ in the limit of large $k$ and $l$, where 
\[
l^{2} \sim k\,. 
\]   
In light of eq. (\ref{assydyn})    
we expect $w_{k,l}$ to behave as  
\[
w_{k,l} 
\sim 
e^{\sigma_{0}k} 
e^{\lambda_{0}l}
f(k,l) 
\]
in this limit, where 
\[
e^{\lambda_{0}} 
= 
c_{+}(g_{0})
\]  
and where $f$ satisfies 
\[
f(\alpha^{2}k, \alpha l) 
= 
\alpha^{\kappa} 
f(k, l)\,. 
\]
We define the continuum disk partition function $W(A,L)$ with fixed area $A$ and fixed length $L$ of the boundary as 
\[
W(A,L) 
= 
a^{\kappa} 
f(A/a^{2},L/a)\,.
\]
We want to approach the continuum limit in a way such that  
the generating function $w(g,z)$ is dominated by triangulations consisting of a large number of triangles $k$
and with a large number of links $l$ on the boundary satisfying that $l^{2} \sim k$. 
Otherwise, we do not obtain any continuum disk partition function $W(\mu,\mu_{B})$ in this limit, 
which we may interpret as a weighed sum over geometries of finite area and with finite boundaries.  
This is obtained by tuning 
\[
g \to g_{0}\,, \qquad z \to c_{+}(g_{0}) = e^{\lambda_{0}}
\] 
in a specific pace  
as we tune $a \to 0$. 
Assuming this is the case we obtain 
\st
w(g,z) 
& = &
\sum_{k=0}^{\infty} 
\sum_{l=0}^{\infty} 
w_{k,l} 
e^{-\sigma k} 
e^{-\lambda (l+1)}
\nonumber\\ 
& \sim &
\frac{1}{c_{+}(g_{0})}
\sum_{k=0}^{\infty} 
\sum_{l=0}^{\infty} 
e^{-(\sigma-\sigma_{0})k} 
e^{-(\lambda-\lambda_{0})l} 
f(k,l)
\nonumber\\
& \approx & 
\frac{a^{-\kappa-3}}{c_{+}(g_{0})} 
\int_{0}^{\infty} d A 
\int_{0}^{\infty} d L 
e^{-\mu A - \mu_{B} L} 
W(A,L)
\nonumber\\
& = &
\frac{a^{-\kappa-3}}{c_{+}(g_{0})} 
W(\mu,\mu_{B}) 
\en
where we have identified    
\[
\frac{\sigma}{a^{2}} 
= 
\frac{\sigma_{0}}{a^{2}} + \mu  
\]
and 
\[
\frac{\lambda}{a} 
= 
\frac{\lambda_{0}}{a} + \mu_{B}\,. 
\]
In light of eqs. (\ref{bareL}) and (\ref{bareZ}) we observe, that both the bulk and the boundary cosmological constants undergo 
additive renormalization in the continuum limit. 
Moreover, the above calculation suggests, 
that the pace, at which $g$ should approach $g_{0}$ and $z$ should approach $c_{+}(g_{0})$ in the continuum limit, is given by 
\[
g = 
e^{-\sigma_{0}-\mu a^{2}} \approx 
g_{0}(1-\mu a^{2}) 
\]  
and 
\[
z = 
e^{\lambda_{0}+\mu_{B}a} \approx 
c_{+}(g_{0}) (1+\mu_{B}a)\,.
\]
Inserting these expressions into eqs. (\ref{c+}), (\ref{c-}), (\ref{c2}) and (\ref{theta}) we may easily show, that 
\[
c_{+} = c_{+}(g_{0}) - \tilde{\alpha} \sqrt{\mu} a + \mathcal{O}(a^{2})\,,
\]
\[
c_{-} = c_{-}(g_{0}) + \mathcal{O}(a^{2}) 
\]
and 
\[
c_{2} = g_{0}c_{+}(g_{0}) + \frac{1}{2}g_{0}\tilde{\alpha} \sqrt{\mu} a + \mathcal{O}(a^{2})\, 
\]
in the limit $a \to 0$, where $\tilde{\alpha}$ is a constant factor.  
Finally, inserting the above expressions into the expression (\ref{tut}) for the generating function we 
obtain\footnote{In the above continuum disk amplitude we have rescaled the boundary cosmological constant $\mu_{B}$ by a strictly positive factor.} 
\[
w(g,z) 
= 
V_{ns} + \gamma W(\mu,\mu_{B}) a^{3/2} + \mathcal{O}(a^{2})\,,
\]
where we consider  
\[\label{condisk}
W(\mu,\mu_{B}) = \left(\mu_{B} - \frac{1}{2}\sqrt{\mu}\right)\sqrt{\mu_{B}+\sqrt{\mu}}
\]
as the continuum disk amplitude with one marked point on the boundary and $\gamma$ is a constant factor. 
Notice, we disregard the leading term 
\[
V_{ns} 
= 
\frac{1}{2}\left\{c_{+}(g_{0})-g_{0}(c_{+}(g_{0}))^{2}\right\}+\frac{1}{2}\left\{c_{+}(g_{0})-2g_{0}(c_{+}(g_{0}))^{2}\right\}\mu_{B}a
\] 
due to the fact, that this term is analytic in both $\mu$ and $\mu_{B}$, 
which implies, that we may not associate this term with any geometries of finite area 
or with finite boundaries.\footnote{If we perform an inverse Laplace transform of $V_{ns}$ with respect to either $\mu$ or $\mu_{B}$ 
we obtain a function related to the Dirac delta function centered at zero by differentiation.}    
Moreover, the leading term is non-universal.  
It is related to the specific class of triangulations, 
which we consider in this discussion.    
From eqs. (\ref{diskamplitude}) and (\ref{cosmolcon}) 
we realize, that the continuum disk amplitude (\ref{condisk}) with one marked point on the boundary 
obtained in the above scaling limit matches the corresponding disk amplitude  
obtained from the Liouville approach to pure 2D euclidean Quantum Gravity. 
As discussed previously in the case of pure 2D euclidean Quantum Gravity the parameter $b$ in Liouville theory 
is given by $b=\sqrt{\frac{p}{q}}=\sqrt{\frac{2}{3}}$.

In the above example we have considered the class of unrestricted triangulations of the disk 
consisting of triangles and double links. 
In the continuum limit we obtained the disk amplitude with one marked point on the boundary in pure 2D euclidean Quantum Gravity.  
Let us proceed and consider the class of unrestricted triangulations consisting of polygons of order $i \leq N$ and double links.   
To each polygon of order $i$ we associate the weight $g_{i}$, which we allow to assume both positive and negative values. 
The generating function for unrestricted triangulations of the disk is defined as  
\[
w(g_{1},\ldots,g_{N},z) 
= 
\sum_{k_{1},\ldots,k_{N},l} 
w_{\{k_{j}\},l}
\:z^{-(l+1)}
\prod_{i=1}^{N}
g_{i}^{k_{i}}
\]
where $w_{\{k_{j}\},l}$ is the number of unrestricted triangulations of the disk 
with $k_{i}$ $i$-sided polygons and 
with one marked link on the boundary consisting of $l$ links. 
As in the case discussed previously the generating function is only convergent for $g_{1},\ldots,g_{N}$ 
belonging to a certain region in weight space and for $\vert z \vert > c_{+}(g_{1},\ldots,g_{N})$. 
The boundary of this region of convergence in weight space is a hyper-surface $M_{2}$ of codimension $1$. 
Embedded in $M_{2}$ we have a hierarchy of hyper-surfaces  
\[
M_{2}\supset M_{3} \supset \ldots \supset M_{N+1}\,,
\]      
where $M_{i}$ is a hyper-surface in weight space of codimension $i-1$. 
The $m^{\textrm{th}}$ multi-critical hyper-surface is given by $M_{m}/M_{m+1}$, 
$2 \leq m \leq N+1$.\footnote{We define $M_{N+2}$ as the empty set.} 
The continuum disk amplitude obtained in the scaling limit, 
in which $(g_{1},\ldots,g_{N})$ approaches a given point $(g^{0}_{1},\ldots,g^{0}_{N})$ on the boundary 
of the region of convergence and $z$ approaches $c_{+}(g^{0}_{1},\ldots,g^{0}_{N})$, depends on, 
which multi-critical hyper-surface $(g^{0}_{1},\ldots,g^{0}_{N})$ belongs to.   
However, the continuum disk amplitude does not depend on the particular point $(g^{0}_{1},\ldots,g^{0}_{N})$ which 
we approach on the given multi-critical hyper-surface.   
The above multi-critical hyper-surfaces are not specific to the generating function of unrestricted 
triangulations of the disk. 
If we consider the generating function of unrestricted triangulations of a given topology 
the same multi-critical hyper-surfaces appear in weight space. 
Each of these multi-critical hyper-surfaces may be associated with 
a particular matter theory coupled to 2D euclidean Quantum Gravity.   
The continuum theory associated with a given multi-critical hyper-surface may be identified from the critical exponents associated with the given multi-critical hyper-surface. 
One of the critical exponents characterizing the different multi-critical hyper-surfaces is the so-called string 
susceptibility $\gamma$ defined by 
\[
Z_{h}(A) \sim A^{\frac{(\gamma-2)\chi(h)}{2}-1}\, 
\] 
where $Z_{h}(A)$ is the continuum partition function, in which 
we sum over all closed geometries of genus $h$ and with fixed area $A$, 
and $\chi(h)$ is the Euler number defined in eq. (\ref{euler}). 
We may determine the string susceptibility associated with a given multi-critical hyper-surface  
from the scaling behaviour of the generating function of closed triangulation of genus $h$.\cite{ADJ} 
For a given conformal field theory coupled to 2D euclidean Quantum Gravity the string susceptibility 
was first determined by Knizhnik, Polyakov and Zamolodchikov in \cite{KPZ} in the genus zero case. 
For higher genus the string susceptibility was first determined by David in \cite{David}. 
Comparing the string susceptibilities and other critical exponents obtained in the different approaches to 2D euclidean Quantum Gravity 
we identify the scaling limit, in which we approach 
the $m^{\textrm{th}}$ multi-critical hyper-surface, as the $(2,2m-1)$ minimal model coupled to 2D euclidean Quantum Gravity. 
In the $(2,2m-1)$ minimal model coupled to 2D euclidean Quantum Gravity we have $m-1$ tachyon operators defined by eq. (\ref{tachyon}),   
which may contribute to the action. 
The background values of the coupling constants of these tachyon operators are determined by the path along 
which we approach a given point on the $m^{\textrm{th}}$ multi-critical hyper-surface. 
Approaching the $m^{\textrm{th}}$ multi-critical hyper-surface along a path such that all coupling constants are set equal to zero 
except the cosmological constants we obtain the so-called conformal background. 
This is the particular scaling limit, which we identify as the tensor product of the $(2,2m-1)$ minimal model, Liouville theory 
and the conformal ghost theory. 
Approaching the given point on the $m^{\textrm{th}}$ multi-critical hyper-surface along a different path 
we obtain a marginal deformation of the conformal background in which case the continuum theory 
does not factorize into a matter part and a Liouville part.\cite{Stau}   
In this thesis we will only consider the conformal background of the $(2,2m-1)$ minimal model coupled to 2D euclidean Quantum Gravity.  

The cylinder amplitude in the $(2,2m-1)$ minimal model coupled to 2D euclidean Quantum Gravity 
was first obtained from dynamical triangulations in \cite{AJM,Stau}.\footnote{Actually, in \cite{AJM,Stau} the cylinder amplitude 
is obtained from the one-matrix model, both the complex and the Hermitian. 
We may view the $1/N$ expansion of the one-matrix model as an ingenious tool to obtain 
the generating function of unrestricted triangulations of a given topology.  
An introduction to the one matrix model is beyond the scope of this thesis.} 
Quite remarkable, the cylinder amplitude obtained from dynamical triangulations is the same in  
all $(2,2m-1)$ minimal models coupled to 2D euclidean Quantum Gravity independently of background. 
The cylinder amplitude obtained from dynamical triangulations with no marked point on either boundary is given by 
\[
Z_{DT} = -\ln\left[\left(\sqrt{\mu_{B_{1}}+\sqrt{\mu}}+\sqrt{\mu_{B_{2}}+\sqrt{\mu}}\right)^{2}a\right] \,,  
\]
where $\mu_{B_{1}}$ and $\mu_{B_{2}}$ are the boundary cosmological constants associated with the two boundaries.  
However, as we have seen in chapter \ref{chaptercyl} there exists a whole family of cylinder amplitudes 
in the $(2,2m-1)$ minimal model coupled to 2D euclidean Quantum Gravity, one for each pair of Cardy matter states. 
The scaling limit of dynamical triangulations in which we approach the $m^{\textrm{th}}$ multi-critical hyper-surface 
corresponds to one concrete realization of the $(2,2m-1)$ minimal model coupled to 2D euclidean Quantum Gravity and to 
one concrete realization of boundary conditions. 
In order to determine which particular Cardy matter condition is realized in this scaling limit 
we compare the cylinder amplitude obtained from dynamical triangulations 
with the cylinder amplitudes  
obtained in eqs. (\ref{cyl1}), (\ref{cyl2}) and (\ref{cyl3}). 
\emph{From this we conclude, that the $(1,1)$ Cardy matter condition is realized in the scaling limit of dynamical triangulations, 
in which we approach the $m^{\textrm{th}}$ multi-critical hyper-surface.}  
As we discussed in case of the Ising model in section \ref{sectionpq} 
the $(1,1)$ Cardy matter state corresponds to a fixed boundary condition in the concrete realizations of the $(p,q)$ minimal models. 
Furthermore, the only open string state, which couples between two $(1,1)$ Cardy boundary conditions, is the state associated 
with the identity operator.

\section{The cylinder amplitude with fixed distance} \label{sectioncyldis}
Formally, we define the cylinder amplitude with fixed distance $D$ as 
\[
N(\mu_{B},\mu'_{B},\mu;D) 
= 
\int \frac{\mathcal{D}g}{\textrm{Voll(Diff)}} 
\exp\left(-\mu_{B}L-\mu'_{B}L'-\mu A\right) 
\prod_{x'  \in \,   c'} \delta\!\left(d(x',c)-D\right) 
\]
where we integrate over all geometries homeomorphic to the cylinder, with one marked point on the entrance loop c 
and no marked points on the exit loop $c'$. 
$\mu_{B}$ denotes the boundary cosmological constant associated with entrance loop, 
while $\mu'_{B}$ denotes the boundary cosmological constant associated with the exit loop. 
The distance $d$ between a given point $x'$ on the exit loop and the entrance loop $c$ is defined as 
\[
d(x',c) = \textrm{min}\{d(x',x) \vert  x \in c \}\,,
\]
where $d(x',x)$ is the distance between the two points $x$ and $x'$. 
The delta functions appearing in the above definition 
impose the constraint, 
that only geometries, that satisfy, that each point $x'$ on the exit loop $c'$ is in a fixed distance $D$ of the entrance loop $c$, contribute to the partition function. 
Notice, this definition is \emph{not} symmetric in $c$ and $c'$. 
Moreover, the points on the exit loop of a given geometry 
satisfying the delta function constraints 
are not necessarily the only points  
in a distance $D$ of the entrance loop. 
\begin{figure}[!b]
\begin{center}
\includegraphics[width=0.25\textwidth]{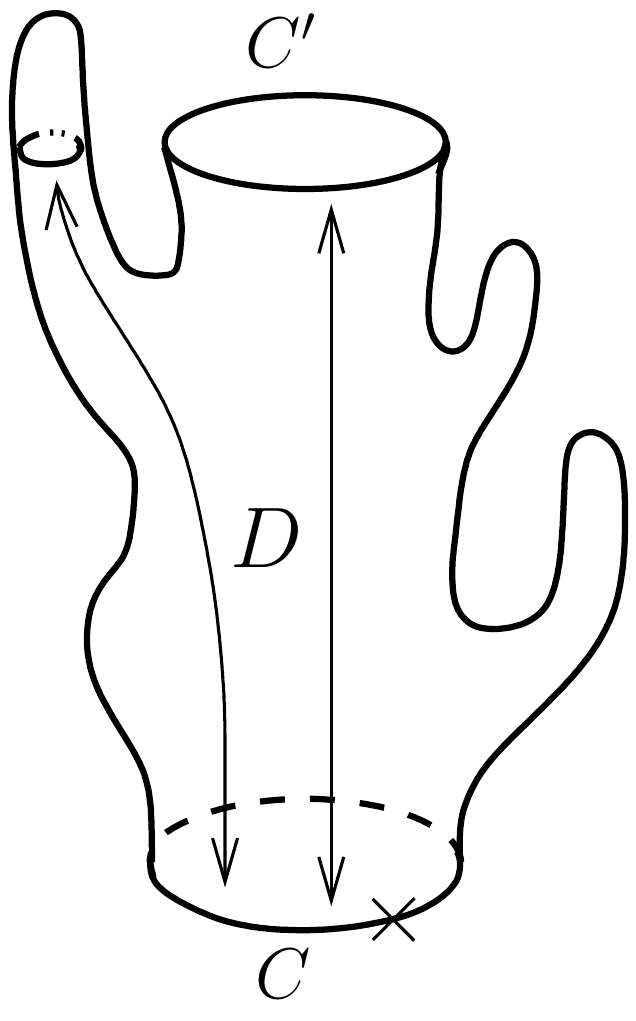}
\end{center}
\caption{The cylinder amplitude with fixed distance}
\end{figure}
In general, the set of points in a given distance $D$ of the entrance loop $c$ consists of several disconnected loops.

The above definition is clearly a formal definition. 
To state a precise and concrete definition, from which one  
may determine the cylinder amplitude with the exit loop in a fixed distance from the entrance loop, 
is an entirely different matter.
The concept of distance is highly non-trivial in Quantum Gravity. 
In Liouville theory the geodesic distance is a derived non-local concept, which is difficult to work with. 
The concept of geodesic distance in Liouville theory is discussed by David in \cite{D2}.   
The fact, that we may introduce a measure of distance in Quantum Gravity from dynamical triangulations, 
illustrates the strength of the combinatorial approach to Quantum Gravity 
offered by dynamical triangulations. 
For a given triangulation we define the dual lattice as the ``fat'' graph obtained by replacing each $i$-sided polygon 
by a vertex with $i$ emerging double lines each crossing a separate link of the $i$-sided polygon. 
We connect the two double lines crossing a link shared by two neighbouring polygons. 
The introduction of double lines instead of single lines on the dual lattice  
allows us to define a notion of orientation on the dual lattice. 
We define the distance between two given links on a given triangulation 
as the minimum number of double lines connecting the two links on the dual lattice.  
Applying this definition Kawai, Kawamoto, Mogami and Watabiki consider the generating function of 
unrestricted triangulations of the cylinder satisfying that the minimum distance from any given link 
on the exit loop to the entrance loop is a fixed positive integer $d$. 
They only consider triangulations consisting of triangles and double links. 
In the so-called transfer matrix formalism they show  
that the cylinder amplitude with fixed distance $D$ and one marked point on the entrance loop 
obtained in the scaling limit corresponding to pure 2D euclidean Quantum Gravity satisfies the PDE \cite{KKMW}
\[\label{PDE}
\frac{\prt}{\prt D} N(\mu_{B},\mu'_{B},\mu;D)  
= 
- \frac{\prt}{\prt \mu_{B}}\left[ W(\mu_{B},\mu) N(\mu_{B},\mu'_{B},\mu;D) \right] 
\]
and the initial condition 
\[\label{initialconcyl}
N(\mu_{B},\mu'_{B},\mu;0) = \frac{1}{\mu_{B}+\mu'_{B}}\,. 
\]
The disk amplitude $W(\mu_{B},\mu)$ is given by (\ref{condisk}). 
Performing an inverse Laplace transform with respect to both $\mu_{B}$ and $\mu'_{B}$ the initial condition reads 
\[\label{inilooplenght}
N(L, L',\mu;0) = \delta(L-L')\,, 
\]
where $L$ is the length of the entrance loop and $L'$ is the length of the exit loop.   
Gubser and Klebanov generalize the calculation done by Kawai, Kawamoto, Mogami and Watabiki and 
show, that the above equations are also valid in the conformal background of the $(2,2m-1)$ minimal model 
coupled to 2D euclidean Quantum Gravity obtained in the scaling limit, 
in which we approach the $m^{\textrm{th}}$ multi-critical hyper-surface.\cite{GK} 
In this case $W(\mu_{B},\mu)$ denotes the disk amplitude with one marked point on the boundary 
in the conformal background of the $(2,2m-1)$ minimal model coupled to 2D euclidean Quantum Gravity. 
In the scaling limit in which we approach the $m^{\textrm{th}}$ multi-critical hyper-surface  
the continuum distance $D$ is defined as  
\[\label{defD}
D = a^{\eta} d\,, 
\]   
where the critical exponent $\eta$ is given by   
\[\label{scaleD}
\eta = m-3/2\,.
\]
The scaling of $D$ in terms of $a$ is fixed by the condition, that  
the two sides of the equation, from which we derive the partial differential equation (\ref{PDE}), 
scale in the same way.    

We may obtain a lot of interesting physics from the above cylinder amplitude. 
In the remaining part of this section 
let us discuss the anomalous dimension of the distance $D$.    
Let us first consider a given geometry homeomorphic to the sphere. 
The area within a distance $D$ of a given point $x$ is given by   
\[\label{de1}
A(x,D) = \int_{0}^{D} dD' \int d^{2}x' \sqrt{g(x')} \:\delta\!\left(d(x',x)-D'\right)\,.  
\]  
With respect to the given geometry 
we define the average area of a disk of radius $D$ as 
\[\label{de2}
A_{g}(D) 
= 
\frac{\int d^{2}x \, \sqrt{g(x)} \, A(x,D)}{\int d^{2}x \, \sqrt{g(x)}}\,.
\]
The corresponding ensemble average for fixed total area $A_{\textrm{tot}}$ is defined as 
\[\label{de3}
\langle A(D) \rangle_{A_{\textrm{tot}}} 
= 
\frac{\int \mathcal{D}g \; \delta\!\left(\int d^{2}x \sqrt{g(x)} - A_{\textrm{tot}}\right) A_{g}(D)}
{\int \mathcal{D}g \; \delta\!\left(\int d^{2}x \sqrt{g(x)} - A_{\textrm{tot}}\right)}\,,
\]
where we integrate over all geometries homeomorphic to the sphere and with area $A_{\textrm{tot}}$. 
The Hausdorff dimension $d_{H}$ of the ensemble of manifolds 
is defined by 
\[\label{hausdorff}
\langle A(D) \rangle_{A_{\textrm{tot}}} 
\approx 
\alpha D^{d_{H}}\,, \qquad D \to 0\,.  
\]
The coefficient $\alpha$ appearing in this definition must be dimensionless. 
The ensemble average $\langle A(D) \rangle_{A_{\textrm{tot}}}$ 
is given in terms of two parameters with a non-trivial dimension, the distance $D$ and the area $A_{\textrm{tot}}$. 
However, in the limit $D \to 0$ the ensemble average $\langle A(D) \rangle_{A_{\textrm{tot}}}$ 
cannot depend on the area $A_{\textrm{tot}}$. 

What is the physics encoded in the Hausdorff dimension $d_{H}$? 
In the following we assume that $D$ measures the geodesic distance. 
Let us consider a given point $x$ on a generic smooth\footnote{Based on the insights obtained from 
studying the paths contributing to the propagator of a free scalar field \cite{Polyakov87,Makeenko} we do not expect the partition function to be dominated by smooth geometries.} 
geometry and let us gradually zoom in on a neighbourhood of $x$.  
On a sufficiently small scale $S$ we expect the geometry to appear flat in a neighbourhood of $x$. 
For $D < S$ the area of the disk of radius $D$ and centered at $x$ is given by the well-known formula
\[
A(x,D) \approx \pi D^{2}\,.
\]
If the ensemble average is dominated by geometries, which all appear flat on a sufficiently small finite scale $S$, 
the Hausdorff dimension is equal to $2$. 
However, if the Hausdorff dimension is different from $2$, then no matter how small a scale $S$ we consider,  
geometries, which are not flat on this scale $S$, contribute significantly to the ensemble average.   

Let us determine the Hausdorff dimension associated with the distance $D$ introduced 
in the $(2,2m-1)$ minimal model coupled to 2D euclidean Quantum Gravity from dynamical triangulations. 
From eqs. (\ref{hausdorff}) and (\ref{defD}) we may determine the dimension of 
the ensemble average\footnote{Notice, if we differentiate 
the ensemble average $\langle A(D) \rangle_{A_{\textrm{tot}}}$ defined by 
eqs. (\ref{de1}), (\ref{de2}) and (\ref{de3}) with respect to $D$ 
we obtain the reparametrization invariant two-point function with fixed area up to a constant factor independent of $D$. 
Moreover, if we shrink both the entrance loop and the exit loop to a point 
in the cylinder amplitude given by eqs. (\ref{PDE}) and (\ref{initialconcyl})  
we obtain the reparametrization invariant two-point function, where we sum over all geometries homeomorphic to the sphere 
independent of their area. This two-point function is related to the two-point function with fixed area by a Laplace transformation.  
Hence, we may obtain the ensemble average $\langle A(D) \rangle_{A_{\textrm{tot}}}$ 
from the cylinder amplitude with fixed distance.} 
\[
[\langle A(D) \rangle_{A_{\textrm{tot}}}] 
= 
[D]^{d_{H}} 
= 
[a]^{\eta d_{H}}\,.
\] 
From the definition (\ref{de3}) of the ensemble average we obtain that 
\[
[\langle A(D) \rangle_{A_{\textrm{tot}}}] 
= 
[A] 
=
[a]^{2}\,.
\] 
We conclude that the Hausdorff dimension associated with the distance $D$ in the $(2,2m-1)$ minimal model coupled to 2D euclidean Quantum Gravity 
is given by 
\[\label{haus1}
d_{H} = \frac{4}{2m-3}\,.
\]
In the semi-classical limit $b \to 0$ corresponding 
to $m \to \infty$ 
we expect, that the partition function with fixed area is dominated by a single smooth geometry, in which case 
the Hausdorff dimension associated with the geodesic distance is equal to $2$. 
The Hausdorff dimension given by the above equation do not approach $2$ in the semi-classical limit \mbox{$m \to \infty$}. 
In \cite{KSW} Kawamoto, Saeki and Watabiki determined the Hausdorff dimension from studying the diffusion equation in Liouville theory.  
They reached the conclusion, that the Hausdorff dimension associated with the geodesic distance is given by 
\[\label{haus2}
d_{H} = 
2 \frac{\sqrt{25-c}+\sqrt{49-c}}{\sqrt{25-c}+\sqrt{1-c}} \,, 
\]
where $c$ is the central charge of the conformal matter theory coupled to 2D euclidean Quantum Gravity.  
Notice, this expression for the Hausdorff dimension indeed approaches $2$ in the semi-classical limit $b \to 0$ 
corresponding to $c \to -\infty$. 
In the case of pure 2D euclidean Quantum Gravity corresponding to $m=2$ the two expressions for the Hausdorff dimension (\ref{haus1}) and (\ref{haus2}) 
match. We conclude, that $D$  
is the geodesic distance  
in the case of pure 2D euclidean Quantum Gravity. 
In the case of the $(2,2m-1)$ minimal model coupled to 2D euclidean Quantum Gravity, $m>2$, 
the two expressions for the Hausdorff dimension do not match. In this case $D$ is \emph{not} the geodesic distance. 
Rather, $D$ is a distance measured in terms of matter excitations. 
This is explicit in some constructions of 2D euclidean Quantum Gravity with matter such as the Ising model 
studied by Ambj\o rn, Anagnostopoulos, Jurkiewics and Kristjansen in \cite{AAJK} 
and the $c=-2$ model formulated as an $O(-2)$ model studied by Ambj\o rn, Kristjansen and Watabiki in \cite{AKW}. 
The fact, that the fractal dimension of pure 2D euclidean Quantum Gravity is equal to $4$ and not $2$ 
is related to the fact, 
that the partition function in pure 2D euclidean Quantum Gravity is dominated by wildly branching geometries.\cite{KKMW} 
From so-called Causal Dynamical Triangulations one may obtain the cylinder amplitude with fixed geodesic distance $D$ 
between the two boundaries in pure 2D Lorentzian Quantum Gravity. 
This amplitude describes the propagation of a closed loop with ``time'' $D$. 
In Causal Dynamical Triangulations we only include geometries in the statistical ensemble, which satisfy, that 
the topology of any given ``time''-slice is $S^{1}$, that is we exclude branching geometries from the statistical ensemble. 
This constraint prevents the propagating loop from splitting up into several loops with ``time'' $D$. 
The Hausdorff dimension obtained from Causal Dynamical Triangulations is $2$.\cite{AL}

\section{ZZ-branes from a worldsheet perspective} \label{worldsheetperspective}
In chapter \ref{chapterZZ} we studied non-compact geometries in the $(p,q)$ minimal model 
coupled to 2D euclidean Quantum Gravity defined in terms of the ZZ-branes. 
In this section we will introduce a different approach to obtaining 
non-compact geometries in the $(2,2m-1)$ minimal model coupled to 2D euclidean Quantum Gravity. 
This section is based upon work done by the author of this thesis 
in collaboration with his academic adviser Prof. Jan Ambj\o rn and 
partly with S. Kawamoto and S. Arianos. 
This work is published in \cite{AG1} and \cite{AG2}. 
The author of this thesis took part in the work published in \cite{AG1} as a part of his master thesis. 
This work concerns the geometry of the disk amplitude with fixed geodesic distance in pure 2D euclidean Quantum Gravity.  
The results published in \cite{AG2} have been obtained 
by the author of this thesis during his time as a Ph.D student 
in collaboration with his academic adviser. 
This work concerns the cylinder amplitudes with fixed distances 
in the $(2,2m-1)$ minimal models coupled to 2D euclidean Quantum Gravity 
including pure 2D euclidean Quantum Gravity and concerns the 
interpretation of the obtained results especially in terms of branes. 

Given the initial condition (\ref{initialconcyl}) 
we may easily solve the partial differential equation (\ref{PDE}) by the method of characteristic curves.  
We first determine the characteristic curve in the $(\tilde{\mu}_{B},\tilde{D})$-plane passing through $(\mu_{B},D)$ 
along which 
\[
G(\tilde{\mu}_{B},\mu'_{B},\mu;\tilde{D}) 
\equiv  
W(\tilde{\mu}_{B},\mu)N(\tilde{\mu}_{B},\mu'_{B},\mu;\tilde{D})
\]
is constant. 
We parametrize the characteristic curve by $\lambda$. 
Applying the partial differential equation (\ref{PDE}) we obtain 
\st
0 
& = & 
\frac{dG}{d\lambda} 
=  
\frac{\prt G}{\prt \tilde{\mu}_{B}} 
\frac{d \tilde{\mu}_{B}}{d\lambda} 
+ 
\frac{\prt G}{\prt \tilde{D}} 
\frac{d \tilde{D}}{d\lambda} 
= 
-\frac{1}{W(\tilde{\mu}_{B},\mu)}
\frac{\prt G}{\prt \tilde{D}}
\frac{d \tilde{\mu}_{B}}{d\lambda}
+
\frac{\prt G}{\prt \tilde{D}} 
\frac{d \tilde{D}}{d\lambda}
\en
from which we derive the characteristic equation 
\[
\frac{d \tilde{D}}{d\lambda} 
= 
\frac{1}{W(\tilde{\mu}_{B},\mu)}
\frac{d \tilde{\mu}_{B}}{d\lambda}\,.    
\]
The solution to the characteristic equation is given by 
\[\label{natureflow}
D = \int_{\hat{\mu}_{B}}^{\mu_{B}} \frac{d\tilde{\mu}_{B}}{W(\tilde{\mu}_{B},\mu)} \,,
\]
where $(\hat{\mu}_{B},0)$ denotes the point in the $(\tilde{\mu}_{B},\tilde{D})$-plane 
at which the characteristic curve crosses the $\tilde{\mu}_{B}$-axis. 
The above equation determines $\hat{\mu}_{B}$ as a function of $\mu_{B}$ and $D$. 
From the initial condition (\ref{initialconcyl}) 
we obtain the cylinder amplitude with fixed distance 
\st
N(\mu_{B},\mu'_{B},\mu;D) 
& = & 
\frac{1}{W(\mu_{B},\mu)}G(\mu_{B},\mu'_{B},\mu;D) 
=  
\frac{1}{W(\mu_{B},\mu)}G(\hat{\mu}_{B},\mu'_{B},\mu;0)
\nonumber\\ 
& = & 
\frac{1}{\hat{\mu}_{B}+\mu'_{B}}\frac{W(\hat{\mu}_{B},\mu)}{W(\mu_{B},\mu)}\,.
\en
Let us begin by considering the cylinder amplitude with fixed distance in pure 2D euclidean Quantum Gravity, 
in which case $D$ measures the geodesic distance. 
In pure 2D euclidean Quantum Gravity the disk amplitude with one marked point on the boundary is given by eq. (\ref{condisk}) 
and we may express the cylinder amplitude with fixed geodesic distance as  
\[\label{cylgeo}
N(\mu_{B},\mu'_{B},\mu;D)  
= 
\frac{1}{\hat{\mu}_{B}+\mu'_{B}}
\frac{\left(\hat{\mu}_{B} - \frac{1}{2}\sqrt{\mu}\right)\sqrt{\hat{\mu}_{B}+\sqrt{\mu}}}
{\left(\mu_{B} - \frac{1}{2}\sqrt{\mu}\right)\sqrt{\mu_{B}+\sqrt{\mu}}}\,,
\]
where $\hat{\mu}_{B}$ is determined by 
\[\label{induced}
D  
= 
\int_{\hat{\mu}_{B}}^{\mu_{B}} \frac{d\tilde{\mu}_{B}}{\left(\tilde{\mu}_{B} - \frac{1}{2}\sqrt{\mu}\right)\sqrt{\tilde{\mu}_{B}+\sqrt{\mu}}}\,.
\] 
Viewing eq. (\ref{PDE}) as a renormalization group equation and $D$ as the scale we see that eq. (\ref{PDE}) 
induces a flow in $\hat{\mu}_{B}$. 
With regard to this flow the real $\mu_{B}$-axis is divided into to two disjoint intervals    
separated by the zero $\sqrt{\mu}/2$ of $W(\mu_{B},\mu)$. 
For $\mu_{B} \in (\sqrt{\mu}/2,\infty)$ the flow of $\hat{\mu}_{B}$ is constrained to the interval $(\sqrt{\mu}/2,\infty)$, while  
for $\mu_{B} \in (-\sqrt{\mu},\sqrt{\mu}/2)$ the flow of $\hat{\mu}_{B}$ is constrained to the interval $(-\sqrt{\mu},\sqrt{\mu}/2)$. 
This follows from eq. (\ref{induced}) and the initial condition $\hat{\mu}_{B}=\mu_{B}$ for $D=0$.
In the following we will consider $\mu_{B} > \sqrt{\mu}/2$ and $\mu'_{B} > -\sqrt{\mu}/2$ in which case 
we have a clear interpretation of $N(\mu_{B},\mu'_{B},\mu;D)$ in terms of geometry. 
For $\mu'_{B} > -\hat{\mu}_{B}$ (which is always the case, if  $\mu_{B} > \sqrt{\mu}/2$ and $\mu'_{B} > -\sqrt{\mu}/2$,)  
we may express the cylinder amplitude with fixed geodesic distance as the Laplace transform  
\[
N(\mu_{B},\mu'_{B},\mu;D)  
= 
\int_{0}^{\infty} dL'
e^{-(\mu'_{B} + \hat{\mu}_{B})L'}
\frac{W(\hat{\mu}_{B},\mu)}{W(\mu_{B},\mu)}
\]   
from which it is clear, that we may view $\hat{\mu}_{B}$ as an induced boundary cosmological constant on the exit loop. 
The induced boundary cosmological constant $\hat{\mu}_{B}$ (or more precisely the factor $e^{-\hat{\mu}_{B}L'}$) is related to 
the entropy of geometries homeomorphic to the cylinder and with an exit loop of fixed length $L'$ 
in a fixed geodesic distance $D$ from a marked entrance loop. Later on when we consider the $(2,2m-1)$ minimal model coupled 
to 2D euclidean Quantum Gravity this is no longer true due to the presence of matter.  

In order to generate a non-compact random geometry we will consider the limit in which the geodesic distance $D$ 
from the exit loop to the entrance loop diverges.   
\emph{In the non-compact limit $D \to \infty$ the induced boundary cosmological constant $\hat{\mu}_{B}$ 
approaches the fixed point $\sqrt{\mu}/2$ independent of the value of $\mu_{B}$.  
The induced boundary cosmological constant $\sqrt{\mu}/2$ is generic to the non-compact limit and 
precisely match the value of the boundary cosmological constant associated with the 
single principal ZZ brane in pure 2D euclidean Quantum Gravity through eq. (\ref{cosmologicalZZ}). 
In this sense we may view the cosmological constant associated with the ZZ-brane and the non-compact 
quantum Lobachevskiy plane in pure 2D euclidean Quantum Gravity as an induced boundary cosmological constant.}  
In order to study the random geometry obtained in the limit $D \to \infty$ 
let us start out by shrinking the entrance loop to a point  
\[\label{diskD}
\widetilde{W}(\mu'_{B},\mu;D) 
= 
\lim_{\mu_{B} \to \infty} \mu_{B}^{3/2} N(\mu_{B},\mu'_{B},\mu;D)  
= 
\frac{W(\hat{\mu}_{B},\mu)}{\mu'_{B}+\hat{\mu}_{B}}\,.
\]
In this limit the induced boundary cosmological constant is obtained eq. (\ref{induced}) with $\mu_{B} = \infty$. 
\[\label{inz}
\hat{\mu}_{B} = \frac{\sqrt{\mu}}{2} \left( 1  + \frac{3}{\sinh^{2}\left(\sqrt{\frac{3}{2}}\mu^{1/4}D\right)} \right)\,.
\]
The disk amplitude $\widetilde{W}(\mu'_{B},\mu;D)$ with fixed geodesic distance 
is similar to the ordinary disk amplitude $W(\mu'_{B},\mu)$ except  
that all points on the boundary are in a fixed geodesic distance $D$ from a point (puncture) in the bulk of the disk. 
Viewing $\widetilde{W}(\mu'_{B},\mu;D) $ as a partition function we may determine 
the average area and the average boundary length of the corresponding random geometry in the limit $D \to \infty$. 
For generic values of $\mu'_{B}$ we obtain from the two above equations   
\st
\langle A \rangle 
& = &  
- \frac{1}{\widetilde{W}(\mu'_{B},\mu;D)}  \frac{\prt}{\prt \mu} \widetilde{W}(\mu'_{B},\mu;D)  
\nonumber\\
& = &
\frac{1}{\mu'_{B}+\hat{\mu}_{B}}
\frac{1}{4\sqrt{\mu}} 
\left(
1 + 
\frac{3\sinh\left(\sqrt{\frac{3}{2}}\mu^{1/4}D\right)
-3\sqrt{\frac{3}{2}}\mu^{1/4}D\cosh\left(\sqrt{\frac{3}{2}}\mu^{1/4}D\right)}
{\sinh^{3}\left(\sqrt{\frac{3}{2}}\mu^{1/4}D\right)}
\right)
\nonumber\\
&& 
+
\frac{\sqrt{3}}{2\sqrt{2}\mu}
\frac{2\cosh^{2}\left( \sqrt{\frac{3}{2}}\mu^{1/4}D \right) + 1}
{\sinh\left(\sqrt{6} \mu^{1/4}D  \right)}
\mu^{1/4}D
-
\frac{3}{4\mu} 
\nonumber\\
& \approx &
\frac{\sqrt{3}}{\sqrt{2}\mu^{3/4}}D + \mathcal{O}(1)\,, \qquad \textrm{for } D \to \infty\,.
\label{avarea}
\en
and 
\st
\langle L' \rangle 
& = & 
- \frac{1}{\widetilde{W}(\mu'_{B},\mu;D)}  \frac{\prt}{\prt \mu'_{B}} \widetilde{W}(\mu'_{B},\mu;D) 
\;= \;
\frac{1}{\mu'_{B}+\hat{\mu}_{B}}
\label{avlength}
\en
from which we see, that the random geometry obtained in the non-compact limit $D \to \infty$ 
for generic values of $\mu'_{B}$ essentially is a 
semi-infinite cylinder of vanishing radius except close to the exit loop, where the radius becomes finite.  
However, precisely for $\mu'_{B} = -\sqrt{\mu}/2$ we obtain 
\[\label{avarea2}
\langle A \rangle 
\approx  
\frac{1}{24\mu}
e^{\sqrt{6}\mu^{1/4}D}\,, \qquad \textrm{for } D \to \infty
\]   
and 
\[\label{avlength2}
\langle L' \rangle 
\approx  
\frac{1}{6\sqrt{\mu}}
e^{\sqrt{6}\mu^{1/4}D}\,, \qquad \textrm{for } D \to \infty\,.
\]
\emph{For $\mu'_{B} = -\frac{\sqrt{\mu}}{2}$ and in the non-compact limit of large $D$ the average area 
and the average boundary length are proportional and they both grow exponentially with the geodesic distance $D$. 
This is exactly the signature of the Lobachevskiy plane discussed in section \ref{Lobachevskiy}.} 
Let us try to understand this result. 
There only exists 
one non-compact solution to classical Liouville theory, the Lobachevskiy plane. 
The negative curvature of the Lobachevskiy plane is determined by the positive value of $\mu$. 
Due to this fact we expect, that (for a fixed value of $\mu$)  
there essentially only exists one non-compact random geometry in pure 2D euclidean Quantum Gravity, a quantum Lobachevskiy plane. 
This is indeed consistent with above the result.\footnote{The infinitely narrow cylinder obtained in the non-compact 
limit $D \to \infty$ for generic values of $\mu'_{B}$ is essentially a one-dimensional geometry.} 
In order to generate the Lobachevskiy plane in the non-compact limit $D \to \infty$ 
we need the boundary of the disk to diverge in this limit. 
This is obtained by tuning the boundary cosmological constant $\mu'_{B}$ to the smallest possible value. 
From the expressions in (\ref{avarea}) and (\ref{avlength}) valid for all $D \geq 0$ 
we may show, that the both the average area and the average boundary length 
are positive for all $D \geq 0$ as long as $\mu'_{B} \geq -\sqrt{\mu}/2$ and $\mu_{B}>\sqrt{\mu}/2$.  
In this case we may view $\widetilde{W}(\mu'_{B},\mu)$ as a partition function in 2D euclidean Quantum Gravity.
For $\mu'_{B} < -\sqrt{\mu}/2$ this interpretation is not valid. 
At a finite geodesic distance $D$ the average boundary length  
becomes negative and any interpretation of $\widetilde{W}(\mu'_{B},\mu)$ in 
terms of geometry is obviously not correct. 
The smallest possible value of $\mu'_{B}$ \emph{viewed as a genuine boundary cosmological constant in 2D euclidean Quantum Gravity} is $-\sqrt{\mu}/2$. 

Let us consider the cylinder amplitude (\ref{cylgeo}) with fixed geodesic 
distance in pure 2D euclidean Quantum Gravity. Retaining a finite entrance loop we may probe 
the boundary condition imposed on the exit loop in the non-compact limit $D \to \infty$. 
For generic values of $\mu'_{B}$ the cylinder amplitude 
vanishes in the non-compact limit $D \to \infty$. 
However, precisely for $\mu'_{B} = -\sqrt{\mu}/2$ the cylinder amplitude remains finite in 
the limit $\D \to \infty$ and is given by
\st
N(\mu_{B},\mu'_{B},\mu;\infty) 
=   
\frac{\sqrt{\frac{3}{2}}\mu^{1/4}}{\left(\mu_{B}-\frac{\sqrt{\mu}}{2}\right)\sqrt{\mu_{B}+\sqrt{\mu}}} 
& = &
-\frac{\prt}{\prt\mu_{B}} 
\ln\left[\frac{\sqrt{\mu_{B}+\sqrt{\mu}}+\sqrt{\frac{3}{2}}\mu^{1/4}}{\sqrt{\mu_{B}+\sqrt{\mu}}-\sqrt{\frac{3}{2}}\mu^{1/4}}\right] 
\nonumber\\
& = &
-\frac{\prt}{\prt\mu_{B}} 
\ln\left[\frac{z + \frac{\sqrt{3}}{2}}{z - \frac{\sqrt{3}}{2}}\right] 
\nonumber\\
& = &
-2\frac{\prt}{\prt\mu_{B}} 
\mathcal{Z}((1,1)_{_{C}},\sigma_{1};(1,1)_{_{C}},(1,1)_{_{ZZ}}) 
\nonumber\\
\label{cylinfty}
\en
where we have introduced the uniformization parameter $z$ defined in eq. (\ref{z}) with 
regard to the boundary cosmological constant on the entrance loop. 
\emph{Up to an overall dimensionless normalization this is exactly the unique FZZT-ZZ cylinder amplitude in pure 
2D euclidean Quantum Gravity obtained in eq. (\ref{FZZTZZcyl}).} 
The differentiation with respect to $-\mu_{B}$ occurs in the above identity due to the fact, that we 
have marked one point on the entrance loop in the cylinder amplitude with fixed geodesic distance $D$. 
Imposing a FZZT boundary condition is equivalent to introducing a boundary cosmological constant. 
For finite values of the geodesic distance $D$ we have imposed a FZZT boundary condition on both the entrance loop and the exit loop. 
\emph{In the transition from compact to non-compact geometry enforced by setting $\mu'_{B} = -\sqrt{\mu}/2$ 
and taking the limit $D \to \infty$ the FZZT-brane imposed on the exit loop transforms into 
the single principal ZZ-brane in pure 2D euclidean Quantum Gravity.} 
The above construction of non-compact geometries in pure 2D euclidean 
Quantum Gravity resolves a puzzle related to the ZZ-branes: 
The world-sheet geometry associated with a ZZ-brane is non-compact. 
However, if we calculate the average area of the random world-sheet  
associated with for instance the FZZT-ZZ cylinder amplitude given by eq. (\ref{FZZTZZcyl}) we obtain a finite value!   
In our construction the resolution to this problem comes from the fact, 
that we cannot determine the correct value of the average area after we set $\mu'_{B} = -\sqrt{\mu}/2$.  
Recall, the average area is given by 
\[
\langle A \rangle = -\frac{1}{N}\frac{\prt N}{\prt \mu}\,.
\]
If we first set $\mu'_{B} = -\sqrt{\mu}/2$ as in eq. (\ref{cylinfty}) and then apply the above formula for the average area 
we do not determine the average area. 
Instead we obtain 
\[
\langle A \rangle - \frac{1}{4\sqrt{\mu}}\langle L' \rangle
\]
which is finite even in the non-compact limit $D \to \infty$.  
It remains to be seen, if a similar conclusion may be reached in the Liouville approach to ZZ-branes. 

Let us consider the cylinder amplitude with fixed distance $D$ in the conformal background of the $(2,2m-1)$ minimal model 
coupled to 2D euclidean Quantum Gravity. 
In the $(2,2m-1)$ minimal model coupled to 2D euclidean Quantum Gravity the ordinary disk amplitude obtained 
from dynamical triangulations is given by \cite{GK}
\[\label{ordinarydisk2m1}
W(\mu_{B},\mu) 
= 
(-1)^{m} \hat{P}_{m}(\mu_{B},\mu)\sqrt{\mu_{B}+\sqrt{\mu}}
=
(-1)^{m} (\sqrt{\mu})^{(2m-1)/2} P_{m}(t) \sqrt{t+1}
\]  
where $t=\mu_{B}/\sqrt{\mu}$ and  
where the polynomial $P_{m}(t)$ in the conformal background is determined 
from eq. (\ref{Chebyshev}) (up to a convenient normalization)            
\[
 P_{m}^{2}(t)(t+1) = 2^{2-2m}(T_{2m-1}(t)+1)\,, 
\]
where $T_{2m-1}(t)$ is the first kind of Chebyshev polynomial of order $2m-1$. 
Explicitly, we may express  
\[
P_{m}(t) = \prod_{n=1}^{m-1}(t - t_{n})\,, \qquad t_{n} = -\cos\left(\frac{2n\pi}{2m-1}\right)\,.
\] 
All the zeros of the polynomial $\hat{P}_{\mu}(\mu_{B},\mu)$  
viewed as a function of $\mu_{B}$ 
lie on the real $\mu_{B}$-axis in between $-\sqrt{\mu}$ and $\sqrt{\mu}$ and divide the real $\mu_{B}$-axis 
into $m$ disjoint intervals 
\[
\big(t_{n}\sqrt{\mu}\, , \,\,t_{n+1}\sqrt{\mu}\big)\,, \qquad 0 \leq n \leq m-1\,,\qquad t_{0} \equiv -1\,, \quad t_{m} \equiv \infty\,.  
\] 
As in the case of pure 2D euclidean Quantum Gravity the PDE (\ref{PDE}) 
induces a flow in $\hat{\mu}_{B}$ with the distance $D$. 
The nature of the flow may easily be determined from eq. (\ref{natureflow}) 
and the initial condition $\hat{\mu}_{B} = \mu_{B}$ for $D = 0$.  
If $\mu_{B}$ belongs to the interval $\big(t_{n}\sqrt{\mu}\, , \,\,t_{n+1}\sqrt{\mu}\big)$ the flow of $\hat{\mu}_{B}$ 
is constrained to the same interval. 
The fixed points of $\hat{\mu}_{B}$ corresponding to $D \to \infty$ are the zeros in the polynomial 
$P(\mu_{B},\sqrt{\mu})$ (viewed as a function of $\mu_{B}$), which precisely match  
the values of the boundary cosmological constant associated with the principal ZZ-branes in the $(2,2m-1)$ minimal model coupled 
to 2D euclidean Quantum Gravity.  
Let us start out by presenting our results concerning the cylinder amplitude $N(\mu_{B},\mu'_{B},\mu;D)$ 
with fixed distance in the conformal background of the $(2,2m-1)$ minimal model coupled to 2D euclidean Quantum Gravity, $m>2$. 
Then we will discuss the interpretation of these results. 
For generic values of the boundary cosmological constant $\mu'_{B}$ on the exit loop 
the cylinder amplitude $N(\mu_{B},\mu'_{B},\mu;D)$ 
vanishes in the limit $D \to \infty$ independent of which fixed point $\hat{\mu}_{B}$ approaches.  
However, as in the case of pure 2D euclidean Quantum Gravity we have a unique situation, 
if we set   
\[
\mu'_{B} = -t_{k}\sqrt{\mu}\,,
\]
where $t_{k}\sqrt{\mu}$ is the fixed point approached 
by  
$\hat{\mu}_{B}$ in the limit $D \to \infty$. 
In this case the cylinder amplitude obtained in the limit $D \to \infty$ 
is given by 
\[
N(\mu_{B},-t_{k}\sqrt{\mu},\mu;D) 
=  
\frac{1}{\sqrt{\mu}}
\frac{1}{\hat{t}-t_{k}}
\frac{\sqrt{\hat{t}+1}\prod_{n=1}^{m-1}(\hat{t} - t_{n})}
{\sqrt{t+1}\prod_{n=1}^{m-1}(t - t_{n})}
\phantom{mapmapmapmapmapmapmap}
\nonumber
\]
\[
\to  
\frac{\alpha_{k}}{\sqrt{\mu}} 
\frac{1}{\sqrt{t+1}\prod_{n=1}^{m-1}(t-t_{n})} 
\phantom{mapmapmapma}
\nonumber
\]
\[
\propto  
\frac{1}{\sqrt{\mu}}
\frac{1}{\sqrt{1+t}}
\sum_{n=1}^{m-1} (-1)^{n} \sin\left(\!\frac{2 \pi n}{2m-1}\!\right)
\left[
\frac{1}{\sqrt{1+t}+\sqrt{1+t_{n}}} 
- 
\frac{1}{\sqrt{1+t}-\sqrt{1+t_{n}}} 
\right]
\label{mellemcyl}
\]
where $\hat{t}=\hat{\mu}_{B}/\sqrt{\mu}$ and where we have omitted a dimensionless constant of proportionality.  
It is natural to assume, that the boundary condition imposed on the finite entrance loop is the boundary 
condition generic to the scaling limit of dynamical triangulations, 
i.e. the $(1,1)$ Cardy matter condition with respect to the matter section. 
As shown in chapter \ref{chapterZZ} the fundamental ZZ boundary condition in Liouville theory
is the basic $(1,1)$ ZZ boundary condition. 
The set of FZZT-ZZ cylinder amplitudes in the conformal background of the 
$(2,2m-1)$ minimal model coupled to 2D euclidean Quantum Gravity with a $(1,1)$ Cardy matter condition imposed on the FZZT boundary 
and the basic $(1,1)$ ZZ boundary condition imposed on the ZZ boundary is determined in eq. (\ref{FZZTZZcyl}).   
\[
\mathcal{Z}((1,1)_{_{C}},\sigma;(1,n)_{_{C}},(1,1)_{_{ZZ}}) 
= 
\frac{1}{2}
\ln \left[ \frac{\sqrt{t+1}+\sqrt{1+t_{n}}}
{\sqrt{t+1}-\sqrt{1+t_{n}}}\right]\,, \qquad 1 \leq n \leq m-1\,, 
\]
where we have expressed the uniformization parameter $z$ defined in eq. (\ref{z}) 
in terms of $t=\mu_{B}/\sqrt{\mu}$. 
Applying the fact that
\[
S_{(1,1);(1,n)} = \frac{2}{\sqrt{2m-1}} (-1)^{m+n+1}\sin\left(\frac{2 \pi n}{2m-1}\right)\,,
\]
where $S$ is the modular S-matrix in the $(2,2m-1)$ minimal model defined in eq. (\ref{mods}), 
we may express the cylinder amplitude obtained in eq. (\ref{mellemcyl}) as 
\st
N(\mu_{B},-t_{k}\sqrt{\mu},\mu;D \to \infty) 
& \propto &
-
\sum_{n=1}^{m-1} 
S_{(1,1);(1,n)} \frac{\prt}{\prt \mu_{B}} 
\mathcal{Z}((1,1)_{_{C}},\sigma;(1,n)_{_{C}},(1,1)_{_{ZZ}}) 
\nonumber\\
& \propto &
-
\frac{\prt}{\prt \mu_{B}} 
\mathcal{Z}((1,1)_{_{C}},\sigma;(1,1)_{_{\textrm{Ishibashi}}},(1,1)_{_{ZZ}}) 
\en
where the omitted constant of proportionality is dimensionless and  
where we have applied the identity 
\[
\sum_{n=1}^{m-1} S_{(1,1);(1,n)} \vert 1,n \rangle_{_{C}} 
= 
\sum_{n=1}^{m-1} \sum_{k=1}^{m-1} 
\frac{S_{(1,1);(1,n)}S_{(1,n);(1,k)}}{\sqrt{S_{(1,1);(1,k)}}} 
\vert\vert \Delta_{1,k} \rangle\rangle 
= 
\frac{1}{\sqrt{S_{(1,1);(1,1)}}} \vert \vert \Delta_{1,1} \rangle \rangle 
\]
derived from eq. (\ref{Cardydef}) and the fact, that the $S^{2}=I$.
\emph{Quite remarkable, independent of the fixed point $t_{k}\sqrt{\mu}$ approached by $\hat{\mu}_{B}$  
we obtain the FZZT-ZZ cylinder amplitude with the basic $(1,1)$ Ishibashi state imposed on the ZZ boundary in the limit $D \to \infty$, 
if we set $\mu'_{B}=-t_{k}\sqrt{\mu}$. 
In this limit we have a transition from a FZZT brane to a ZZ brane on the exit loop.  
The FZZT-ZZ cylinder amplitude with the $(1,1)$ Ishibashi state imposed on the ZZ boundary is 
characterized by the fact, that the only physical closed string states, 
which propagate in this cylinder amplitude, are the closed string states   
with a trivial matter part corresponding to the identity operator.}\footnote{These are the physical 
closed string states constructed from the BRST cohomology groups 
\[
\mathcal{H}_{\ast}\left(\Delta_{1,1},P\right) 
= 
\bigoplus_{n}
\frac{\textrm{Ker}\left( Q : C_{n}(\Delta_{1,1},P) \longmapsto C_{n+1}(\Delta_{1,1},P)    \right)}
{\textrm{Im}\left( Q : C_{n-1}(\Delta_{1,1},P) \longmapsto C_{n}(\Delta_{1,1},P)    \right)}
\]
where
\[
P = \pm i \frac{4(2m-1)t + 2m-1 \pm 2}{2\sqrt{2}\sqrt{2m-1}}\,, \qquad t \in \mathbf{Z}  
\]
and where 
\[
C_{n}(\Delta_{1,1},P) 
\equiv 
\bigg\{ \vert \sigma \rangle \in  
L(c_{2,2m-1},\Delta_{1,1})  
\otimes
\mathcal{F}(c_{L},P)  
\otimes \mathcal{H}^{L}_{\textrm{Ghost}}
\bigg\vert \: 
b_{0} \vert \sigma \rangle =  
L_{0}^{\textrm{tot}} \vert \sigma \rangle = 0, \: 
N_{g} \vert \sigma \rangle = n \vert \sigma \rangle 
\bigg\} 
\]
Recall, the irreducible matter representation $L(c_{2,2m-1},\Delta_{1,1})$ only contains the state corresponding to identity operator.}  
This conclusion follows from the following argument: 
As discussed previously 
we may express any given FZZT-ZZ cylinder amplitude involving a principal ZZ brane 
as the difference between two FZZT-FZZT cylinder amplitudes. 
The FZZT-FZZT cylinder amplitude with the $(1,1)$ Ishibashi state imposed on one of the boundaries 
is given by eq. (\ref{truecyl}) (up to an overall normalization), 
if we simply leave out the sum over the different matter Virasoro characters    
and only include the contribution from the matter Virasoro character $\chi_{1,1}(q_{c})$. 
It follows from eq. (\ref{remark}) and the following discussion that the only poles, which contribute to this particular 
FZZT-FZZT cylinder amplitude, are the poles associated with the physical closed string states 
with a trivial matter part corresponding to the identity operator and the poles, 
which we cannot associate with any physical closed string states in the BRST formalism. 
The contribution from the latter poles cancels in the difference between 
the two FZZT-FZZT cylinder amplitudes giving us the FZZT-ZZ cylinder amplitude. 

The limit $D \to \infty$ plays an instrumental role in the above transition from a FZZT brane  
to a ZZ brane on the exit loop and we would like to address two important aspects with regard to this limit. 
Firstly, as discussed in section \ref{sectionTheFZZTandZZbranes} Seiberg and Shih advocates, that target space in 
$(p,q)$ minimal string theory is given by the Riemann surface $\mathcal{M}_{p,q}$ defined in eq. (\ref{mpq}). 
In the case of $(2,2m-1)$ minimal string theory this Riemann surface is a double sheeted cover of the complex $\mu_{B}$-plane except 
at the singularities, 
which are the zeros in the polynomial $\hat{P}_{m}(\mu_{B},\mu)$ appearing in the ordinary disk 
amplitude.\footnote{$\mathcal{M}_{p,q}$ 
is defined as the set of points $(\mu_{B},W(\mu_{B},\mu))$. 
Due to the square root appearing in the ordinary disk amplitude given by eq. (\ref{ordinarydisk2m1}) 
$\mathcal{M}_{2,2m-1}$ is a double sheeted cover of the complex $\mu_{B}$-plane. 
However, we cannot distinguish between the two sheets for values of $\mu_{B}$ for which the ordinary disk amplitude vanishes.  
The singularities of $\mathcal{M}_{2,2m-1}$ are therefore given by the zeros in the polynomial $P(\mu_{B},\mu)$.} 
The zeros in the polynomial $\hat{P}_{m}(\mu_{B},\mu)$ precisely match the values of the boundary cosmological 
constant associated with the principal ZZ branes and Seiberg and Shih advocates, 
that we should associate each singularity with a particular principal ZZ-brane. 
One is also led to this extended target space in the above construction. 
As mentioned previously the fixed points of the 
induced boundary cosmological constant $\hat{\mu}_{B}$ are precisely the zeros in the polynomial $\hat{P}_{m}(\mu_{B},\mu)$. 
We want $\hat{\mu}_{B}$ to be able to approach any of the fixed points 
in the limit $D \to \infty$ i.e. we want all the fixed points to be attractive. 
This is only possible, if we consider the running boundary cosmological constant $\hat{\mu}_{B}$ as a function and $\mu_{B}$ as a variable 
taking values on the Riemann surface $\mathcal{M}_{2,2m-1}$. 
The reason is that every other fixed point on a given sheet is repulsive. 
The attractive fixed points on one sheet are repulsive on the other sheet and vice versa. 
The flow of $\hat{\mu}_{B}$ in the $(2,2m-1)$ minimal model coupled to 2D euclidean Quantum Gravity, $m$ odd, also 
hints at this extended target space. 
In the case $m$ odd the largest fixed point $t_{m-1}\sqrt{\mu}$ is  
repulsive on the first sheet. 
If we consider $\mu_{B} > t_{m-1}\sqrt{\mu}$ the induced boundary cosmological constant 
gradually approaches infinity along the real axis as we increase $D$ 
and reaches infinity for a finite value of $D$. 
If we keep on increasing $D$ the induced boundary cosmological constant $\hat{\mu}_{B}$ flows down the real axis on the second sheet 
and gradually approaches the attractive fixed point $t_{m-1}\sqrt{\mu}$ on the second sheet.   

Secondly, our construction also adds to the understanding of the
relation (\ref{FZZTZZb2irr}) discovered by Martinec. 
As shown in section \ref{sectionZZ} in Liouville theory for generic values of $b$ 
there is a one-to-one correspondance between the ZZ boundary states 
labeled by $(m,n)$ and the degenerate primary operators $V_{\alpha_{m,n}}$. 
This correspondance completely determines the Liouville 
cylinder amplitude with two ZZ boundary conditions: The spectrum of
states flowing in the open string channel between two ZZ boundary
states is obtained from the fusion algebra of the corresponding
degenerate operators. Similarly, there is a one-to-one
correspondance between the FZZT boundary states labeled by $\sigma>0$
and the non-local primary operators
$V_{Q/2+i\sigma/2}$. 
The conformal dimension of the spin-less degenerate primary operator
$V_{\alpha_{m,n}}$ is given by 
\[
\Delta(\alpha_{m,n})=\frac{Q^{2}-(m/b+nb)^{2}}{4}\,, 
\] 
while the conformal
dimension of the spin-less non-local primary operator $V_{Q/2+i\sigma/2}$ is
given by 
\[
\Delta(Q/2+i\sigma/2) 
= 
\frac{Q^{2}+\sigma^{2}}{4}\,. 
\]
Since $\Delta(\alpha_{m,n})=\Delta(Q/2+i\sigma/2)$ for $\sigma=i(m/b+nb)$, one is
naively led to the wrong conclusion, that a FZZT boundary state
turns into a ZZ boundary state, if one tunes $\sigma=i(m/b+nb)$.
However, the operator $V_{\alpha_{m,n}}$ is degenerate and in addition to
setting $\sigma=i(m/b+nb)$ we therefore have to truncate the spectrum
of open string states, that couple to the FZZT boundary state, in
order to obtain a ZZ boundary state. This is precisely captured in
the relation \ref{FZZTZZb2irr} with regard to 
the principal ZZ boundary states in Liouville theory for $b^{2}$ 
rational.
The world-sheet geometry characterizing the FZZT brane is compact,
while the world-sheet geometry of the ZZ-brane is non-compact.
Hence, truncating the spectrum of open string states induces a
transition from compact to non-compact geometry. In our concrete
realization of this transition this truncation is obtained by first
setting the boundary cosmological constant 
$\mu'_{B}=-t_{k}\sqrt{\mu}$ on the exit loop and then taking $D \to \infty$. 
In this limit the induced boundary cosmological constant approaches 
the fixed point $t_{k}\sqrt{\mu}$ associated with one of the principal ZZ boundary conditions. 

\chapter{Conclusion and discussion}
Let us shortly summarize the main results obtained in this thesis 
and discuss the main questions arising from this work.  
In chapter \ref{chaptercyl} we determined the FZZT-FZZT cylinder amplitude 
for all pairs of Cardy matter states 
in the $(p,q)$ minimal model coupled to 2D euclidean Quantum Gravity. 
We showed, that to any given physical closed string state appearing in the BRST formalism 
we may associate a pole in the integrand in the expression (\ref{truecyl2}) 
for the FZZT-FZZT cylinder amplitude. 
We interpret the corresponding residue as the amplitude of the given physical closed string state 
propagating between the two FZZT branes.  
In addition to these poles, which are in a one-to-one correspondance 
with physical closed string states appearing in the BRST formalism, a different 
set of poles appeared in the integrand in the expression (\ref{truecyl2}) for the FZZT-FZZT cylinder amplitude. 
The nature of these poles is clearly different 
from the nature of the poles associated with physical closed string states. 
These additional poles appear in the product of the two FZZT boundary wave functions,  
while the poles associated with physical closed string states appear in the 
trace of the operator $\frac{1}{L^{tot}_{0}+\bar{L}^{tot}_{0}}$. 
One of the main questions arising from our calculation of the FZZT-FZZT cylinder amplitude in the $(p,q)$ minimal 
model coupled to 2D euclidean Quantum Gravity is the nature of these additional poles, which we cannot associate 
with any physical closed string states in the BRST formalism. 
The nature of these poles is clearly related to the compact nature of 
the FZZT-FZZT cylinder amplitude. 
This follows from the fact, that    
these additional poles do not appear in the integrand in expression (\ref{truecylzz})
for the non-compact FZZT-ZZ cylinder amplitude. 
Quite remarkable, we may associate some of these additional poles 
with scaling operators, which appear in the matrix model approach to 2D euclidean Quantum Gravity  
but not in the BRST formalism discussed in section \ref{physoperator}.     

Comparing the complete set of FZZT-FZZT cylinder amplitudes obtained in chapter 
\ref{chaptercyl} with the cylinder amplitude obtained from dynamical triangulation/the one-matrix model 
we concluded in section \ref{sectionDT}, that the particular matter boundary condition realized in the 
scaling limit of dynamical triangulations, in 
which we approach the $m^{\textrm{th}}$ multi-critical hyper-surface, 
is the $(1,1)$ Cardy matter condition. 

In chapter \ref{chapterZZ} we argued and provided some evidence, that 
there essentially only exists one ZZ boundary state in pure Liouville theory, the basic $(1,1)$ ZZ boundary state. 
All the other principal ZZ boundary states correspond to effective boundary conditions in Liouville theory obtained 
by integrating out the matter degrees of freedom. With regard to the one point function evaluated on the 
quantum Lobachevskiy plane and the FZZT-ZZ and ZZ-ZZ 
cylinder amplitudes we showed explicitly, that the dressing factor, 
which distinguishes the $(m,n)$ ZZ boundary state from the basic $(1,1)$ ZZ boundary state, 
appears when we impose the $(m,n)$ Cardy matter condition 
on the ZZ boundary and integrate out the matter degress of freedom. 
A truly existing future line of research would concern the ZZ branes from a boundary perspective. 
The interpretation of the principal ZZ boundary states proposed 
in this thesis suggests, that there exists an isomorphism 
between the set of physical open string states, that couple to the ZZ brane on the left hand side of eq. (\ref{id2}) 
and the set of physical open string states, that couple to the ZZ brane on the right hand side of eq. (\ref{id2}). 
As a starting point for studying the ZZ branes from a boundary perspective 
one may consider the FZZT-ZZ and the ZZ-ZZ cylinder amplitudes in the \emph{open} string channel. 
Given our experience with the cylinder amplitudes in the closed string channel 
we may expect, that a given cylinder amplitude involving a ZZ brane and evaluated in the open string channel 
is given by a sum over residues each associated with a particular physical \emph{open} string state. 
This analysis may shed some light on the spectrum of physical open string states, that couple to a given ZZ brane 
or flow between a ZZ brane and another given brane. 
This research project may be interesting from an AdS/CFT point of view.

In section \ref{worldsheetperspective} we studied the cylinder amplitude with fixed distance in the conformal background of 
the $(2,2m-1)$ minimal model coupled to 2D euclidean Quantum Gravity and the disk amplitude with fixed geodesic distance 
in pure 2D euclidean Quantum Gravity.  
In the case of pure 2D euclidean Quantum Gravity we showed, that if we set $\mu'_{B}=-\sqrt{\mu}/2$
then in the non-compact limit $D \to \infty$ we have a transition from compact to 
non-compact worldsheet geometry, that is the Lobachevskiy plane, 
and we have a transition from a FZZT brane to a ZZ brane on the exit loop. 
In the case of the $(2,2m-1)$ minimal model coupled to 2D euclidean Quantum Gravity, $m>2$, we gave some evidence, 
that we have a similar transition from a FZZT brane to a ZZ brane on the exit loop in the limit $D \to \infty$, 
if we set $\mu'_{B}=-t_{k}\sqrt{\mu}$, where $t_{k}\sqrt{\mu}$ 
is the fixed point approached by the induced boundary cosmological constant $\hat{\mu}_{B}$ in the limit $D \to \infty$. 
The particular matter boundary condition realized in this limit on the ZZ boundary is the $(1,1)$ Ishibashi state  
and the FZZT-ZZ cylinder amplitude obtained in this limit is characterized by the fact, that the only physical closed string states, 
which propagate in the closed string channel, are the physical states with a matter part corresponding to the identity operator. 
A better understanding of this fact may shed some light on the nature of the distance $D$. 

\appendix
\chapter{The one and two point functions in the upper half plane}
\label{apptwopoint}
Let us consider the two point function of two primary operators $V_{\alpha'}(z_{1},\bar{z}_{1})$ and 
$V_{\alpha}(z_{2},\bar{z}_{2})$ inserted in the upper plane. Moreover, let us use conformal gauge. 
The group of conformal transformation mapping the upper half plane to the upper half plane is given by 
$PSL(2,\mathbf{R})$, 
which is a three parameter group. 
Hence, we may fix the position of three real coordinates by a conformal transformation. 
In the following we will construct a conformal transformation mapping $z_{2} \to i$ and $\textrm{Re}(z_{1}) \to 0$. 
The conformal transformation is obtained in the following way. 
We first map the upper half plane to the unit disk with the conformal transformation
\[
z \to 
\omega(z) = \frac{i z + 1}{z + i} \, .
\]
We then perform a rotation of the disk 
\[
\omega \to 
\omega' = i \frac{\vert\omega_{2}\vert}{\omega_{2}} \omega
\]
where $\omega_{2} = \omega(z_{2})$. 
(The indices $1$ and $2$ refer to the position of the operators $V_{\alpha'}$ and $V_{\alpha}$ throughout this argument.) 
Notice, this conformal transformation maps $\omega_{2}$ to the imaginary axis. 
The disk is now mapped to the upper half plane by the conformal transformation 
\[
\omega' \to 
z' = \frac{\omega'+i}{i\omega' + 1} \,.
\]
Notice, $\omega'_{2}$ is mapped to the imaginary axis. 
Applying the conformal transformation 
\[
z' \to 
\tilde{z} = \frac{1}{\vert z'_{2} \vert} z' 
\] 
we map $z'_{2} \to i$. 
We now map the upper half plane to the unit disk by the conformal transformation 
\[
\tilde{z} \to 
\tilde{\omega} = \frac{i \tilde{z} + 1}{\tilde{z} + i} \, .
\] 
Under this transformation $\tilde{z}_{2} = i$ is mapped to the center of the unit disk.
We then perform a rotation of the disk 
\[
\tilde{\omega} 
\to 
\hat{\omega} 
= 
- i \frac{\vert \tilde{\omega}_{1} \vert}{\tilde{\omega}_{1}} \tilde{\omega} \, .
\]
and finally, we map the unit disk to the upper half plane by the conformal transformation 
\[
\hat{\omega} \to 
\hat{z} = \frac{\hat{\omega}+i}{i\hat{\omega} + 1} \,.
\]
Notice, $z_{2}$ is mapped to $i$ and $z_{1}$ is mapped to $it$, $0<t<1$, under the composition of these conformal transformations. 
The value of $t$ may be determined from the fact that the parameter 
\[
\eta = \frac{(z_{1}-z_{2})(\bar{z}_{1}-\bar{z}_{2})}{(z_{1}-\bar{z}_{2})(\bar{z}_{1}-z_{2})}
\]
is invariant under conformal transformations. 
Hence, we may express $\eta$ as
\[
\eta = \frac{(i-it)(-i+it)}{(i+it)(-i-it)} 
\]
from which we may determine $t$ in terms of $\eta$
\[
t = \frac{1-\sqrt{\eta}}{1+\sqrt{\eta}} \, .
\]
Applying the transformation law (\ref{conftransprim}) governing the primary operators under a given conformal transformation successively  
we obtain 
\[
\bigg\langle
V_{\alpha'}(z_{1},\bar{z}_{1})V_{\alpha}(z_{2},\bar{z}_{2})
\bigg\rangle 
\nonumber
\]
\[
=
\frac{1}{\vert z_{2}-\bar{z}_{2} \vert^{2(\Delta(\alpha)-\Delta(\alpha'))} \vert z_{1}-\bar{z}_{2} \vert^{4\Delta(\alpha')}}
\frac{2^{2(\Delta(\alpha)+\Delta(\alpha'))}}{(1+\sqrt{\eta})^{4\Delta(\alpha')}}
\bigg\langle
V_{\alpha'}(it,-it)V_{\alpha}(i,-i)
\bigg\rangle 
\]
The form of the one-point function is obtained from the above two-point function by setting $V_{\alpha'}$ equal to the identity operator. 
\[\label{onepointupperappendix}
\bigg\langle 
V_{\alpha}(z,\bar{z}) 
\bigg\rangle
= 
\frac{U(\alpha)}{\vert z-\bar{z} \vert^{2\Delta(\alpha)}}
\] 
where 
\[
U(\alpha) 
= 
2^{2\Delta(\alpha)} 
\bigg\langle 
V_{\alpha}(i,-i) 
\bigg\rangle
\]

\chapter{Identities used in the derivation of the cylinder amplitude} \label{app2}
In this appendix we derive some of the identities used in the derivation of the cylinder amplitude in chapter \ref{chaptercyl}. 
Let us begin by performing the integral given in eq. (\ref{remark}). 
From eq. (\ref{virchaP}), (\ref{virchamin}) and (\ref{qclosed}) we obtain
\st
&&
\int_{0}^{\infty} \!\!d\tau
\,\chi_{m,n}(q_{c})  
\,\chi_{P}(q_{c})
\,\eta(q_{c})^{2} 
\nonumber\\
& = & 
\sum_{t \in \mathbf{Z}} 
\int_{0}^{\infty} \!\!d\tau
\left\{
e^{-2\pi\tau\left(P^{2}+\frac{(2pqt+mq-np)^{2}}{4pq}\right)}
-
e^{-2\pi\tau\left(P^{2}+\frac{(2pqt+mq+np)^{2}}{4pq}\right)}
\right\}
\nonumber\\
& = &
\frac{1}{2\pi}
\sum_{t \in \mathbf{Z}} 
\left\{
\frac{1}{P^{2}+\frac{(2pqt+mq-np)^{2}}{4pq}}
-
\frac{1}{P^{2}+\frac{(2pqt+mq+np)^{2}}{4pq}}
\right\}
\label{B1}
\en
In order to proceed we first need to prove the identity 
\[
\sum_{n=1}^{\infty} 
\frac{1}{an^{2}+bn+c}
= 
\frac{1}{a(z_{2}-z_{1})}
\bigg(
\psi(1-z_{1})
- 
\psi(1-z_{2})
\bigg)\,,
\label{identitypsi}
\]
valid for $\vert\arg(z_{1})\vert<\pi$ and $\vert\arg(z_{2})\vert<\pi$,   
where $\psi(z)=\Gamma'(z)/\Gamma(z)$ is the Psi function defined in eq. (6.3.1) in \cite{Ab} 
and where $z_{1}$ and $z_{2}$ are the two roots in the second order polynomial 
\[
g(z) 
= 
az^{2}+bz+c\,.
\]
In order to prove this identity we first notice, that 
\[
\frac{1}{an^{2}+bn+c} 
= 
\frac{1}{a(n-z_{1})(n-z_{2})} 
= 
\frac{1}{a(z_{2}-z_{1})}
\left[
\frac{1}{n-z_{2}} 
- 
\frac{1}{n-z_{1}}
\right]\,.
\]
From eq. (6.3.6) in \cite{Ab} we obtain 
\[
\sum_{n=1}^{k} 
\frac{1}{an^{2}+bn+c} 
= 
\frac{1}{a(z_{2}-z_{1})}
\left[
\sum_{n=1}^{k} 
\frac{1}{n-z_{2}} 
- 
\sum_{n=1}^{k} 
\frac{1}{n-z_{1}}
\right] 
\nonumber
\]
\[
= 
\frac{1}{a(z_{2}-z_{1})}
\bigg[
\psi(k+1-z_{2}) - \psi(1-z_{2}) - \psi(k+1-z_{1}) + \psi(1-z_{1})
\bigg]\,.
\]
From eq. (6.3.18) in \cite{Ab} we realize, that 
\[
\lim_{k \to \infty} 
\bigg(
\psi(k+1-z_{2}) - \psi(k+1-z_{1}) 
\bigg)
= 
\lim_{k \to \infty} 
\bigg(
\ln(k+1-z_{2}) - \ln(k+1-z_{1}) 
\bigg)
= 
0
\,,
\]
which completes the proof of the identity (\ref{identitypsi}).  
Applying the identity (\ref{identitypsi}) we obtain from eq. (\ref{B1})
\[ 
\int_{0}^{\infty} \!\!d\tau
\,\chi_{m,n}(q_{c})  
\,\chi_{P}(q_{c})
\,\eta(q_{c})^{2} 
\nonumber
\]
\[
=  
\frac{2pq}{\pi}
\left\{
\sum_{t = 1}^{\infty} 
\frac{1}{4pqP^{2}+(2pqt+mq-np)^{2}}
+
\sum_{t = 1}^{\infty} 
\frac{1}{4pqP^{2}+(-2pqt+mq-np)^{2}} 
\right.
\nonumber
\]
\[
\left.
\phantom{\sum_{t = 1}^{\infty} 
\frac{1}{4pqP^{2}+(-2hghghghjhfgg)^{2}} }
+ 
\frac{1}{4pqP^{2}+(mq-np)^{2}} 
- 
(n \to -n) 
\right\}
\nonumber
\]
\[
= 
\frac{1}{2\pi pq} 
\Bigg\{
\frac{\sqrt{pq}}{2iP}
\left[
\psi\left(1+\frac{(mq-np)+2i\sqrt{pq}P}{2pq}\right) 
- 
\psi\left(1+\frac{(mq-np)-2i\sqrt{pq}P}{2pq}\right)
\right.
\nonumber
\]
\[
\phantom{mapmap}
\left.
+ 
\psi\left(1-\frac{(mq-np)-2i\sqrt{pq}P}{2pq}\right) 
- 
\psi\left(1-\frac{(mq-np)+2i\sqrt{pq}P}{2pq}\right)
\right]
\nonumber
\]
\[
+ 
\frac{4p^{2}q^{2}}{(mq-np)^{2}+4pqP^{2}}
- 
(n \to -n) 
\Bigg\}\phantom{map}
\]
Finally, applying the formula 
\[
\psi(1+z) - \psi(1-z) 
= 
\frac{1}{z} - \pi \cot(\pi z)
\]
derived from eqs. (6.3.5) and (6.3.7) in \cite{Ab} we get, that 
\[
\int_{0}^{\infty} \!\!d\tau
\,\chi_{m,n}(q_{c})  
\,\chi_{P}(q_{c})
\,\eta(q_{c})^{2} 
\nonumber
\] 
\[
= 
\frac{1}{2\pi pq} 
\left\{
\frac{\sqrt{pq}}{2iP}
\left[
\frac{2pq}{(mq-np)+2i\sqrt{pq}P}
- 
\pi
\cot\left(
\pi\frac{(mq-np)+2i\sqrt{pq}P}{2pq}
\right)
\right.\right.
\nonumber
\]
\[
\phantom{mapmap}
\left.
-
\frac{2pq}{(mq-np)-2i\sqrt{pq}P}
+  
\pi
\cot\left(
\pi\frac{(mq-np)-2i\sqrt{pq}P}{2pq}
\right)
\right]
\nonumber
\]
\[
\left.
+ 
\frac{4p^{2}q^{2}}{(mq-np)^{2}+4pqP^{2}}
- 
(n \to -n) 
\right\}
\nonumber
\]
\[
= 
\frac{\sinh\left(\frac{2 \pi P}{\sqrt{pq}}\right)}{2\sqrt{pq}P}
\left\{ 
\frac{1}{\cosh\left(\frac{2 \pi P}{\sqrt{pq}}\right) - \cos\left(\frac{\pi(mq-np)}{pq} \right)}
- 
\frac{1}{\cosh\left(\frac{2 \pi P}{\sqrt{pq}}\right) - \cos\left(\frac{\pi(mq+np)}{pq} \right)}
\right\}
\label{remark2}
\]
Let us now turn our attention to the function $f$ defined in eq. (\ref{ffunc})
\st
f(z) 
& = &
\begin{array}{c}
\displaystyle\sum_{m=1}^{p-1} \,\,\, 
\sum_{n=1}^{q-1}
\\
\textrm{\scriptsize{$mq\!-\!np>0$}}
\end{array}  
(-1)^{1+m(s+l+1)+n(r+k+1)} 
\frac{ \sin\left(\frac{\pi q r m}{p}\right) \sin\left(\frac{\pi q k m}{p}\right) }{\sin\left(\frac{\pi q m}{p}\right)}
\frac{ \sin\left(\frac{\pi p s n}{q}\right) \sin\left(\frac{\pi p l n}{q}\right) }{\sin\left(\frac{\pi p n}{q}\right)}
\nonumber\\
&&
\phantom{mapma}
\times 
\left\{ 
\frac{1}{\cos\left(z \right) - \cos\left(\frac{\pi(mq-np)}{pq} \right)}
- 
\frac{1}{\cos\left(z\right) - \cos\left(\frac{\pi(mq+np)}{pq} \right)}
\right\}
\label{ffunc2}
\en
and let us try to derive a simpler expression for $f(z)$ than the above expression.  
In order to proceed we first need to determine the set of poles of $f(z)$. 
Given the fact, that $\textrm{gcd}(p,q)=1$, we may easily show, 
that if one of the terms in the above sum has a pole at $z_{0}$, 
then none of the other terms have a pole at $z_{0}$.
From this fact and the above definition of $f(z)$ 
we realize, that the set of points, at which $f(z)$ has a pole, is given by 
\[\label{defsetA}
\tilde{\mathcal{A}}
= 
\bigg\{
\frac{\pi(2tpq \pm mq \pm np)}{pq}\bigg\vert\; 
t\in\mathbf{Z}\,,\; 
1 \leq m \leq p-1\,,\;
1 \leq n \leq q-1\,,\;
mq-np>0 
\bigg\}\,.
\]     
Furthermore, all the poles of $f(z)$ are simple poles. 
\\
Let $z$ be given by 
\[\label{du}
z 
= 
\frac{\pi(iq+jp)}{pq}\,,\quad i,j\in\mathbf{Z}\,,\;p \ndiv i\,,\;q \ndiv j\,. 
\]
We may express $z$ as 
\st
z 
& = & 
\frac{\pi((i\,\textrm{mod }p)q+(j\,\textrm{mod }q)p+kpq)}{pq}
\nonumber\\
& = &
\frac{\pi(-\{p - (i\,\textrm{mod }p)\}q+(j\,\textrm{mod }q)p+(k+1)pq)}{pq}\,, 
\label{zexpression}
\en
where 
\[
i\,\textrm{mod }p \in \{1, \ldots, p-1\}\,,
\quad
j\,\textrm{mod }q \in \{1, \ldots, q-1\}\,,
\quad k \in \mathbf{Z}\,.
\]
From this we realize, that $z \in \tilde{\mathcal{A}}$. 
Moreover, let $z \in \tilde{\mathcal{A}}$ be given. 
Then we obviously express $z$ as in eq. (\ref{du}). 
We conclude, that we may parametrize the set $\tilde{\mathcal{A}}$ by  
\[\label{de}
\tilde{\mathcal{A}} 
= 
\bigg\{
\frac{\pi(iq+jp)}{pq} 
\bigg\vert 
i,j\in\mathbf{Z}\,,\;p \ndiv i\,,\;q \ndiv j 
\bigg\}\,.
\] 
Let us consider the following ideal in $\mathbf{Z}$  
\[
\mathcal{I} 
\equiv 
\bigg\{
iq+jp 
\bigg\vert 
i,j\in\mathbf{Z} \bigg\}\,.
\]
Due to the fact, that $\mathbf{Z}$ is a principal ideal domain, we know, that the ideal $\mathcal{I}$ 
is generated by a single integer, that is $\exists n\in \mathbf{N}$ such, that 
\[
\mathcal{I} 
= 
\bigg\{
kn 
\bigg\vert 
k\in\mathbf{Z} \bigg\}\,.
\
\]
In order to determine $n$ we have to determine the smallest positive integer belonging to $\mathcal{I}$. 
According to B\`ezout's theorem this integer is given by 
\[
n = \textrm{gcd}(p,q) = 1\,.
\]
We conclude, that 
\[
\mathcal{I} = \mathbf{Z}\,.
\]
In the light of this result we realize, that we may express $\tilde{\mathcal{A}}$ as 
\[
\tilde{\mathcal{A}} 
= 
\bigg\{
\frac{\pi i}{pq} 
\,\bigg\vert\,
i\in\mathbf{Z}\,,\;p \ndiv i\,,\;q \ndiv i
\bigg\}\,.
\label{setA}
\]
Let us consider the function $g(z)$ defined by  
\[
g(z) 
\equiv 
\frac{\sin(pz)\sin(qz)}{\sin(pqz)\sin(z)}\,.
\]
Notice, the set of point, at which $g(z)$ has a pole, is equal to the set of points $\tilde{\mathcal{A}}$, 
at which $f(z)$ has a pole. 
Moreover, all the poles of $g(z)$ are simple poles as the poles of $f(z)$. 
The residue of $g(z)$ at 
\[
z_{i}=\frac{\pi i}{pq}\,,\quad i\in\mathbf{Z}\,,\;p \ndiv i\,,\;q \ndiv i
\]
is given by 
\[
\textrm{Res}(g(z),z_{i}) 
= 
\lim_{z \to z_{i}} (z-z_{i})g(z) 
= 
\frac{(-1)^{i}\sin(pz_{i})\sin(qz_{i})}{pq\sin(z_{i})}\,. 
\]
Let us determine the residue of $f(z)$ at $z_{i}$. 
From our previous discussion we realize, that we may express $z_{i}$ as 
\st
z_{i} 
& = & 
\frac{\pi(\tilde{m}q-(q-\tilde{n})p+jpq)}{pq}
\nonumber\\
& = & 
-
\frac{\pi((p-\tilde{m})q-\tilde{n}p-jpq)}{pq}\,,
\en
where
\[
\tilde{m} \in \{1,\ldots,p-1\}\,,\quad
\tilde{n} \in \{1,\ldots,q-1\}\,.
\]
Moreover, let us consider the case $\tilde{m}q-(q-\tilde{n})p>0$, $j$ even. 
In this case the residue of $f(z)$ at $z_{i}$ is given by 
\[
\textrm{Res}(f(z),z_{i}) 
=  
\lim_{z \to z_{i}} (z-z_{i})f(z) 
\nonumber
\]
\[
=  
(-1)^{1+\tilde{m}(s+l+1)+(q-\tilde{n})(r+k+1)} 
\frac{ \sin\left(\frac{\pi q r \tilde{m}}{p}\right) \sin\left(\frac{\pi q k \tilde{m}}{p}\right) }
{\sin\left(\frac{\pi q \tilde{m}}{p}\right)}
\frac{ \sin\left(\frac{\pi p s (q-\tilde{n})}{q}\right) \sin\left(\frac{\pi p l (q-\tilde{n})}{q}\right) }
{\sin\left(\frac{\pi p (q-\tilde{n})}{q}\right)} 
\phantom{m}
\nonumber
\]
\[
\lim_{z\to z_{i}}\frac{z-z_{i}}{\cos\left(z \right) - \cos\left(\frac{\pi(\tilde{m}q-(q-\tilde{n})p)}{pq} \right)}
\phantom{mapmapmapmapmapmapmapmapmapmapm}
\nonumber
\]
\[
=  
(-1)^{1+\tilde{m}(s+l+1)+(q-\tilde{n})(r+k+1)} 
\frac{ \sin\left(q r z_{i} + \pi r (q-\tilde{n}-j q)\right) \sin\left(q k z_{i} + \pi k (q-\tilde{n}-j q)\right) }
{\sin\left(q z_{i} + \pi (q-\tilde{n}-j q)\right)}
\nonumber
\]
\[
\frac{ \sin\left(- p s z_{i} + \pi s (\tilde{m} + j p)\right) \sin\left(- p l z_{i} + \pi l (\tilde{m} + j p)\right) }
{\sin\left(- p z_{i} + \pi (\tilde{m} + j p)\right)} 
\lim_{z\to z_{i}}\frac{z-z_{i}}{\cos\left(z \right) - \cos\left(z_{i}\right)}
\nonumber
\]
\[\label{resf}
=
-
\frac{\sin( q r z_{i})\sin( q k z_{i})}{\sin( q z_{i})}
\frac{\sin( p s z_{i})\sin( p l z_{i})}{\sin( p z_{i})}
\frac{1}{\sin{z_{i}}}
\phantom{mapmapmapmapmapmapmapmapma}\,\,
\]
\[
= 
pq
\frac{\sin( q (p - r) z_{i})\sin( q k z_{i})}{\sin^{2}( q z_{i})}
\frac{\sin( p s z_{i})\sin( p l z_{i})}{\sin^{2}( p z_{i})}
\textrm{Res}(g(z),z_{i})\,.
\phantom{mapmapmapmapmapm}\,\,\,
\]
\[
= 
pq
\frac{\sin( q r z_{i})\sin( q k z_{i})}{\sin^{2}( q z_{i})}
\frac{\sin( p (q - s) z_{i})\sin( p l z_{i})}{\sin^{2}( p z_{i})}
\textrm{Res}(g(z),z_{i})\,.
\phantom{mapmapmapmapmapm}\,\,\,
\]
By a similar calculation we may determine the residue of $f(z)$ at $z_{i}$ in the three remaining cases 
$\tilde{m}q-(q-\tilde{n})p>0$, $j$ odd, $\tilde{m}q-(q-\tilde{n})p<0$, $j$ even and $\tilde{m}q-(q-\tilde{n})p<0$, $j$ odd. 
In all the remaining cases we get the same result as in the above equation. 
Thus, if we multiply $g(z)$ with the two entire functions 
\[
pq
\frac{\sin( q (p - r) z)\sin( q k z)}{\sin^{2}( q z)}
\frac{\sin( p s z)\sin( p l z)}{\sin^{2}( p z)}\,
\]
or
\[
pq
\frac{\sin( q r z)\sin( q k z)}{\sin^{2}( q z)}
\frac{\sin( p (q - s) z)\sin( p l z)}{\sin^{2}( p z)}
\]
we obtain two functions 
\[\label{hfunc}
h_{1}(z) 
= 
pq
\frac{\sin( q (p - r) z)\sin( q k z)}{\sin( q z)}
\frac{\sin( p s z)\sin( p l z)}{\sin( p z)}
\frac{1}{\sin(pqz)\sin(z)}
\]
and 
\[
h_{2}(z) 
= 
pq
\frac{\sin( q r z)\sin( q k z)}{\sin( q z)}
\frac{\sin( p (q - s) z)\sin( p l z)}{\sin( p z)}
\frac{1}{\sin(pqz)\sin(z)}\,,
\]
which have the same pole structure as $f(z)$, and which have the same residue at a given pole as $f(z)$, that is 
both $f(z)-h_{1}(z)$ and $f(z)-h_{2}(z)$ are entire functions.

Let us consider $h_{1}(z)$. 
In order to prove, that $f(z)$ is equal to $h_{1}(z)$ we need to show, 
that the difference between the two functions is bounded in the entire complex plane. 
If this is the case Liouville's theorem tells us, that the two functions differ by a constant. 
For generic values of the Cardy indices $f(z)-h_{1}(z)$ is not bounded in the complex plane. 
This follows from the fact, that 
\[\label{Cconstant}
\lim_{t \to \infty}
\bigg(
f(it) - h_{1}(it)
\bigg) 
= 
\lim_{t \to \infty} \left[
\alpha \exp(-t)
-
pq \exp\left(-\left\{ 
rq - sp + (1-k)q + (1-l)p + 1\right\}t
\right)\right]\,,
\]
where $\alpha$ is a constant. 
Let us consider the case $k=l=1$ and $mq-np>0$, 
in which the difference between $f(it)$ and $h_{1}(it)$ vanishes in the limit $t \to \infty$, 
and let us prove, that $f(z)-h_{1}(z)$ is indeed bounded in the entire complex plane in this special case. 
It follows from eq. (\ref{ffunc2}) and (\ref{hfunc}), that 
\[
f(z+2\pi) - h_{1}(z+2\pi) 
= 
f(z) - h_{1}(z) 
\]
and 
\[
f(-z) - h_{1}(-z) 
= 
f(z) - h_{1}(z)\,.  
\]
Hence, we only need to show, that $f(z)-h_{1}(z)$ is bounded on the set 
\[
\bigg\{
x+iy \in \mathbf{C}
\bigg\vert
\,
0 \leq x \leq 2\pi\,,\;
y \geq 0
\bigg\}\,.
\]
Due to the fact, that the set  
\[
\mathcal{D} 
= 
\bigg\{x+iy \in \mathbf{C} 
\bigg\vert\,
0 \leq x \leq 2\pi\,,\;
0 \leq y \leq 2\ln(2) 
\bigg\}
\]
is closed and bounded and $\vert f(z) - h_{1}(z) \vert$ is continuous, 
there exists an $a \geq 0$ such that
\[
\vert f(z) - h_{1}(z) \vert \leq a\,,\; \forall z \in \mathcal{D}\,.
\]  
Let us now consider any given $z=x+iy \in \mathbf{C}$, $y \geq 2\ln(2)$.  
For $t \geq 1$ we obtain the following estimate 
\st
\vert \sin(t z) \vert^{2} 
& = &
\vert \sin(t x)\cosh(t y) + i\cos(t x) \sinh(t y) \vert^{2} 
\nonumber\\
& = & 
\sin^{2}(t x) \cosh^{2}(t y) + \cos^{2}(t x) \sinh^{2}(t y)
\nonumber\\
& = &
\cosh^{2}(t y) - \cos^{2}(t x) 
\geq 
\cosh^{2}(t y) - 1 
\nonumber\\
& = &
\sinh^{2}(t y) 
= 
\frac{1}{2}\bigg(\cosh(2 t y) - 1 \bigg) 
\nonumber\\  
& \geq & 
\frac{1}{4} e^{2 t y} - \frac{1}{2} 
\geq  
\frac{1}{8} e^{2 t y}\,,   
\label{estimate1}
\en
where we used the fact, that $y \geq 2\ln(2)$ in the last estimate. 
Furthermore, 
\[\label{estimate2}
\vert \sin(tz) \vert^{2} 
= 
\cosh^{2}(t y) - \cos^{2}(t x) 
\leq  
\cosh^{2}(t y)
\leq 
e^{2ty} 
\]
and 
\st
\vert \cos(z) \vert^{2}
& = & 
\vert \cos(x)\cosh(y) - i\sin(x)\sinh(y) \vert^{2} 
\nonumber\\
& = &
\cos^{2}(x)\cosh^{2}(y) + \sin^{2}(x)\sinh^{2}(y) 
\nonumber\\
& = &
\cosh^{2}(y) - \sin^{2}(x) 
\geq 
\cosh^{2}(y) - 1 
\nonumber\\
& \geq & 
\frac{1}{8} e^{2y} \geq 1\,, 
\label{estimate3}
\en
where in the two last estimates we have used some of the intermediate result 
in (\ref{estimate1}) and the fact, that $y \geq 2\ln(2)$. 
From the estimate (\ref{estimate3}) and eqs. (\ref{ffunc2}) with $k=l=1$ we obtain for $y \geq 2\ln(2)$  
\st
\vert f(z) \vert 
& \leq &
\!\!\!
\begin{array}{c}
\displaystyle\sum_{m=1}^{p-1} \,\,\, 
\sum_{n=1}^{q-1}
\\
\textrm{\scriptsize{$mq\!-\!np>0$}}
\end{array}  
\left\{ 
\frac{1}{\big\vert \vert\!\cos\left(z \right)\!\vert - \vert\!\cos\left(\frac{\pi(mq-np)}{pq}\right)\!\vert \big\vert}
+ 
\frac{1}{\big\vert \vert\!\cos\left(z\right)\!\vert - \vert\!\cos\left(\frac{\pi(mq+np)}{pq}\!\vert \right)\big\vert}
\right\}
\nonumber\\
& \leq &
(p-1)(q-1)
\frac{1}{\vert\cos\left(z \right)\vert - 1}
\leq
(p-1)(q-1)
\frac{1}{\frac{1}{\sqrt{8}}e^{y} - 1}
\nonumber\\
& \leq &
(p-1)(q-1)
\frac{1}{\frac{1}{\sqrt{8}}e^{2\ln(2)} - 1}
\nonumber\\
& = &
\frac{(p-1)(q-1)}{\sqrt{2}-1}\,.
\en
Moreover, for $y \geq 2\ln(2)$ we obtain from (\ref{hfunc}) and the two estimates (\ref{estimate1}) and (\ref{estimate2}), 
that 
\[
\vert h_{1}(z) \vert 
\leq 
8 p q 
\exp\left(-\{rq-sp +1\}y\right)
\leq
8 p q \,.
\]
Thus, for $y \geq 2 \ln(2)$ we obtain the estimate 
\[
\vert f(x+iy) - h_{1}(x+iy) \vert 
\leq 
\vert f(x+iy)\vert + \vert h_{1}(x+iy) \vert 
\leq 
\frac{(p-1)(q-1)}{\sqrt{2}-1} + 8 pq\,. 
\]
Hence, we conclude, that 
\[
\vert f(x+iy) - h_{1}(x+iy) \vert 
\leq 
\textrm{max}\big\{a , \frac{(p-1)(q-1)}{\sqrt{2}-1} + 8 pq  \big\}\,,\quad \forall z \in \mathbf{C}\,.
\]
From Liouville's theorem we obtain, that $f(z)$ and $h_{1}(z)$ only differ by a constant $C$ in the case $k=l=1$ and $rq-sp>0$. 
We may determine this constant from eq. (\ref{Cconstant}) 
\[
C = \lim_{t \to \infty}
\bigg(
f(it) - h(it)
\bigg) 
= 0\,.
\]
Thus, 
\[
f(z) = h_{1}(z)\,,\quad \textrm{for } k=l=1\,,\;rq-sp>0\,.  
\]
By a similar argument we may show, that 
\[
f(z) = h_{2}(z)\,,\quad \textrm{for } k=l=1\,,\;rq-sp<0\,.
\]

\end{document}